\documentclass[aps,superscriptaddress,groupedaddress,12pt]{article}
\usepackage[utf8]{inputenc}
\usepackage[margin=1.25in]{geometry}
\usepackage{datetime2}
\usepackage{graphicx}
\usepackage{xspace}
\usepackage{amsmath}
\usepackage{setspace}
\usepackage{hyperref}
\usepackage{multirow}
\usepackage[super]{nth}
\usepackage{color}
\usepackage{authblk}

\hypersetup{
    colorlinks=true,
    linktoc=all,
    linkcolor=blue,
}

\begin{document}

\title{Analysis note: jet reconstruction, energy spectra, and substructure analyses with archived ALEPH data}

\author[1]{Yi Chen}
\author[1]{Yen-Jie Lee}
\author[2]{Marcello Maggi}
\author[3]{Paoti Chang}
\author[4]{Yang-Ting Chien}
\author[5]{Christopher McGinn}
\author[5]{Dennis Perepelitsa}

\affil[1]{Massachusetts Institute of Technology, Cambridge, Massachusetts, USA}
\affil[2]{INFN Sezione di Bari, Bari, Italy}%
\affil[3]{National Taiwan University, Taipei, Taiwan}%
\affil[4]{Georgia State University, Atlanta, Georgia, USA}%
\affil[5]{University of Colorado Boulder, Boulder, Colorado, USA}%

\date{\today}

\maketitle

\begin{abstract}
The first measurements of anti-$k_{T}$ jet energy spectrum and substructure in hadronic $Z$ decays are presented. The archived $e^+e^-$ annihilation data at a center-of-mass energy of 91 GeV were collected with the ALEPH detector at LEP in 1994. The jet substructure was analyzed as a function of jet energy. The results are compared with the perturbative QCD calculations and predictions from the {\sc pythia} v6.1, {\sc sherpa}, and {\sc herwig} v7.1.5 event generators. In this note, jet reconstruction procedure, jet energy calibration and the performance with archived ALEPH data and Monte Carlo simulations are also documented.
\end{abstract}

\newcommand{\ak}{anti-k$_\text{T}$\xspace}
\newcommand{\ee}{$e^+e^-$\xspace}
\newcommand{\fastjet}{\textsc{FastJet}\xspace}
\newcommand{\sd}{{soft drop }}
\newcommand{\zg}{\ensuremath{z_G}\xspace}
\newcommand{\Rg}{\ensuremath{R_G}\xspace}
\newcommand{\ME}{\ensuremath{M/E}\xspace}
\newcommand{\MgE}{\ensuremath{M_G/E}\xspace}
\newcommand{\Bayes}{\texttt{BayesUnfold}\xspace}
\newcommand{\SVD}{\texttt{SVDUnfold}\xspace}
\newcommand{\includegraphicsfour}[1]{\includegraphics[width=0.24\textwidth]{#1}}
\newcommand{\includegraphicsthree}[1]{\includegraphics[width=0.32\textwidth]{#1}}
\newcommand{\includegraphicstwo}[1]{\includegraphics[width=0.45\textwidth]{#1}}
\newcommand{\includegraphicsone}[1]{\includegraphics[width=0.80\textwidth]{#1}}
\newcommand{\includegraphicsonewide}[1]{\includegraphics[width=0.95\textwidth]{#1}}
\newcommand{\includegraphicsonesmall}[1]{\includegraphics[width=0.60\textwidth]{#1}}
\newcommand{\HybridE}{\ensuremath{E_\text{sum}^\text{hybrid}}\xspace}
\newcommand{\fixme}[1]{{\color{red} \textbf{#1}}}
\newcommand{\followup}[1]{{\color{magenta} {#1}}}
\clearpage

\tableofcontents
\clearpage

\section{Introduction}\label{Section:Introduction}


Jets, collimated sprays of particles originating from fast-moving quarks or gluons, are some of the most useful tools for studing Quantum Chromodynamics (QCD) and searching for new physics beyond the Standard Model (SM) in high energy colliders. Since the end of the Large Electron Positron Collider (LEP) operation, significant progress has been made in jet definitions, jet algorithms, and jet substructure observables~\cite{Larkoski:2017jix}. However, those techniques, which were widely explored in the data analyses of the proton-proton~\cite{Kogler:2018hem} and heavy-ion collisions~\cite{Cao:2020wlm}, are not yet used in the cleanest $e^+e^-$ collision system. Monte Carlo (MC) simulations, such as \textsc{pythia}~\cite{Sjostrand:2000wi}, \textsc{sherpa}~\cite{Gleisberg:2008ta}, and \textsc{herwig}~\cite{Reichelt:2017hts}, are tuned with hadron spectra and hadronic event shape observables. They are then employed to predict the jet spectra and jet substructures in more complicated hadron-hadron collision environments.

Studies of jets in electron-positron ($e^+e^-$) annihilation using identical algorithms as those used in high-energy hadron colliders such as Large Hadron Collider (LHC) and Relativistic Heavy-Ion Collider (RHIC) are of great interest.
Unlike hadron-hadron collisions, $e^+e^-$ annihilation does not have beam remnants, gluonic initial state radiation, or the complications of parton distribution functions.
Therefore, electron-position annihilation data provide the cleanest test for perturbative QCD and phenomenological models that are tuned with hadronic event shapes.
Moreover, fully reconstructed jets provide us an opportunity to inspect the quark and gluon fragmentation in great detail on a shower-by-shower basis.
Finally, studies of jet substructure and their comparison to modern event generators are of great interest since jet substructure observables are novel tools for jet flavor identification, electroweak boson/top tagging~\cite{Thaler:2008ju,Abdesselam:2010pt}, and studies of the Quark-Gluon Plasma properties at hadron colliders~\cite{Andrews:2018jcm}. 

In this analysis note, the first measurement of anti-$k_{T}$ jet~\cite{Cacciari:2008gp} momentum spectrum, jet splitting functions, and subjet opening angle distribution in hadronic $Z$ boson decays are presented, with two types of jet selections. The inclusive observables include all the reconstructed jet above the jet energy threshold and inside a defined acceptance. They are sensitive to higher-order corrections of jet spectra in perturbative QCD.
They include low momentum jets in the events which are typically associated with soft gluon radiation or random combinatorial jets with hadrons from different partons. On the other hand, the leading dijet observables consider the leading and subleading jet in the event. This type of observable focuses more on the dominant energy flow and is less sensitive to soft radiation.

The analysis note is organized in the following way: The obsersables are defined in Section~\ref{Section:Observable}.  The data and Monte Carlo samples are documented in Section~\ref{Section:Samples}. Hadronic event selection and background rejection criteria are described in Section~\ref{Section:Selection}. Jet reconstruction and calibration procedures are documented in Section~\ref{Section:JetReconstruction} and \ref{Section:JetCalibration}. The simulated jet resolution and data-driven correction to the resolution function are documented in Section~\ref{Section:JetResolution}. Leading jet selection criteria are described in Secion~\ref{Section:LeadingJet}. Resolution unfolding, systematics and the results are summarized in Section~\ref{Section:Unfolding},~\ref{Section:Systematics} and \ref{Section:Result}.  Additional cross checks are documented in Section~\ref{Section:CrossCheck}.

\clearpage

\section{Observable Definition}\label{Section:Observable}

In this analysis, seven different observables are considered, and they can be grouped into two categories: inclusive observables, and leading dijet observables. The inclusive jet observables are the ones which are closest to the jet analyses in the hadron colliders. The leading dijet observables are also described.  

\subsection{Inclusive Observables}

In the inclusive jet analysis, all jets within the detector acceptance, $0.2\pi < \theta_\text{jet} < 0.8\pi$, are considered. The following observables include a wide range of jet spectra and jet substructure analyses:
\begin{enumerate}
    \item Jet energy, from 10 GeV up to 50 GeV
    \item Jet mass to jet energy ratio, in 5 GeV bins of jet energy
    \item Groomed jet mass to jet energy ratio, in bins of jet energy
    \item Groomed momentum sharing $\zg$ defined in Sec.~\ref{Subsection:JetGrooming}, in bins of jet energy
    \item Groomed jet radius $\Rg$ defined in Sec.~\ref{Subsection:JetGrooming}, in bins of jet energy
\end{enumerate}
The grooming procedure, which is based on the soft drop algorithm, is outlined in Sec.~\ref{Section:JetReconstruction} and described in detail in~\cite{Larkoski:2014wba}.  Jet mass is defined as the invariant mass, which is based on the sum of the 4-momenta of all jet constituents.

We present the ratio of the jet mass ($M$) to its energy ($E$) instead of the jet mass itself since there is a strong correlation between the jet mass and the jet energy.  The measurement of the $M/E$ ratio decouples the jet energy related systematic uncertainties from other systematic effects that affect the jet mass.

For the observables other than jet energy, the spectra are normalized (with area equal to 1) for each jet energy range in order to minimize the effects from the overall jet energy migration which affects the normalization.

\subsection{Leading Dijet Observables}

Another set of observables is considered for leading dijets, which are the leading and subleading jets ranked by jet energy in the event~\cite{Dasgupta:2014yra} (see for instance, calculation in~\cite{Neill:2021std}).  The acceptance requirement of the leading dijets is the same as that in the inclusive jet analyses: $0.2\pi < \theta_\text{jet} < 0.8\pi$.  Since we are interested in measuring the event-wide leading dijet, a functional definition of what we measure is the spectra of the leading dijets when both of the two jets are inside the acceptance. Some complications arise when the leading jet(s) coincides with the beam line and is reconstructed with a lower energy. In this analysis, if the leading or subleading jet is out of the acceptance, the event is rejected in the generator level. In the data analysis, the acceptance effect correction is applied based on Monte Carlo simulation. The details of the special selection are designed and described in Sec.~\ref{Section:LeadingJet}.

For the leading dijets, the following observables are measured:
\begin{enumerate}
    \item Dijet energy.  This is equivalent to the sum of the leading jet energy spectrum and subleading jet energy spectrum. 
    \item Dijet total energy.  The spectra of the total energy of the leading dijets.
\end{enumerate}

\clearpage

\section{ALEPH Detector and Data Samples}
\label{Section:Samples}

\subsection{ALEPH Detector}
The ALEPH detector is described in ~\cite{Decamp:1990jra}. The central part of the detector is designed for efficient reconstruction of charged particles. The trajectories of them are measured by a two-layer silicon strip vertex detector, a cylindrical drift chamber and a large time projection chamber (TPC). Those tracking detectors are inside a 1.5 T axial magnetic field generated by a superconducting solenoidal coil. The charged particle transverse momenta are reconstructed with a resolution of $\delta p_t/p_t = 6\times 10^{-4} p_t \oplus 0.005$ (GeV/c). 

Electrons and photons are identified in the electromagnetic calorimeter (ECAL) situated between the TPC and the superconducting coil. The ECAL is a sampling calorimeter, made by lead plates and proportional wire chambers segmented in $0.9^\circ\times 0.9^\circ$ projective towers. They are read out in three sections in depth and have a total thickness of around 22 radiation lengths. Isolated photons are reconstructed with a relative energy resolution of $0.18/\sqrt{E}+0.009$ (GeV).

The iron return yoke constructed with 23 layers of streamer tubes is also used as the hadron calorimeter (HCAL) for the detection of charged and neutral hadrons. The relative energy resolution for hadrons is $0.85/\sqrt{E}$. Muons are identified by their pattern in HCAL and by the muon chambers, made by two double-layers of streamer tubes outside the HCAL. 

The information from trackers and calorimeters is combined in an energy-flow algorithm~\cite{ALEPH:1994ayc}. This algorithm provides a set of charged and neutral particles, call energy-flow objects. They are used in the jet reconstruction in this analysis.

\subsection{Data Samples}

This study is performed with hadronic $Z$ decays. The archived $e^+e^-$ annihilation data at a center-of-mass energy of 91 GeV were collected with the ALEPH detector at LEP~\cite{Decamp:1990jra} in 1994. To analyze these data, an MIT Open Data format was created~\cite{Tripathee:2017ybi} and was validated and used in the two-particle correlation function analysis~\cite{Badea:2019vey}. Currently, only the data taken in 1994 is analyzed because of the availability of the archived Monte Carlo simulation. In the future, more data could be added when the archived Monte Carlo samples from other years become available.

\subsection{Simulation Samples}

Archived $\textsc{pythia}$ 6.1~\cite{Sjostrand:2000wi} Monte Carlo (MC) simulation samples, which was produced with the 1994 run detector condition by the ALEPH collaboration, was the only available archived MC sample at the time of this analysis. The MC samples are used for the derivation of jet energy correction factors, event selection efficiency and corrections and acceptance effect corrections.

A set of new $\textsc{pythia}$ events are generated with $\textsc{pythia}$8 version 8.303 at a center-of-mass energy of 91.2~GeV.  The Monash 2013 tune is used, with the weak boson exchange and weak single boson processes turned on.  Pure electroweak events are rejected by filtering the outgoing particles in the hard process to contain at least one hadron.

Sherpa samples are generated with version 2.2.5, with electron positron events generating 2--5 outgoing partons, which are then showered into jets. The strong coupling constant $\alpha_s (M_Z)$ is set to 0.1188.

A set of $\textsc{herwig}$ samples\cite{Bellm:2015jjp} is generated using version 7.2.2.  2--3 outgoing partons are specified in the hard process, and leptonic decays of $Z$ boson are turned off to increase the fraction of hadronic events.  The order of $\alpha$ coupling is set to 2, whereas the colored $\alpha_s$ order is set to 0.

In order to understand effects from potential quenching effects~\cite{Bjorken:1982tu,ATLAS:2010isq,CMS:2011iwn}, a sample is generated with the $\textsc{pyquen}$ generator~\cite{Lokhtin:2005px} (version 1.5.3 with hard scattering generated by $\textsc{ pythia}$ 6.428).  The strength of the quenching is set to be equivalent to a minimum bias sample of PbPb collisions at 5.02 TeV.  Two subsamples are generated, with and without explicit wide-angle radiation of partons.  The default spectra shown in the results are without wide-angle radiation.

\clearpage

\section{Event Selection}\label{Section:Selection}

\subsection{Hadronic event selection}

Hadronic $Z$ boson decay events are selected by requiring the sphericity axis to have a polar angle in the laboratory reference frame ($\theta_{\text{lab}}$) between $7\pi/36$ and $29\pi/36$ to ensure that the event is well contained within the detector. At least five tracks having minimum total energy of 15 GeV are also required to suppress electromagnetic interactions~\cite{Barate:1996fi}. 

The residual contamination from processes such as $e^+e^-\rightarrow\tau^+\tau^-$ is expected to be less than 0.26\% for these event selections based on the studies documented in ~\cite{Barate:1996fi}.

\subsection{Mercedes-Benz event rejection}

\begin{figure}[htp!]
    \centering
    \includegraphicstwo{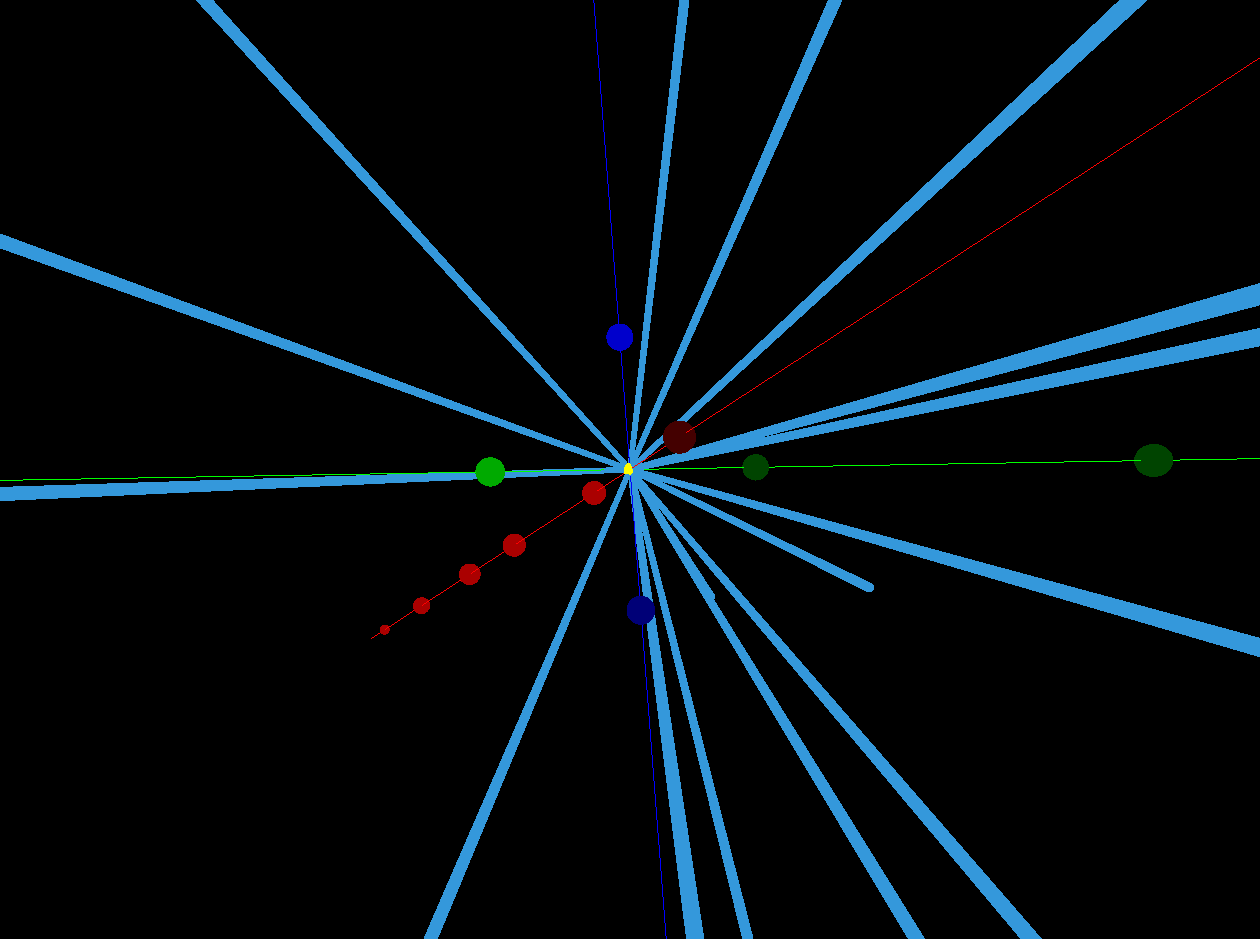}
    \includegraphicstwo{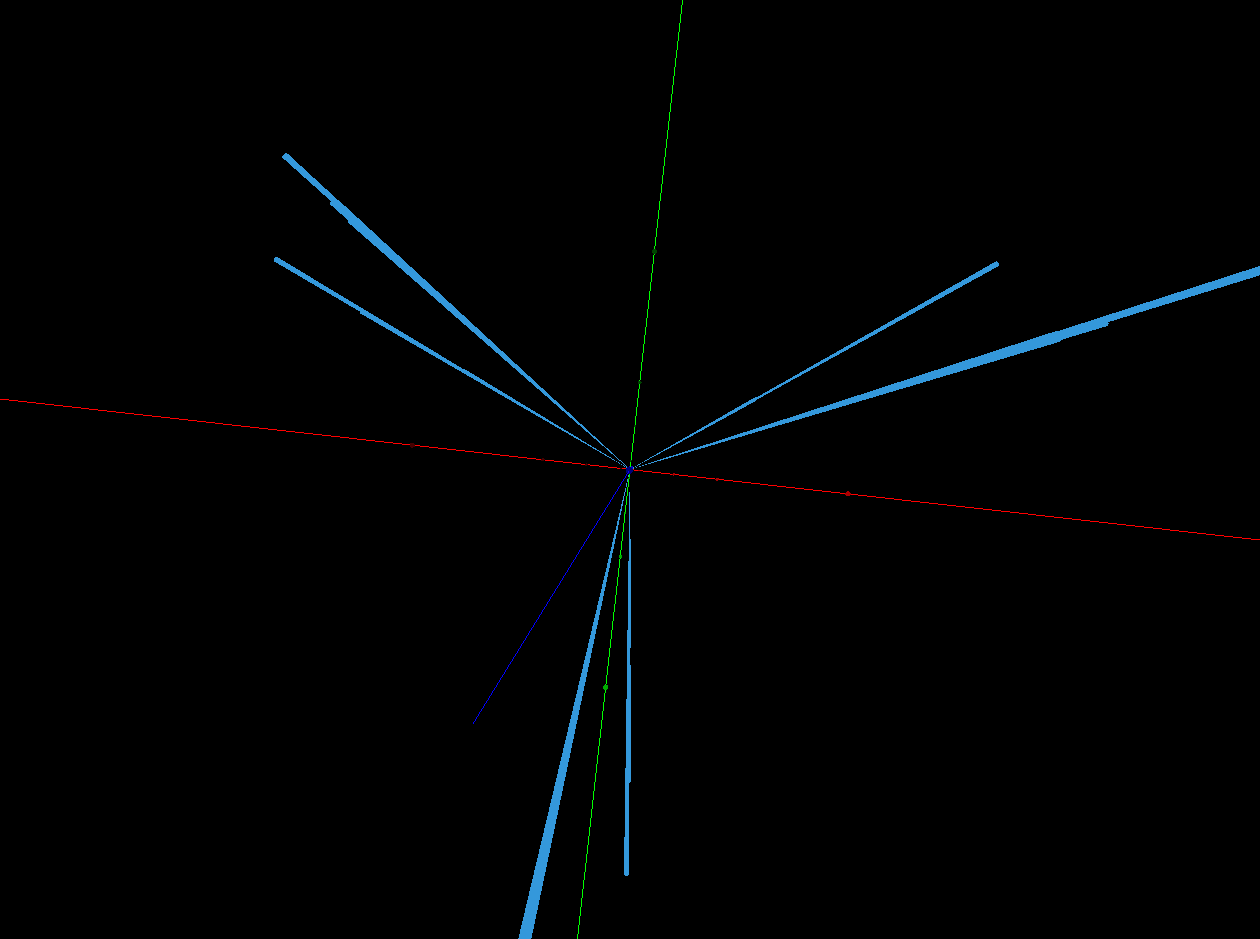}
    \caption{Example of the ``Mercedes-Benz'' events.  The thin lines indicate the axes ($x$ = red, $y$ = green, $z$ = blue).  Light blue lines are the particles, with the length proportional to the momentum of the particle.  The particles are all around 40 GeV.  The right panel show the view from the $-z$ direction.  Each of the three branches typically has 4-7 particles.}
    \label{Figure:Selection-MercedesBenz}
\end{figure}

Through an application of multivariate analysis, we discovered that there is a type of pathological event where we have a number of 40-45 GeV particles arranged in a Mercedes-Benz-like (MBl) pattern.  One example event display is shown in Fig.~\ref{Figure:Selection-MercedesBenz}.

\begin{figure}[htp!]
    \centering
    \includegraphicsonesmall{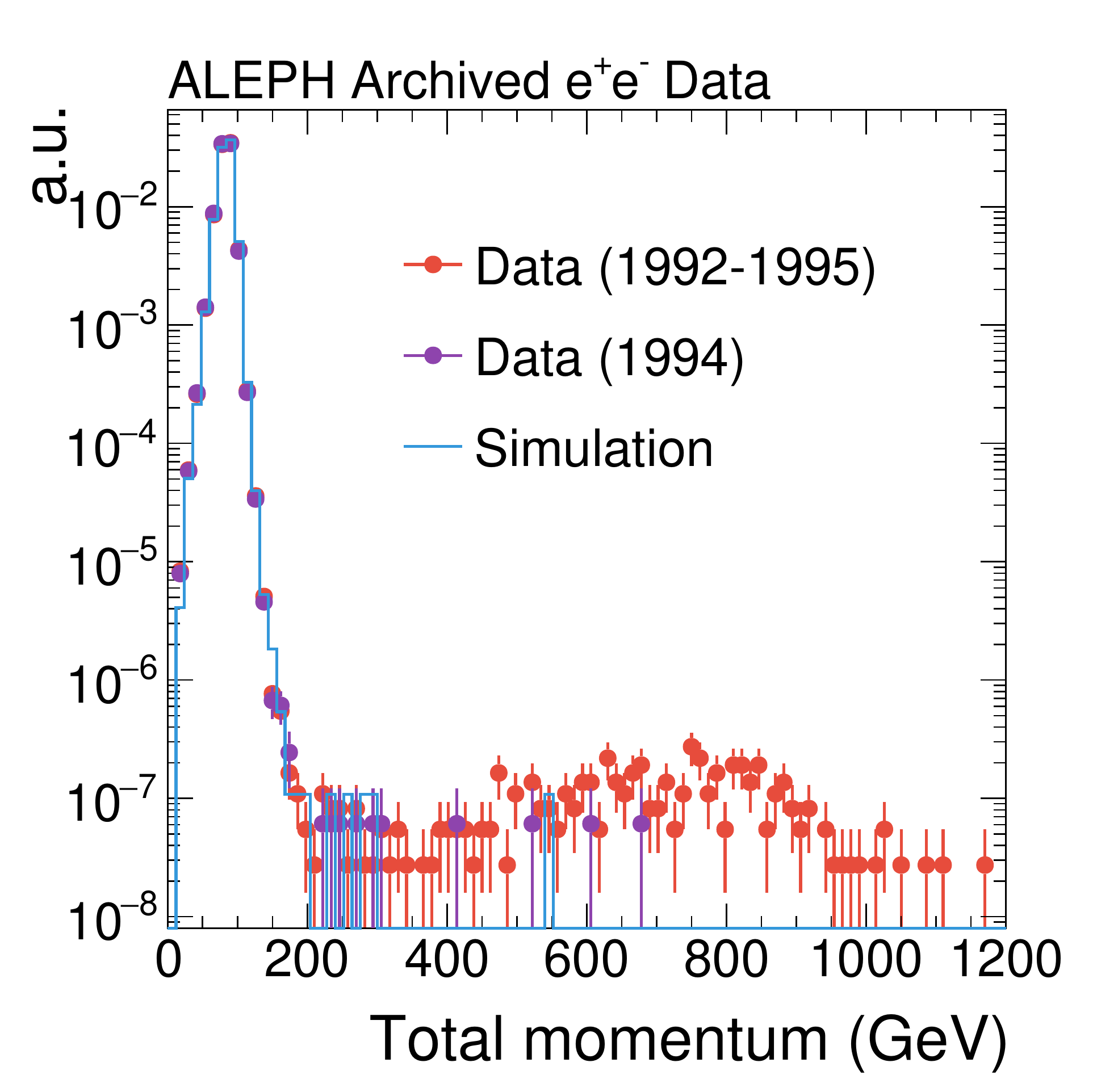}
    \caption{Total visible momentum distribution for data (taken between 1992 and 1995) and simulation 
    }
    \label{Figure:Selection-MercedesBenzRejection}
\end{figure}

The source of this type of event is mainly from laser calibration events and the pattern is an artifact of the low level reconstruction algorithm. Due to the feature of having multiple high-energy particles, we can easily sieve out this type of event by considering the total visible momentum, defined as the sum of the magnitude of the momenta for all reconstructed particles, shown in Fig.~\ref{Figure:Selection-MercedesBenzRejection}.  A selection cut at 200 GeV is seen to be efficient in rejecting this type of event.  Note that most of the MBl events are from data taken in 1995.

\clearpage

\section{Jet Reconstruction}\label{Section:JetReconstruction}

\subsection{Jet Clustering}

Jets are constructed using the \ak algorithm with the \ee variant, which is described below. In this analysis, a resolution parameter of $R_0 = 0.4$ is considered. This resolution parameter is chosen since it is widely used in the jet analyses in proton-proton and heavy-ion collisions carried out at the hadron colliders. Moreover, the chosen value also gives us an opportunity to examine the shower from quarks in detail. According to the \fastjet user's manual~\cite{Cacciari:2005hq,Cacciari:2011ma}, the distance metric is set to be
\begin{align}
    d_{ij} &= \min\left(E_i^{-2}, E_j^{-2}\right) \dfrac{1 - \cos \theta_{ij}}{1 - \cos R_0}\\
    d_{iB} &= E_i^{-2},
\end{align}
where $d_{ij}$ is the distance between two pseudojets $i$ ad $j$, $E_i$ is the energy, and $\theta_{ij}$ is the opening angle.  The termination is defined by the distance to beam-pipe, $d_{iB}$. This is different from the use of transverse momenta and pseudorapidities of constituents in algorithm for hadron-hadron colliders.  The option used is as follows:
\begin{verbatim}
JetDefinition Definition(ee_genkt_algorithm, 0.4, -1);
\end{verbatim}

In the data analysis, energy flow objects which are reconstructed using the tracker and calorimeter information, are used for jet reconstruction. Generator-level jets are clustered by considering all visible final state particles by the ALEPH detector (i.e., excluding neutrinos).  Reconstructed-level jets are clustered with all energy-flow candidates reconstructed by the tracker and calorimeters.

To avoid jets which are partially outside the detector acceptance (around the beam pipe), in this analysis, we only consider jets within $0.2\pi < \theta_\text{jet} < 0.8\pi$, where $\theta_\text{jet}$ is the angle between the jet 3-momentum vector and electron beam direction.

\begin{figure}
    \centering
    \includegraphicsone{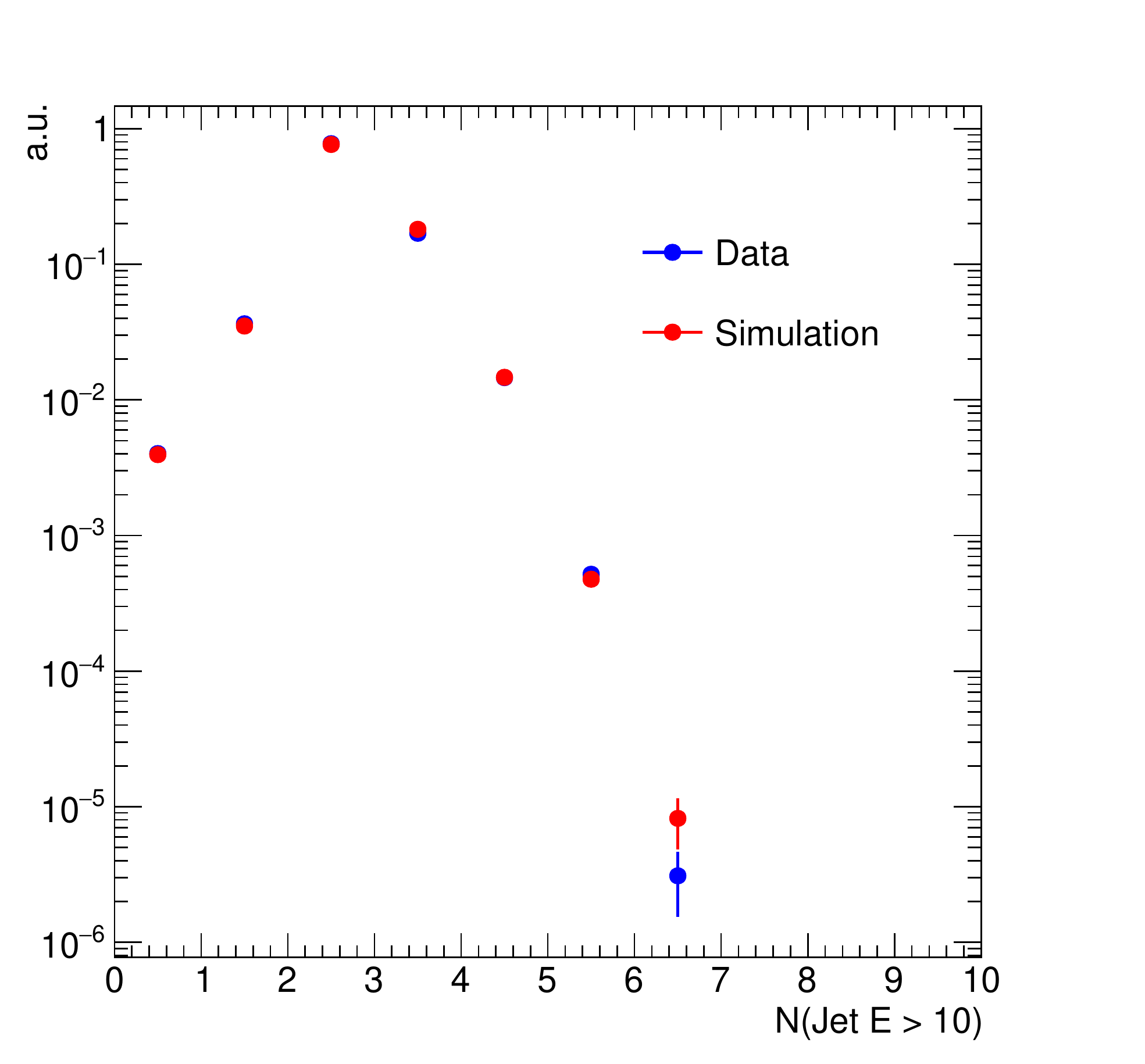}
    \caption{Distribution of number of jets above 10 GeV inside the acceptance in data and simulation.}
    \label{Figure:JetReconstruction-NJet}
\end{figure}

The number of jets above 10 GeV in the acceptance in an event is shown in Figure~\ref{Figure:JetReconstruction-NJet}.

\subsection{Jet Grooming}
\label{Subsection:JetGrooming}

To study the hard part of the jet and to suppress the contribution from soft radiation, we also considered groomed jet quantities by a version of the \sd algorithm~\cite{Larkoski:2014wba,Larkoski:2017bvj} for the \ee collisions.  The \sd algorithm proceeds by first reclustering all jet constituents using the Cambridge-Aachen algorithm, again modified using opening angle and constituent energy in the metric.

The clustering history can be represented as a binary tree.  The tree is then traced, and a declustering procedure is carried out, starting from the root node and following the branches at each step with higher energy.  At each node of the tree, the \sd condition is examined:
\begin{align}
    z \equiv \dfrac{\min(E_1, E_2)}{E_1 + E_2} \leq z_\text{cut} \left(\dfrac{\theta_{12}}{R_0}\right)^\beta,
\end{align}
where indices 1 and 2 represent the two branches originating from the node, and the $\theta_{12}$ is the opening angle.  The parameters $z_\text{cut}$ and $\beta$ are free parameters, and in this work we choose $z_\text{cut} = 0.1, \beta = 0.0$.  If the condition is met, the algorithm stops, and the $z$ and $\theta$ are denoted as $\zg$ and $\Rg$.  If the condition is not met, we go to the next node in the tree.  The choice of jet grooming algorithm and the associated parameters is motivated by the earlier measurements from the LHC~\cite{Larkoski:2017bvj,CMS:2017qlm,CMS:2018ypj,CMS:2018fof,ALICE:2019ykw}, and later at RHIC~\cite{STAR:2020ejj}.  Other grooming parameters and methods are planned in an upcoming followup study of jets using ALEPH data.

The groomed mass $M_G$ is defined as the invariant mass of the two branches of the node where the algorithm terminates.

\clearpage

\section{Jet Calibration}\label{Section:JetCalibration}

\subsection{Jet Calibration Strategy}

Jets are calibrated with a multi-stage strategy.  In the first stage, a Monte Carlo simulation-based jet energy calibration (Section~\ref{Subsection:MCCalibration}) is derived using the archived \textsc{pythia} 6.1 samples.  This correction is then applied on both the simulated sample and the data. There are two additional stages of correction, aimed at correcting the difference between data and simulation.  These ``residual'' corrections correct first the jet energy scale dependence on the jet direction (Section~\ref{Subsection:RelativeSide}), and then the absolute scale using the multijet peak (Section~\ref{Subsection:AbsoluteScale}).  The relative scale derivation described in Section~\ref{Subsection:RelativeScale} serves as an important cross check of the analysis and is not used as the nominal result due to the larger uncertainties.

\subsection{Simulation-based calibration}
\label{Subsection:MCCalibration}

The simulation based calibration compares the reconstructed jet energy with the generated jet energy.  The mean of the response, defined as the ratio of the reconstructed jet energy to the generated jet energy, is used as the metric for the calibration.
Jets are matched with an angular requirement where the opening angle is required to be less than half the size of the jet distance parameter.

The calibration is carried out in bins of the reconstructed jet $\theta$, and for each bin a \nth{4} order polynomial function is fitted to the jet response, the ratio between reconstructed jet energy and the generated jet energy, as a function of generated jet energy.  The function is then inverted to obtain the actual correction factors, to be applied to the reconstructed jets.

\begin{figure}[htp!]
    \centering
    \includegraphicsfour{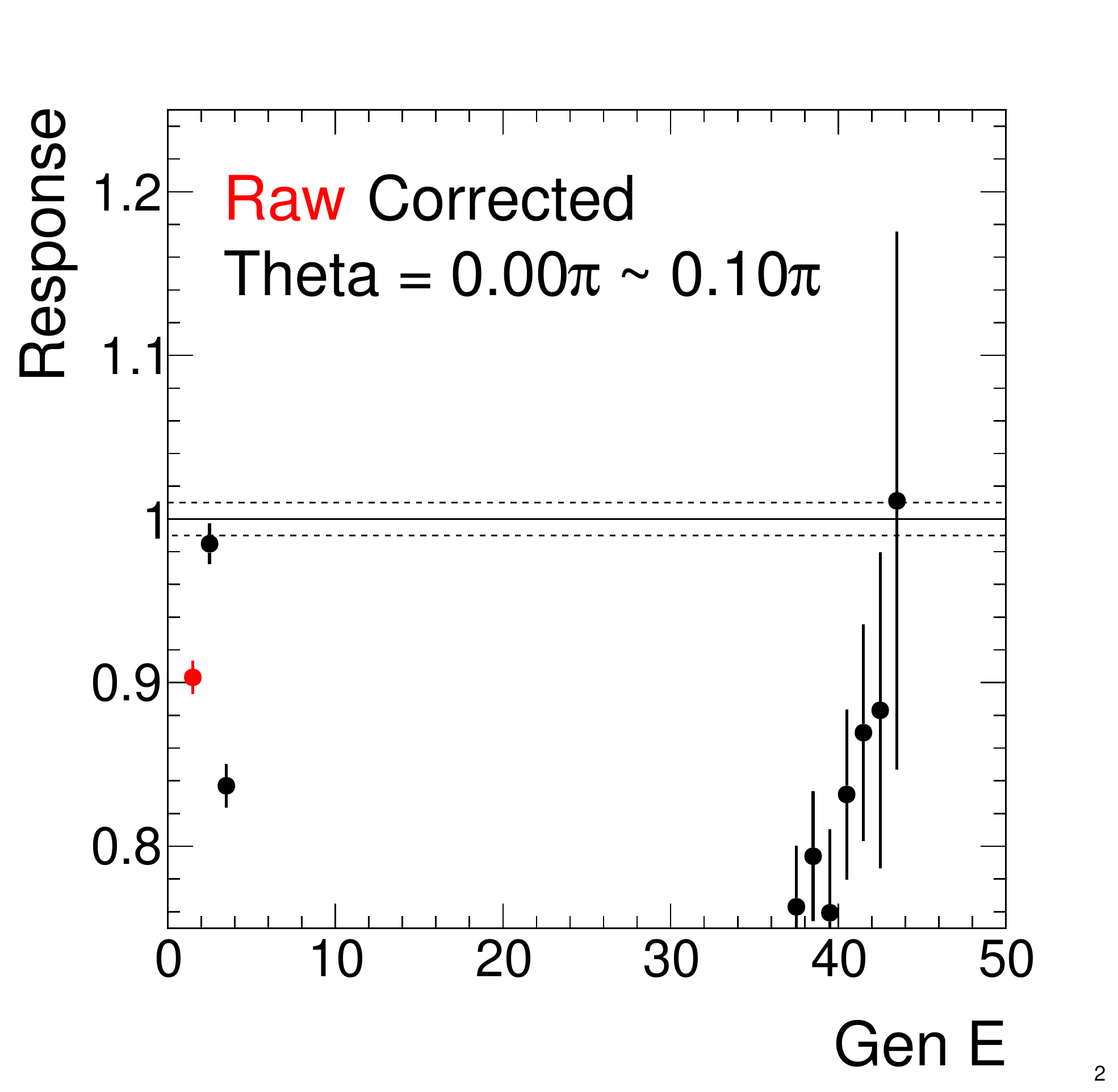}
    \includegraphicsfour{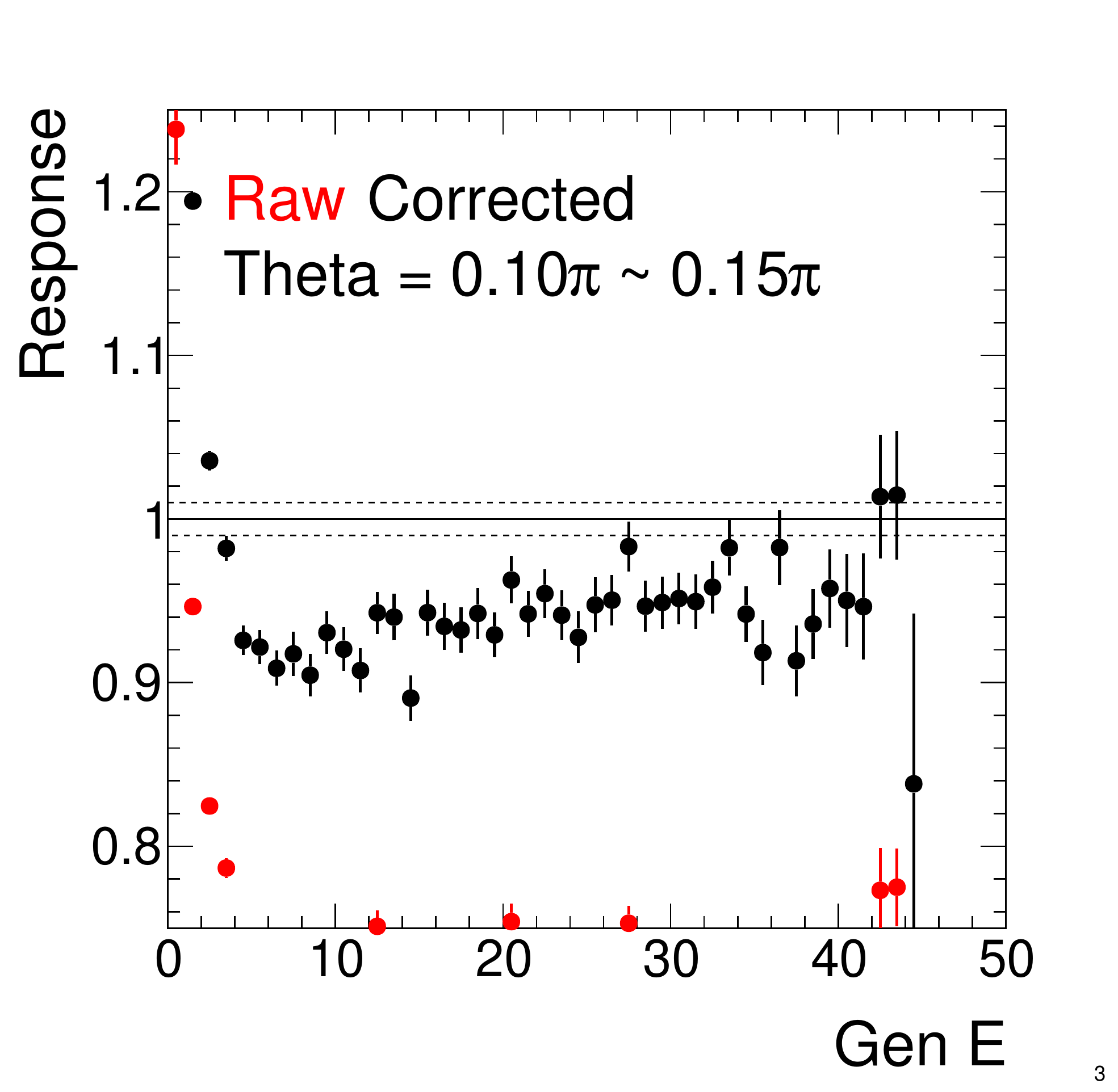}
    \includegraphicsfour{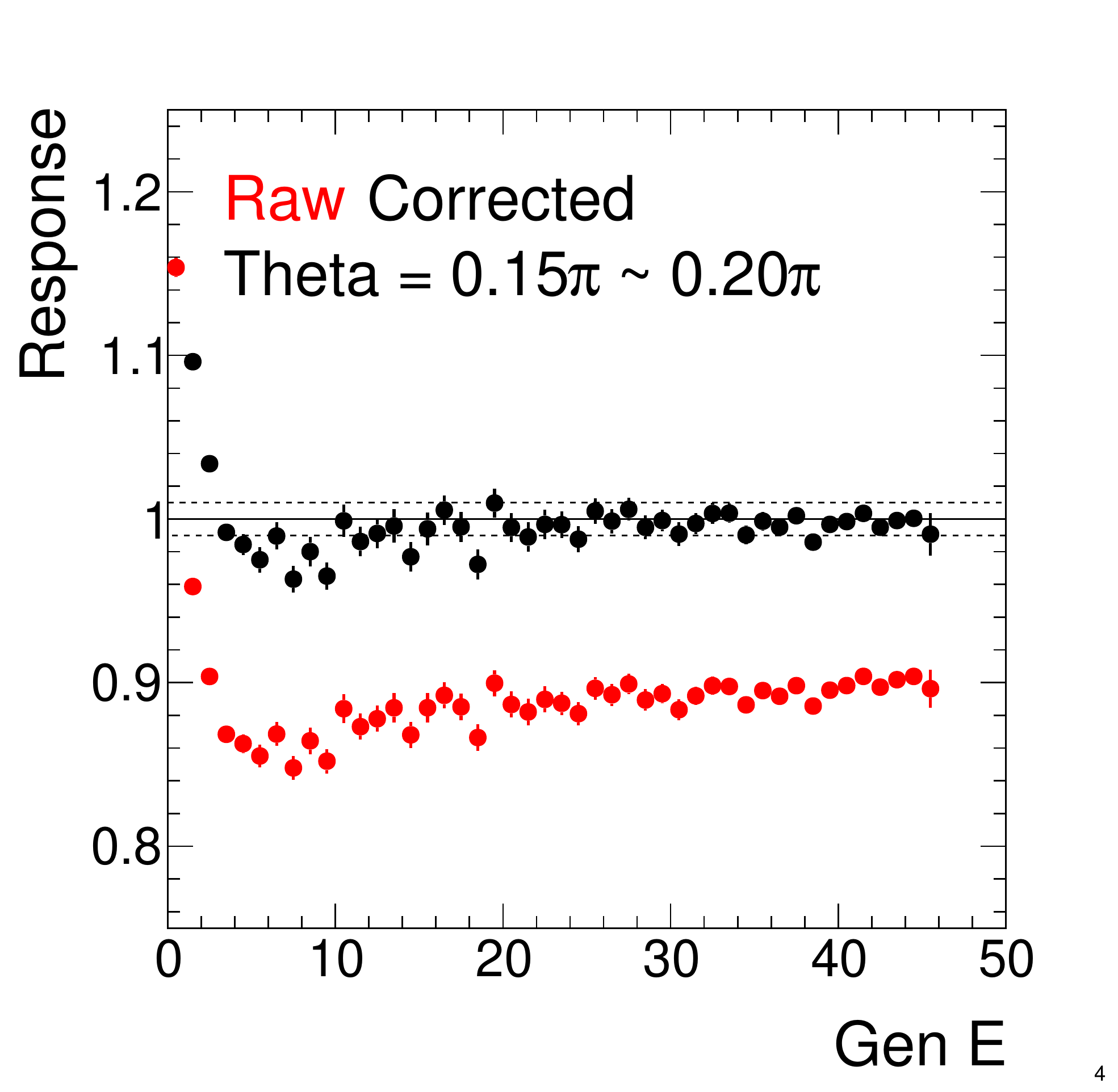}
    \includegraphicsfour{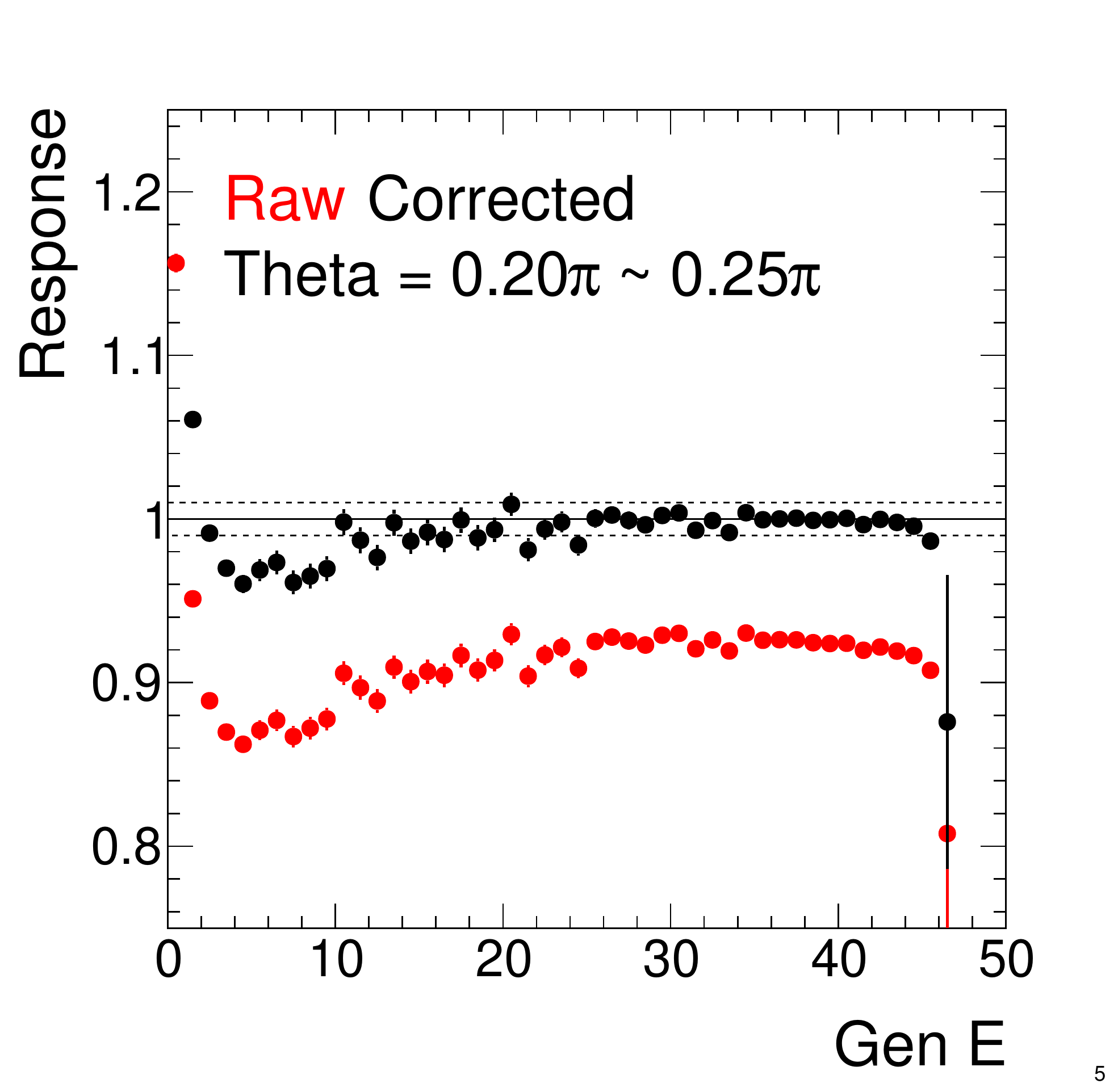}
    \includegraphicsfour{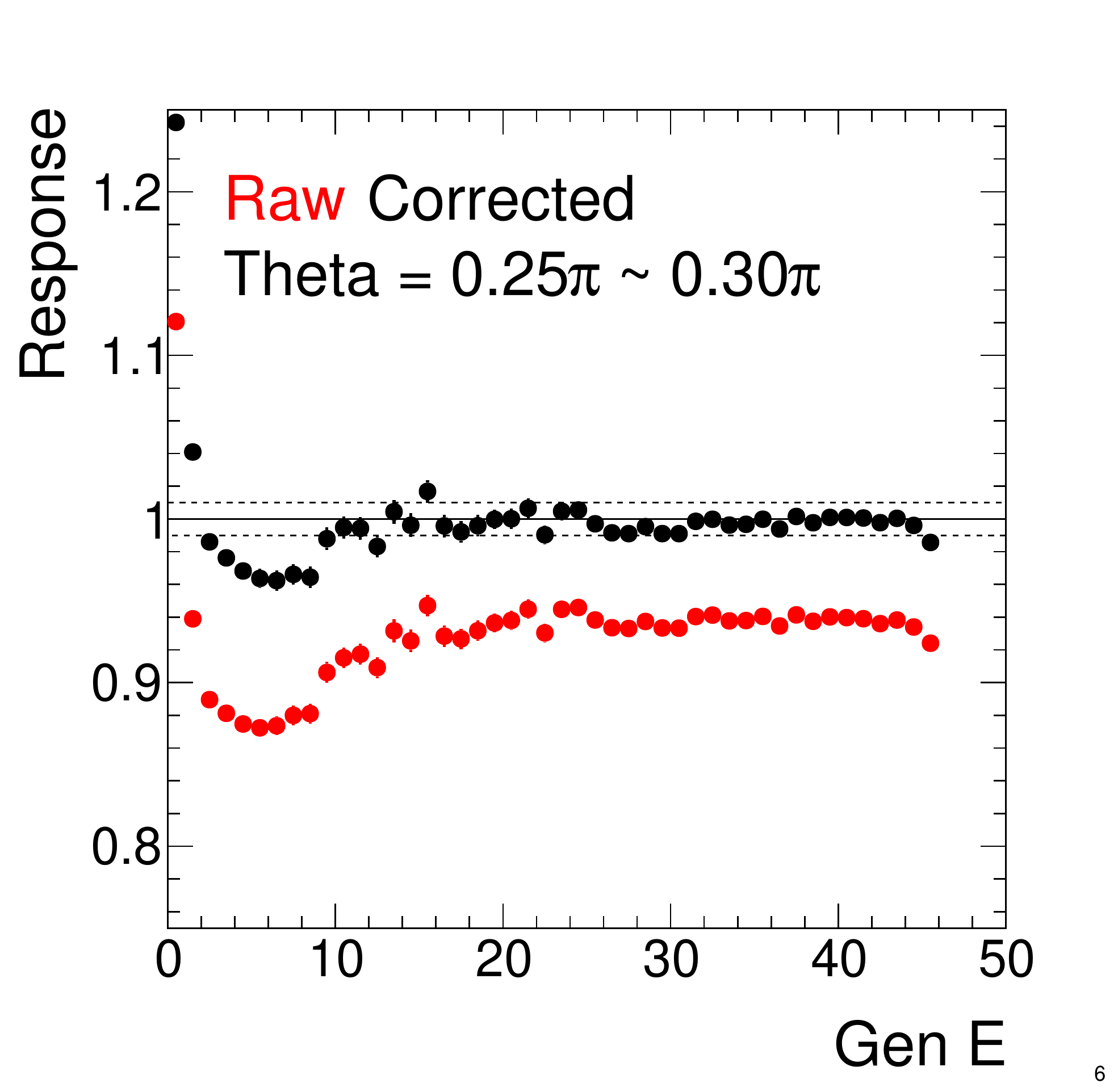}
    \includegraphicsfour{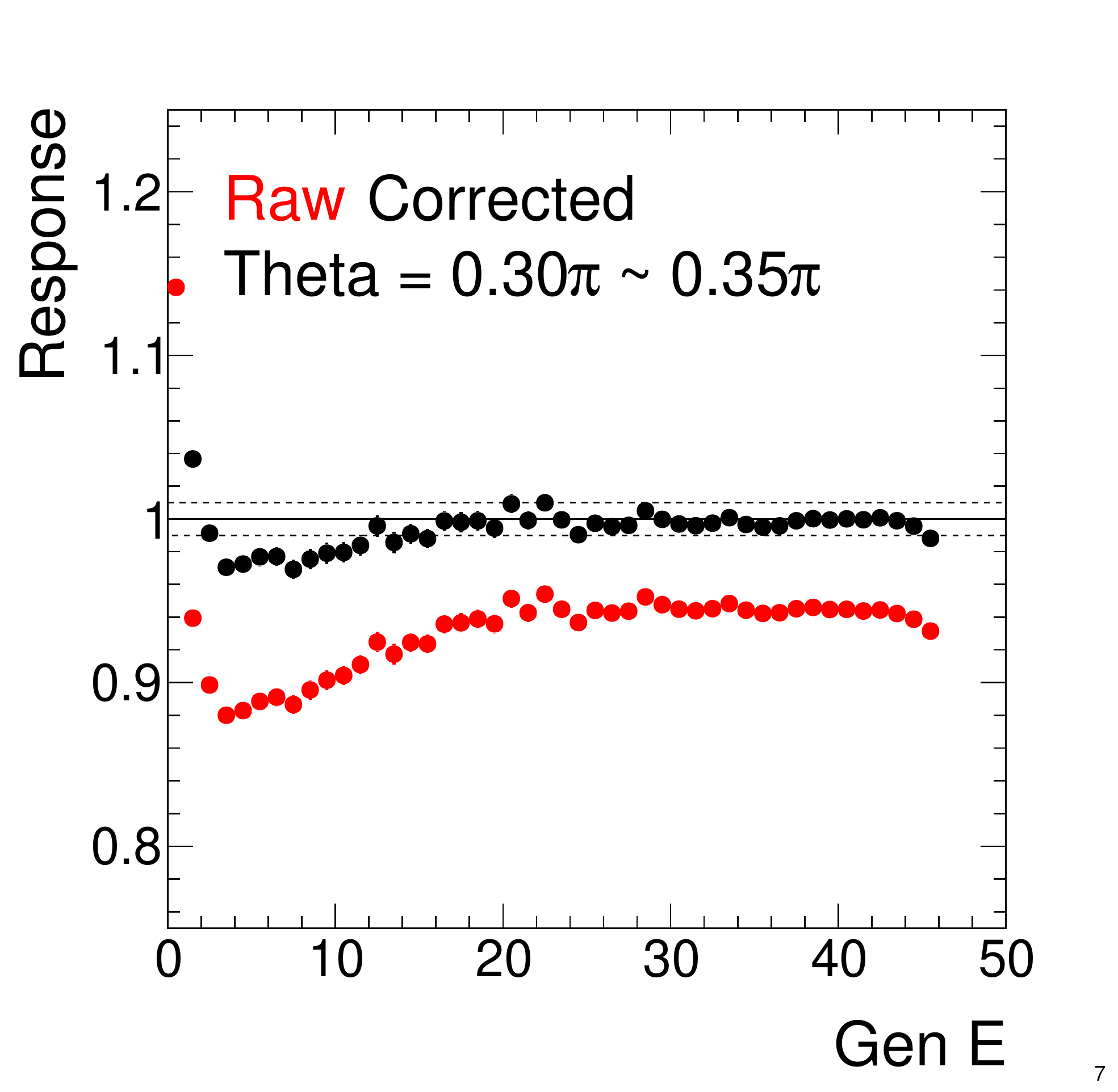}
    \includegraphicsfour{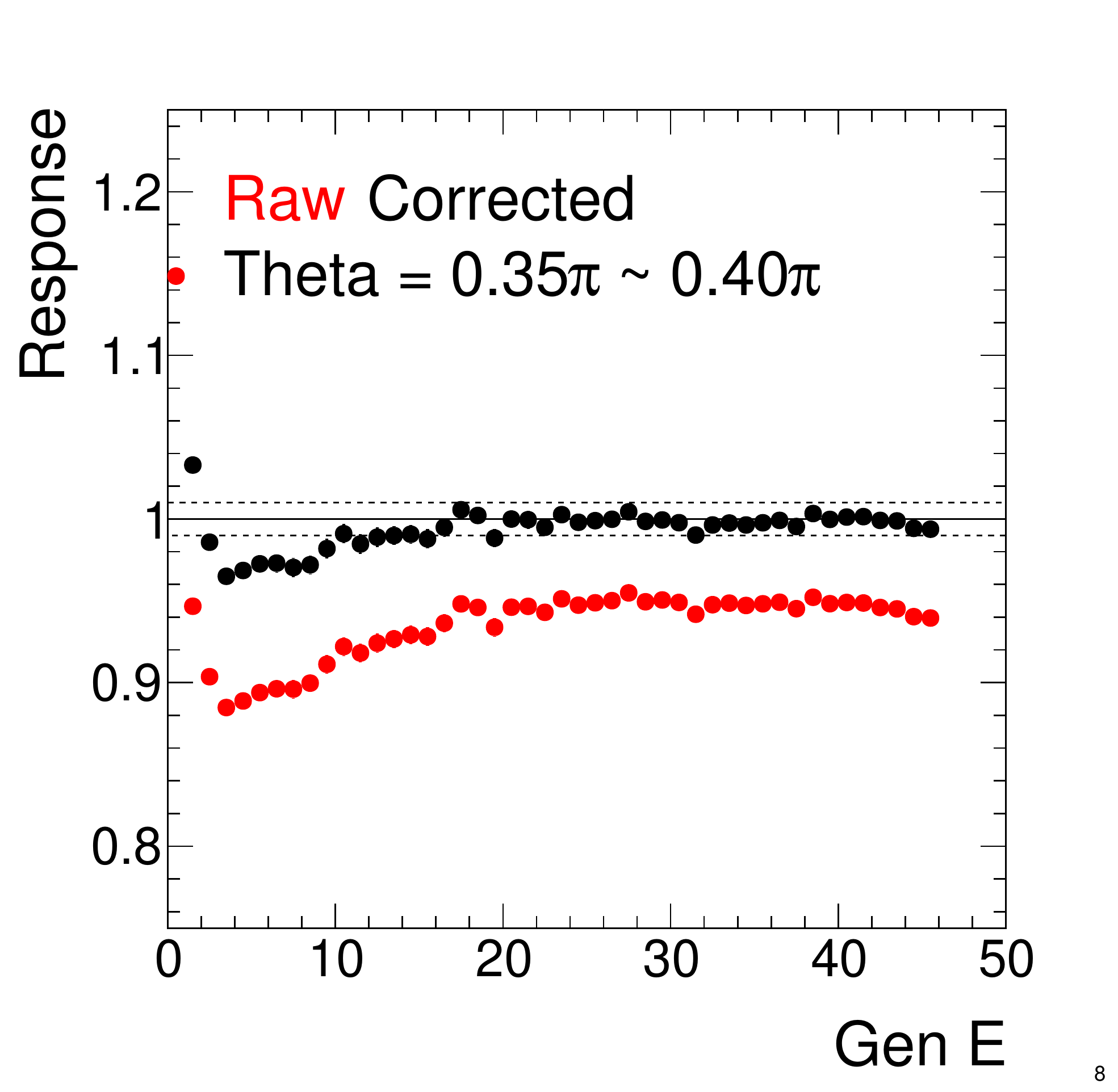}
    \includegraphicsfour{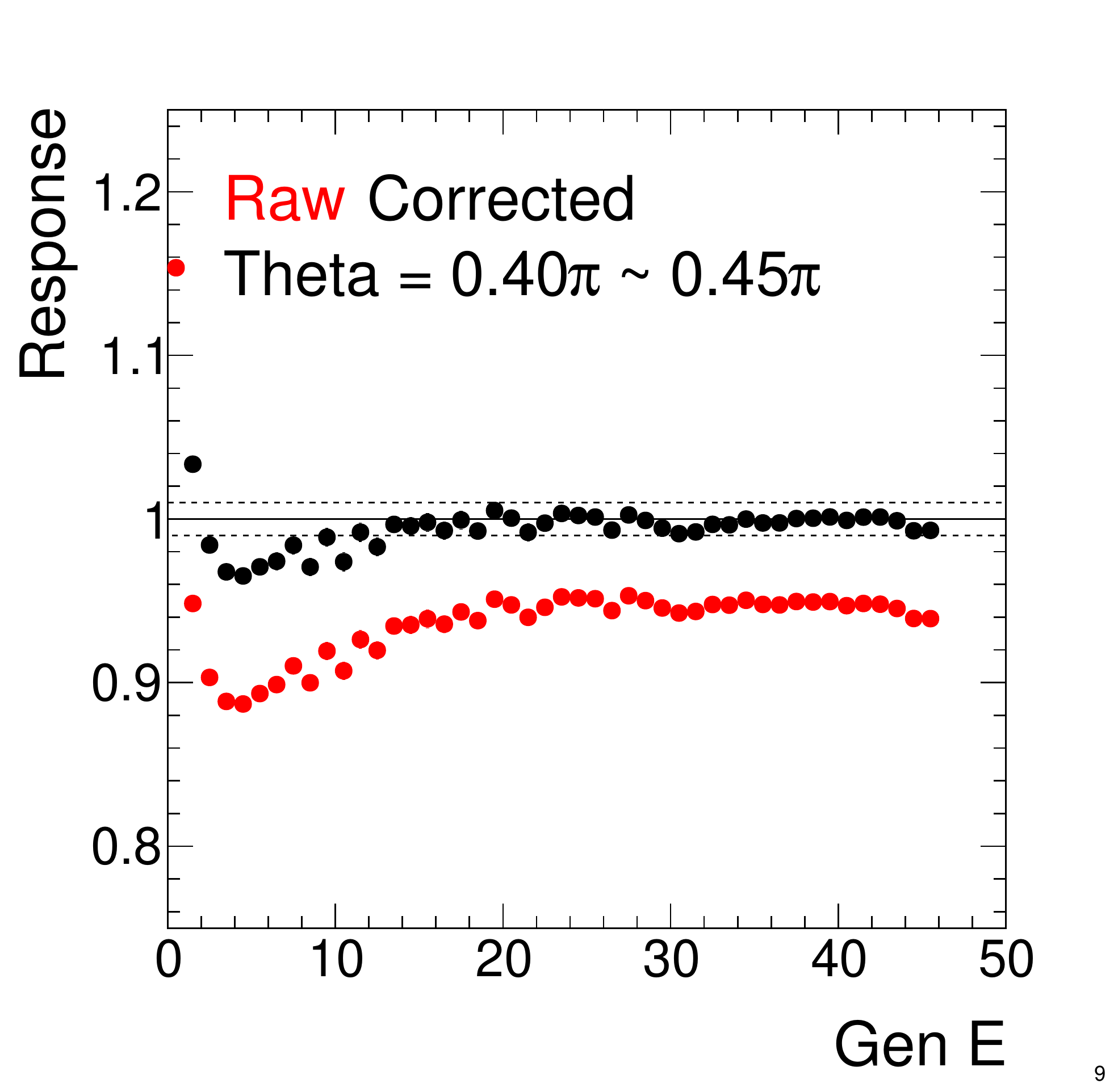}
    \includegraphicsfour{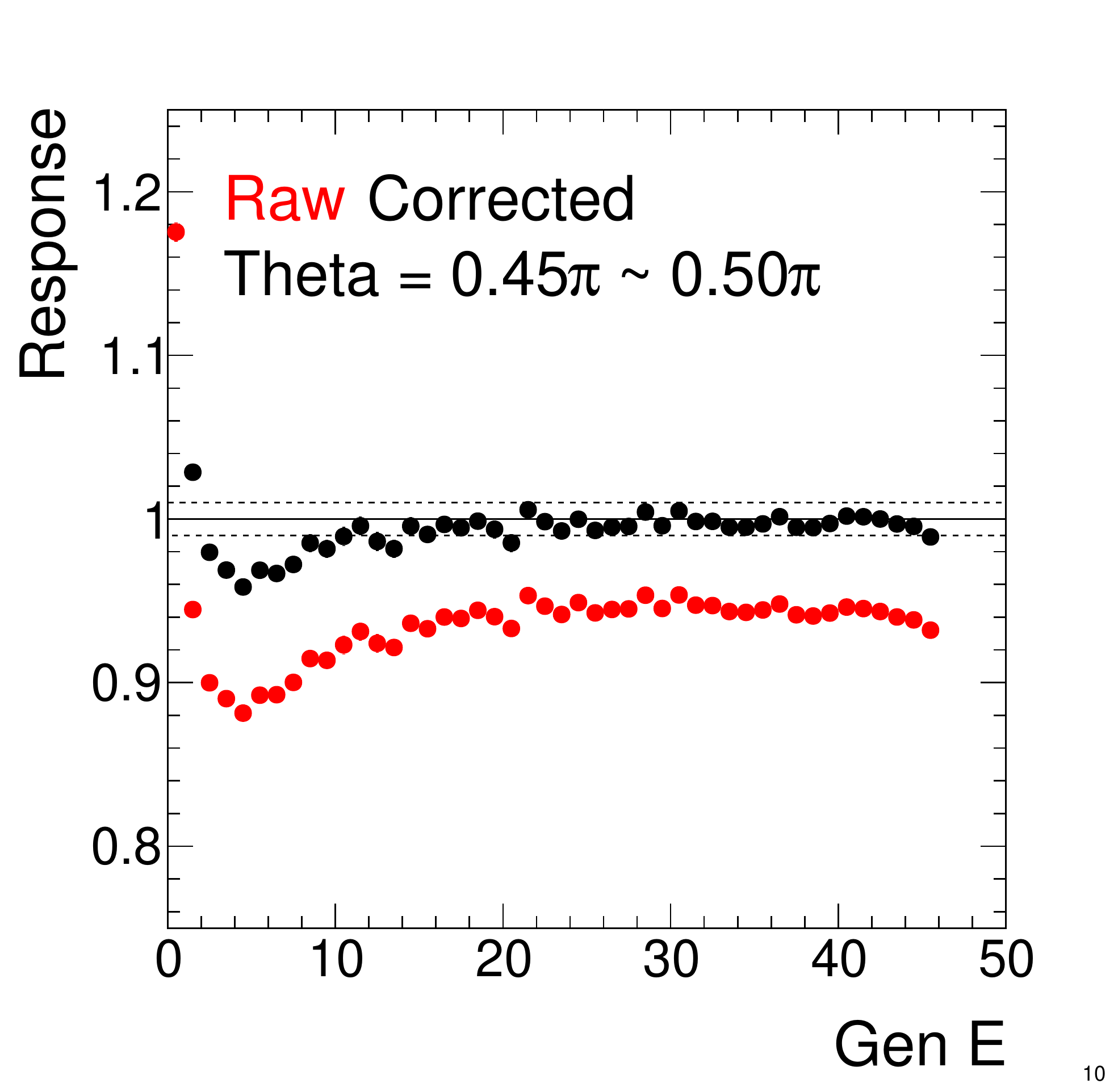}
    \includegraphicsfour{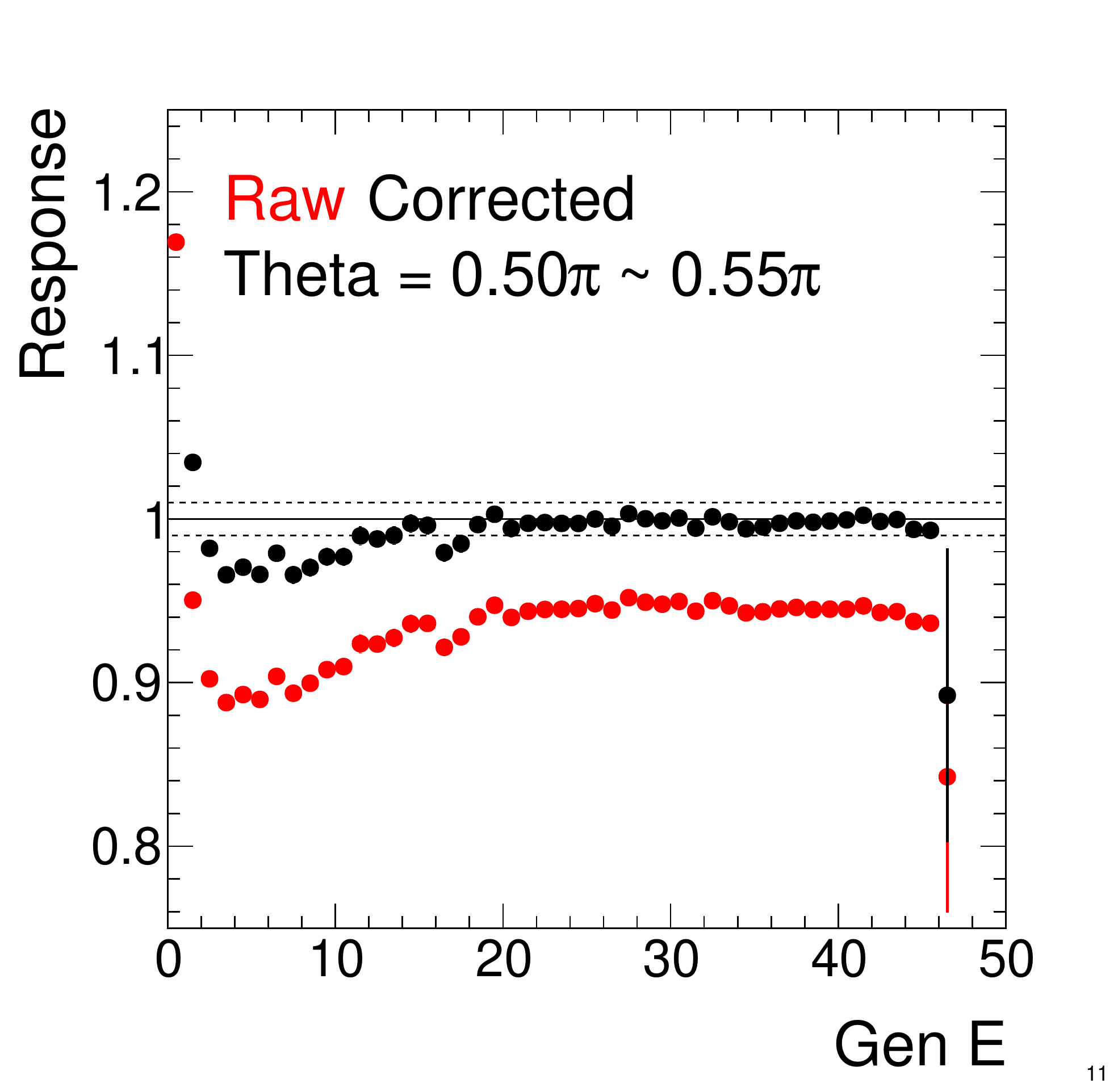}
    \includegraphicsfour{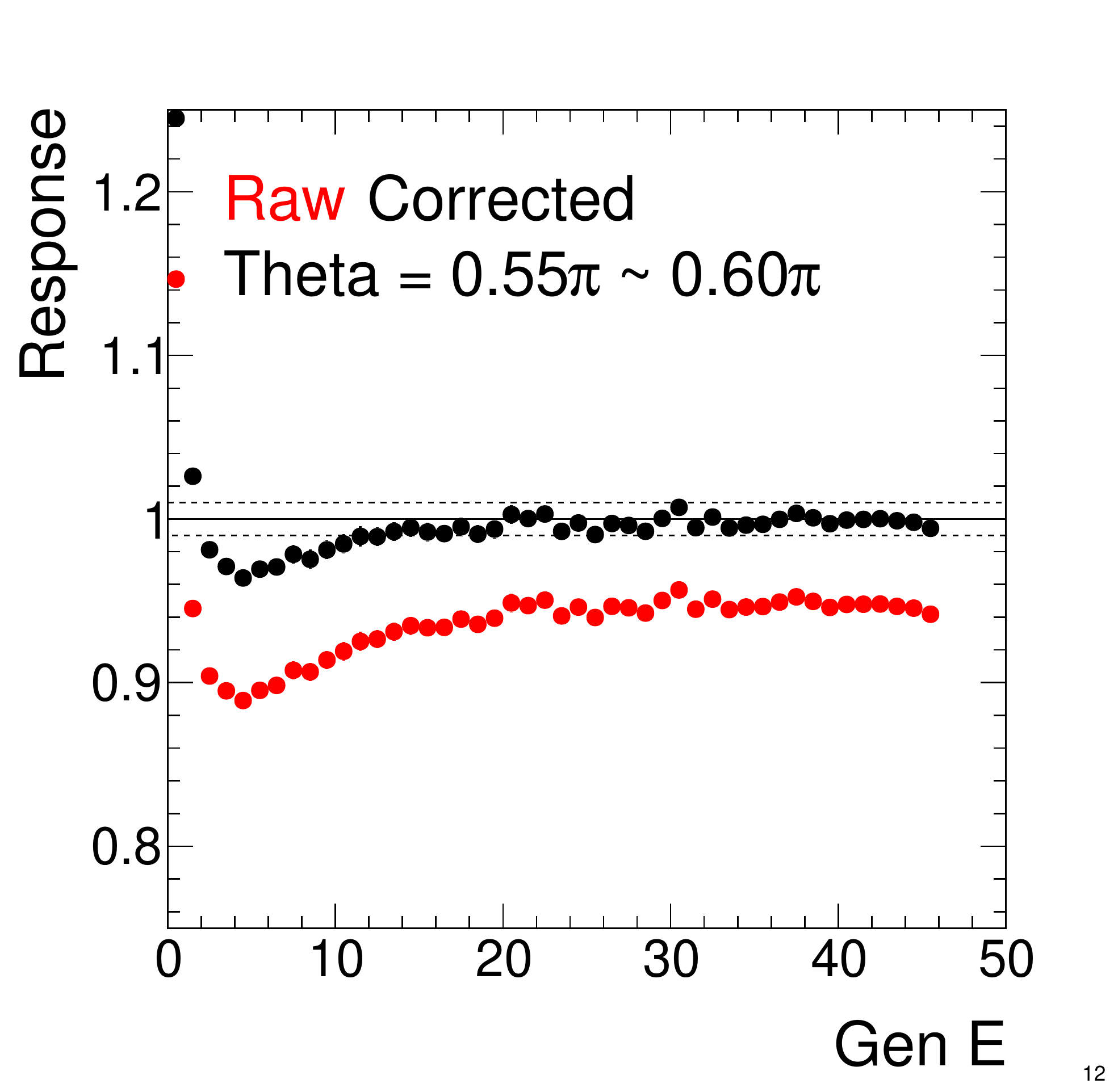}
    \includegraphicsfour{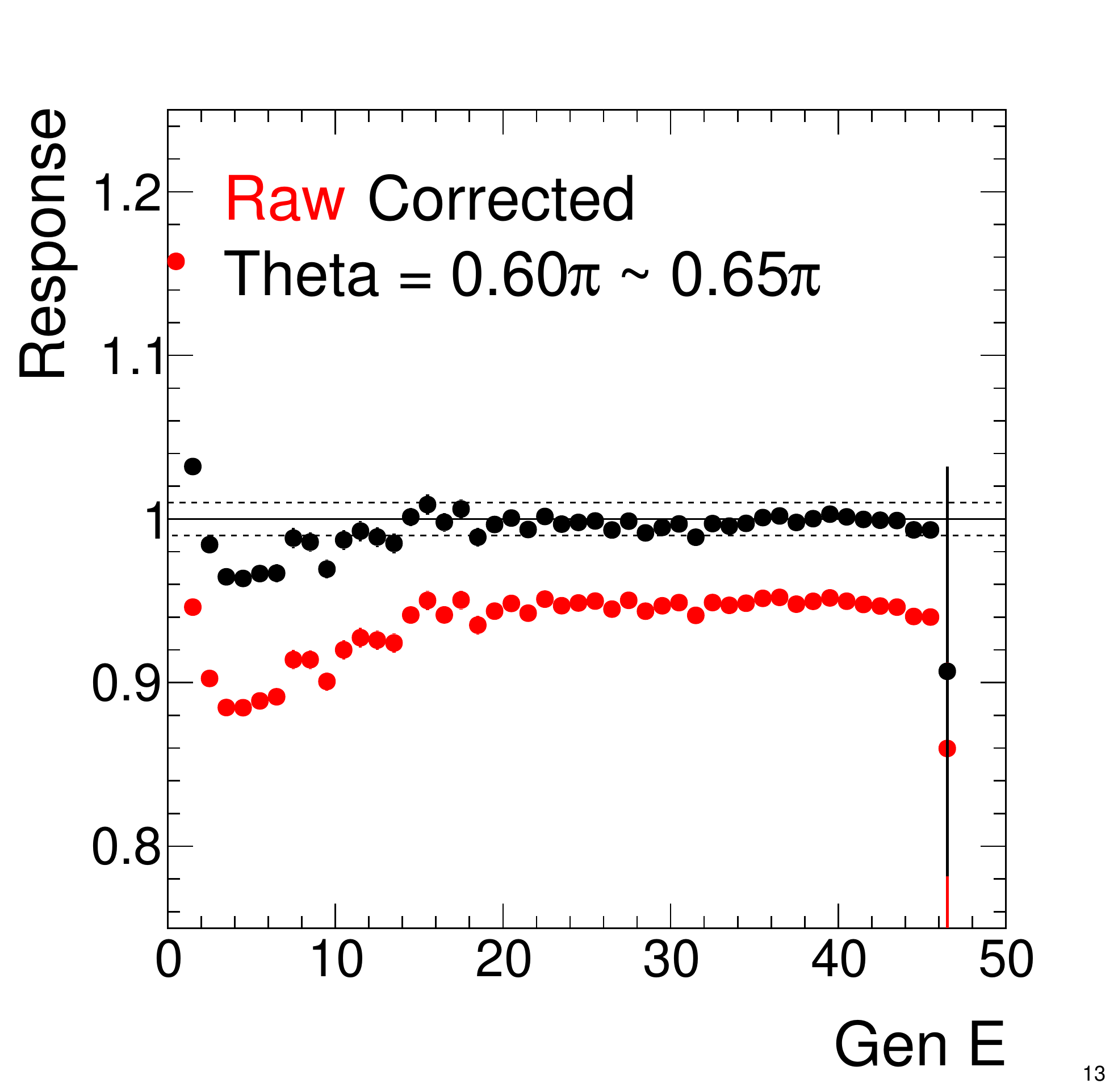}
    \includegraphicsfour{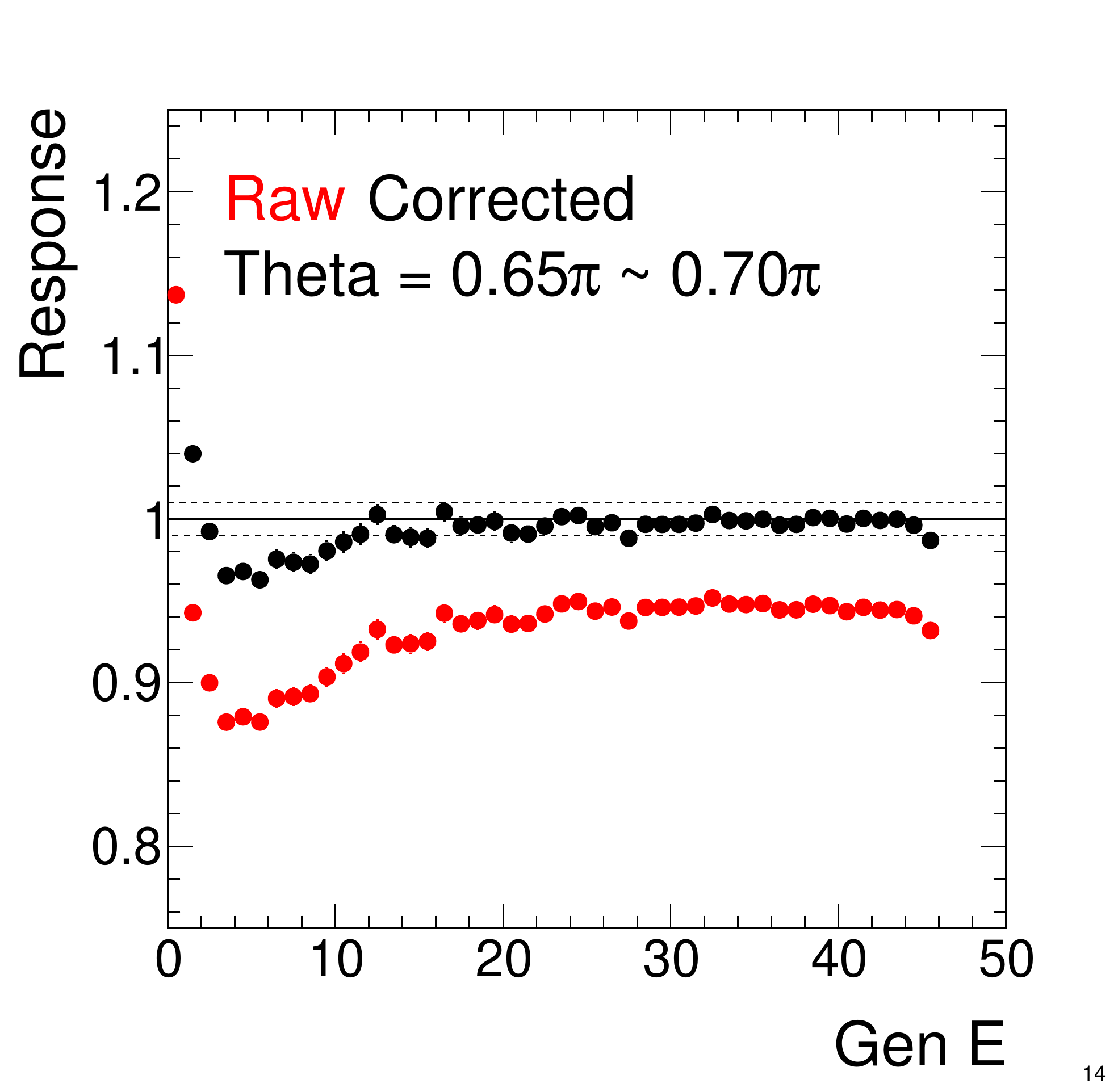}
    \includegraphicsfour{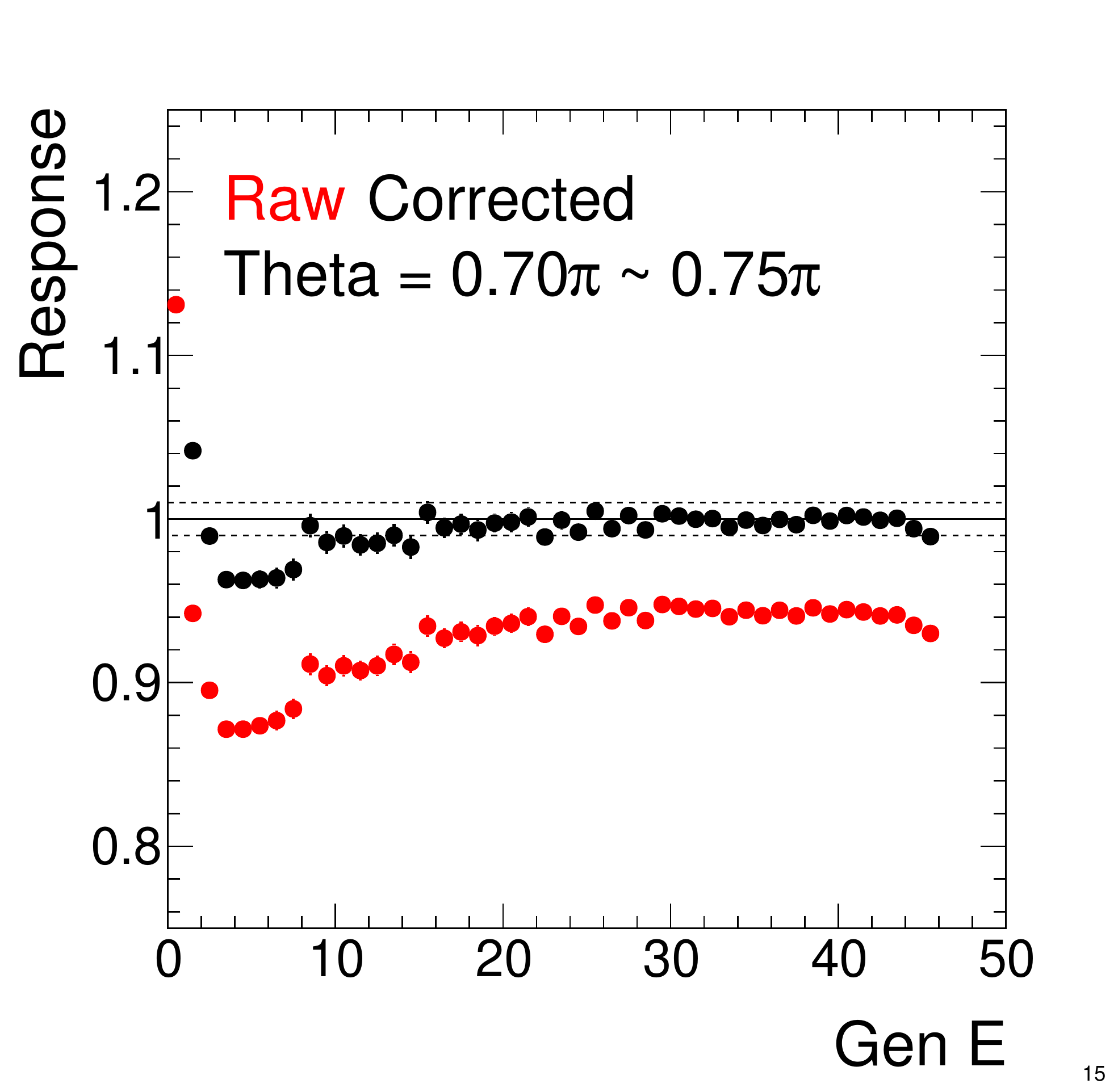}
    \includegraphicsfour{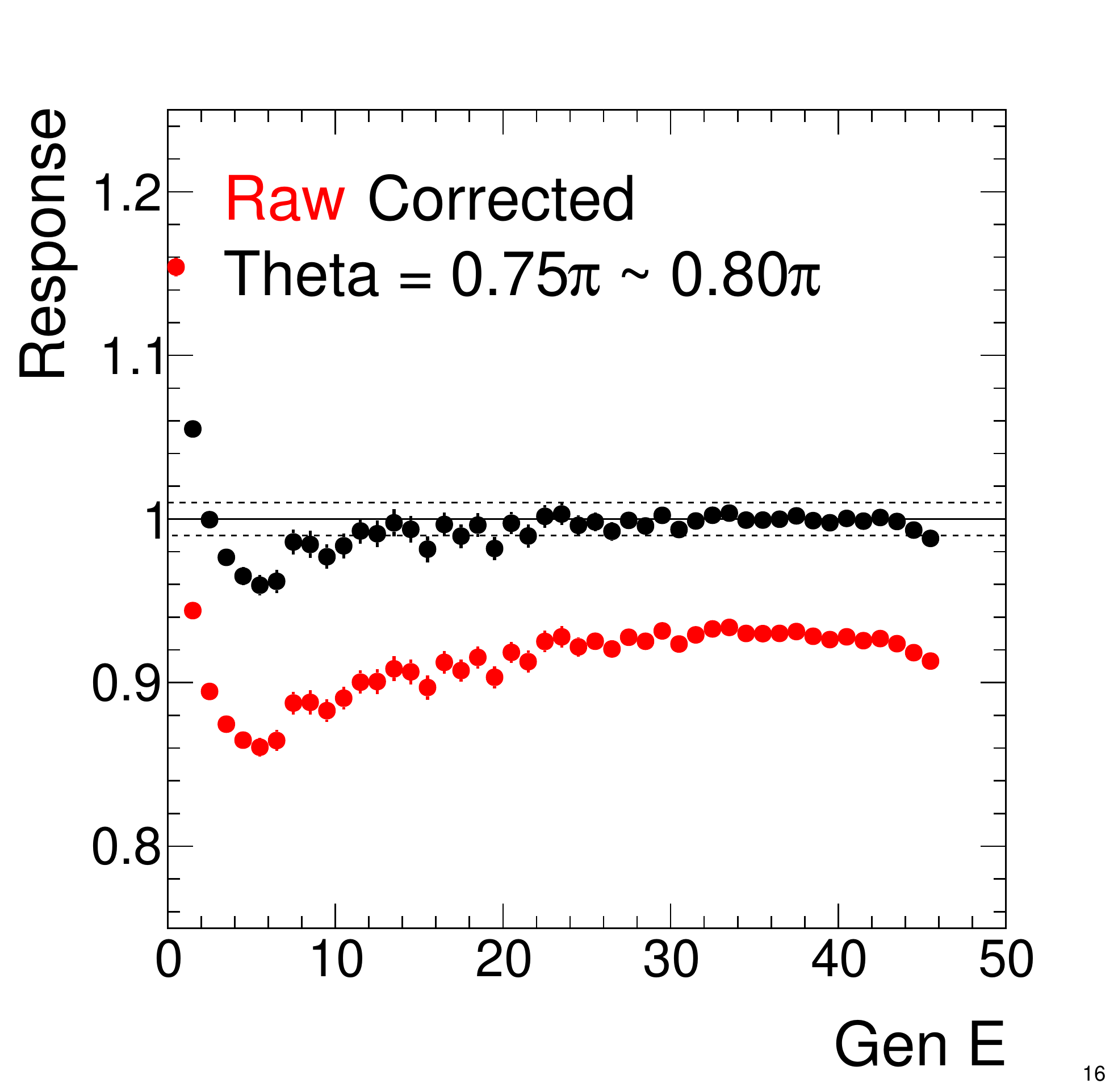}
    \includegraphicsfour{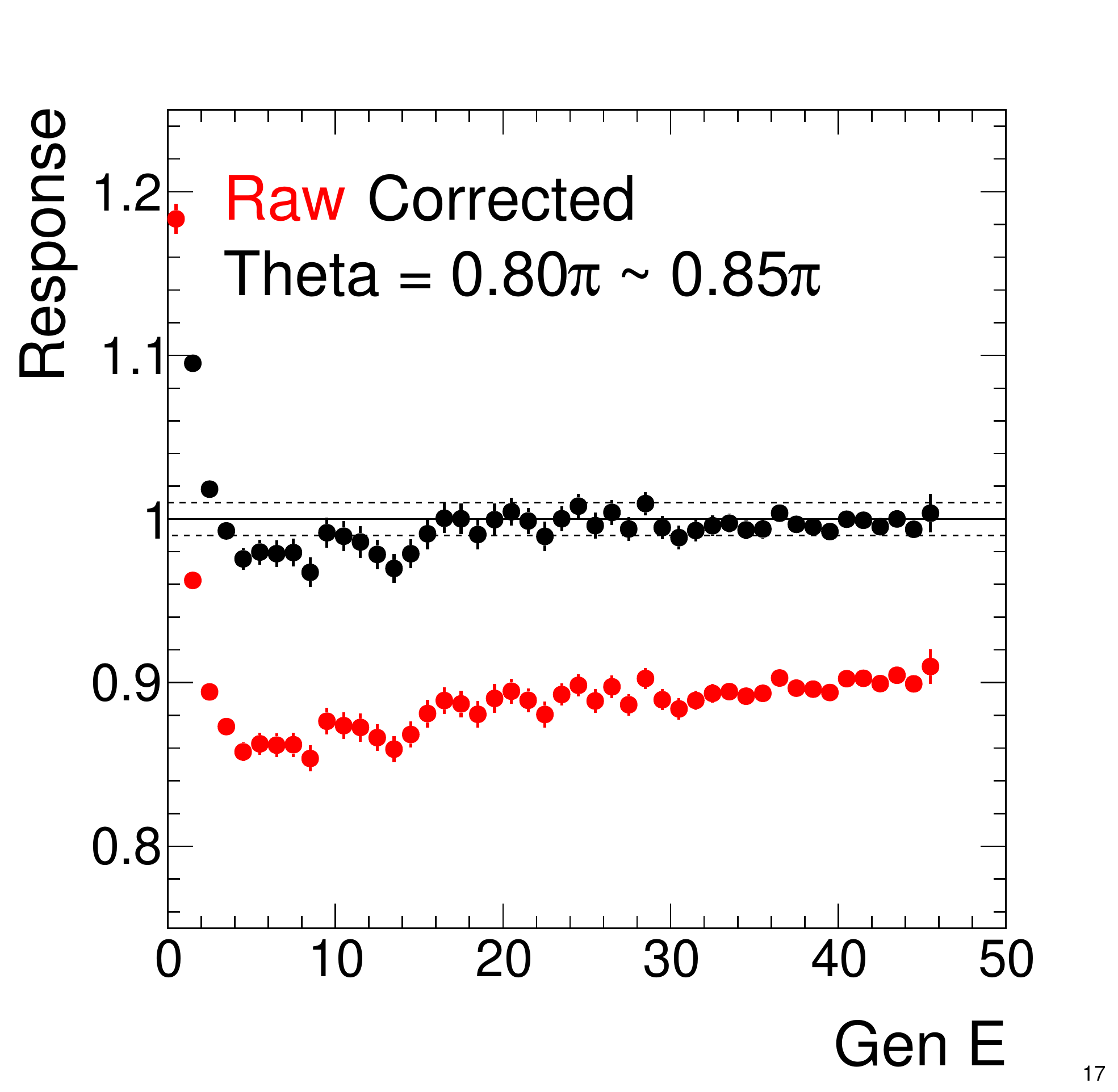}
    \includegraphicsfour{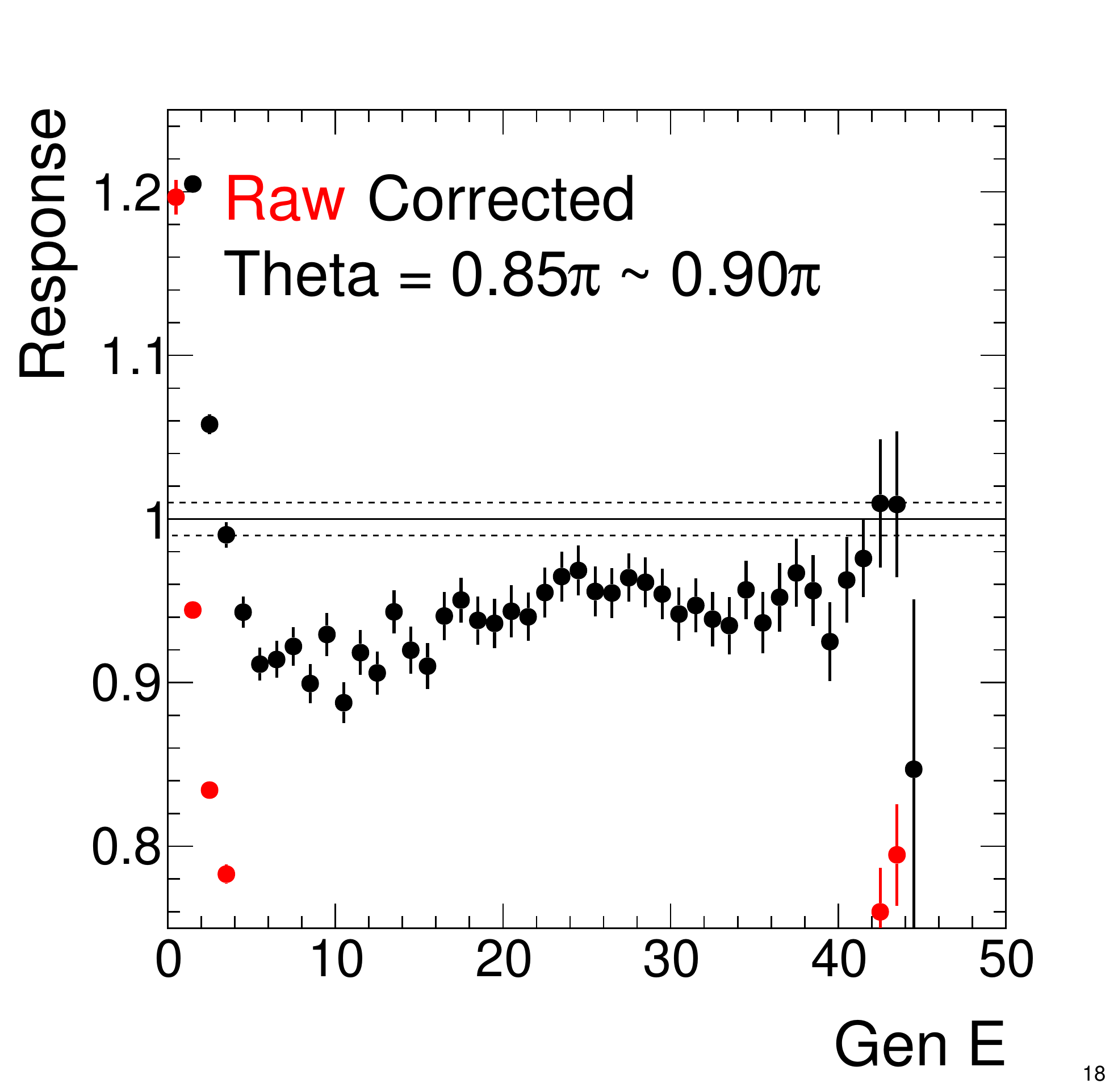}
    \includegraphicsfour{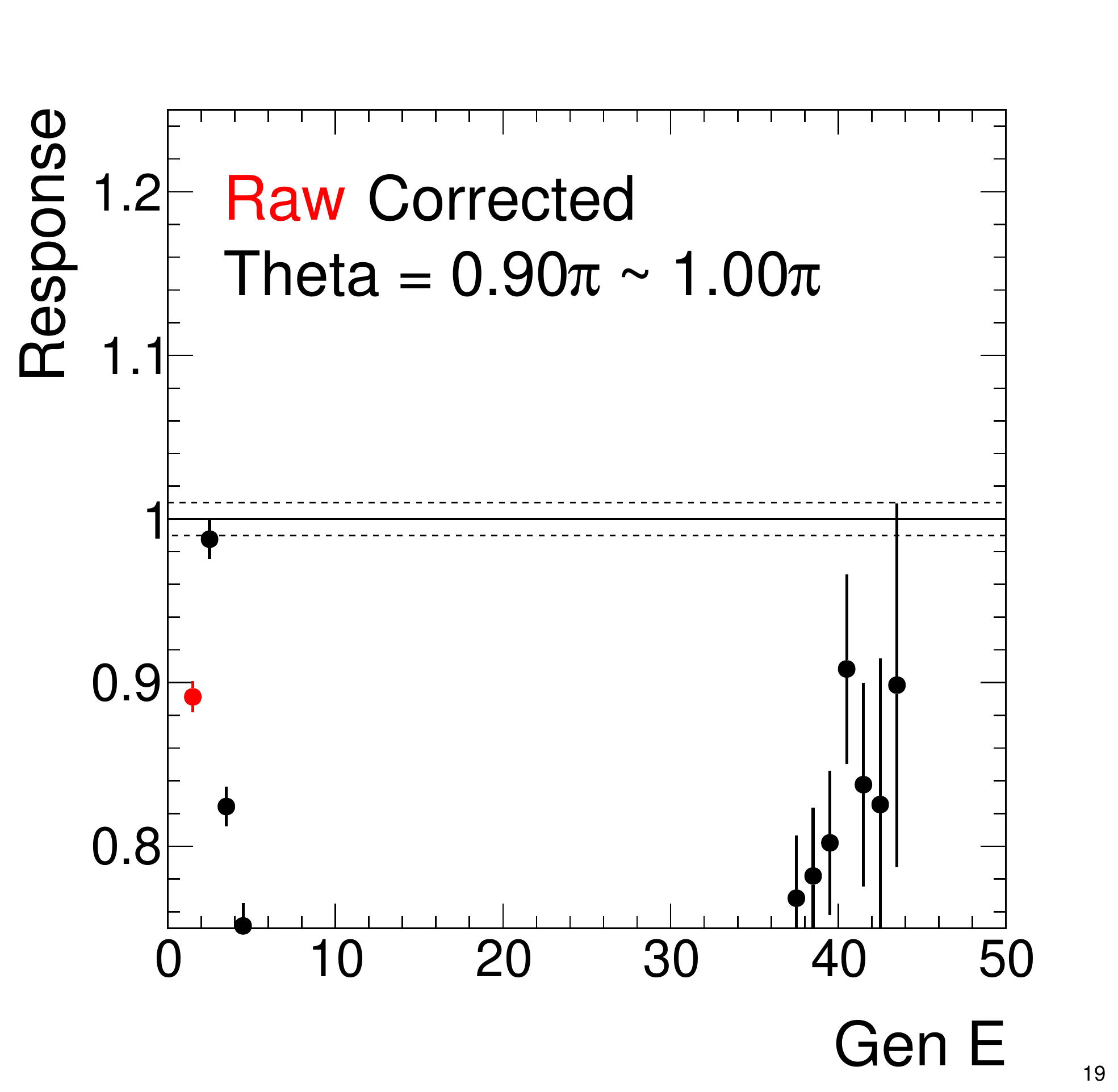}
    \caption{Jet energy response before (red) and after (black) applying corrections, as a function of jet energy in different bins of jet $\theta$.  A decent closure is seen for jets between $0.15\pi$ and $0.85\pi$ and with energy above 10 GeV.}
    \label{Figure:JetCalibration-JECThetaBin}
\end{figure}

The raw response and the corrected response as a function of jet energy in different bins of jet $\theta$ is shown in Fig.~\ref{Figure:JetCalibration-JECThetaBin}. These ``closure tests'', the inspection of the jet energy correction factors, focuses on the differences between the corrected jet energies and the generated jet energies in the MC samples. A perfect closure refers to zero average difference between corrected jet energy and the generated jet energy. The performance of jet energy correction is stable for the jets that are within our analysis selection $0.2\pi < \theta_\text{jet} < 0.8\pi$. Non-closure is observed for those jets which are falling (partially) outside of the ALEPH detector acceptances.

\begin{figure}[htp!]
    \centering
    \includegraphicsthree{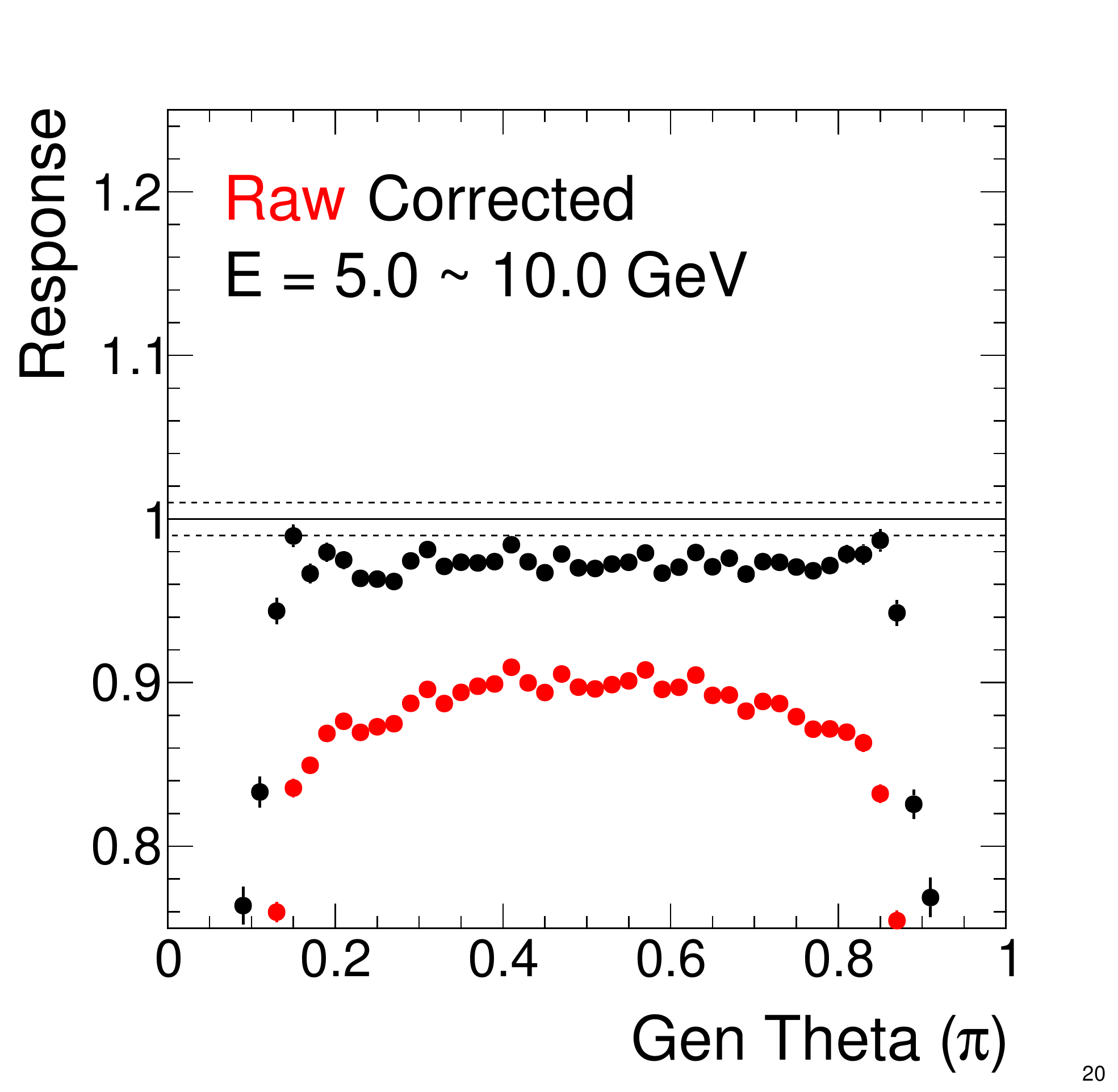}
    \includegraphicsthree{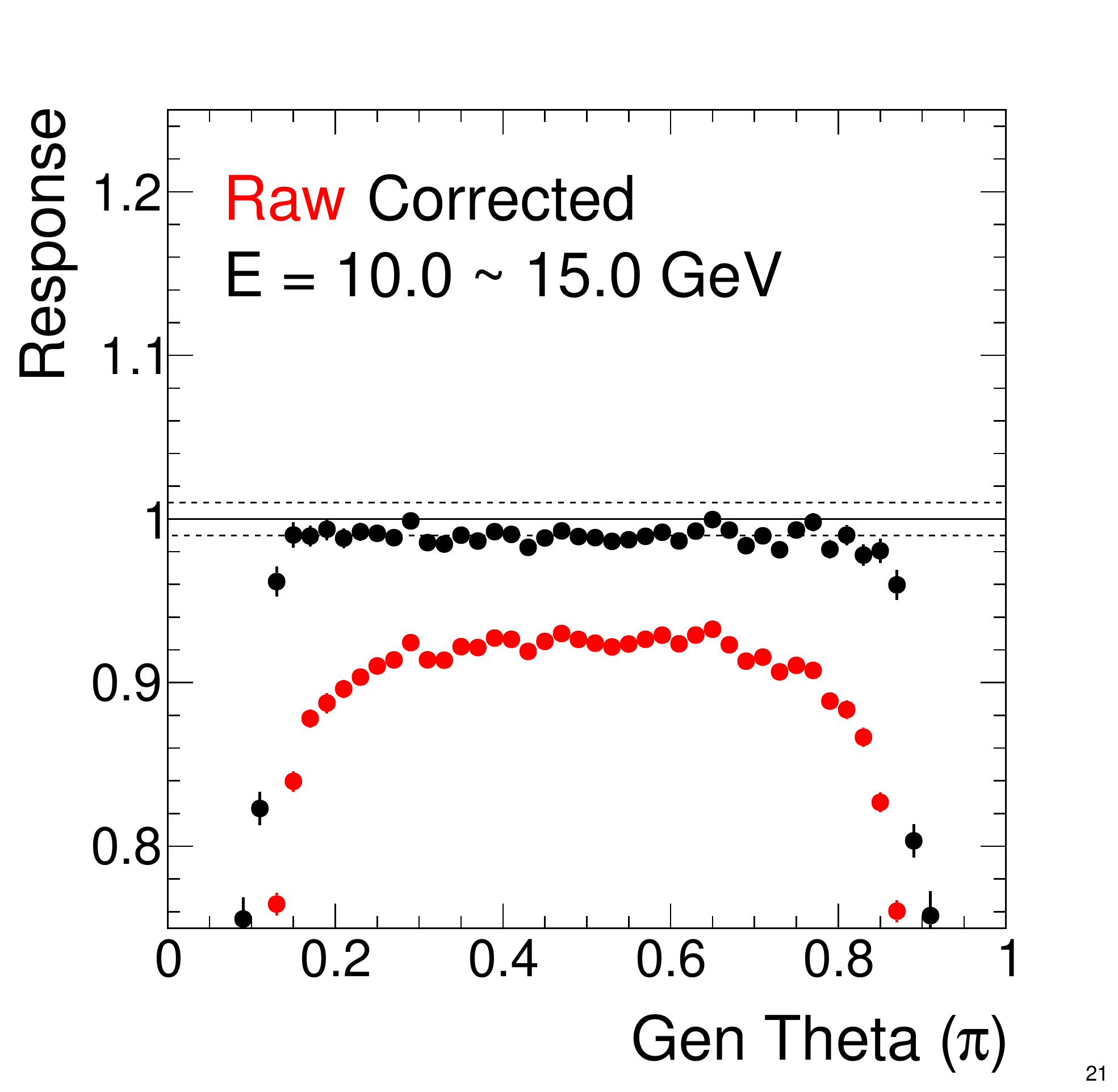}
    \includegraphicsthree{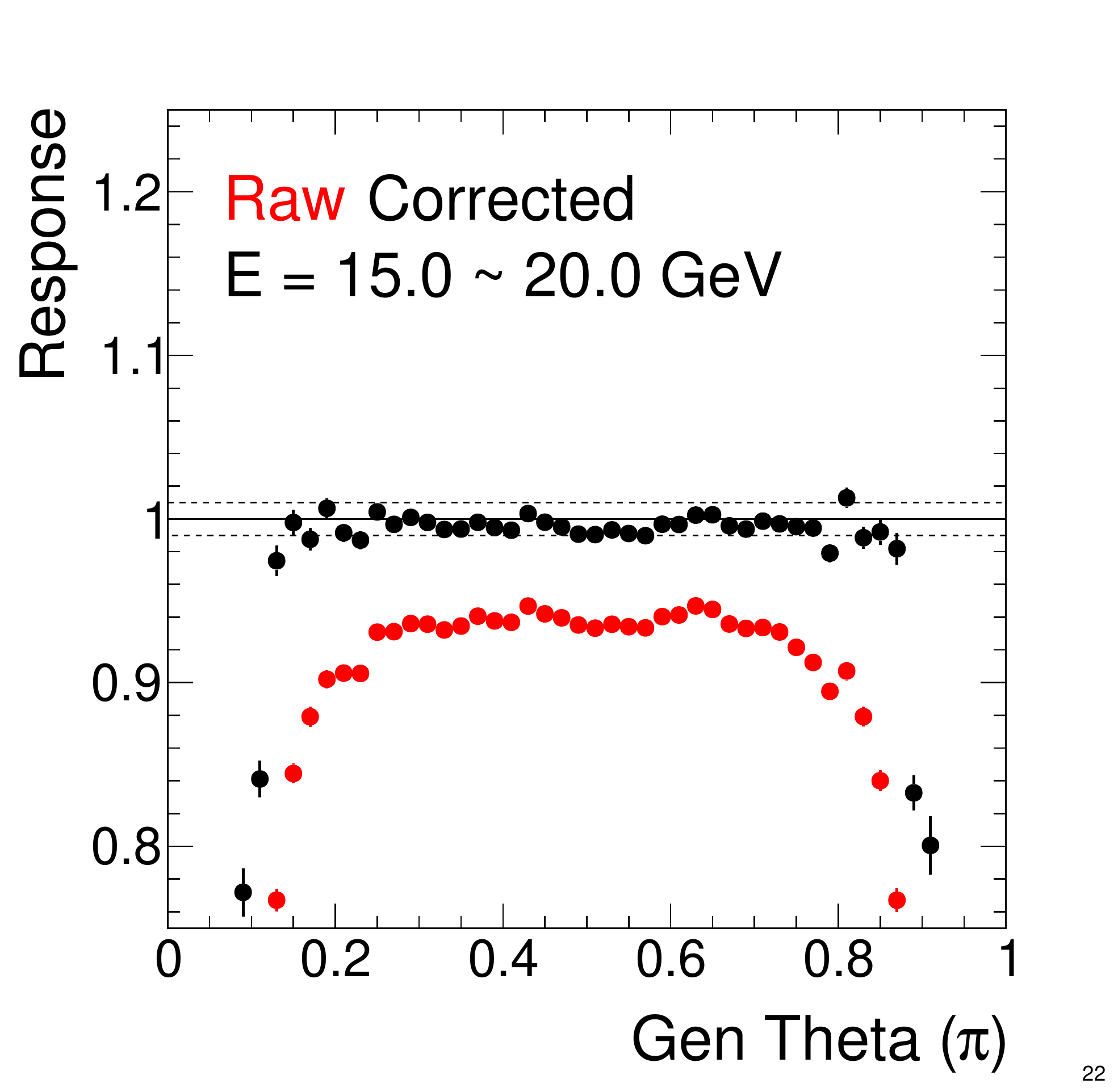}
    \includegraphicsthree{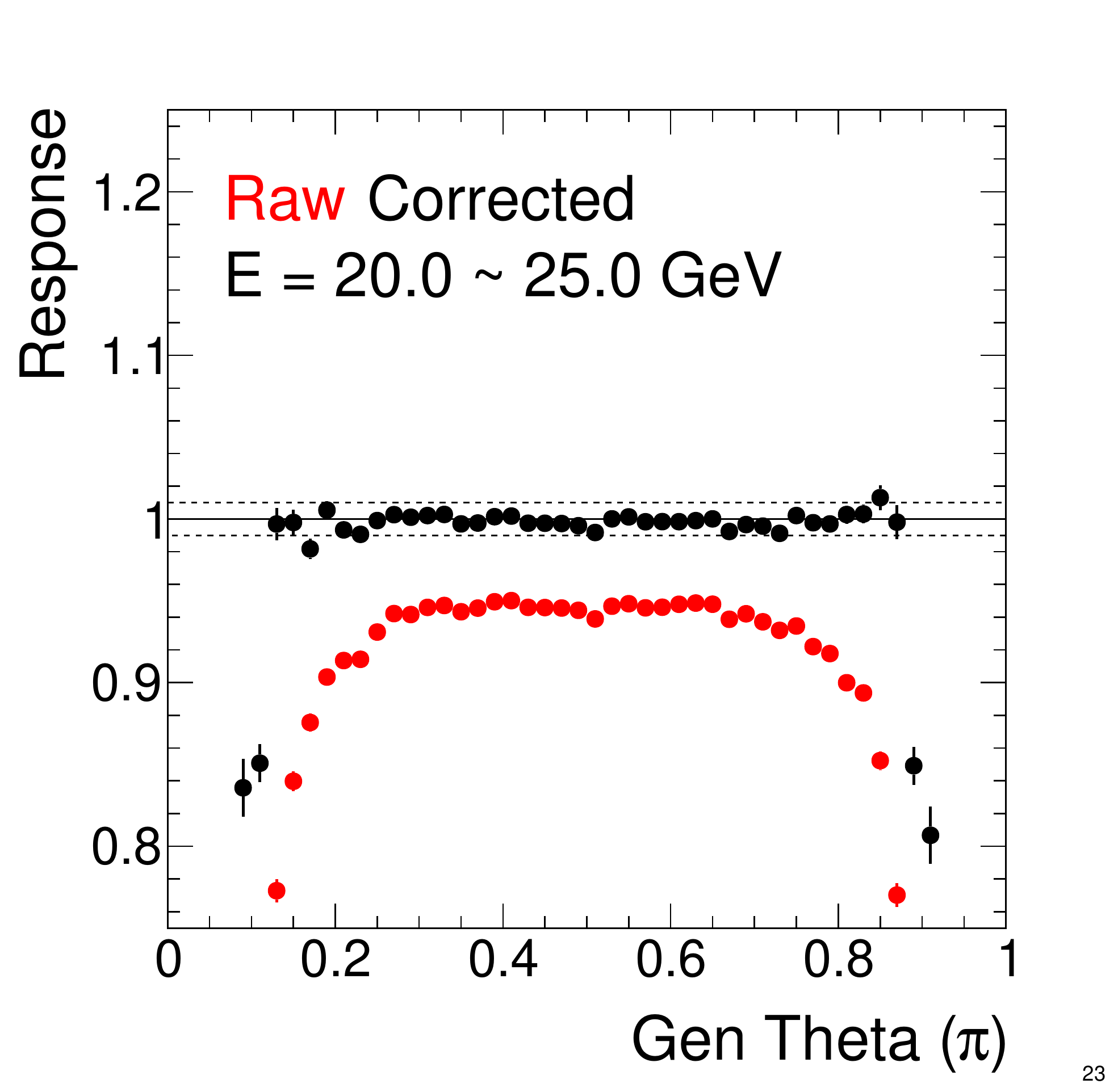}
    \includegraphicsthree{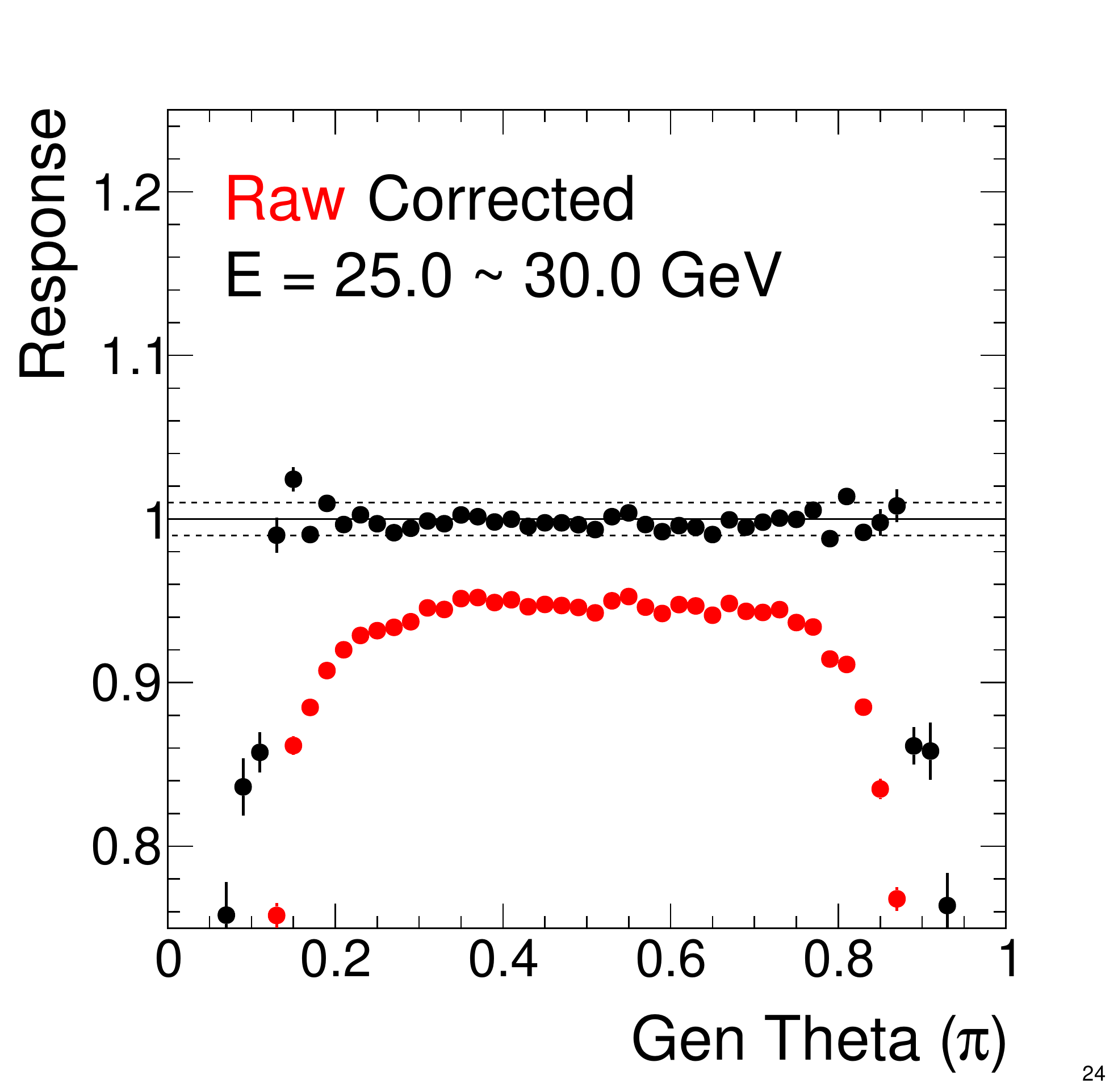}
    \includegraphicsthree{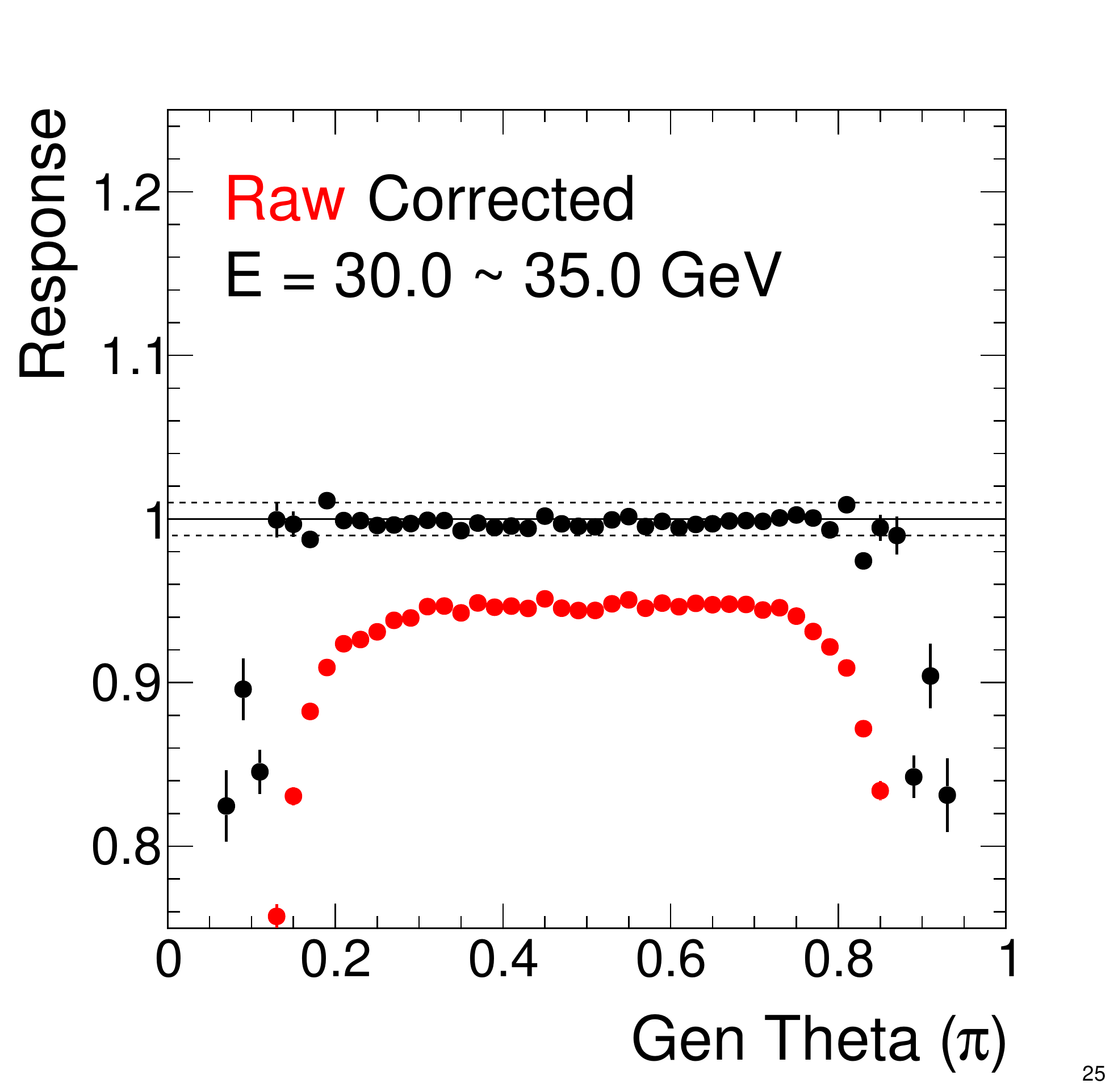}
    \includegraphicsthree{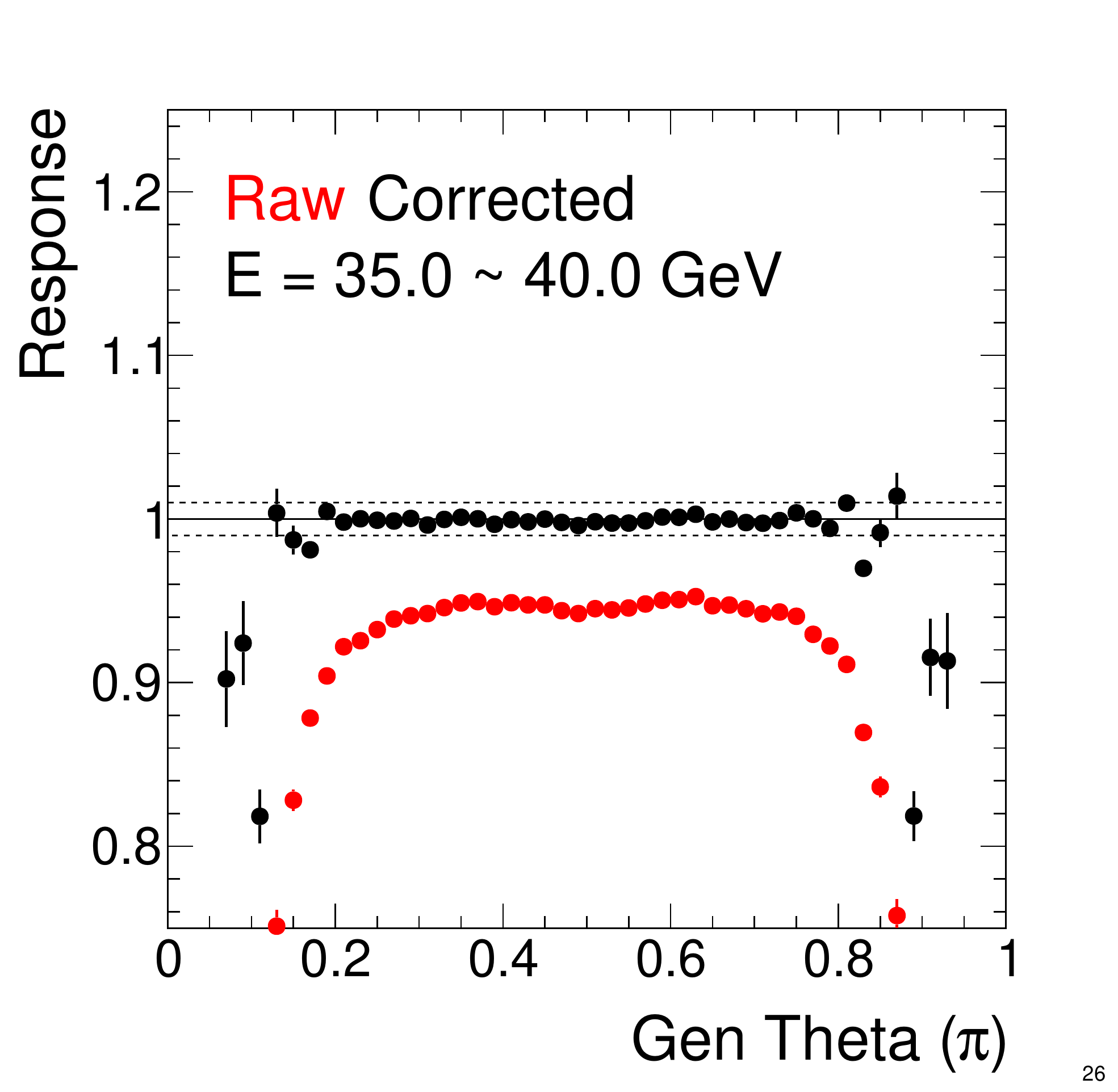}
    \includegraphicsthree{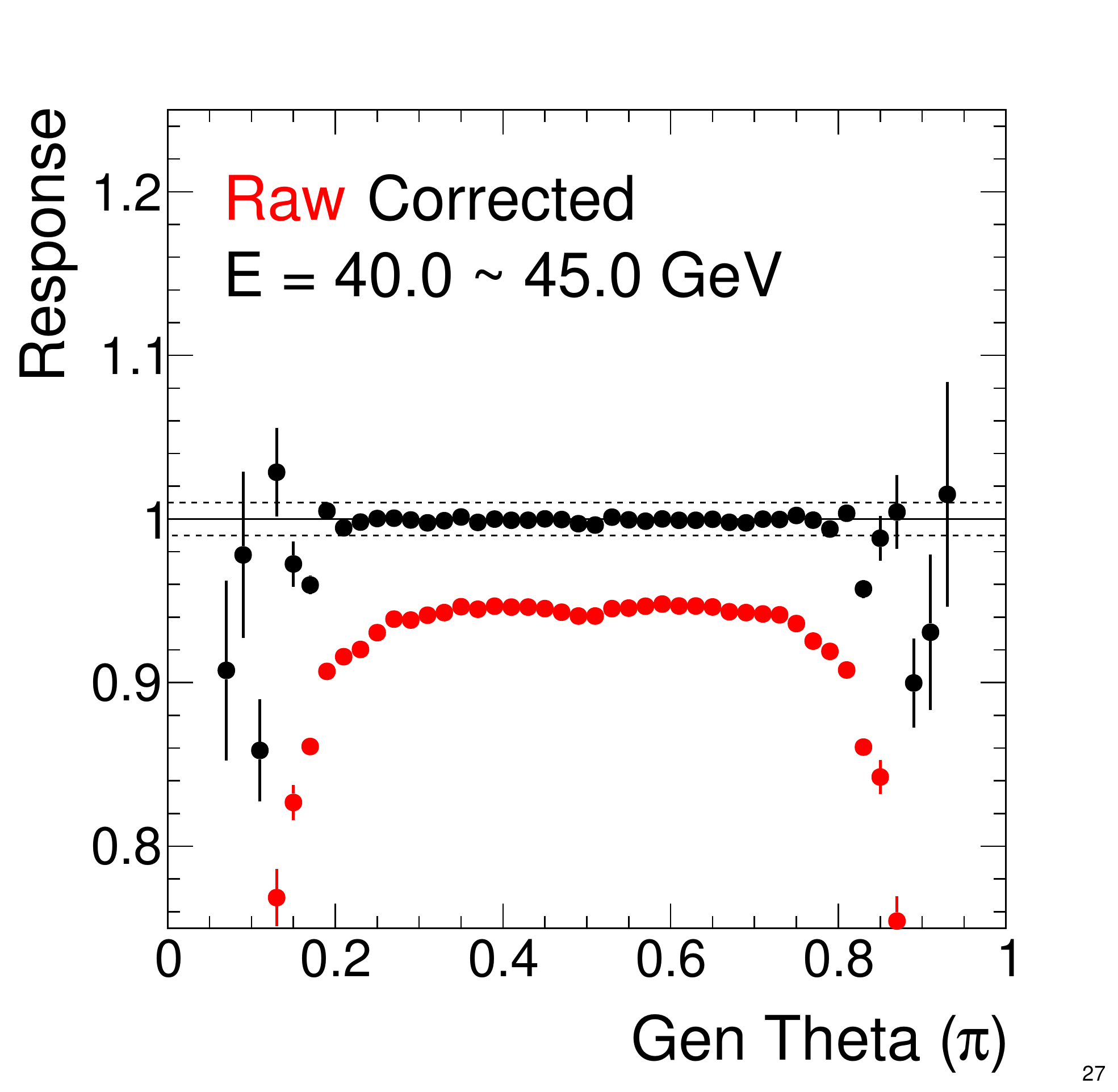}
    \includegraphicsthree{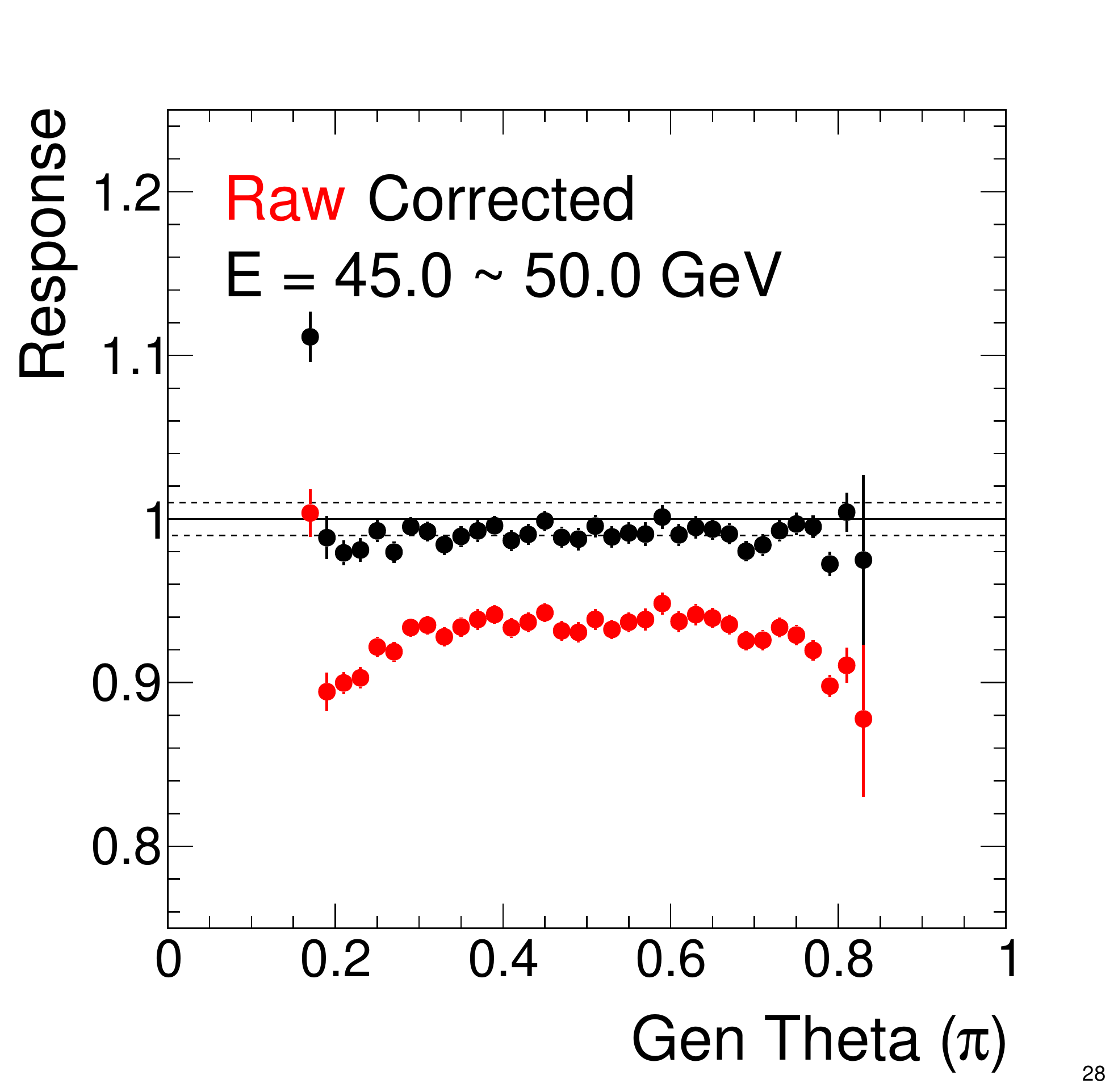}
    \caption{Jet energy response before (red) and after (black) applying corrections, as a function of jet $\theta$ in different bins of jet energy.  A decent closure is seen for jets between $0.15\pi$ and $0.85\pi$.}
    \label{Figure:JetCalibration-JECPBin}
\end{figure}

The same studies are also performed in bins of jet $\theta$ for different energy intervals in Fig.~\ref{Figure:JetCalibration-JECPBin}.
A good closure (1\%) is achieved for jets with energy around 10 GeV and within $0.15\pi < \theta < 0.85\pi$.  As shown in those performance plots, the closure is better for higher energy jets.
For lower energy jets, the response distribution becomes progressively non-Gaussian. Moreover,  a closer look into the jet matching is also needed if we would like to extend the analysis to very low jet energy.
When jet direction is close to the beam line, the raw response drops fast and the response also becomes non-Gaussian.  A significant effort will be needed to correct and have all uncertainties under control.  From Fig.~\ref{Figure:JetCalibration-JECPBin} we can observe that while the corrected jets look reasonable for jets close to $0.15\pi / 0.85\pi$, the raw response already changes quite rapidly.  It indicates that the jet resolution and jet calibration-related uncertainties will also vary rapidly close to the values.  Therefore in this work we place the boundary at $0.20\pi$--$0.80\pi$.

\subsection{Data-based calibration: relative scale}
\label{Subsection:RelativeScale}

Here we present the ``relative scale'' calibration, similar to the ``L2Residual'' or ``$\eta$-intercalibration'' method~\cite{CMS:2011shu,ATLAS:2011lgt} which was used in many of the jet analysis at the LHC experiment. In this analysis, we used this method as an important cross-check of the main analysis. The relative residual calibration aims to equalize jet response difference as a function of jet direction in data.  Since the simulation-based calibration is done in bins of jet $\theta$, the simulated corrected jet response as a function of jet $\theta$ is assumed to be flat in this step.

The calibration proceeds by calibrating jet response for different jet $\theta$ to the jets in the reference region $0.45\pi < \theta < 0.55\pi$.  We look at leading dijet energy balance where one of the leg is in the reference region, and another is in the target region defined by some interval in $\theta$.  And the mean of the balance, defined as
\begin{align}
    R = \dfrac{E_{\text{target}}}{E_{\text{reference}}},
\end{align}
is compared between data and simulation to derive the calibration factors.

Since the leading dijet balance depends on the activity of the third-leading (and softer) jets, which may be different between data and simulation and will incur bias, an extrapolation using the magnitude of the third-leading jet is performed.  The size of the third jet is quantified by $\alpha$, which is defined as follows:
\begin{align}
    \alpha \equiv \dfrac{E_{3}}{E_{1} + E_{2}}
\end{align}
where the index number indicate the jets (1 = leading, 2 = subleading, 3 = third-leading).

Events where the reference jet energy is less than 20 GeV are not considered since a larger non-closure is observed based on studies with Monte Carlo simulation, as well as ones where $\alpha > 0.15$.

\begin{figure}[htp!]
    \centering
    \includegraphicsthree{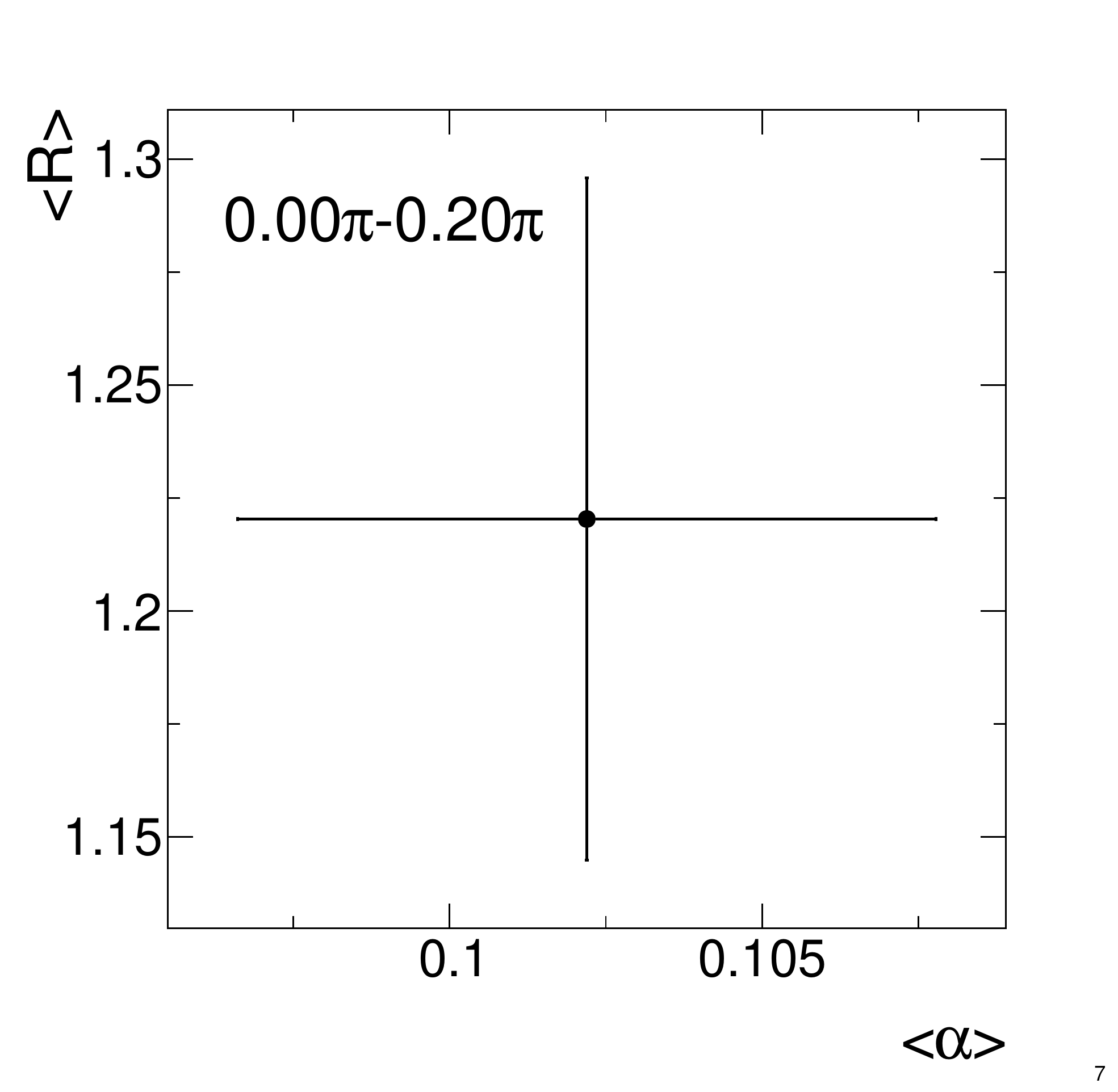}
    \includegraphicsthree{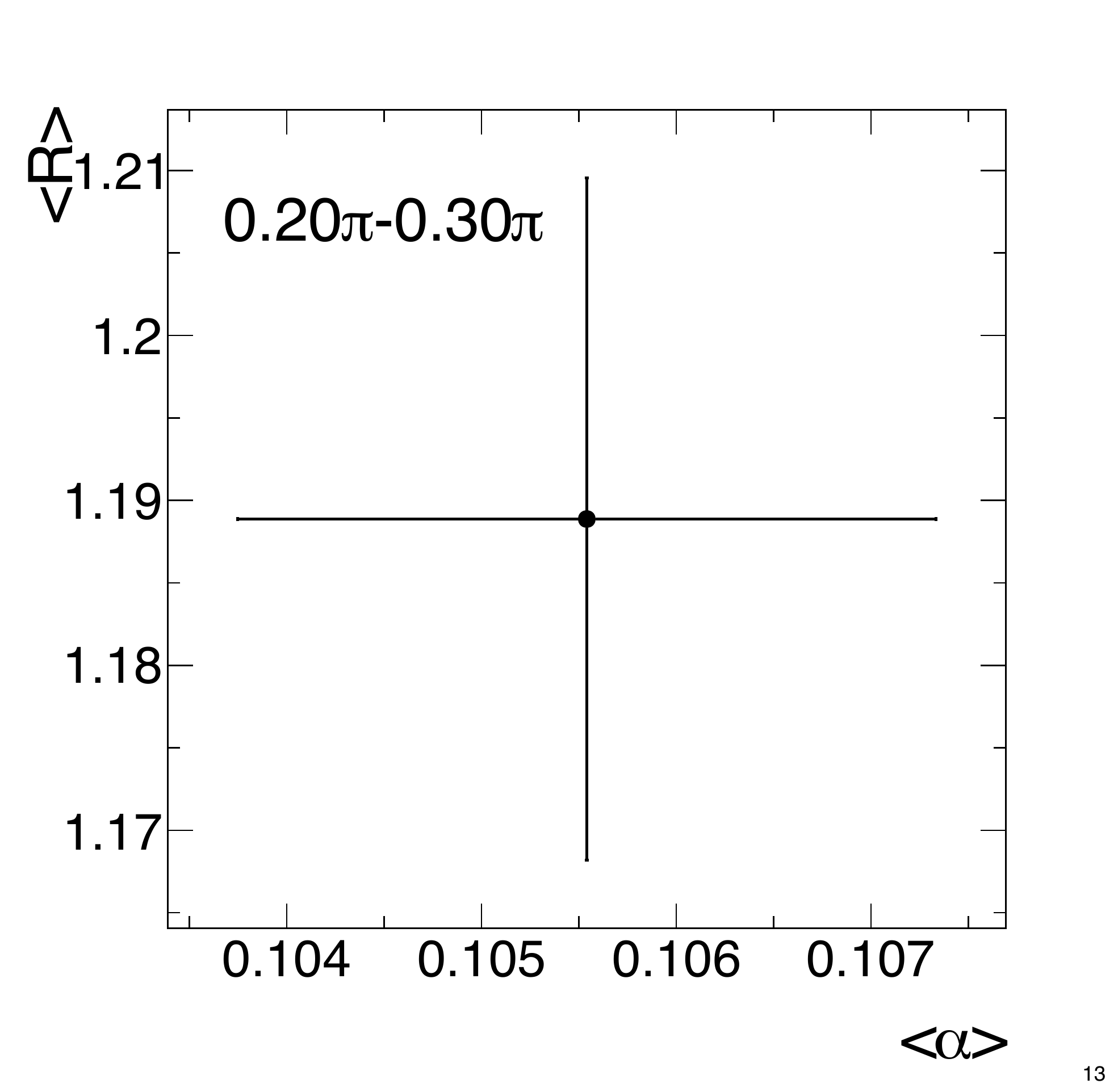}
    \includegraphicsthree{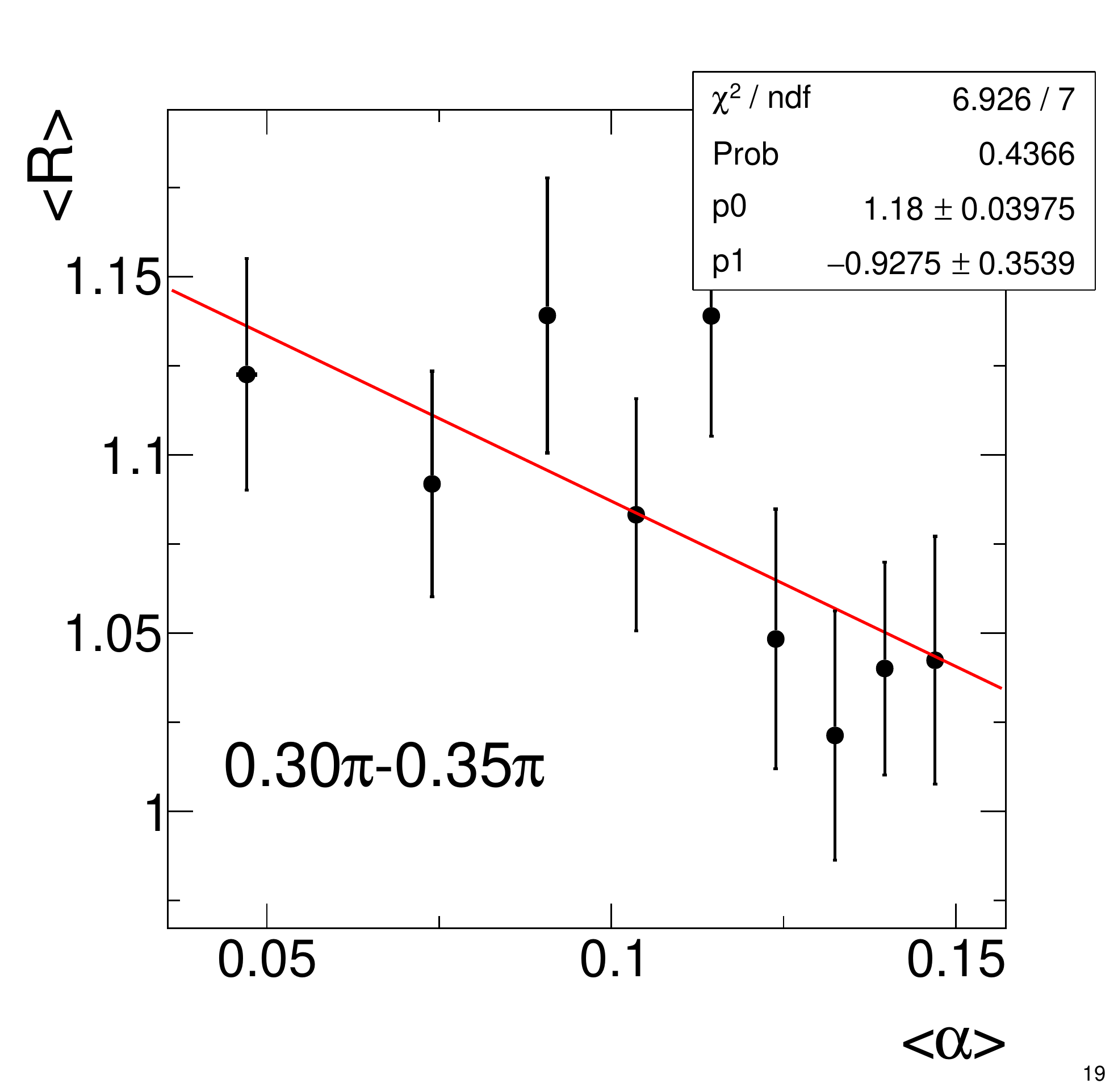}
    \includegraphicsthree{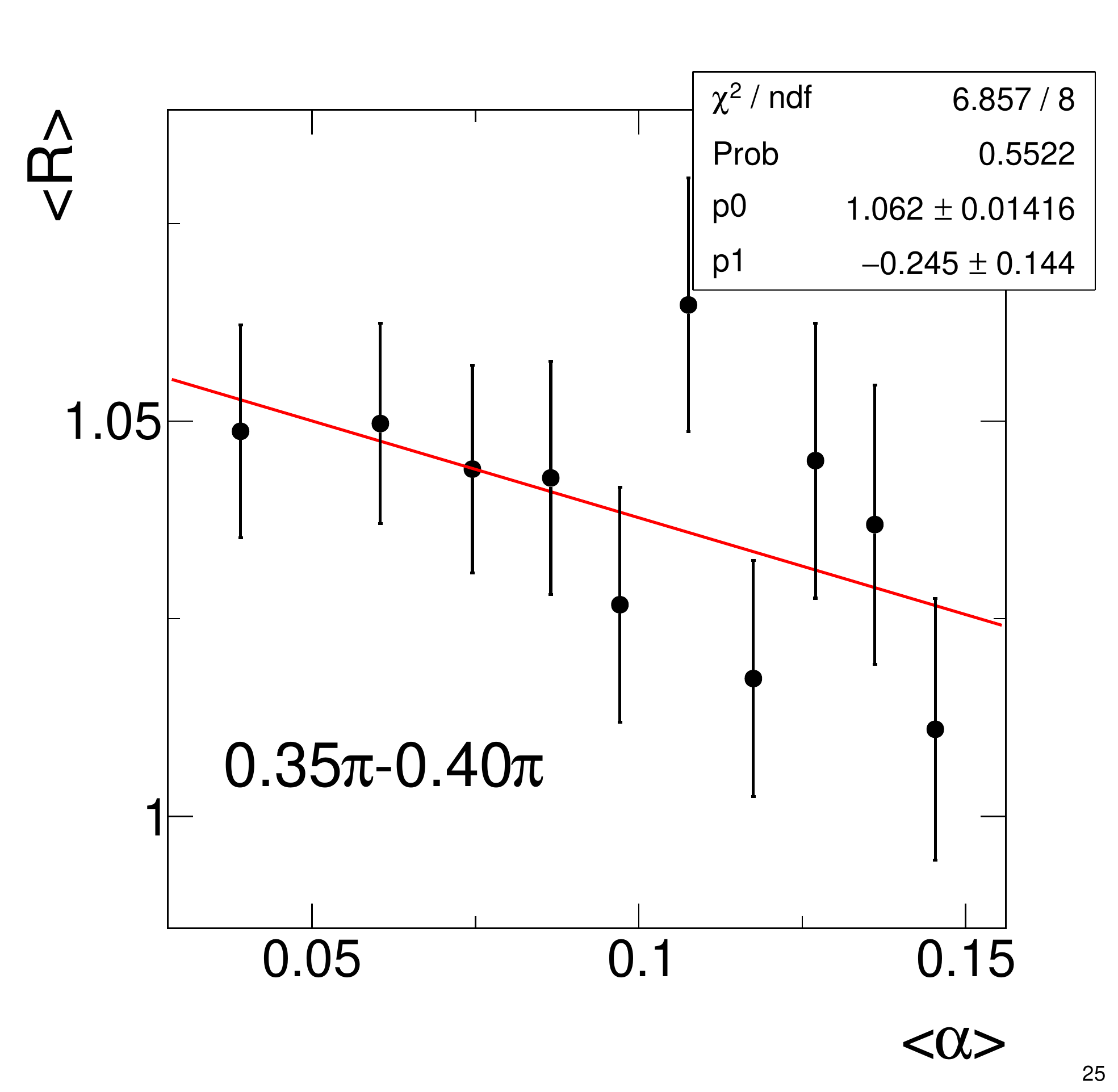}
    \includegraphicsthree{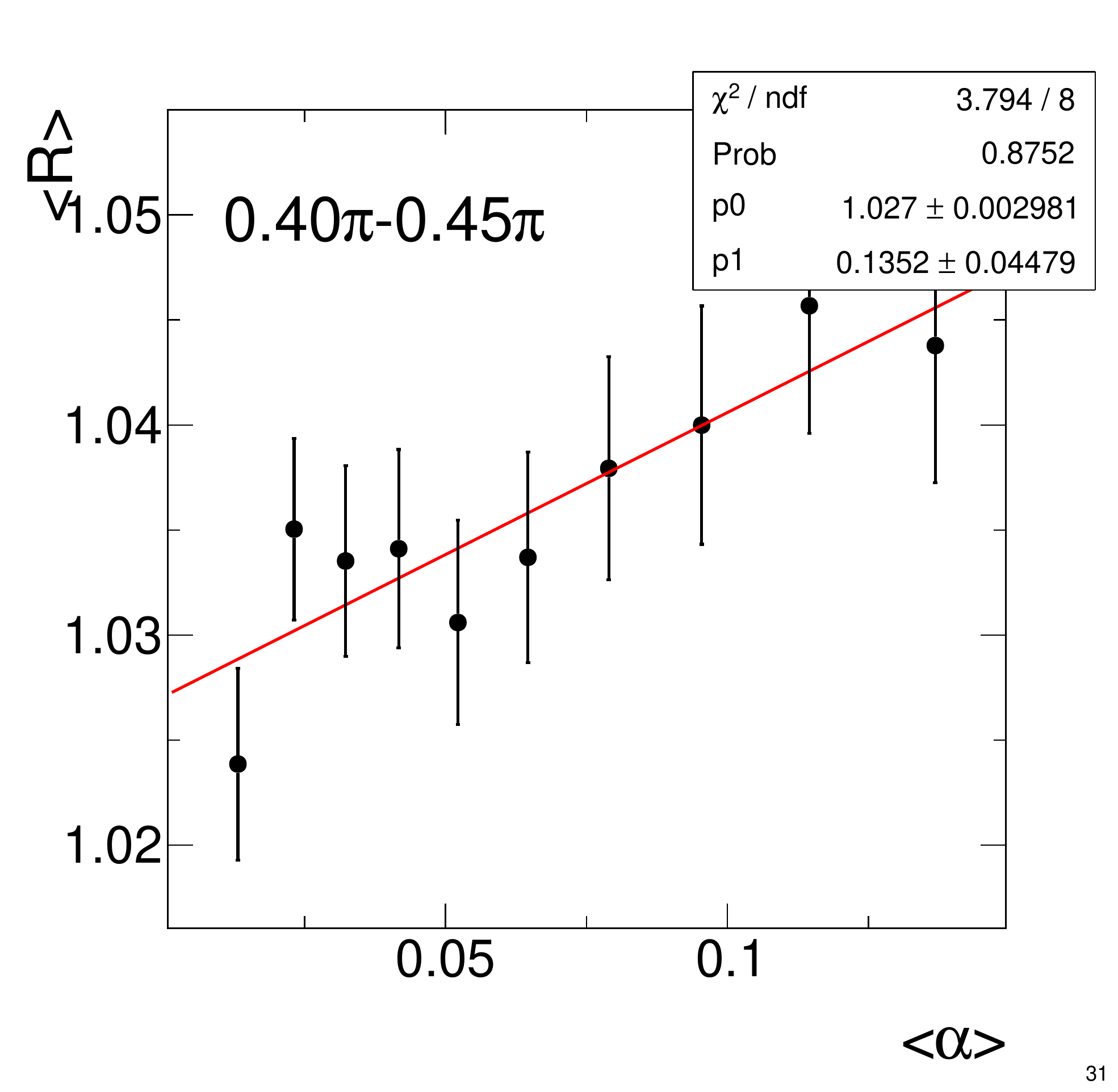}
    \includegraphicsthree{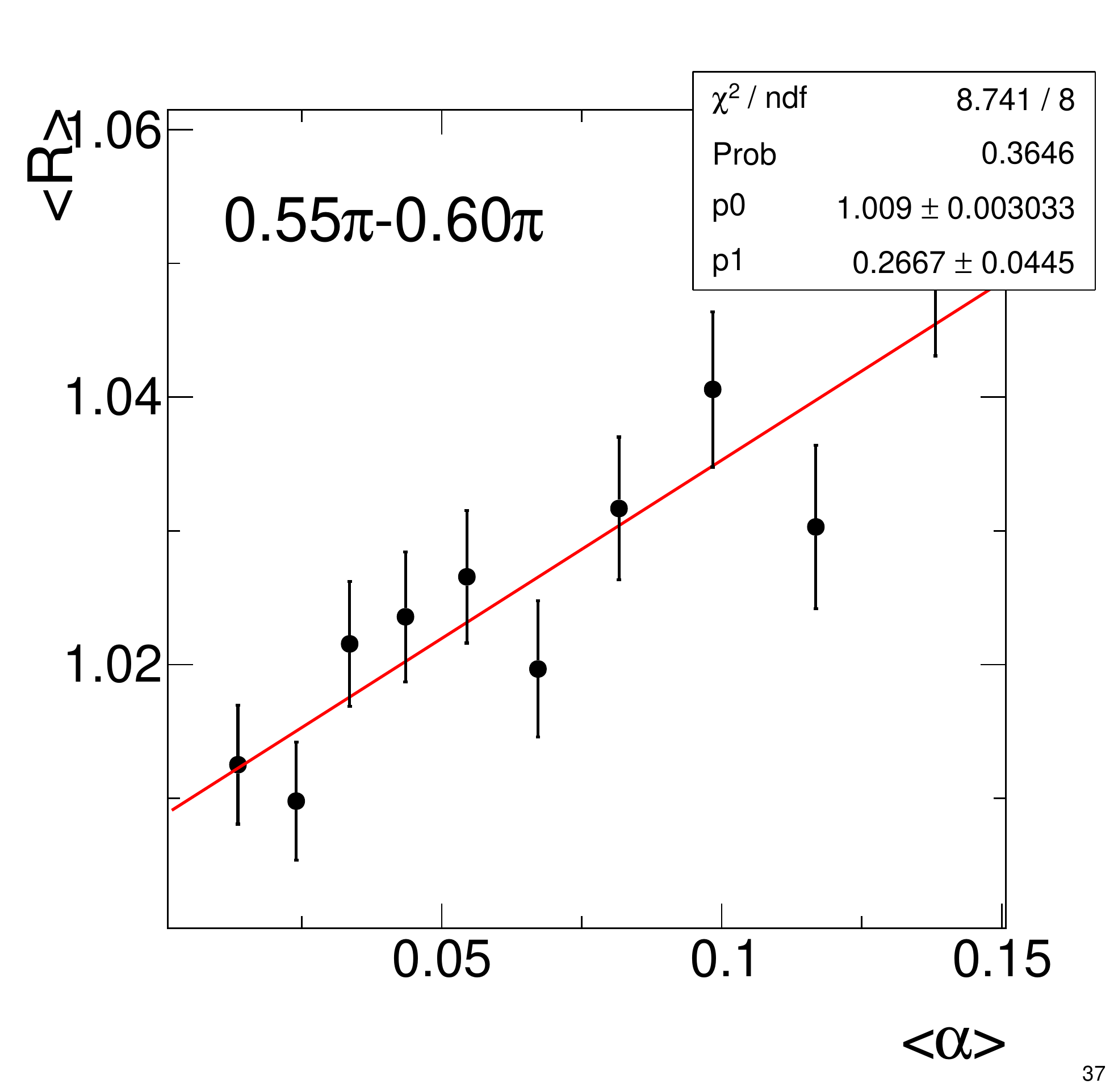}
    \includegraphicsthree{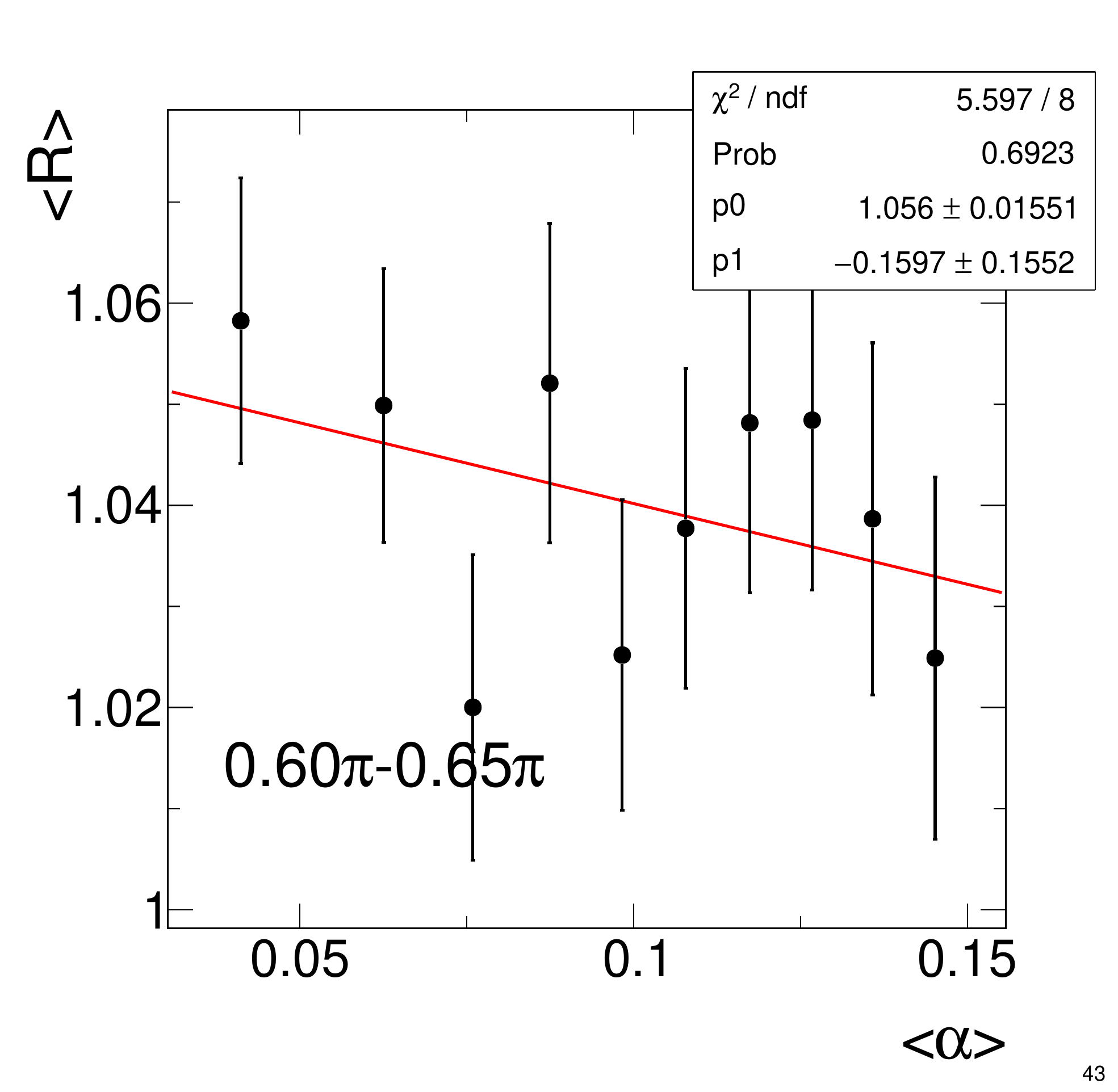}
    \includegraphicsthree{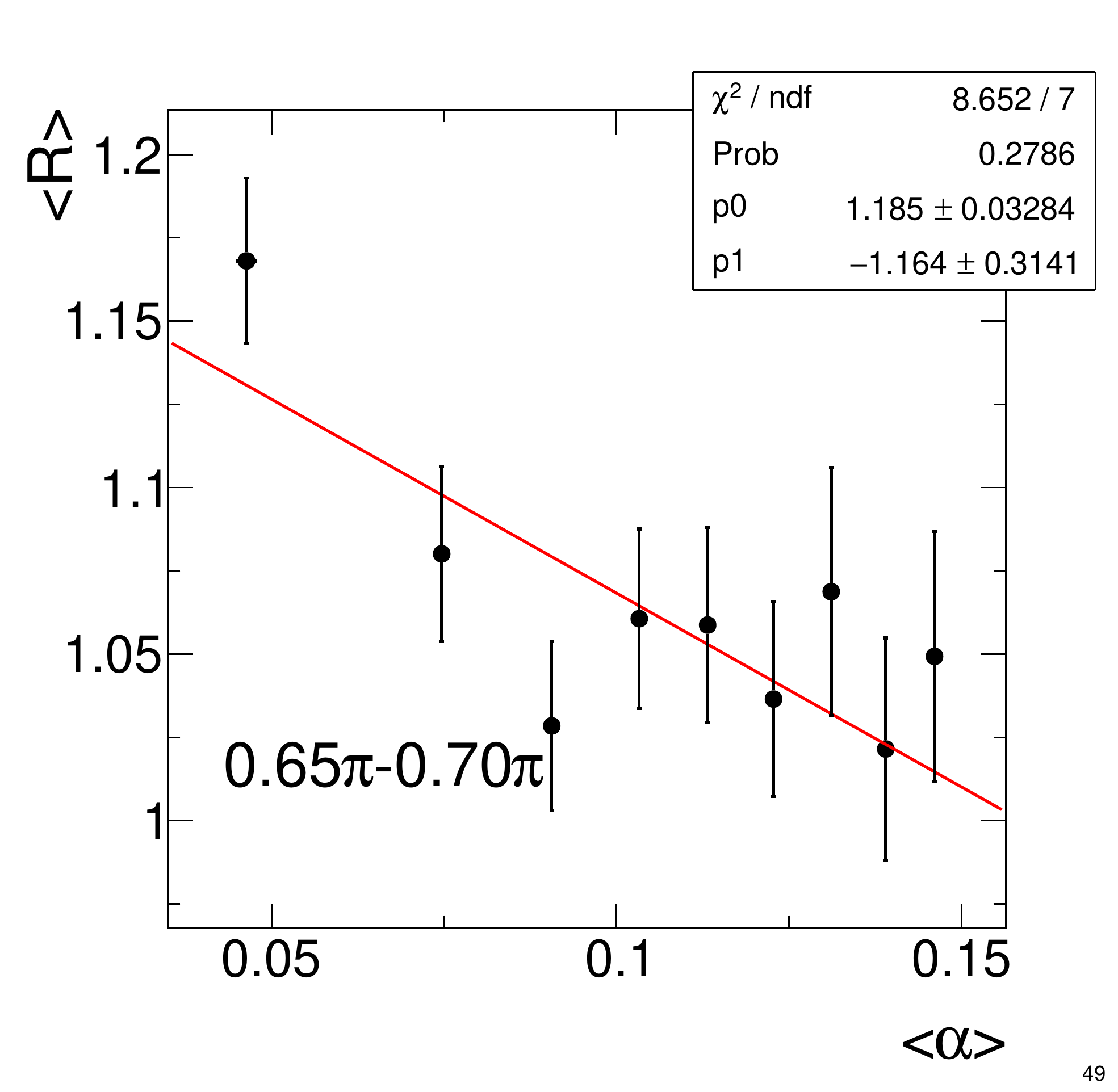}
    \includegraphicsthree{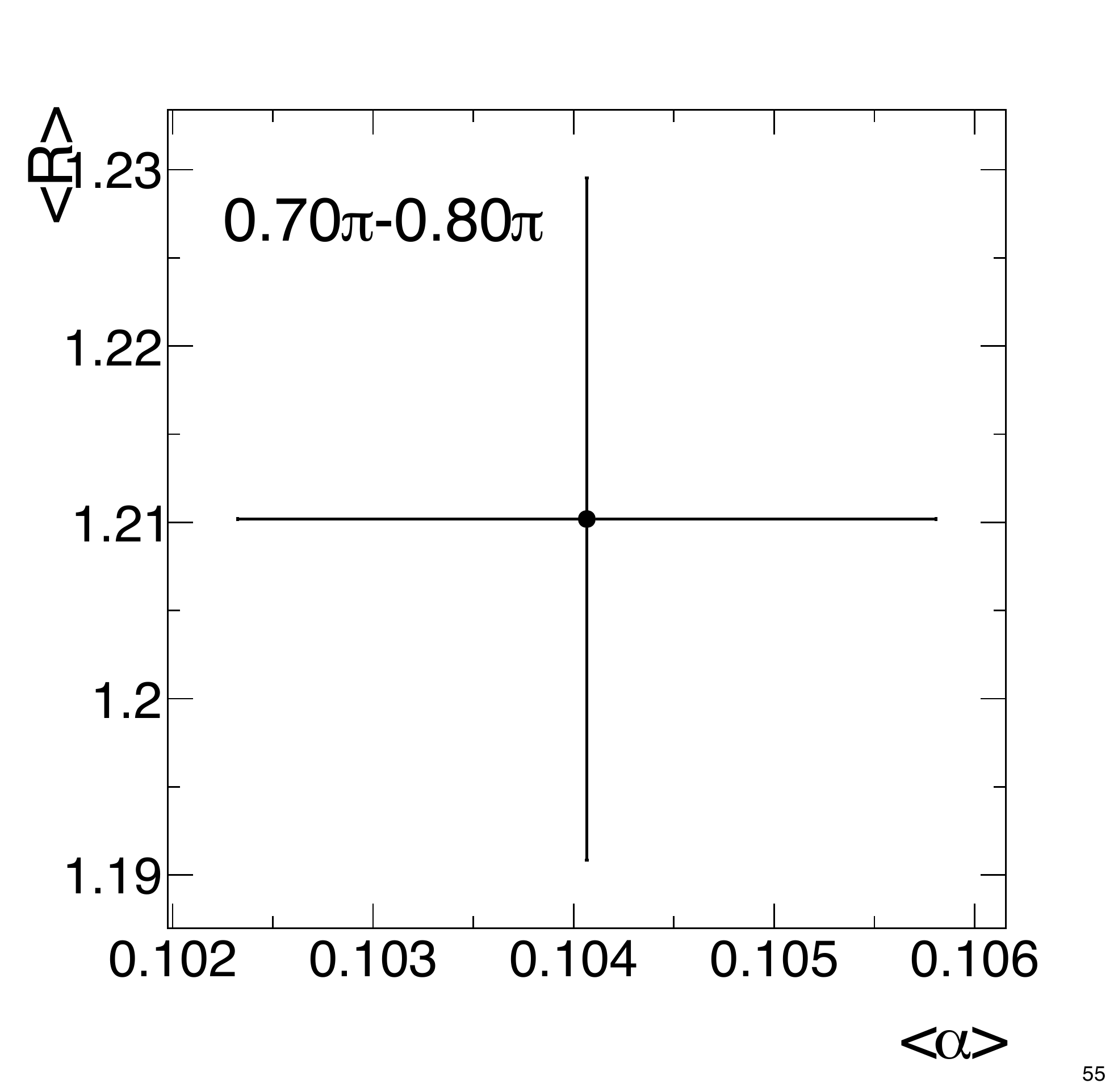}
    \includegraphicsthree{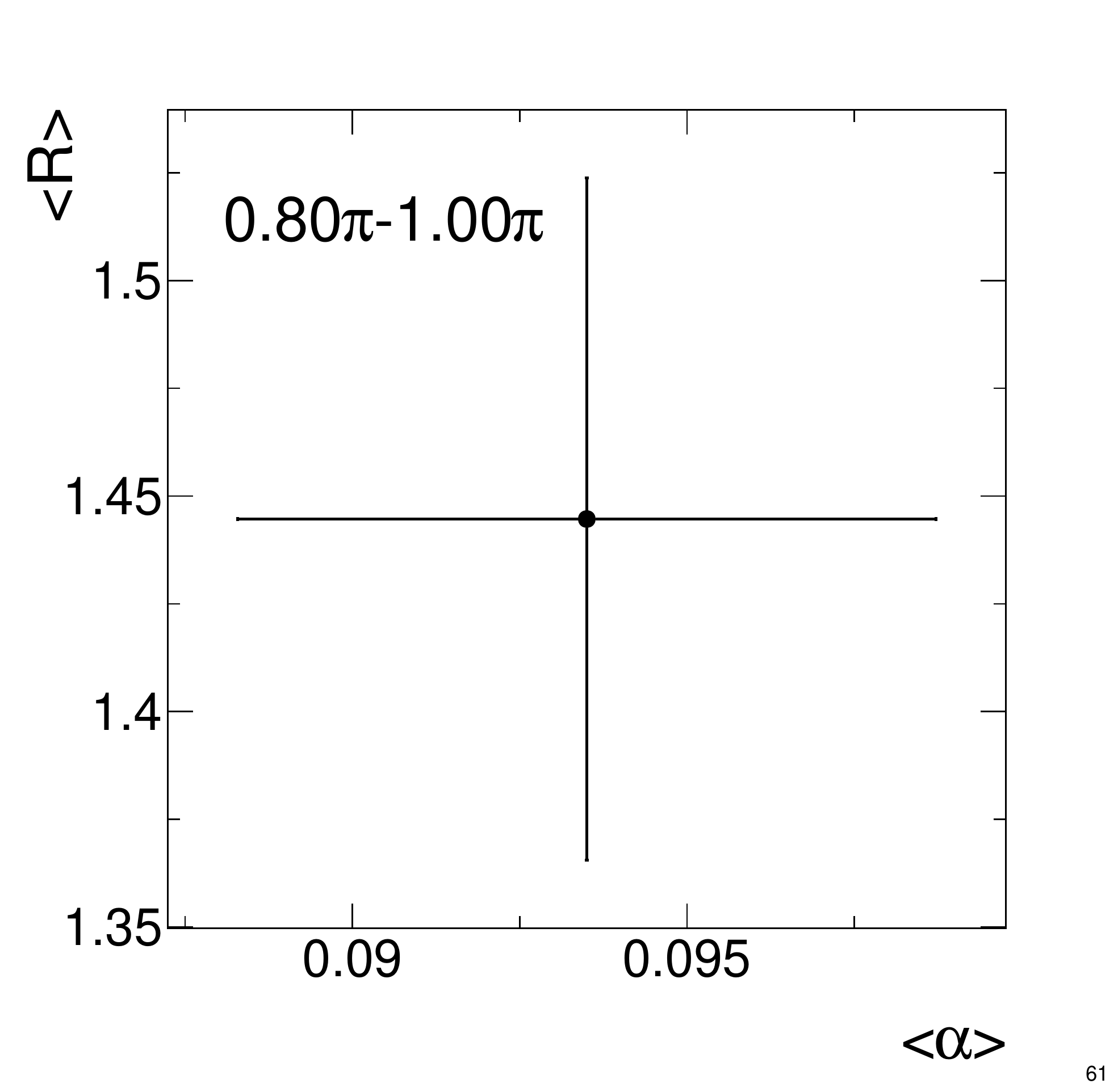}
    \caption{Jet response in data as a function of third jet activity $\alpha$.  The extrapolated value (to $\alpha = 0$) is taken as the representative response in any given bin.}
    \label{Figure:JetCalibration-RelativeResidualData}
\end{figure}

\begin{figure}[htp!]
    \centering
    \includegraphicsthree{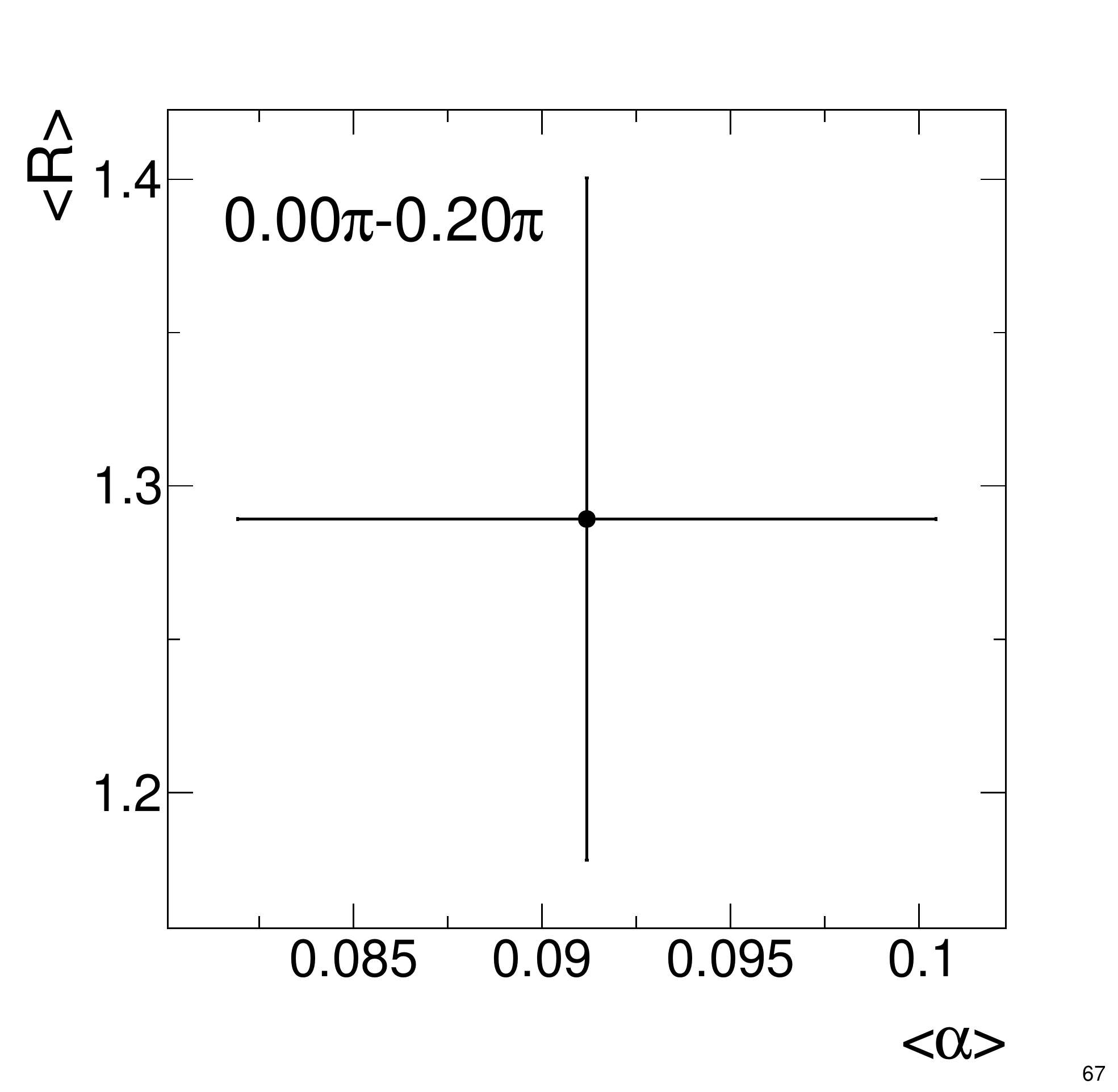}
    \includegraphicsthree{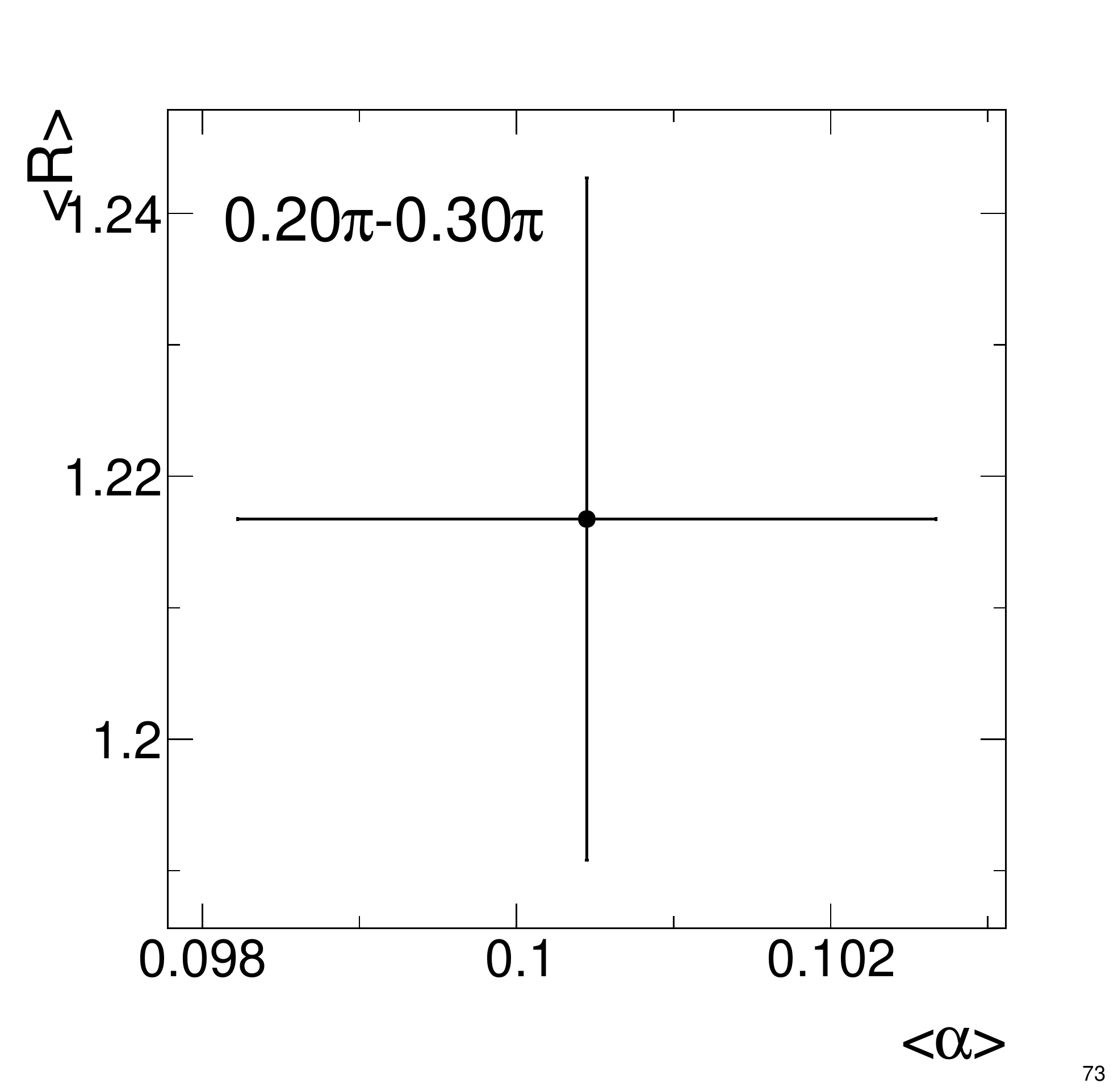}
    \includegraphicsthree{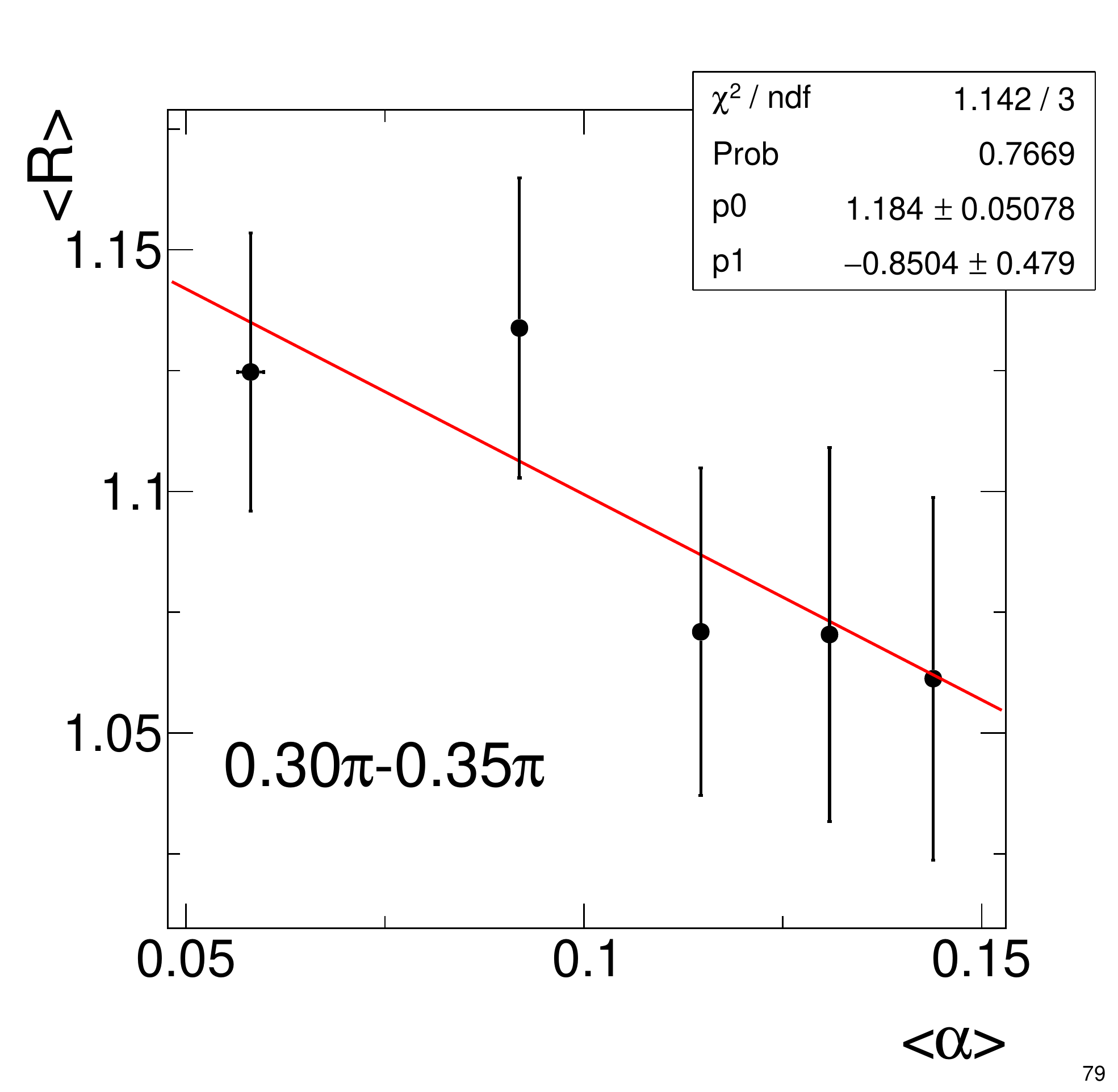}
    \includegraphicsthree{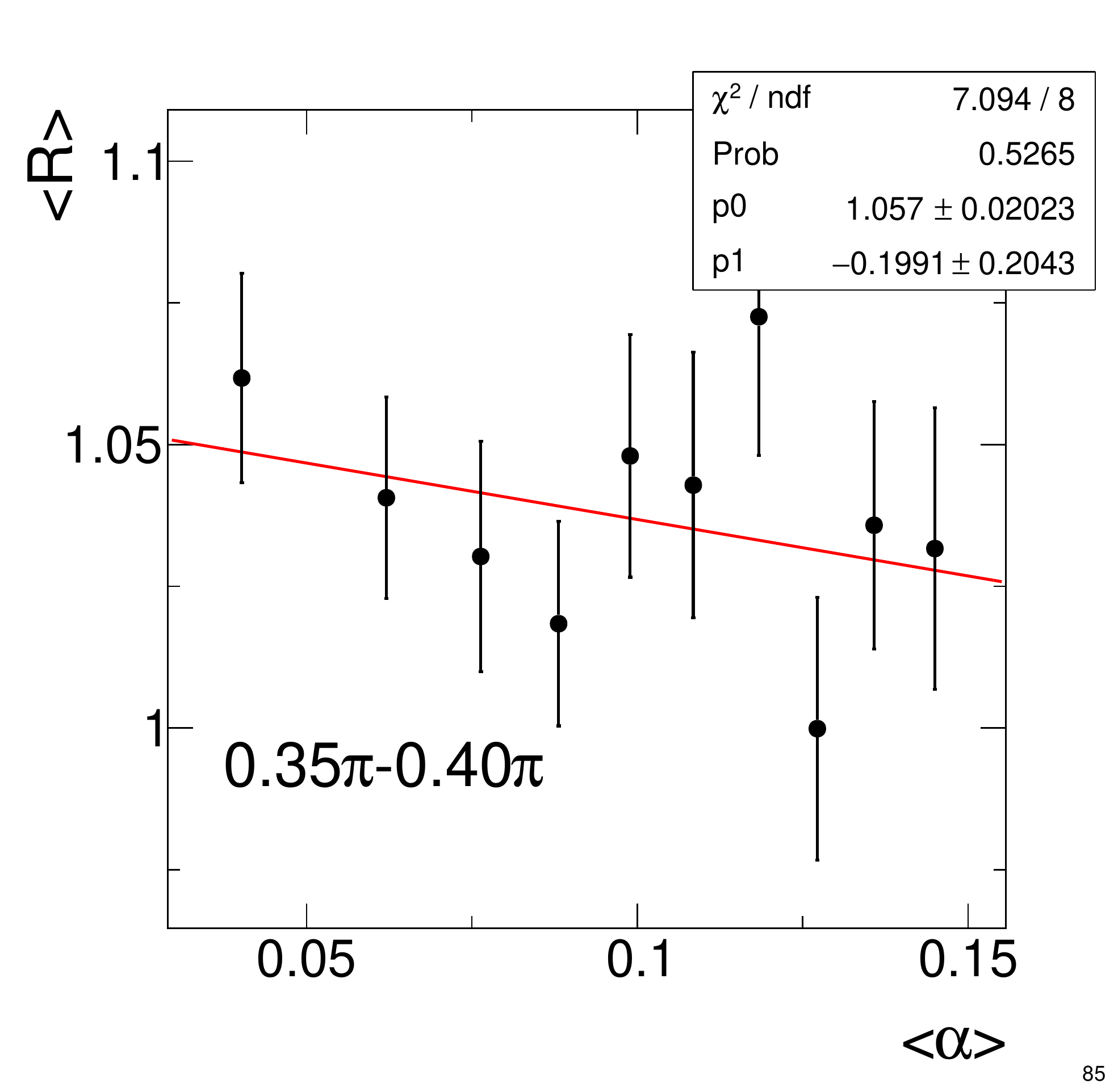}
    \includegraphicsthree{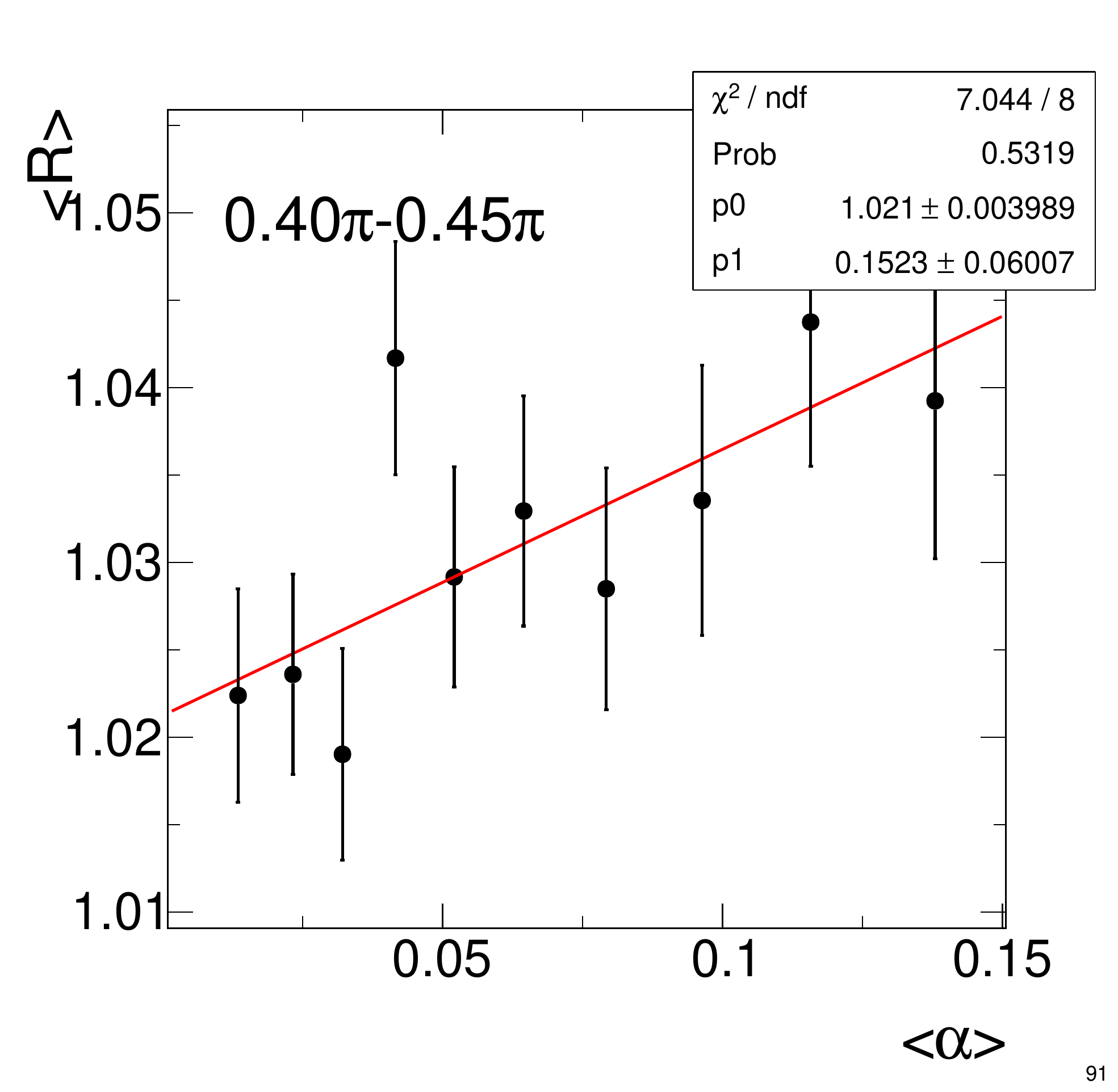}
    \includegraphicsthree{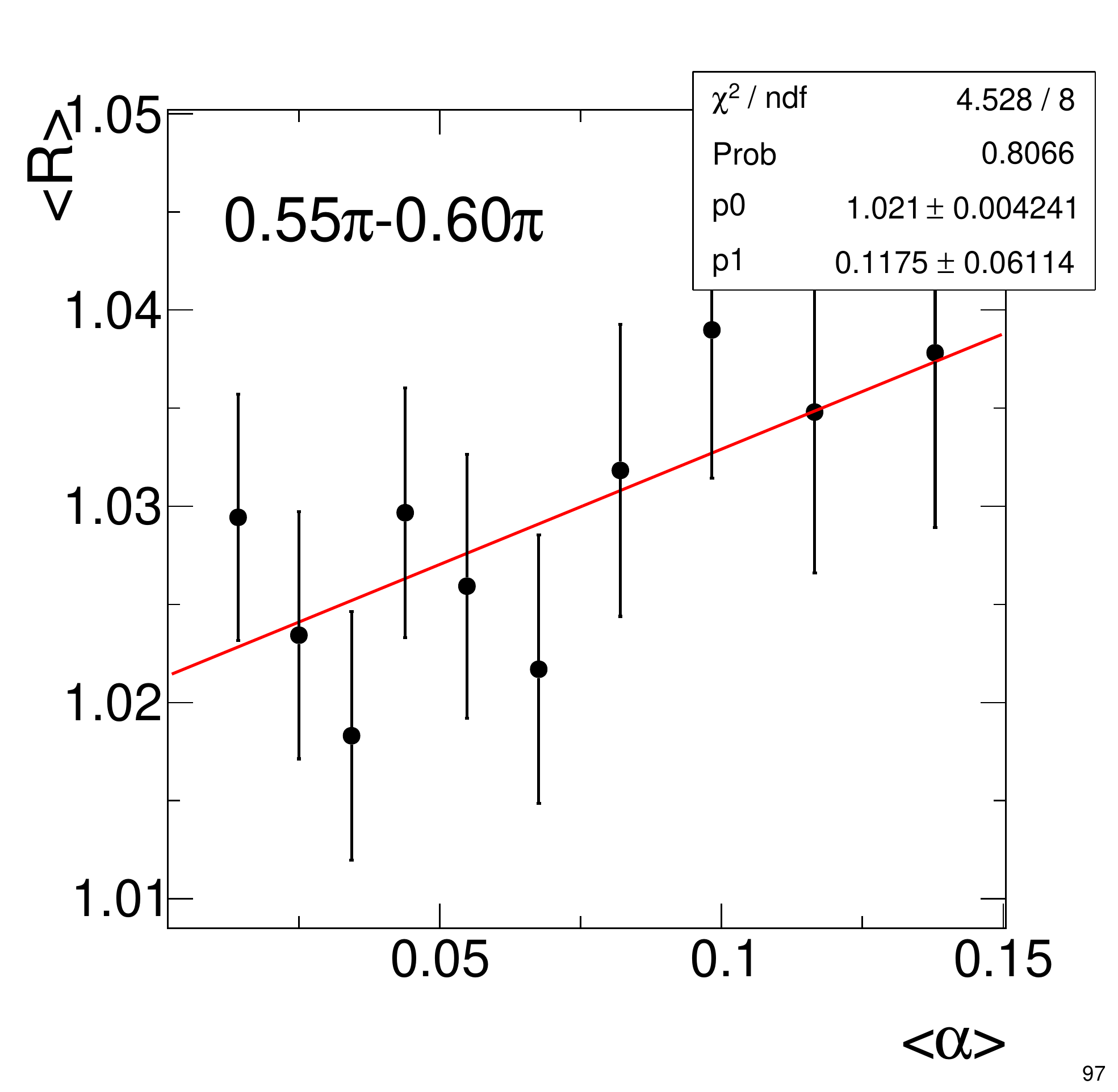}
    \includegraphicsthree{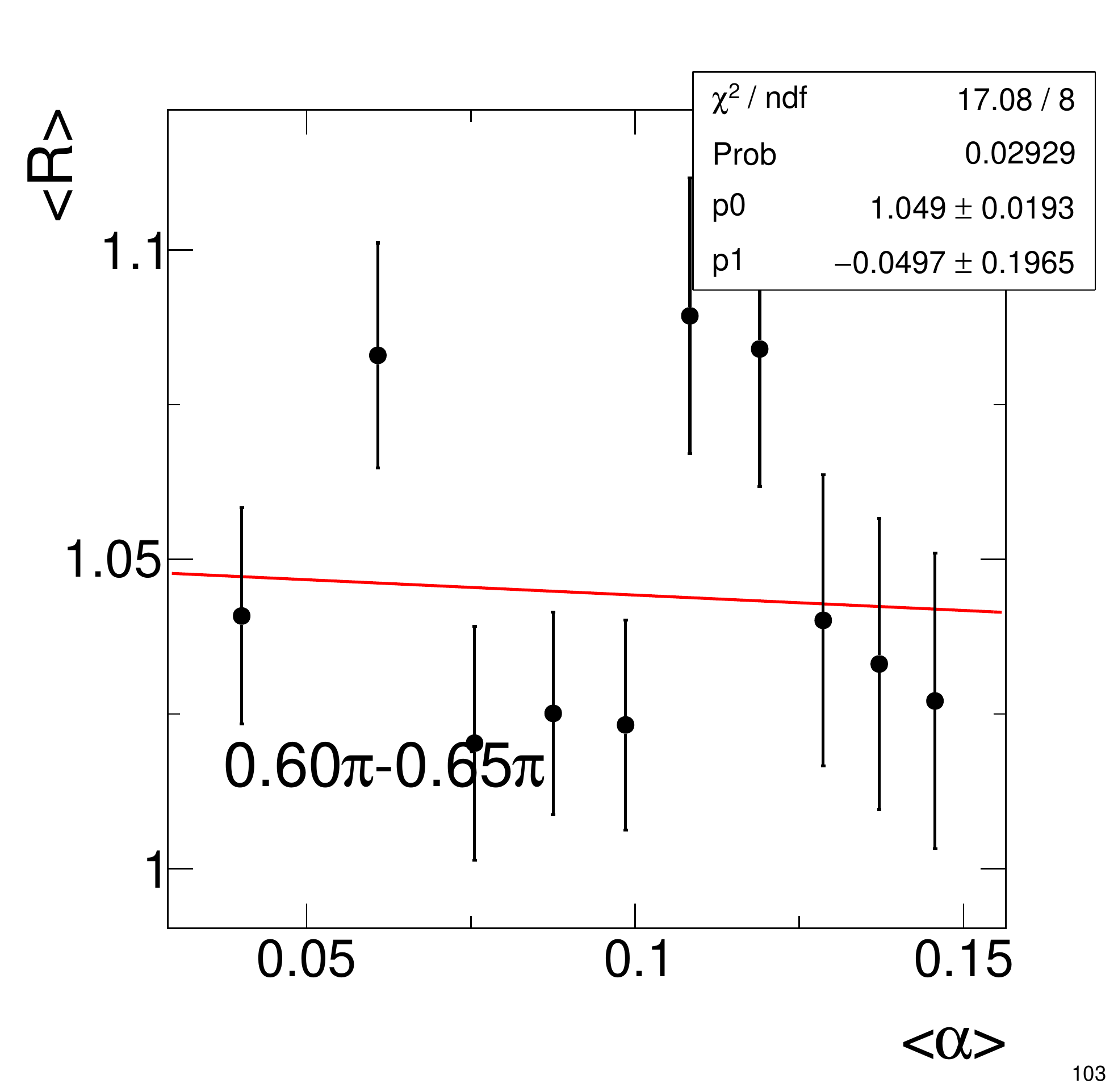}
    \includegraphicsthree{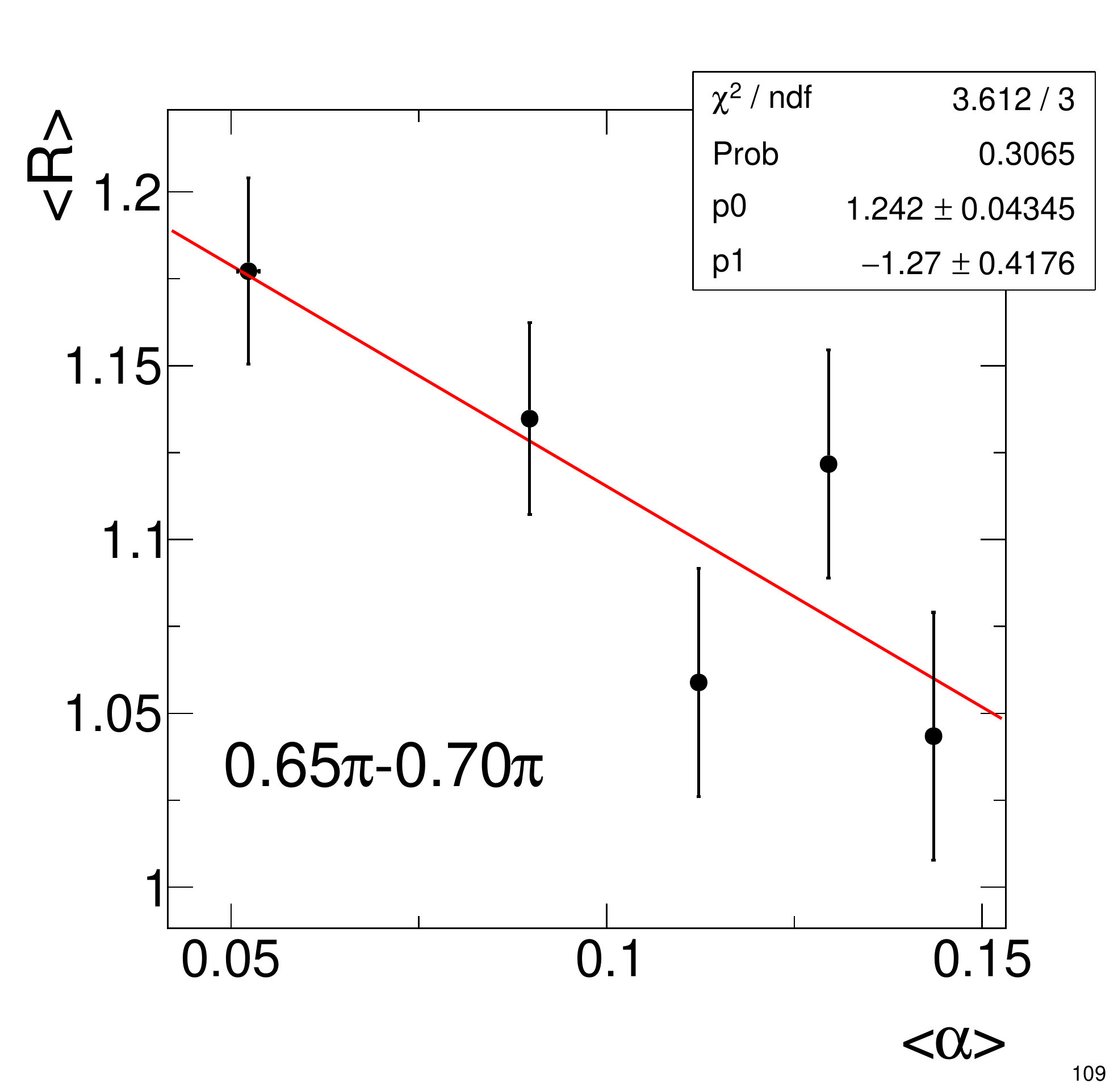}
    \includegraphicsthree{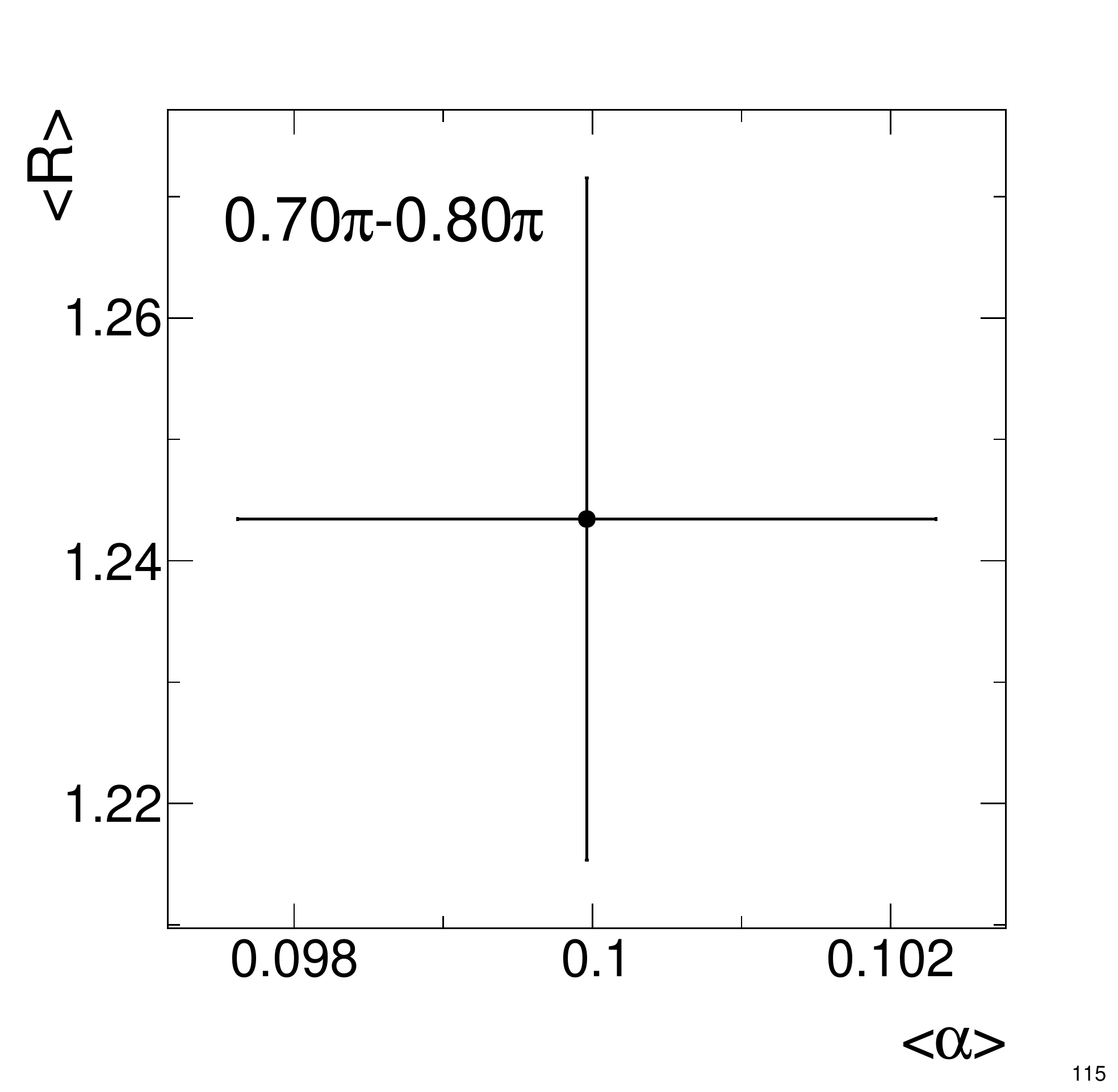}
    \includegraphicsthree{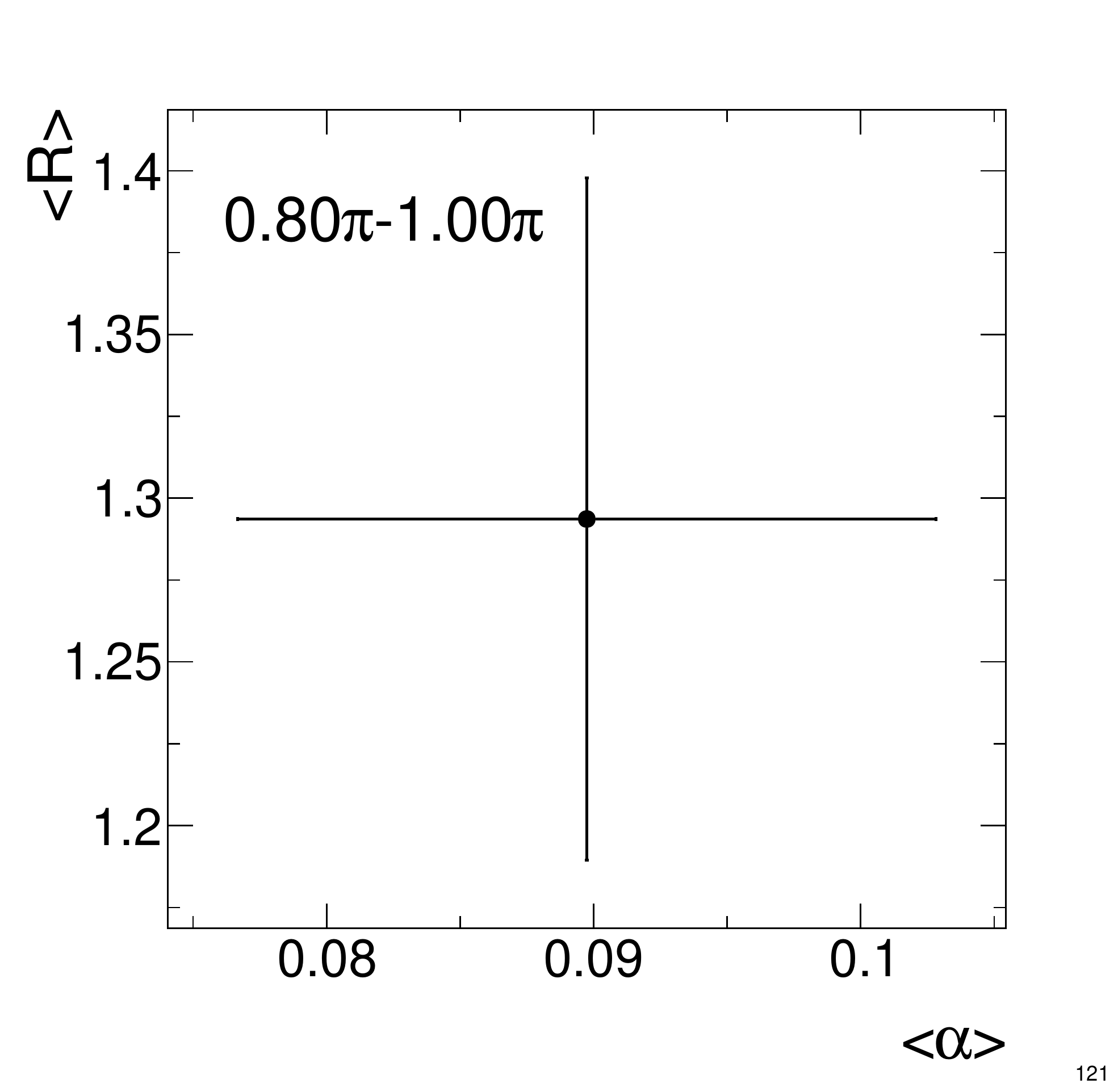}
    \caption{Jet response in simulation as a function of third jet activity $\alpha$.  The extrapolated value (to $\alpha = 0$) is taken as the representative response in any given bin.}
    \label{Figure:JetCalibration-RelativeResidualMC}
\end{figure}

The mean dijet balance for data in different bins of target jet $\theta$ as a function of the third jet activity $\alpha$ is shown in Fig.~\ref{Figure:JetCalibration-RelativeResidualData}, and in Fig.~\ref{Figure:JetCalibration-RelativeResidualMC} for simulation.  For jets close to the beam line, there are not enough statistics for a proper extrapolation, and the average value is taken without extrapolation.

\begin{figure}[htp!]
    \centering
    \includegraphicsthree{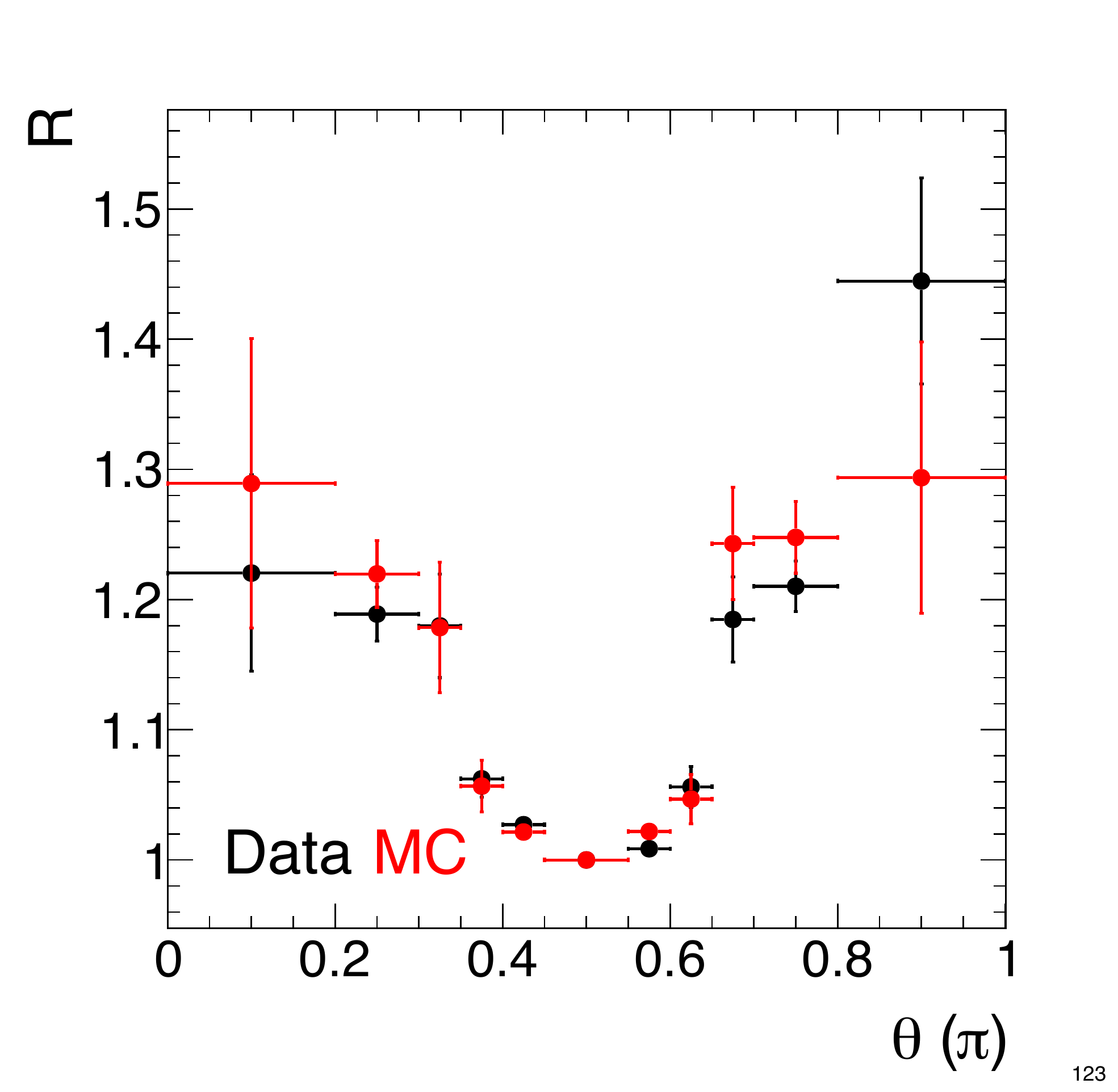}
    \includegraphicsthree{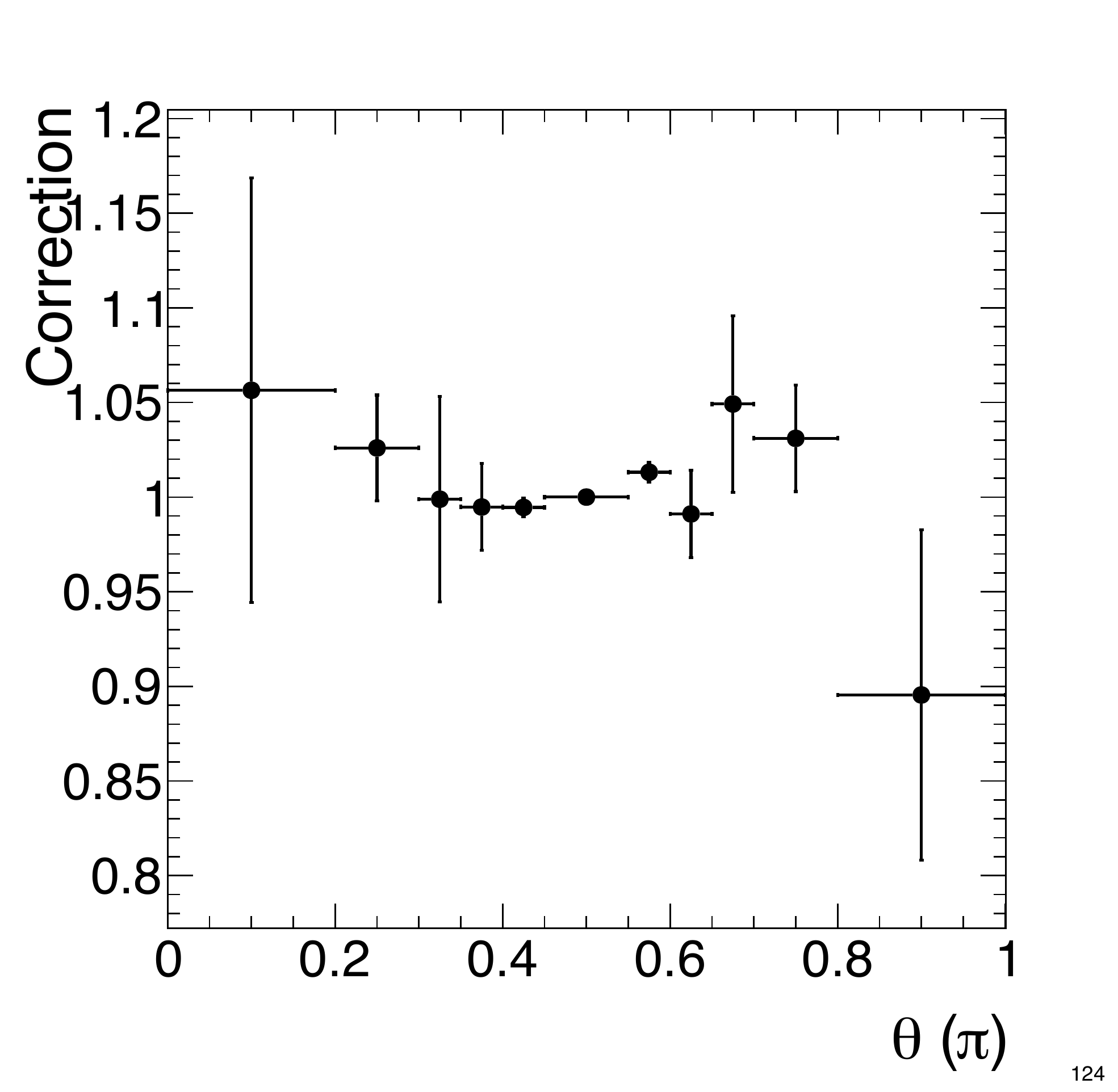}
    \includegraphicsthree{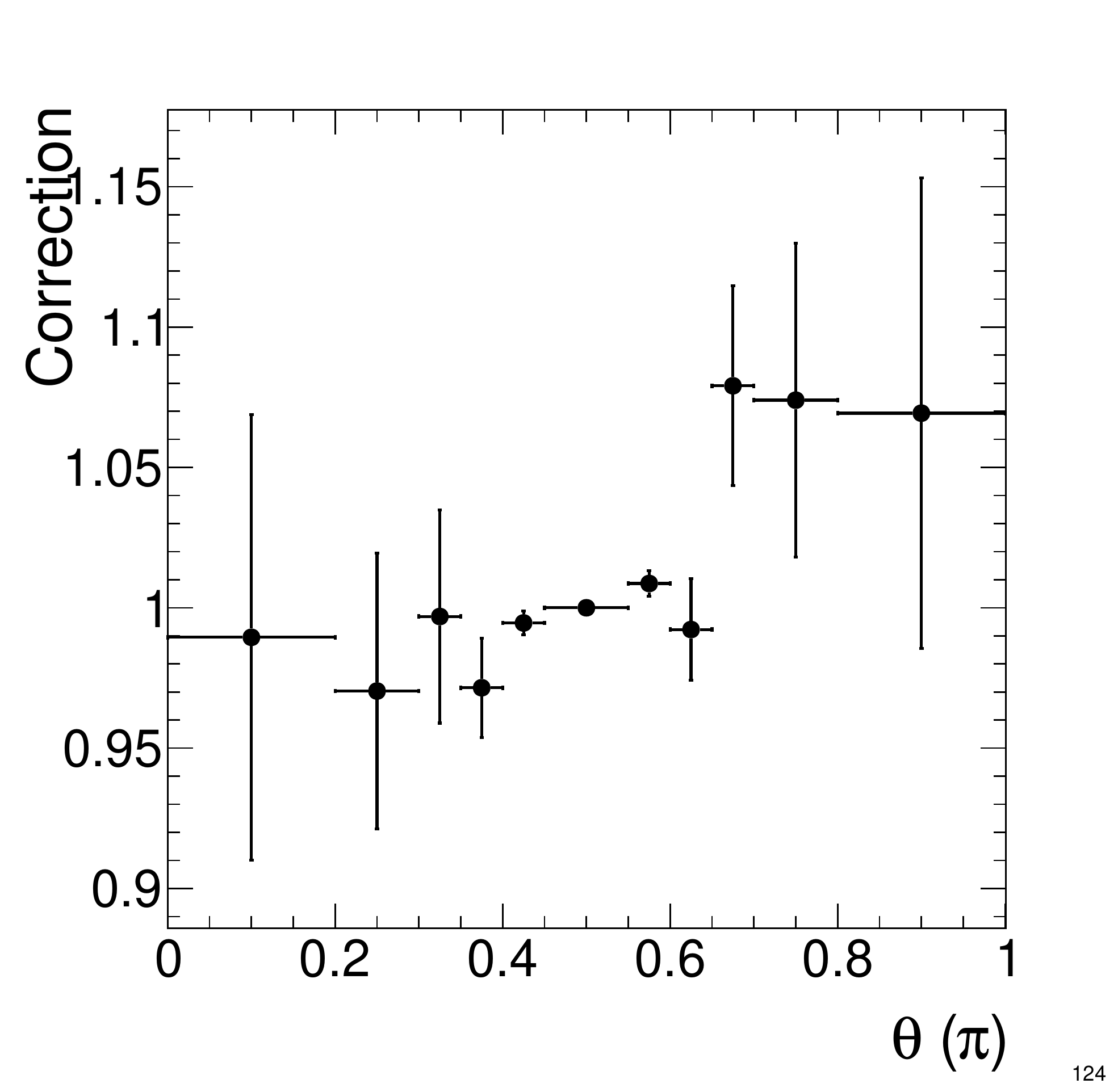}
    \caption{Left: summary of extrapolated response in data and simulation as a function of jet direction.  Middle: Ratio of extrapolated responses of simulation to data.  Right: ratio of extrapolated response using only $\alpha < 0.1$.}
    \label{Figure:JetCalibration-RelativeResidual}
\end{figure}

The extrapolated response (to $\alpha = 0$) is summarized in the left panel of Fig.~\ref{Figure:JetCalibration-RelativeResidual}, and the correction (ratio of data to simulation) is shown in the middle panel of Fig.~\ref{Figure:JetCalibration-RelativeResidual}.  A variation of the derivation where the maximum $\alpha$ is limited to 0.10 is done, and the result is shown in the right panel of Fig.~\ref{Figure:JetCalibration-RelativeResidual}.  The two versions are consistent within statistical uncertainty.

\clearpage

\subsection{Data-based calibration: relative plus/minus difference}
\label{Subsection:RelativeSide}

The previous approach ``relative scale'' works, but it results in large uncertainties.  Therefore an alternative strategy, ``relative plus/minus difference'' method, is adopted in the end as the main analysis. In this first step, the scale difference between two sides of the detector is derived.  Hereafter the side 0--$0.5\pi$ is referred to as the the ``$-$'' side, and $0.5\pi$--$1.0\pi$ is referred to as the ``$+$'' side.  The relative difference between different $\theta$ bins is combined into the next step during the fits for the absolute scale.

\begin{figure}[htp!]
    \centering
    \includegraphicstwo{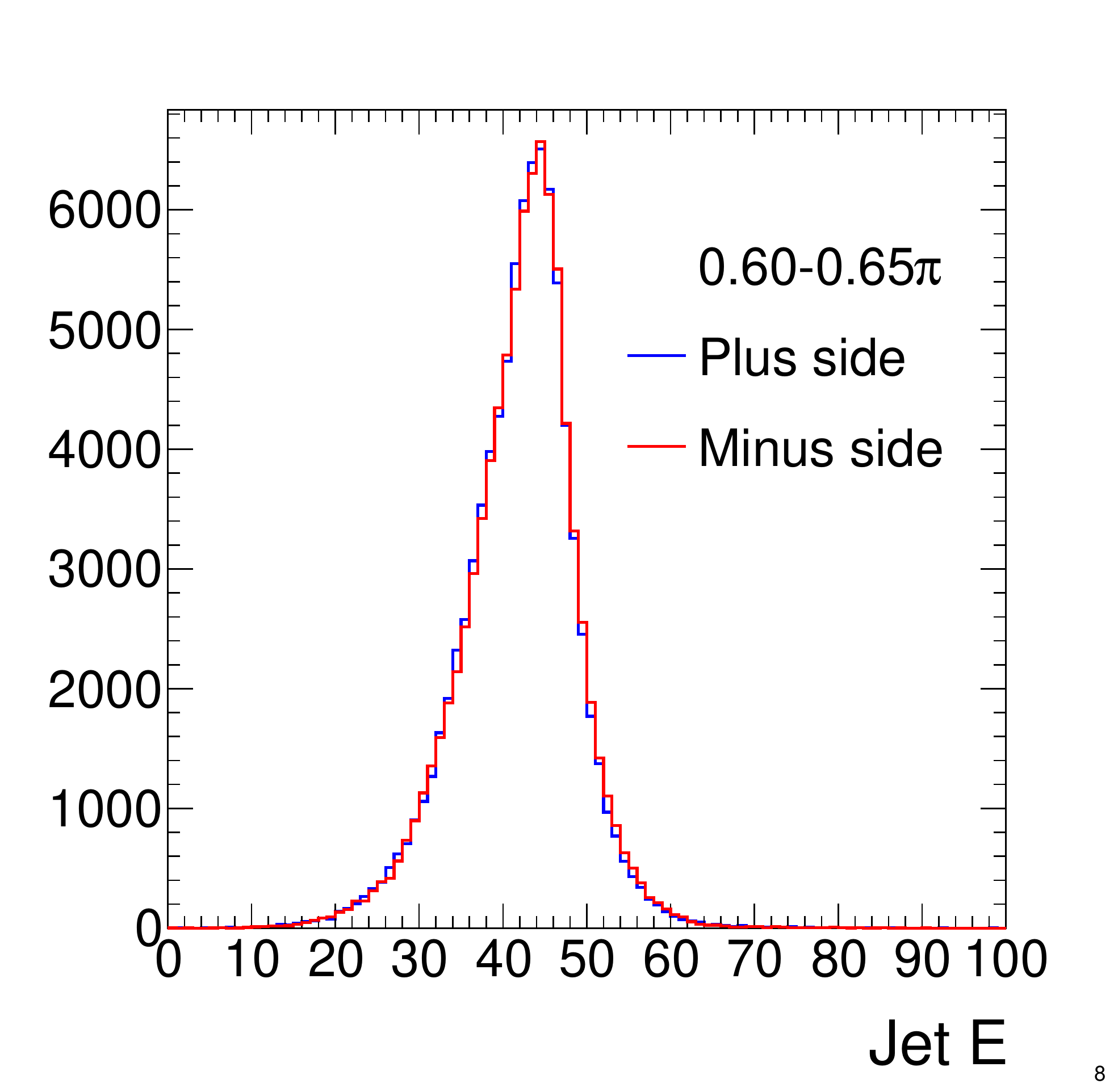}
    \includegraphicstwo{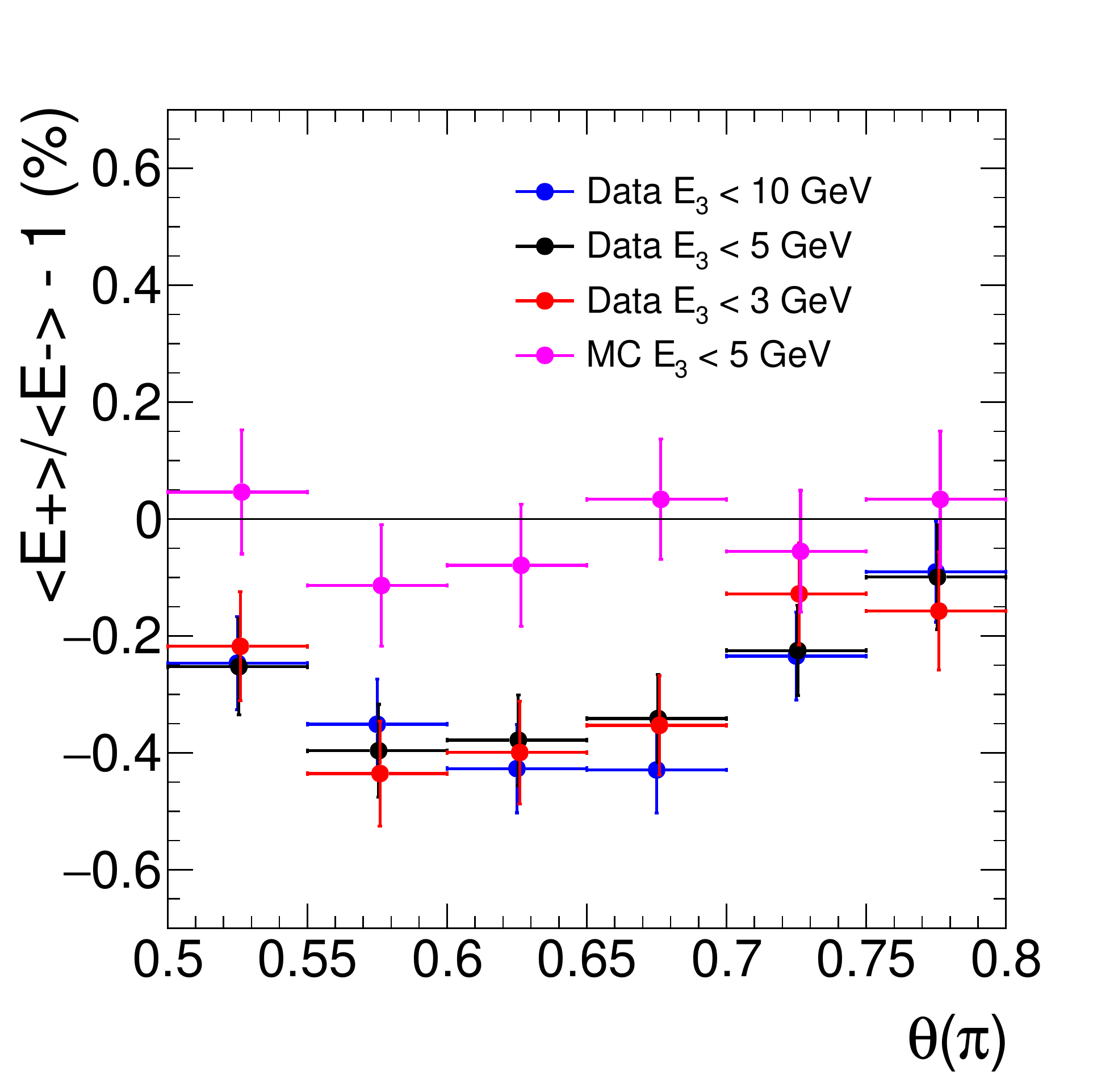}
    \caption{Left: Leading dijet energy distribution for the plus (blue) and the minus (red) sides.  An overall scale difference can be seen.  Right: Shifted ratio of the scale between the plus and the minus sides for data (blue, black, red) and simulation (pink), as a function of jet $\theta$.  The ratio is stable in data regardless of third jet activity.  It is consistent with zero for simulation.}
    \label{Figure:JetCalibration-ResidualSide}
\end{figure}

A set of histograms are filled using the leading dijet in the event, limiting the third-leading jet energy to be below $X$ GeV.  $X$ is varied between 3 and 10 GeV to assess potential dependence of the result on the soft jet activity.  The mean jet energy between the positive and the negative side is used to derive this correction.  An example is shown in the left panel of Fig.~\ref{Figure:JetCalibration-ResidualSide}.  The minus side is consistently higher than the positive side.  The result in bins of $\theta$ is summarized in the right panel.  The dependence on third jet activity is small, and the results are consistent with each other.

The same procedure is also carried out in simulated events.  We observe that the positive/negative scale difference is consistent with 1.

\clearpage

\subsection{Data-based calibration: absolute scale}
\label{Subsection:AbsoluteScale}

The final step of the jet calibration is the absolute scale calibration, and it is done with the multi-jet invariant mass.  Both the simulation-based calibration and the relative correction are applied before deriving the corrections for this step.  Since the collision energy is 91.2 GeV, event-wide jet invariant mass is highly correlated with the $Z$ rest mass.

The calibration proceeds by fitting the parameters of the jet energy correction as a function of jet energy in order to minimize the difference in mean of the invariant mass of the (up to) $N$ leading jets between data and simulation.  The size of the $N$+1-th leading jet is required to be at most $X$ GeV in order to control potential effects from soft jets.  Both $N$ and $X$ are varied to study the sensitivity to soft(er) jets.  The overall scale for each jet $\theta$ bin is also floated.

The fit is set up with all individual events as input.  A $\chi^2$ function is set up to
\begin{enumerate}
    \item Correct all jets in data event by event by a set of JEC parameters as input;
    \item Remove jets out of the acceptance;
    \item Filter using $N$ and $X$ after correction;
    \item Calculate the mean of the invariant mass between the 10\% and the 90\% quantile for both simulation and data, in 2\% quantile ranges (i.e., 10-12\%, 12-14\%, ...), in order to avoid outliers ruining the fit;
    \item Calculate the total $\chi^2$ from the square of the difference of the mean for each quantile range,
\end{enumerate}
and the function is passed to the \textsc{Minuit} minimizer in the \textsc{ROOT} library and fit for the parameters of the jet energy correction.

For each $N$ and $X$ combination, the fit is repeated for different orders of polynomial function from 0-th order up to 5-th order.  Various combinations are studied, and they are summarized in Table~\ref{Table:JetCalibration-AbsoluteResidualNXCombination}.

\begin{table}[htp!]
    \centering
    \begin{tabular}{|c|c|}
        \hline
        N & X (GeV) \\\hline
        2,3,4,5 & 3,5 \\\hline
        3,4 & 6,7,8,9,10 \\\hline
        9 & 3,5,6,8,10 \\\hline
    \end{tabular}
    \caption{Combinations of $N$ and $X$ explored in deriving the residual corrections.}
    \label{Table:JetCalibration-AbsoluteResidualNXCombination}
\end{table}

Due to the shape of the jet energy spectrum (peak at 40-45 GeV and lower in the intermediate energy range, and rising again as jet energy gets lower), and also due to the amount of available statistics, it is determined that fits with a polynomial with order larger than 1 is not stable.  There are only two ``anchor points'', one at high energy and one at low energy.  Having too many degrees of freedom also risks morphing the $Z$ peak shape in data to that of simulation, which is not ideal.

\begin{figure}[htp!]
    \centering
    \includegraphicstwo{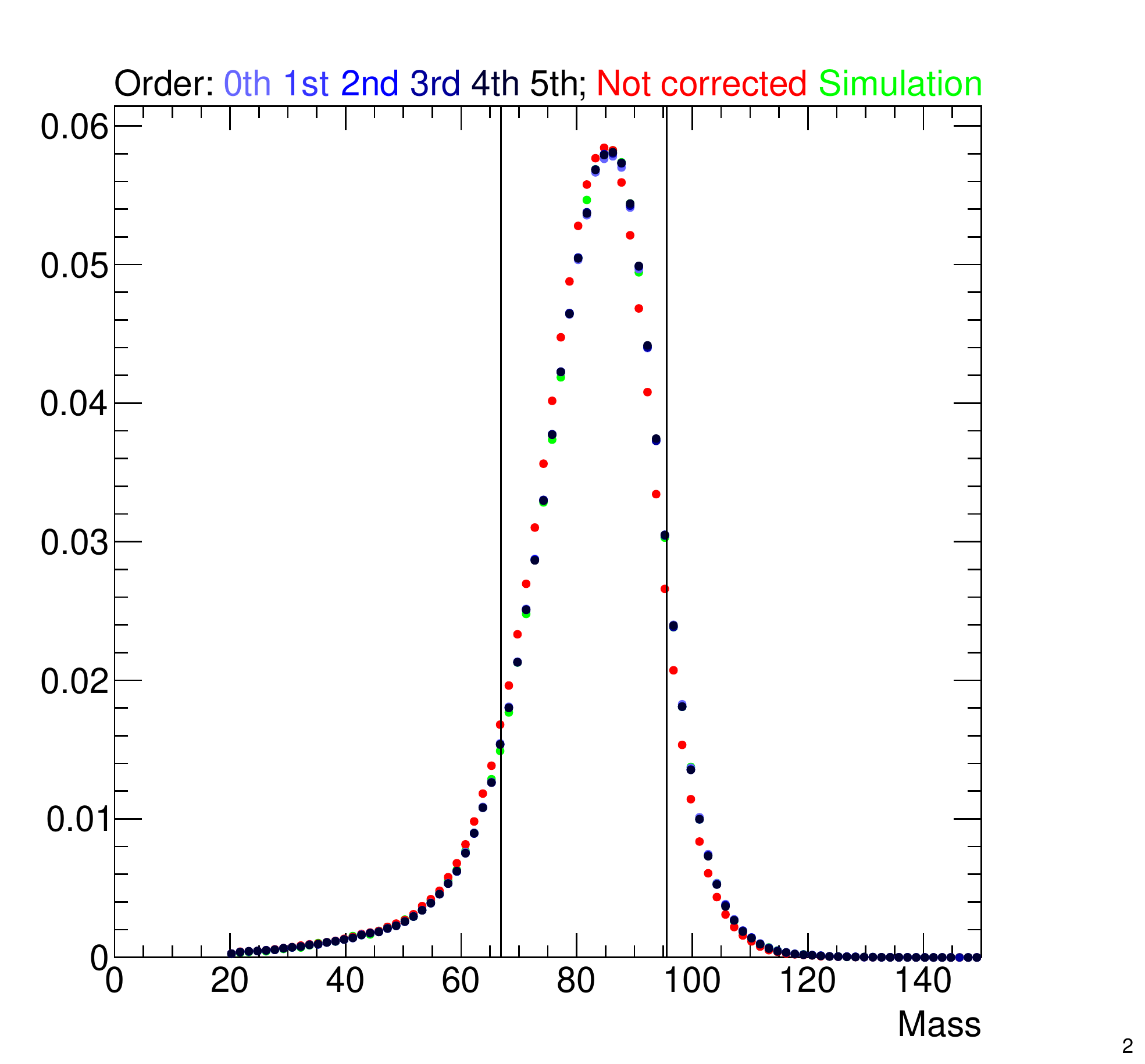}
    \includegraphicstwo{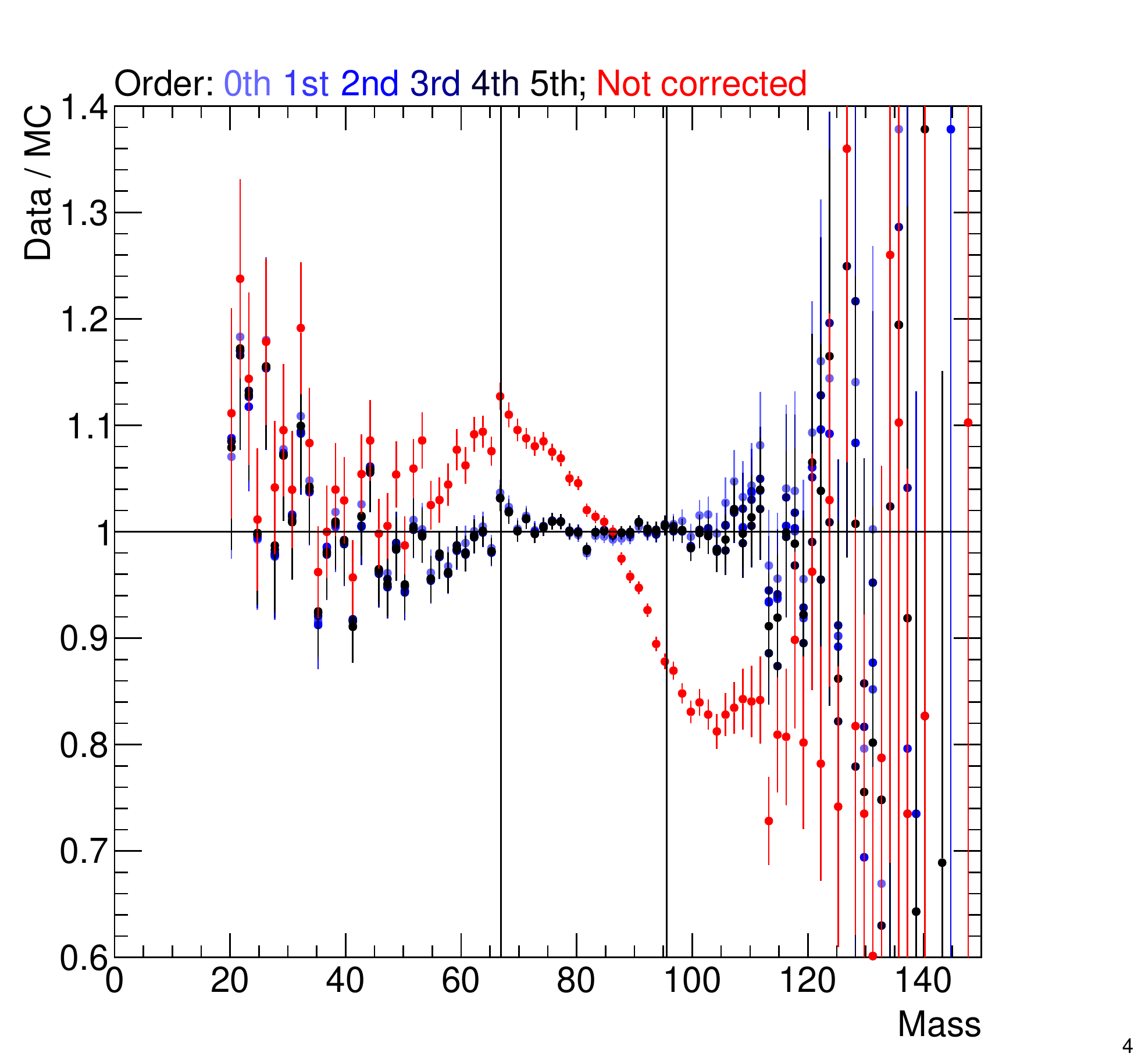}
    \caption{An example of the fit to the multijet invariance mass spectrum.  Left: uncorrected (red) and corrected (various shades of blue) spectra are compared to the simulated spectrum (green).  After correction the spectra match better.  Right: ratio to the simulated spectrum.}
    \label{Figure:JetCalibration-AbsoluteResidualExample}
\end{figure}

An example result is shown in the left panel of Fig.~\ref{Figure:JetCalibration-AbsoluteResidualExample} for $N = 9, X = 3$ GeV.  The red points shows the spectrum for data before correction, and the points in various shades of blue are the corrected spectra done with different order polynomial.  Green is the simulated spectrum.  In the right panel, the ratio to simulated spectrum is shown.  The vertical lines indicate the location of the 10\% and the 90\% percentile in the simulated spectrum.  We do not observe any significant improvement with higher order correction.

\begin{figure}[htp!]
    \centering
    \includegraphicstwo{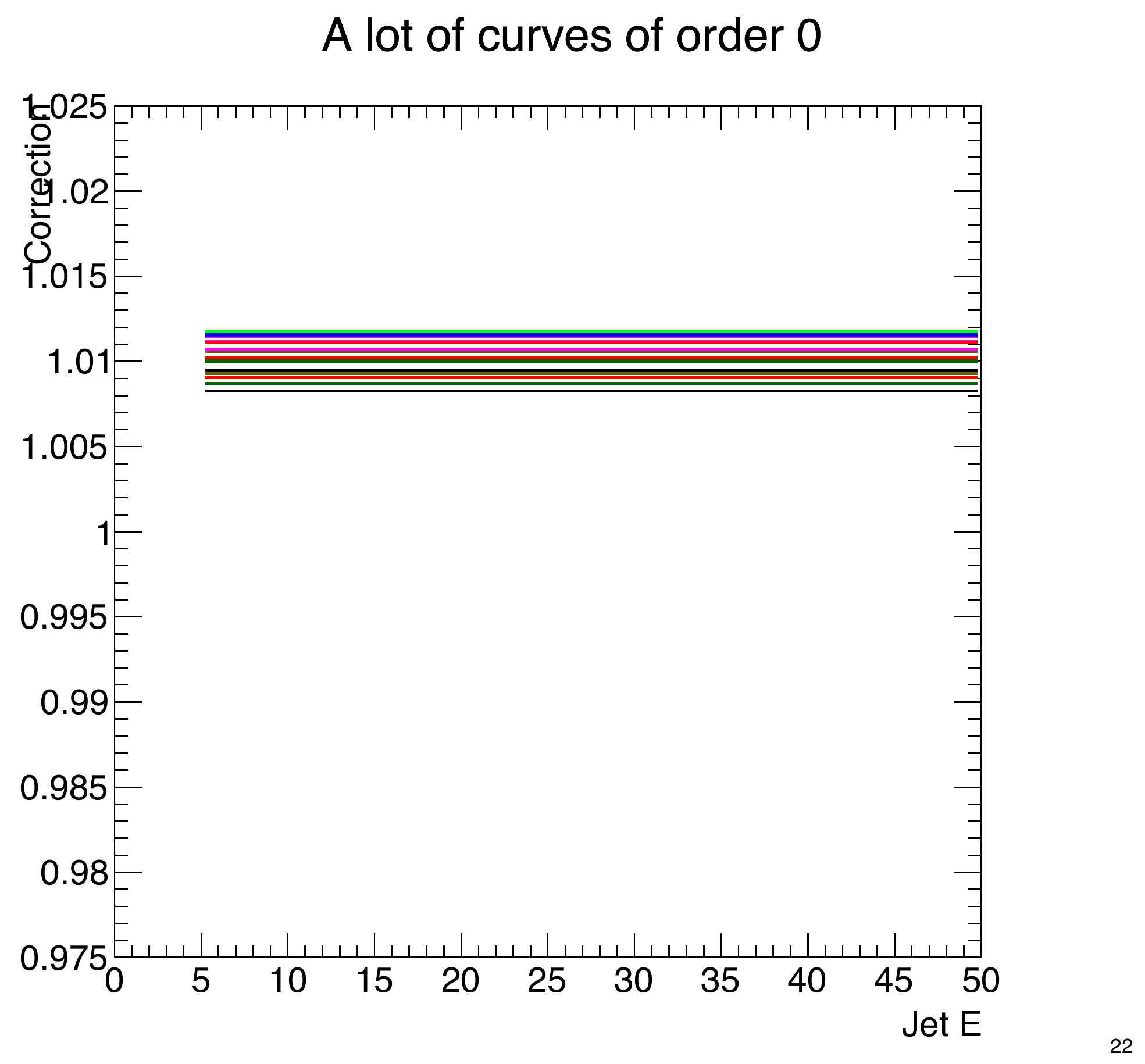}
    \includegraphicstwo{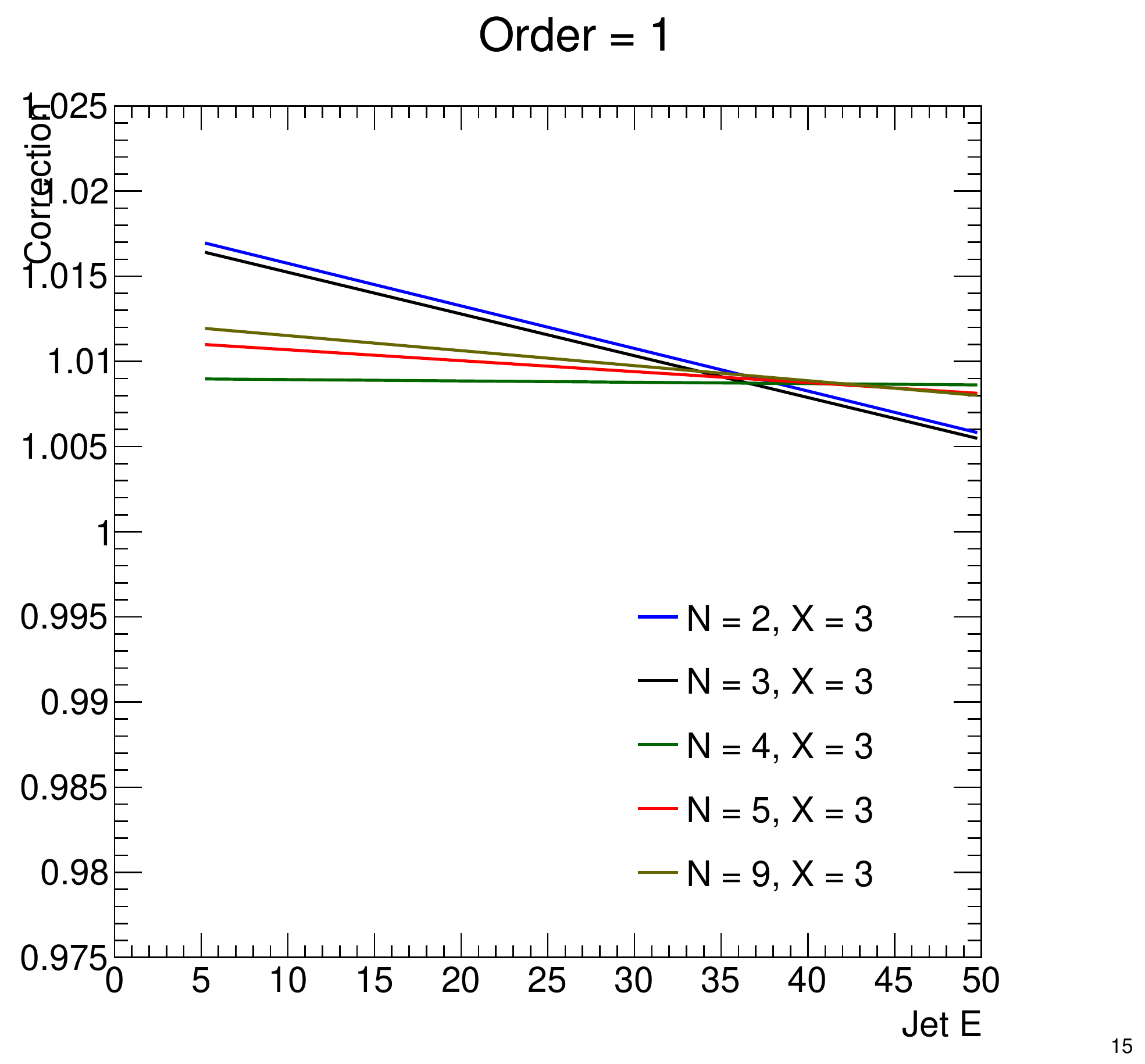}
    \caption{Example of fitted corrections with order 0 (left) and 1 (right).  All variations of $N$ and $X$ agree for 0-th order fits, and there is a dependence on number of jets $N$ at low energy.}
    \label{Figure:JetCalibration-AbsoluteResidualFitExample}
\end{figure}

The fitted corrections for all the combinations are shown in Fig.~\ref{Figure:JetCalibration-AbsoluteResidualFitExample} for 0-th order polynomial in the left panel.  The correction is not very sensitive to the choice of $N$ and $X$.  A scan of different threshold for $X = 3$ is shown in the right panel for polynomial order 1.  With too small $N$, the leading jets are all high energy, and the constraining power at low jet energy is limited.  The high energy end around 40-45 GeV however, is always well constrained.  For a linear function, larger $N$ is preferred.

In the final fits, an overall factor for each $\theta$ bin is also floated to account for potential scale difference as a function of $\theta$, since we only corrected the plus/minus side difference in the previous step.  For the final correction used in this analysis, the first order polynominal ($O(1)$) function is chosen to allow potential scale difference between higher energy jets and lower energy jets.

\clearpage

\section{Jet Resolution}\label{Section:JetResolution}

\subsection{Simulated Jet Resolution}

\begin{figure}[htp!]
    \centering
    \includegraphicsfour{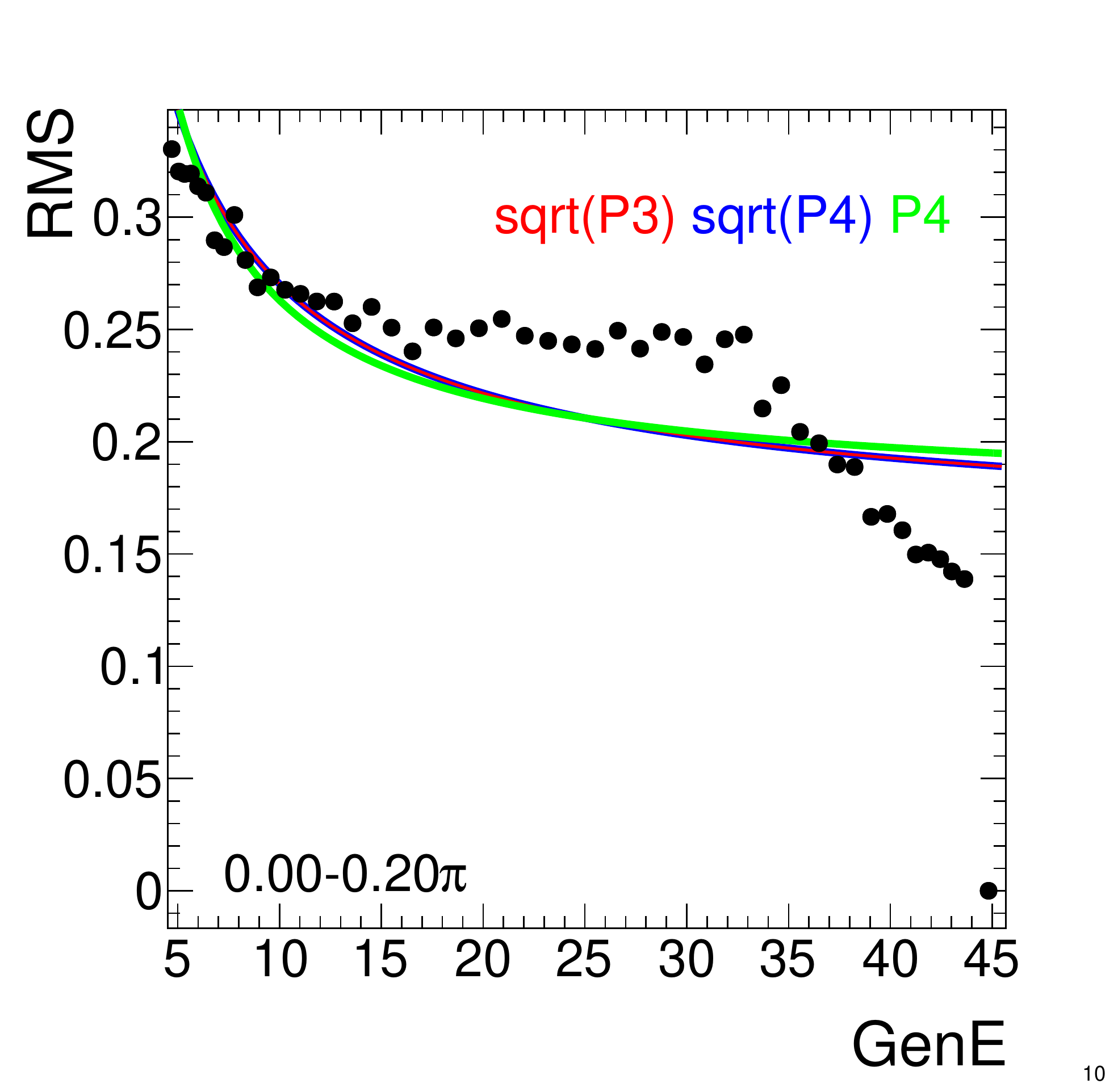}
    \includegraphicsfour{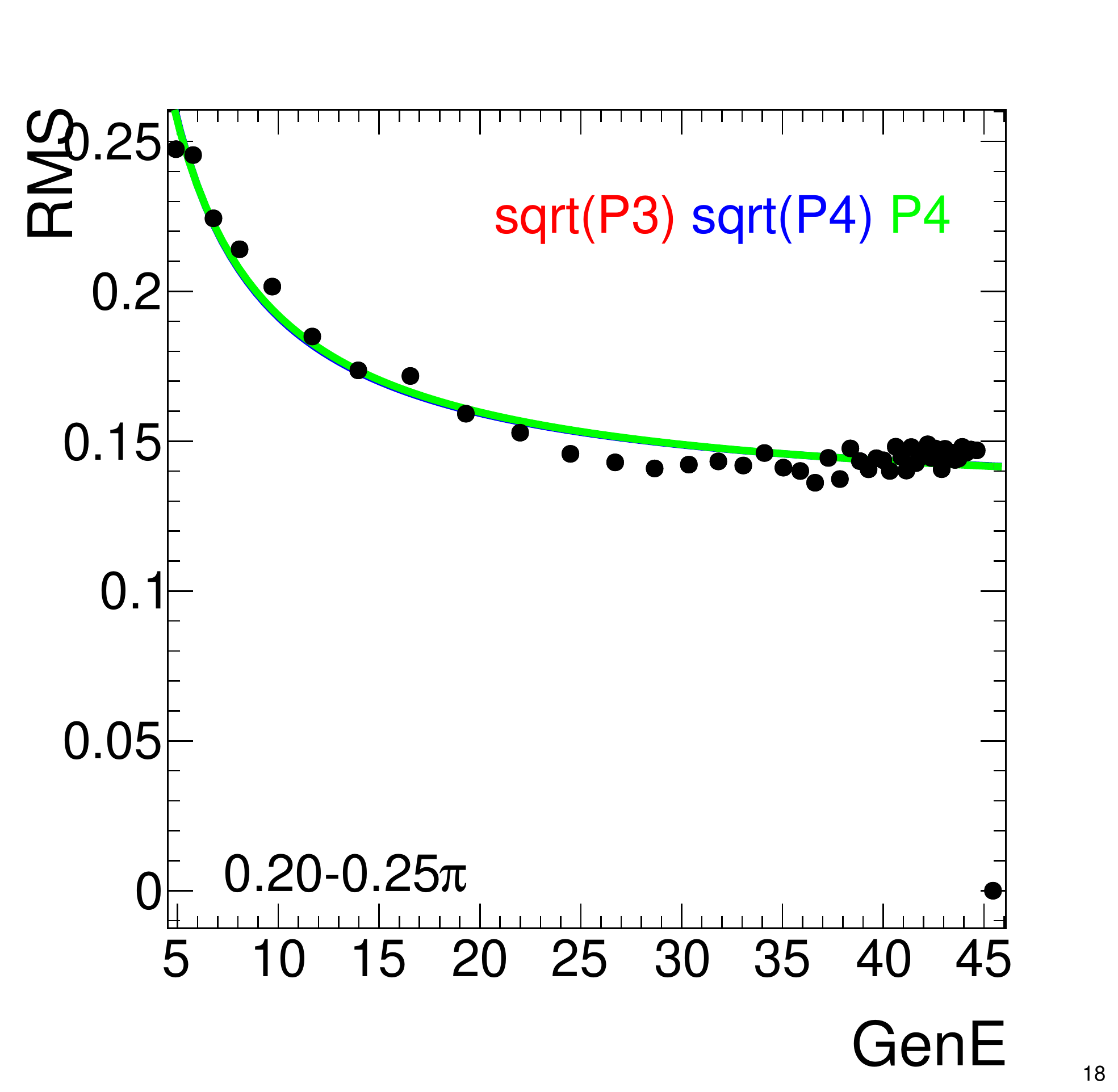}
    \includegraphicsfour{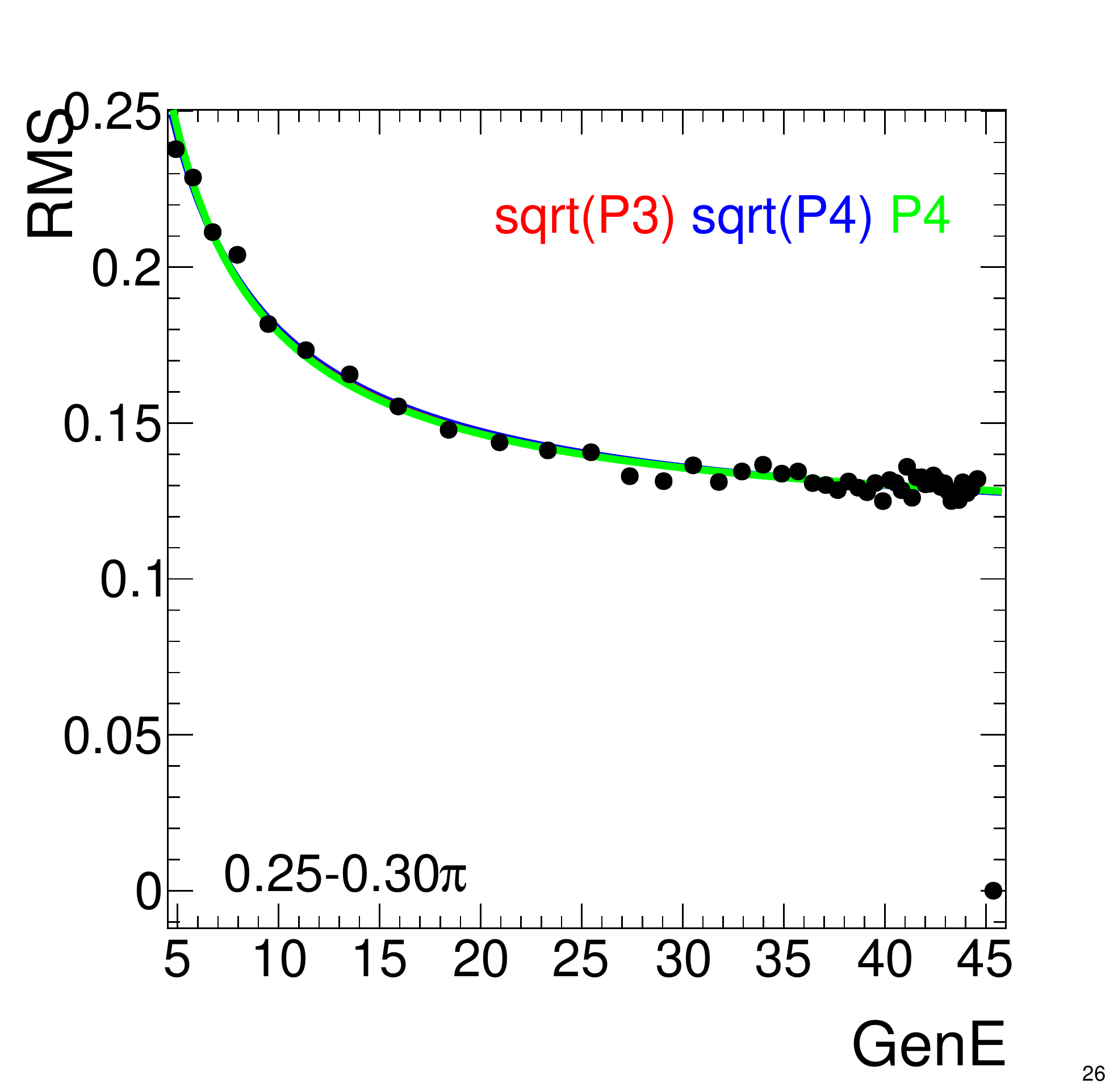}
    \includegraphicsfour{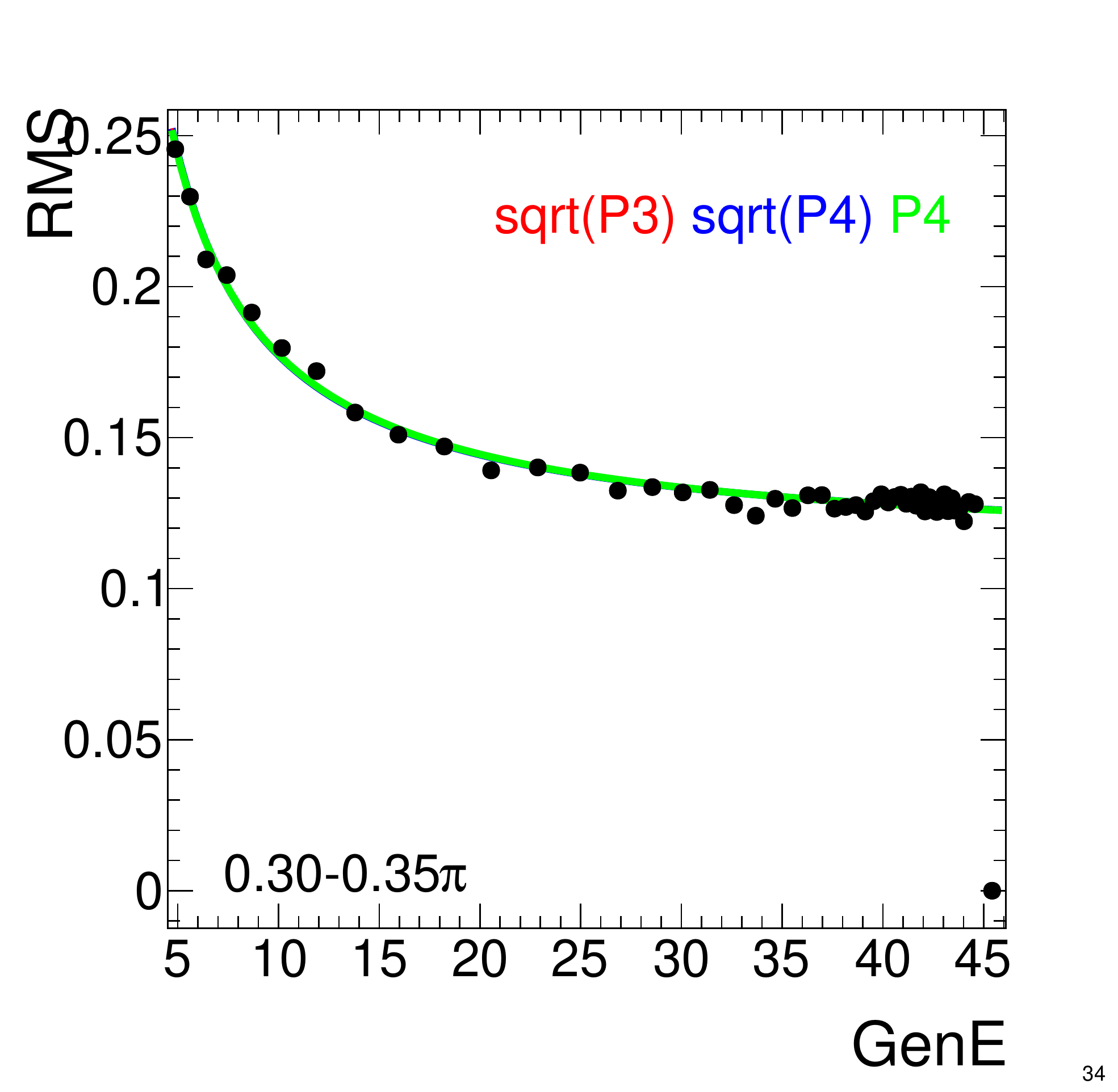}
    \includegraphicsfour{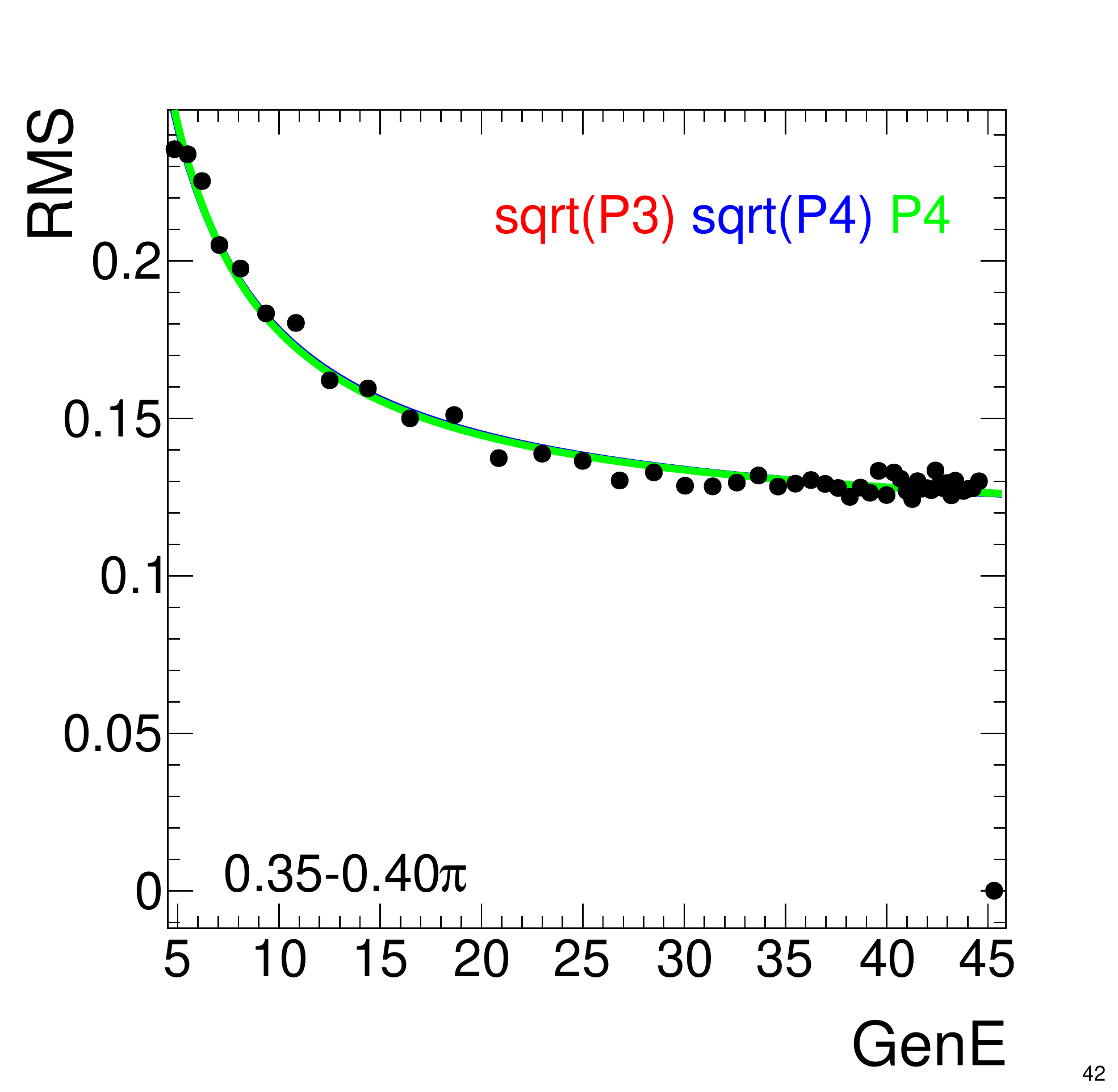}
    \includegraphicsfour{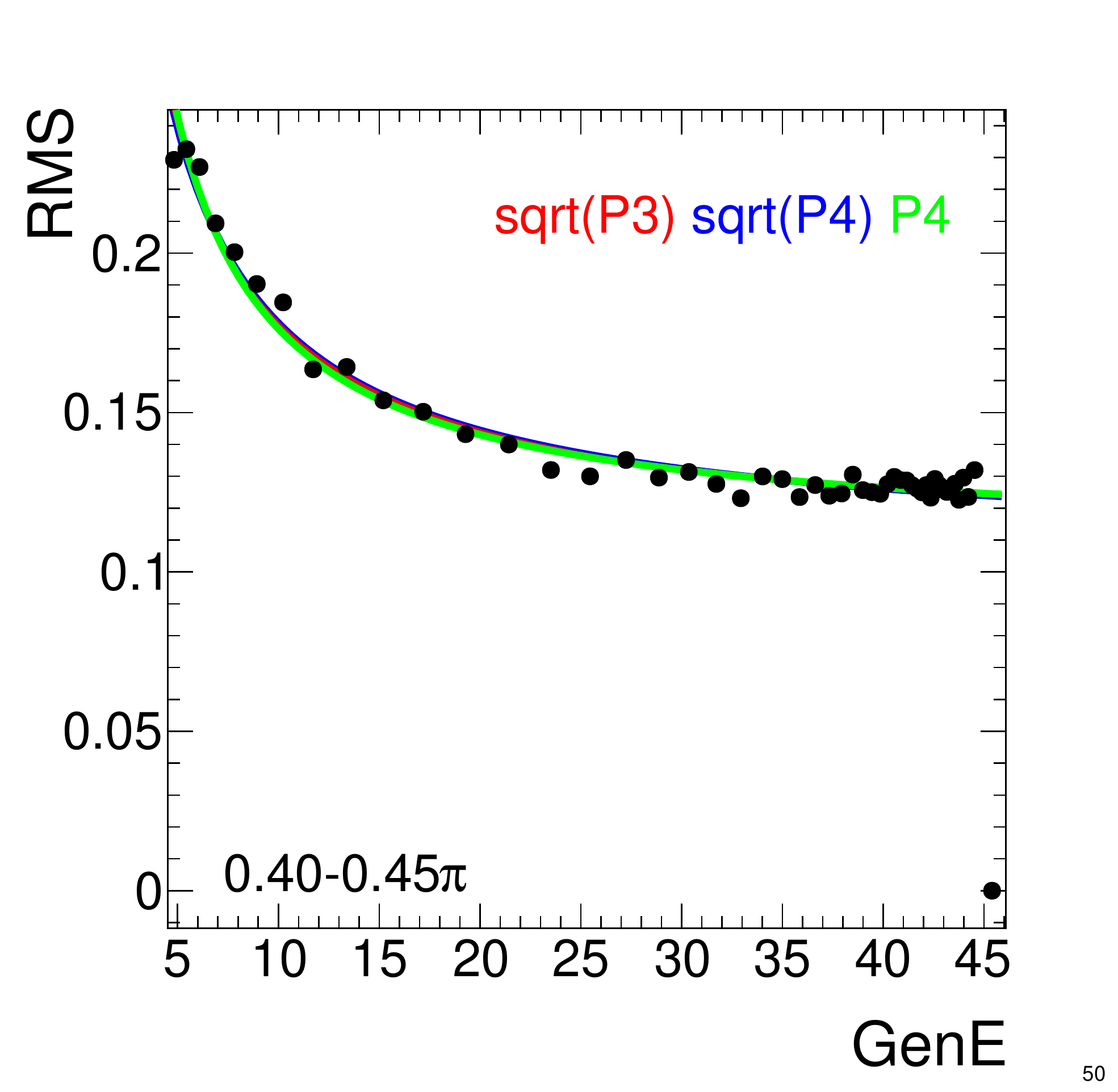}
    \includegraphicsfour{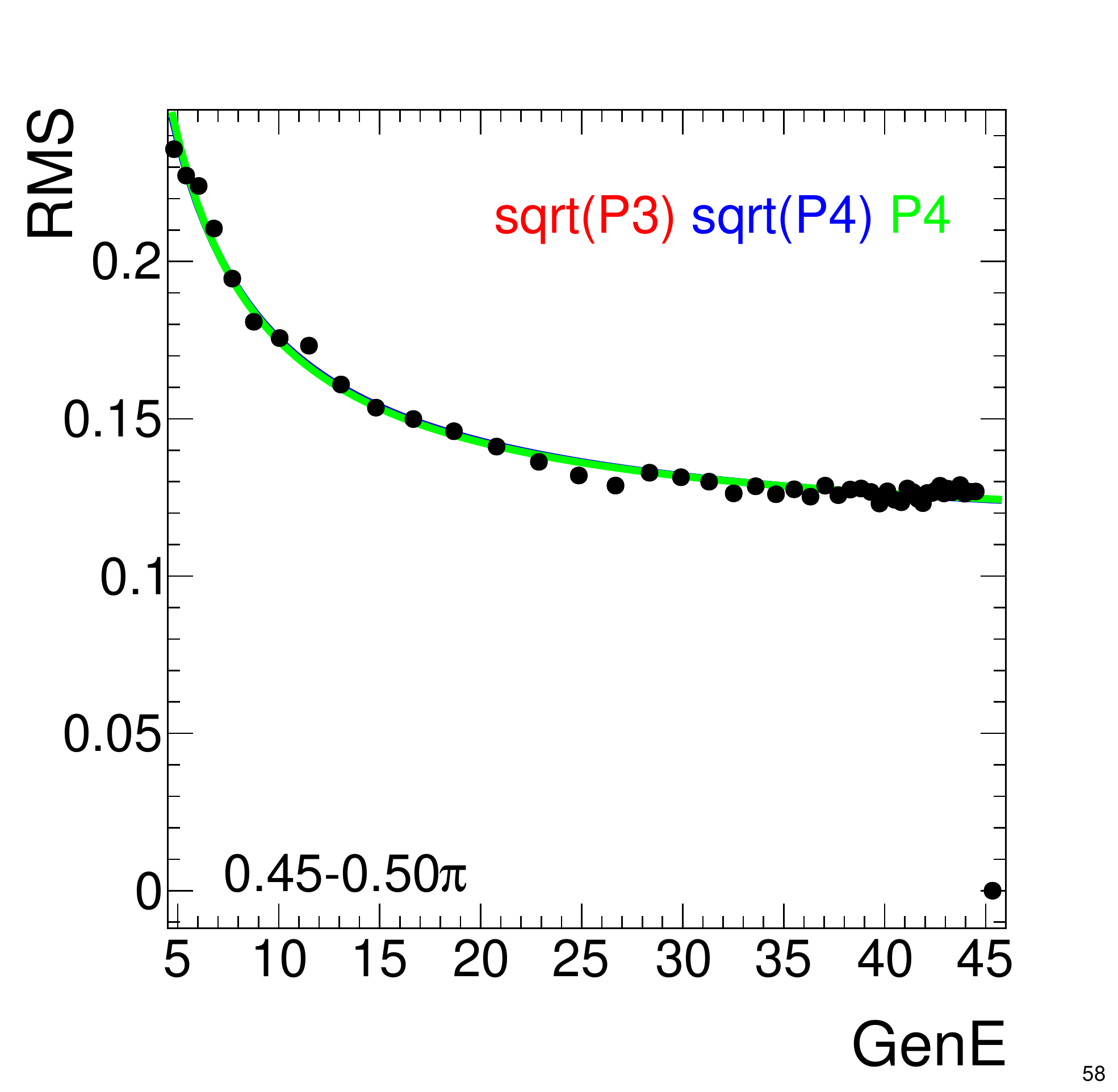}
    \includegraphicsfour{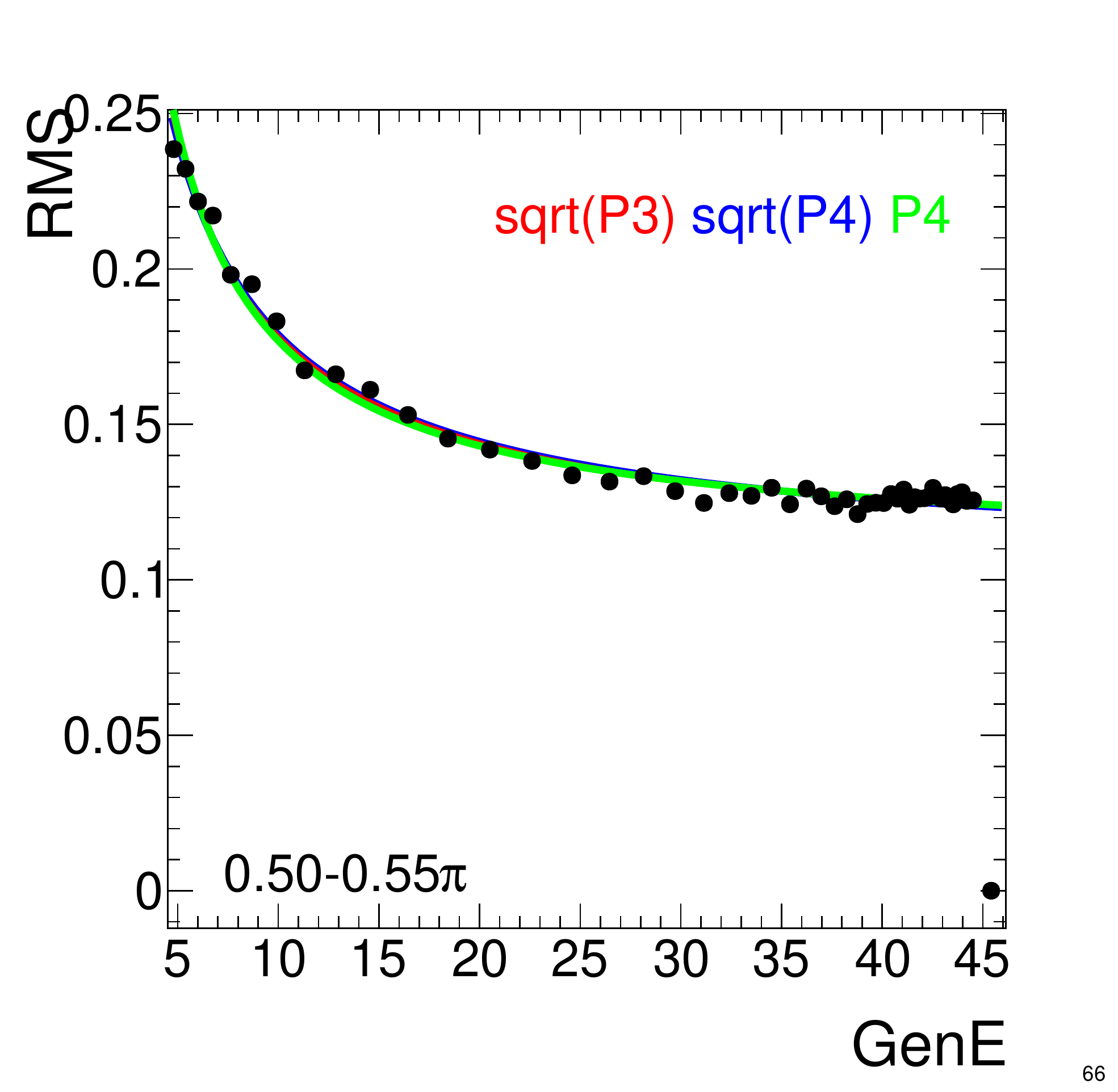}
    \includegraphicsfour{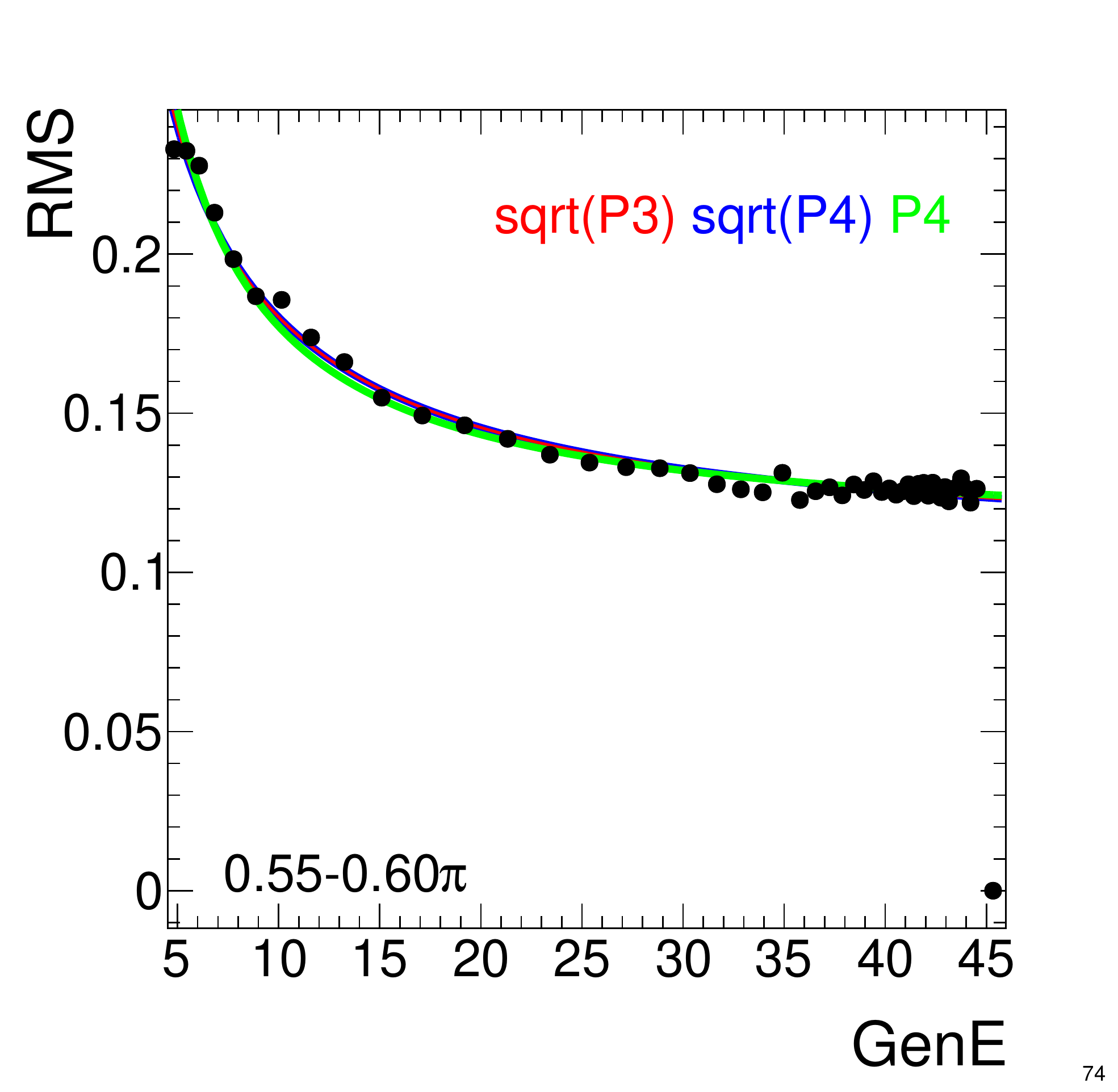}
    \includegraphicsfour{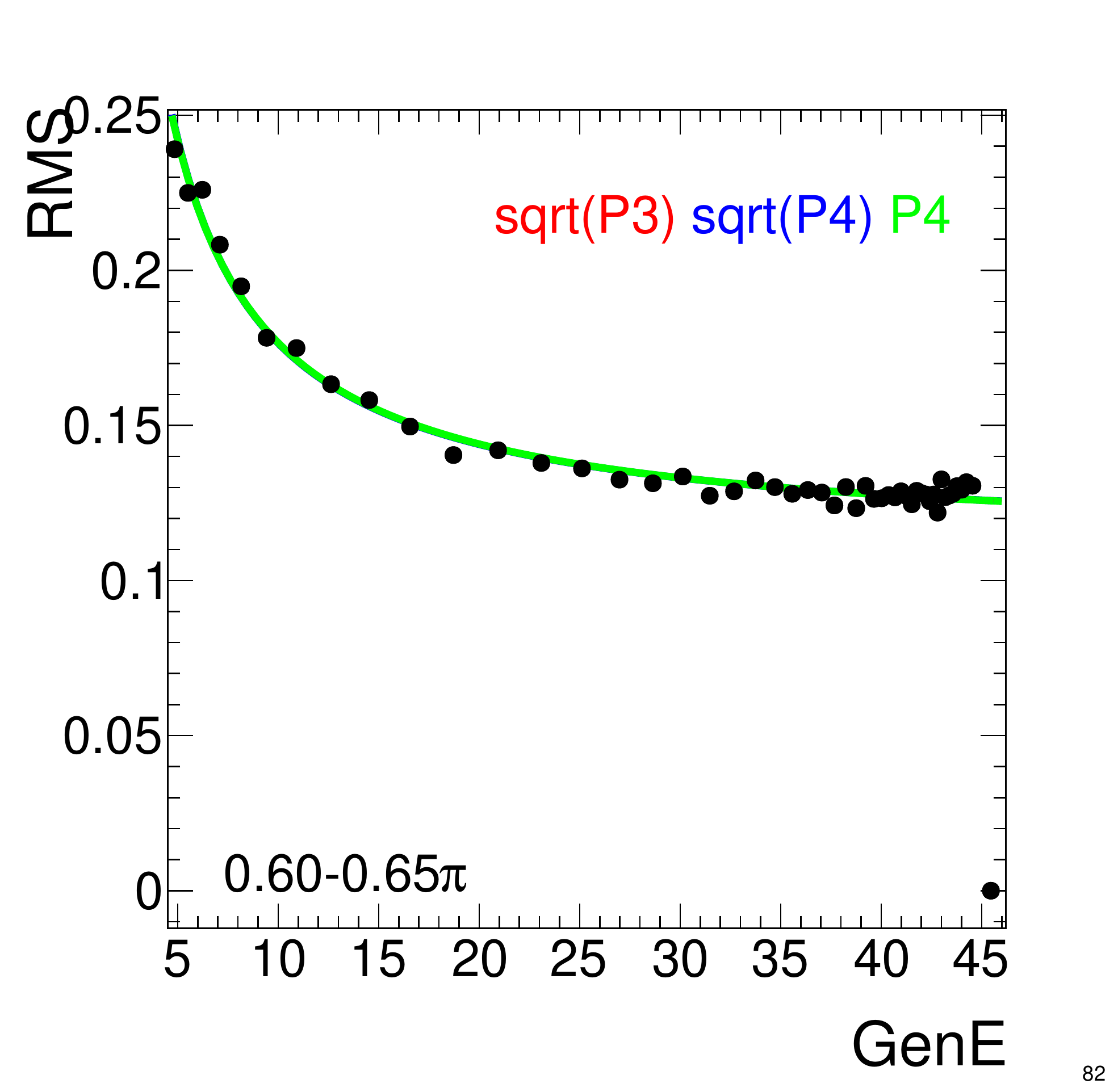}
    \includegraphicsfour{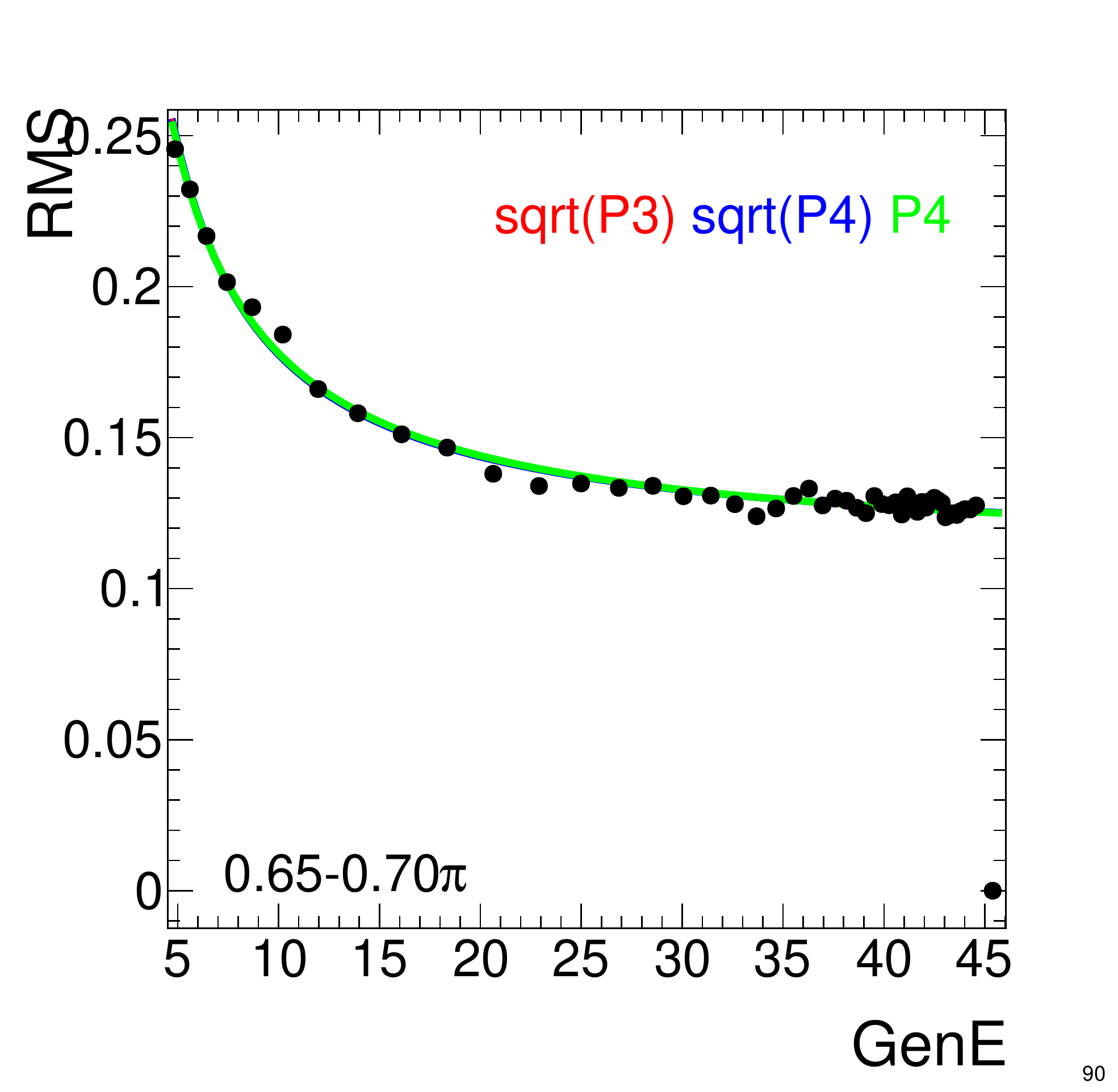}
    \includegraphicsfour{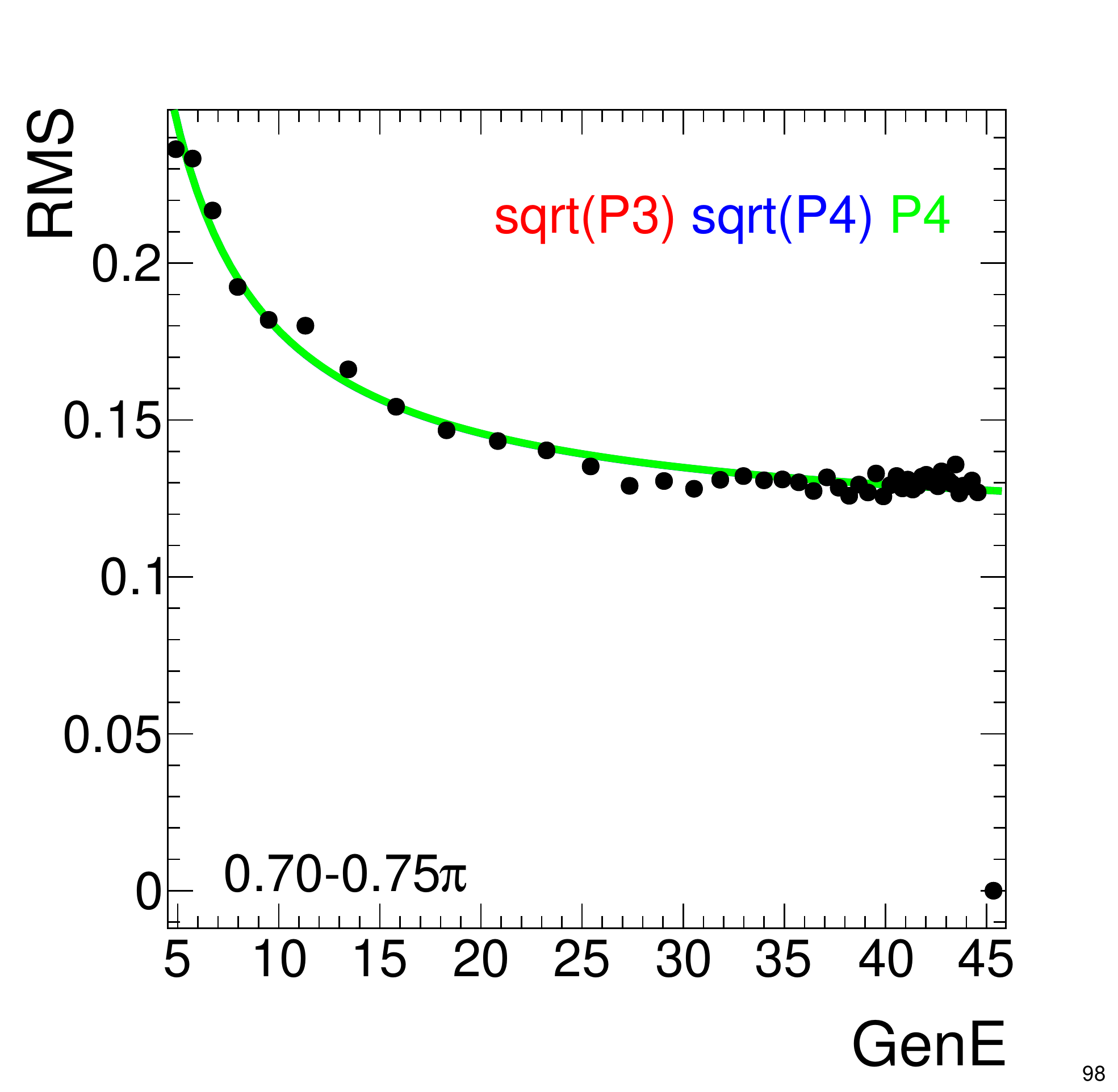}
    \includegraphicsfour{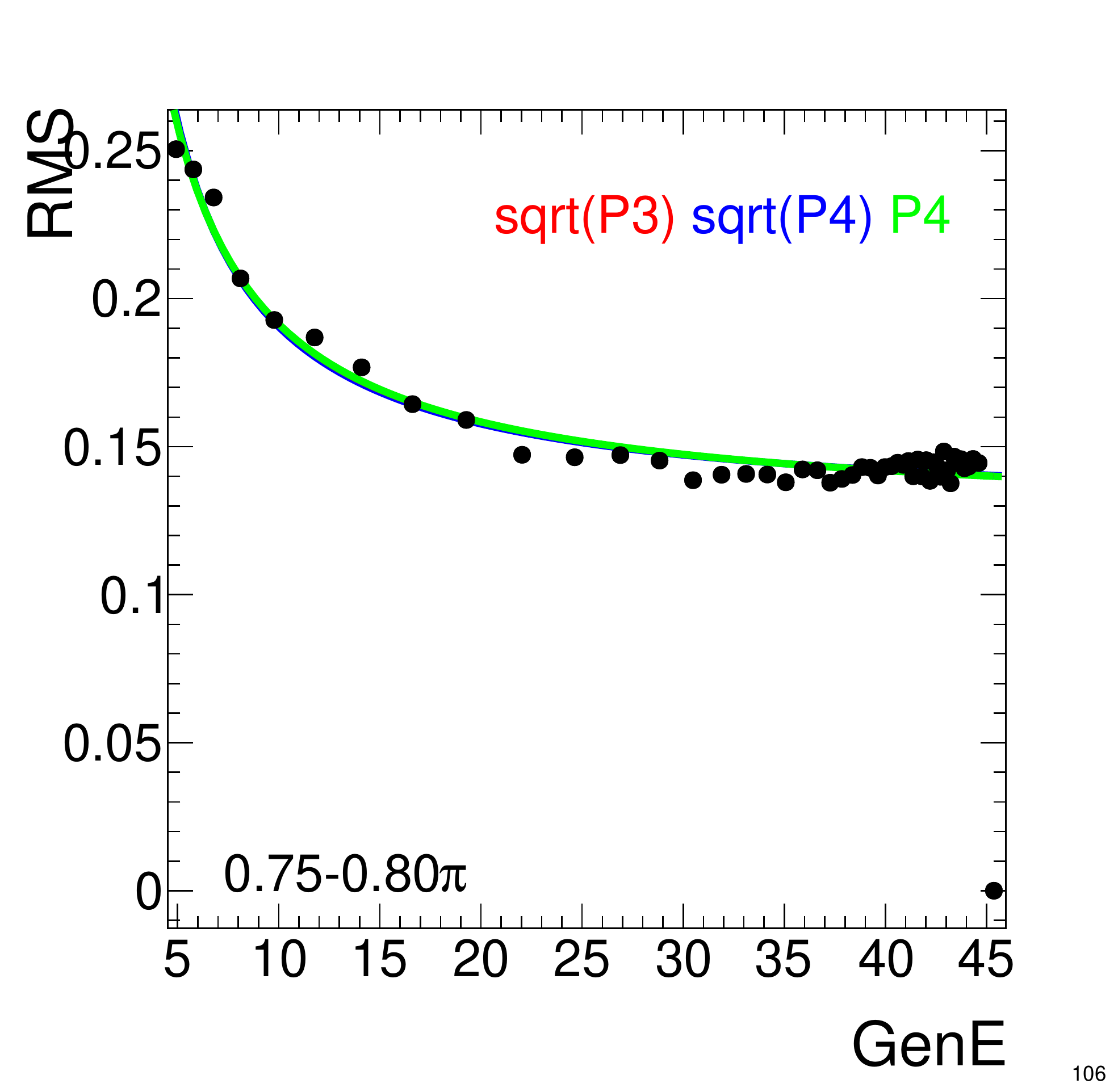}
    \includegraphicsfour{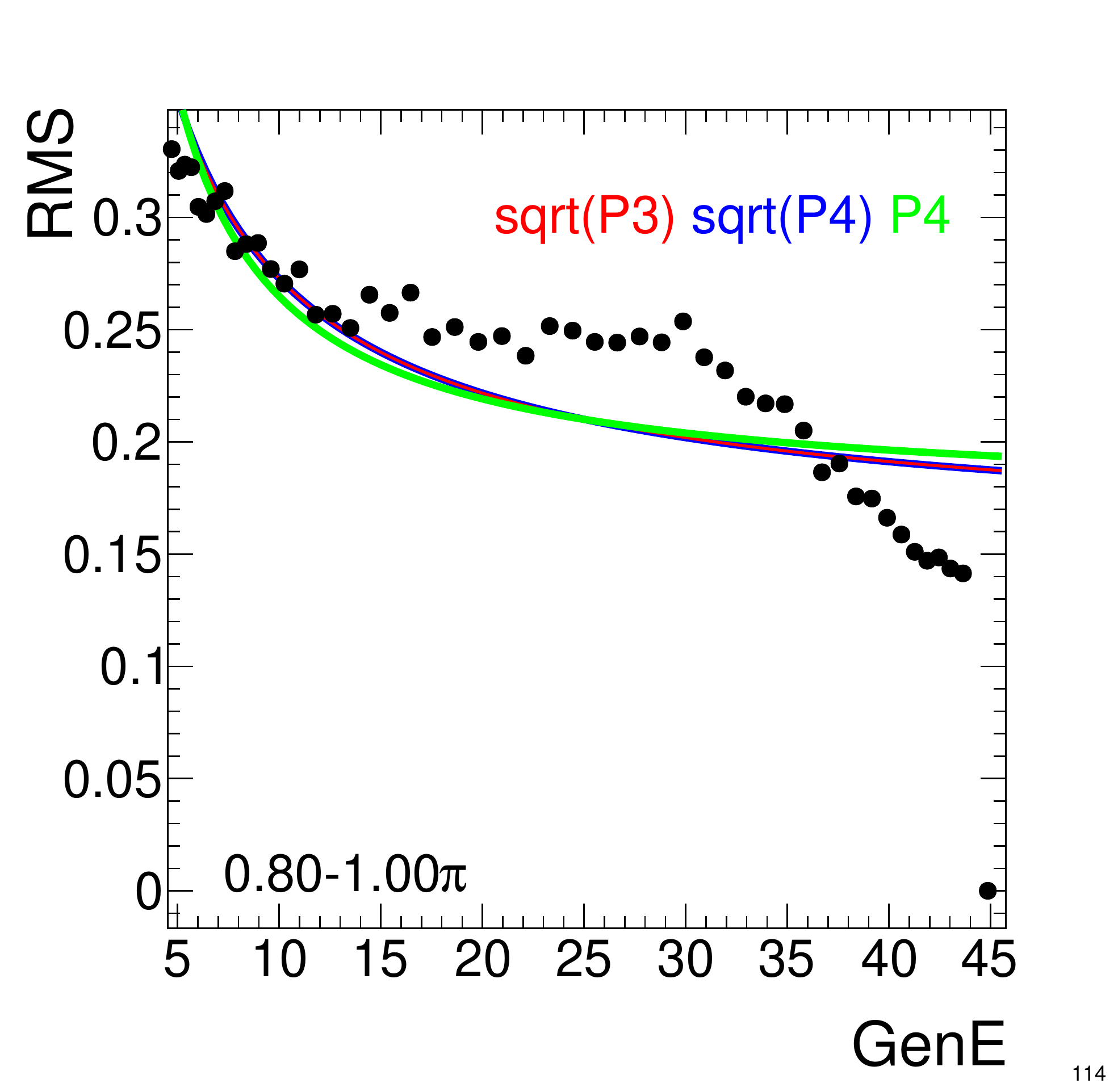}
    \caption{Jet momentum resolution in simulated samples as a function of generated jet momentum in different bins of jet $\theta$.  They are fitted with different functional forms, all of which decently describes the resolution.}
    \label{Figure:JetResolution-MCPResolution}
\end{figure}

The energy, $\theta$, and $\phi$ resolution of the simulated jet is fitted in bins of jet direction $\theta$. The extracted resolution functions could be provided as a useful supplementary information for smearing the theoretical curves or samples from event generators.
 The fitted energy resolution is shown in Fig.~\ref{Figure:JetResolution-MCPResolution} for all different $\theta$ bins.  Three different empirical functions are tried to fit for the resolution:
\begin{align}
    f_\texttt{sqrt(P3)}(p) &= \sqrt{a_0^2 + \dfrac{a_1^2}{E} + \dfrac{a_2^2}{E^2} + \dfrac{a_3^2}{E^3}}\\
    f_\texttt{sqrt(P4)}(p) &= \sqrt{a_0^2 + \dfrac{a_1^2}{E} + \dfrac{a_2^2}{E^2} + \dfrac{a_3^2}{E^3} + \dfrac{a_4^2}{E^4}}\\
    f_\texttt{P4}(p) &= a_0^2 + \dfrac{a_1^2}{E} + \dfrac{a_2^2}{E^2} + \dfrac{a_3^2}{E^3} + \dfrac{a_4^2}{E^4},
\end{align}
where $a_i$'s are free parameters to be fitted.  The three functions are labeled as \texttt{sqrt(P3)}, \texttt{sqrt(P4)} and \texttt{P4}, respectively, in the plots.  All three functions fit equally well, and a more traditional function (the first one) is chosen as the parameterization of the energy resolution.  The coefficient with the highest power, $a_3$, is seen to be negligible.

When the jets are close to the beam line, the resolution deviates from the functional forms.  It is expected as a significant portion of the jet energy leaks into the dead region.

\begin{figure}[htp!]
    \centering
    \includegraphicsfour{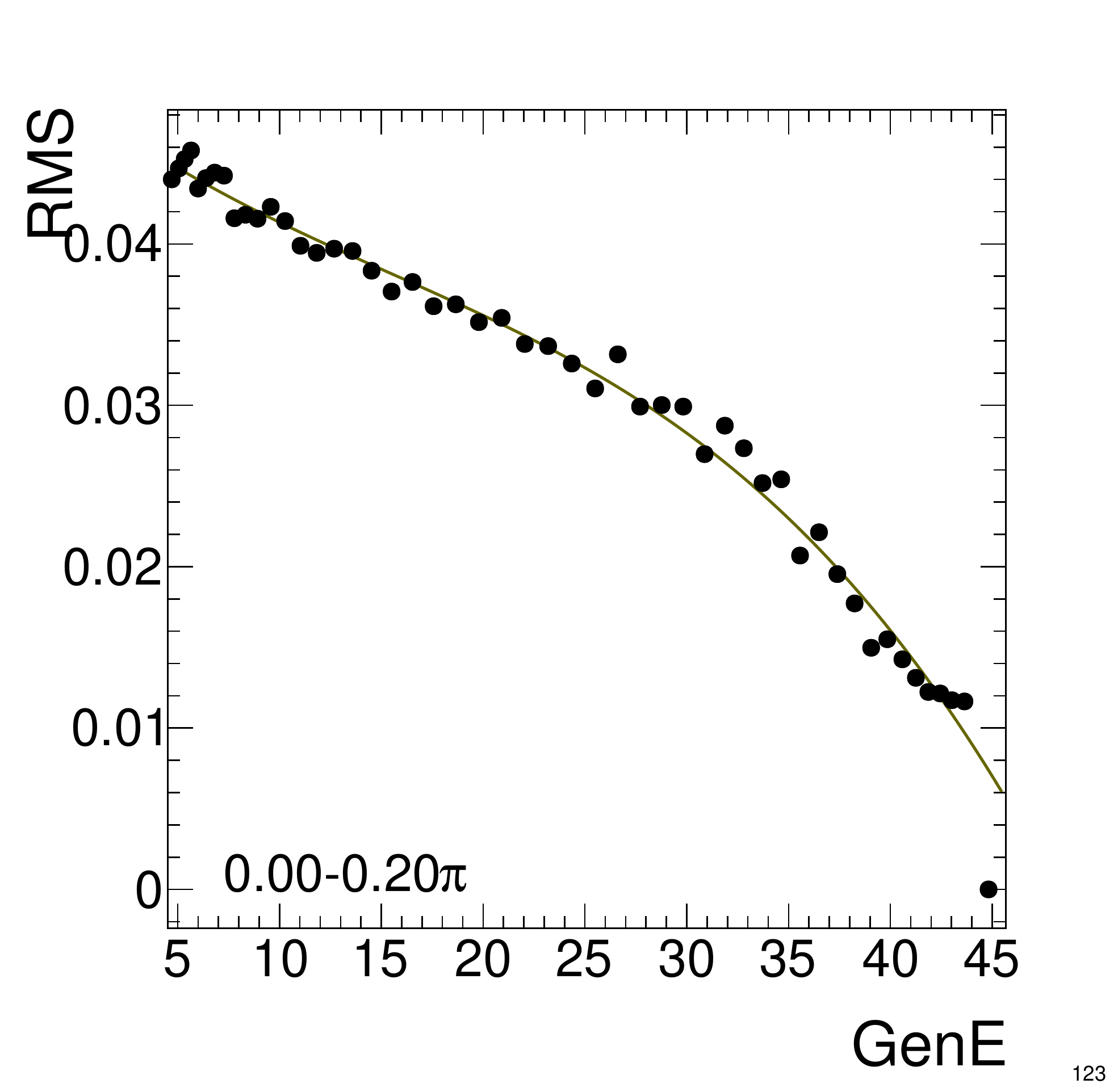}
    \includegraphicsfour{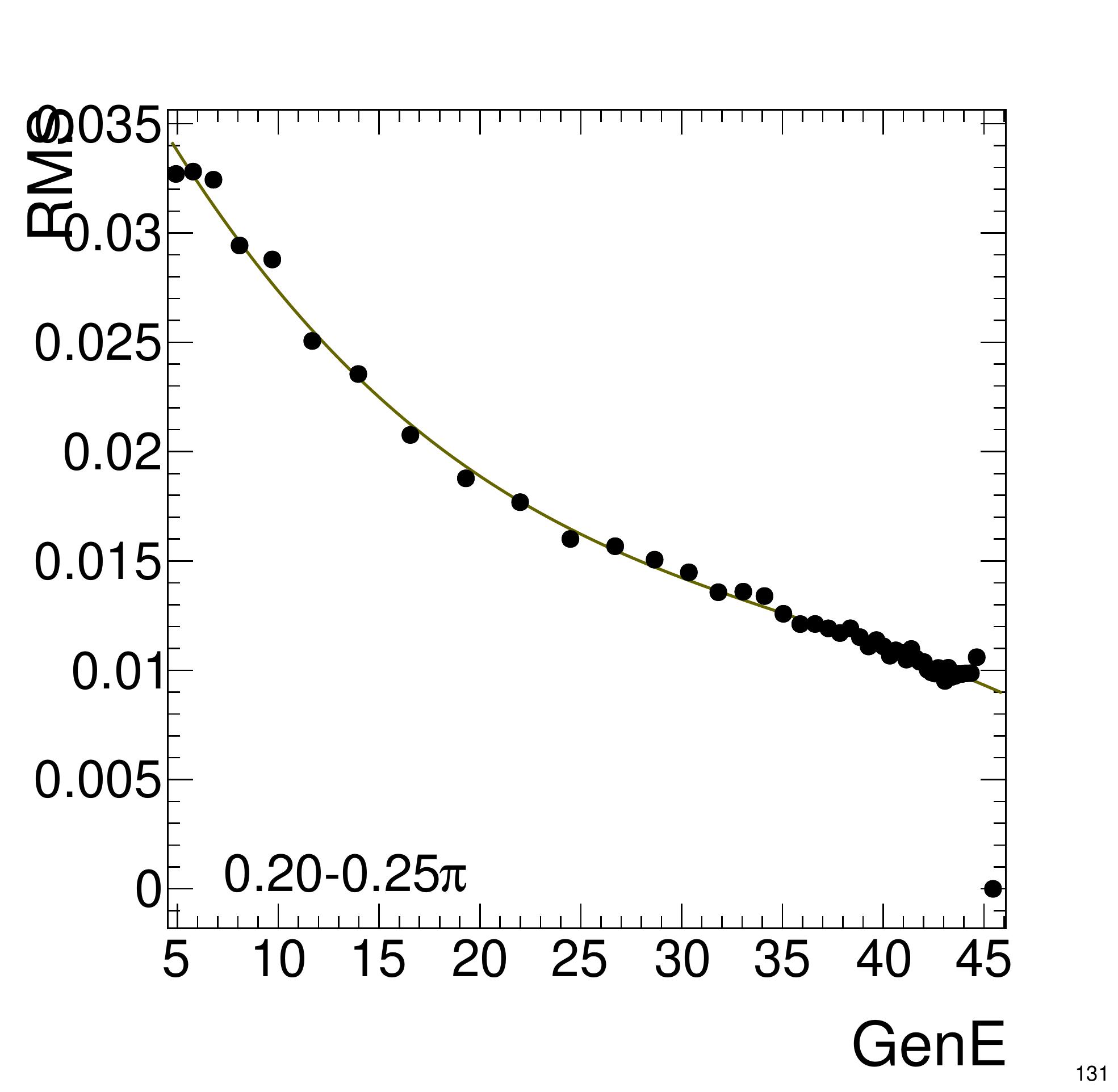}
    \includegraphicsfour{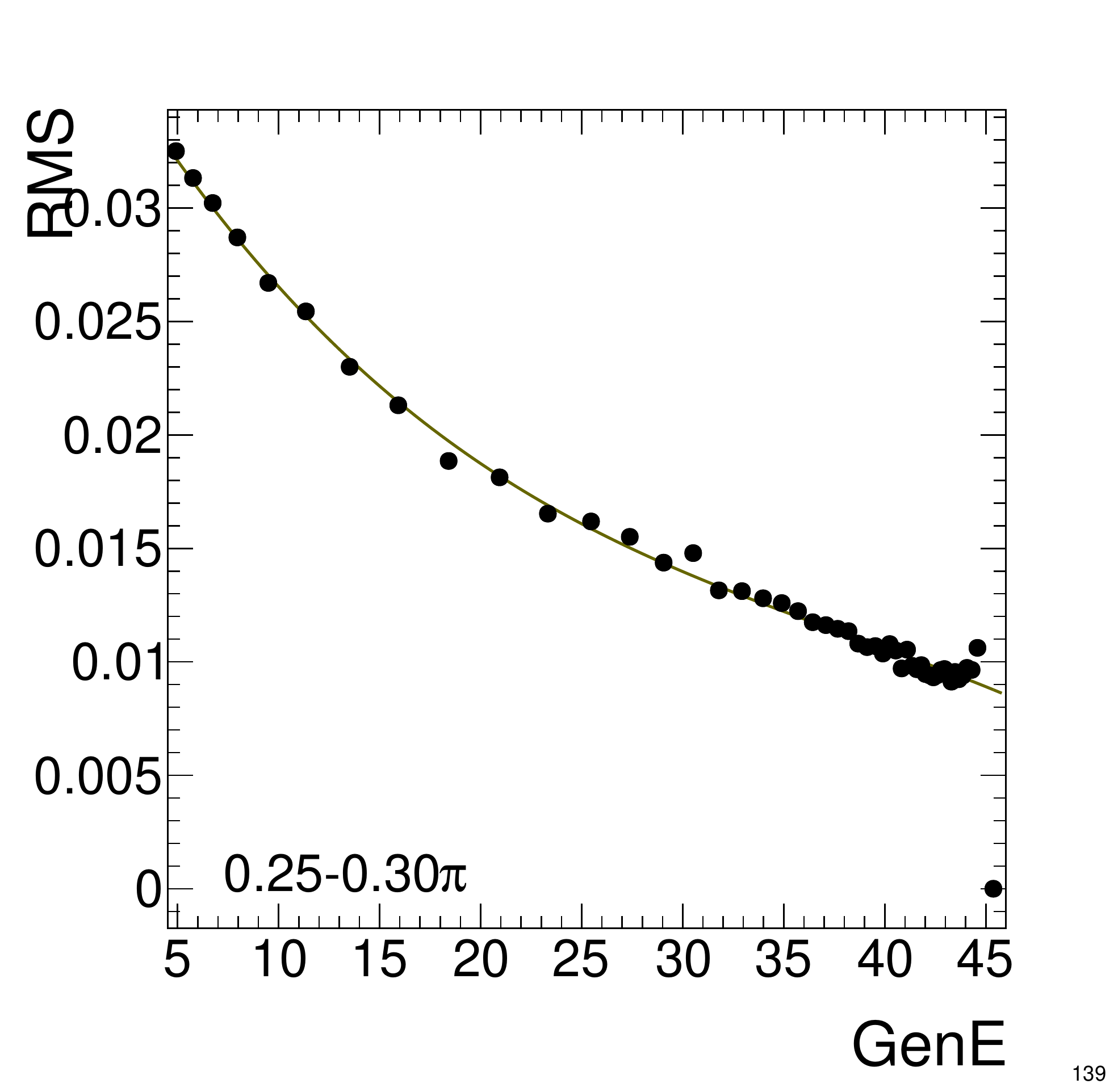}
    \includegraphicsfour{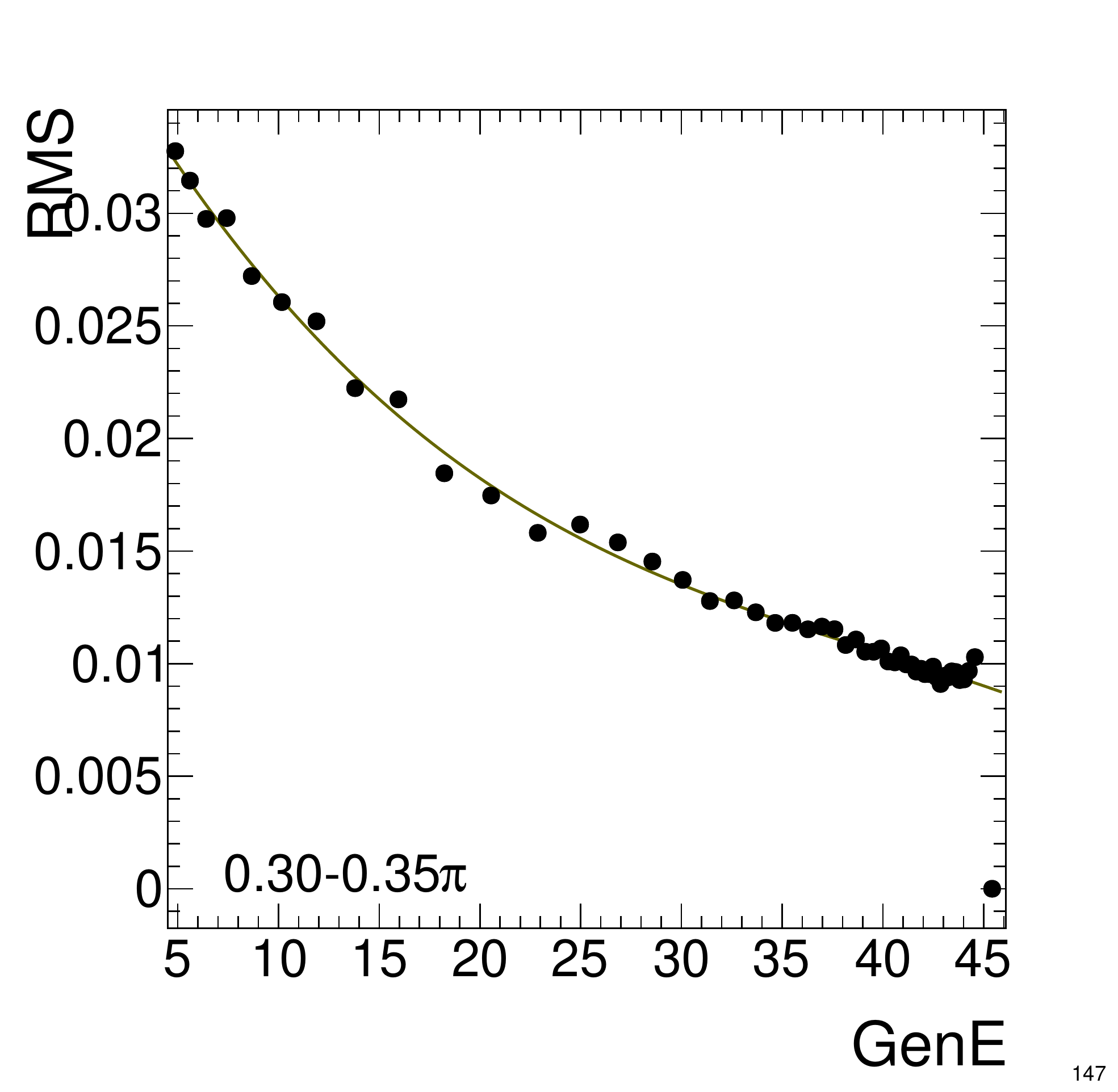}
    \includegraphicsfour{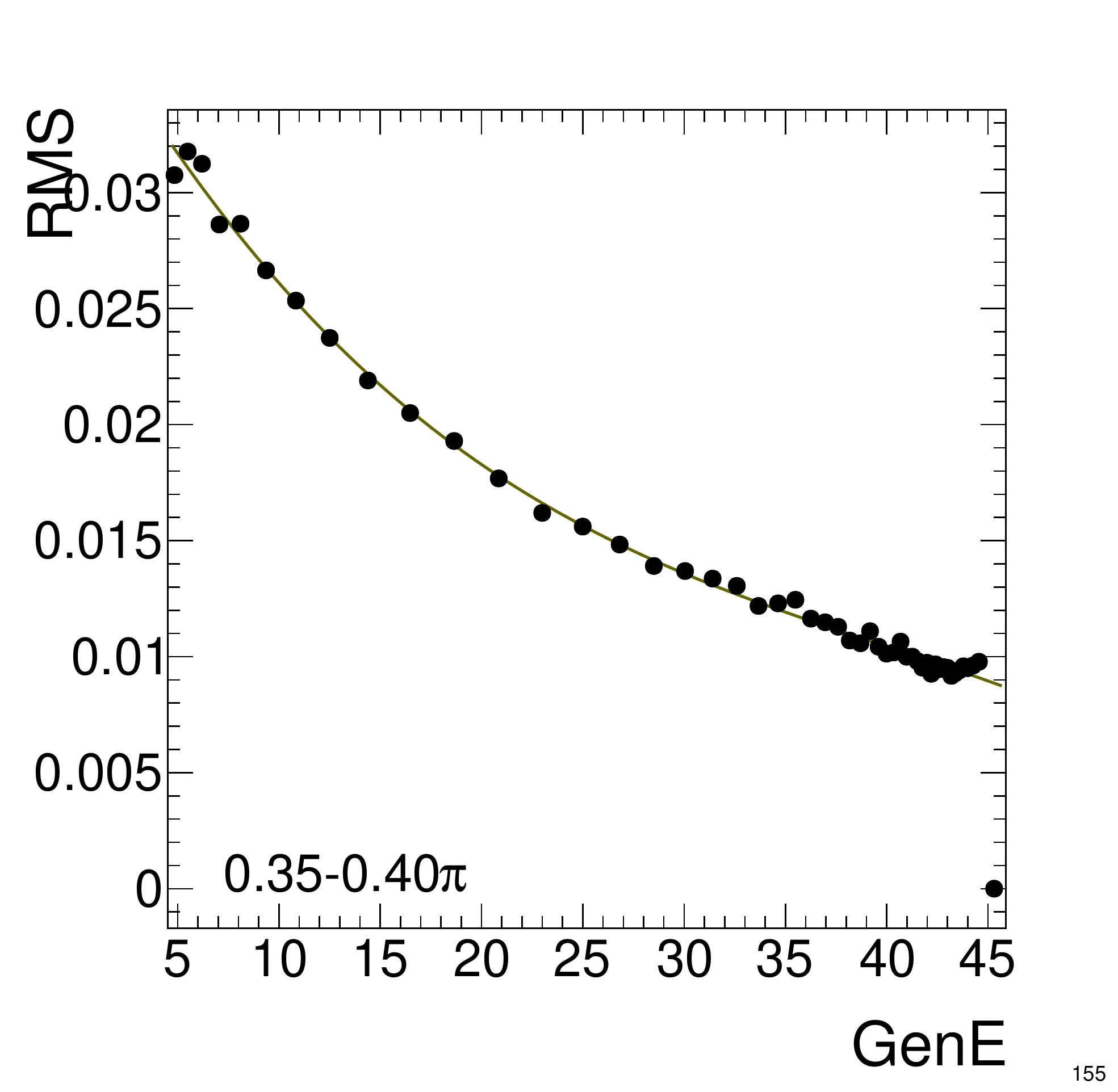}
    \includegraphicsfour{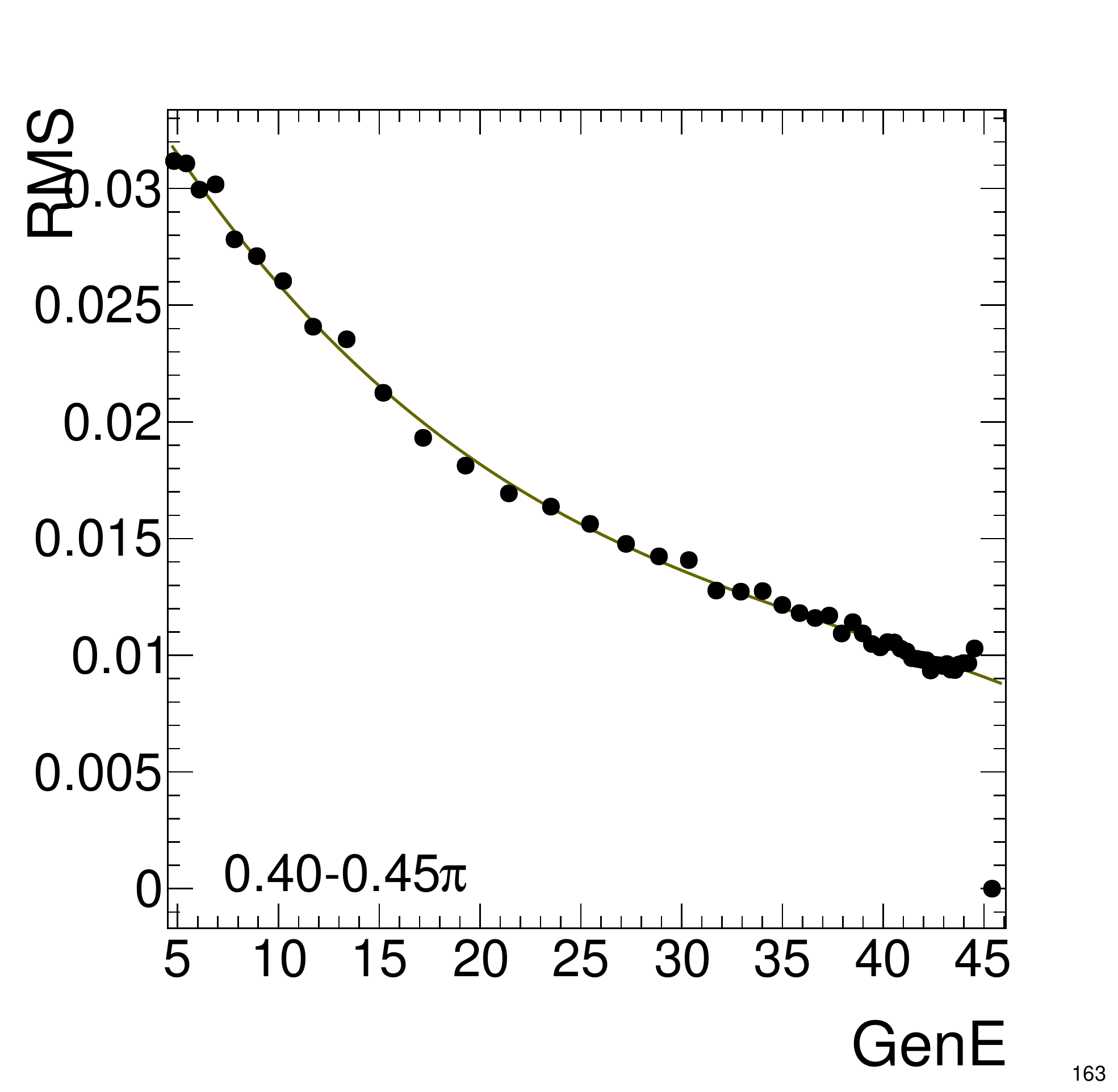}
    \includegraphicsfour{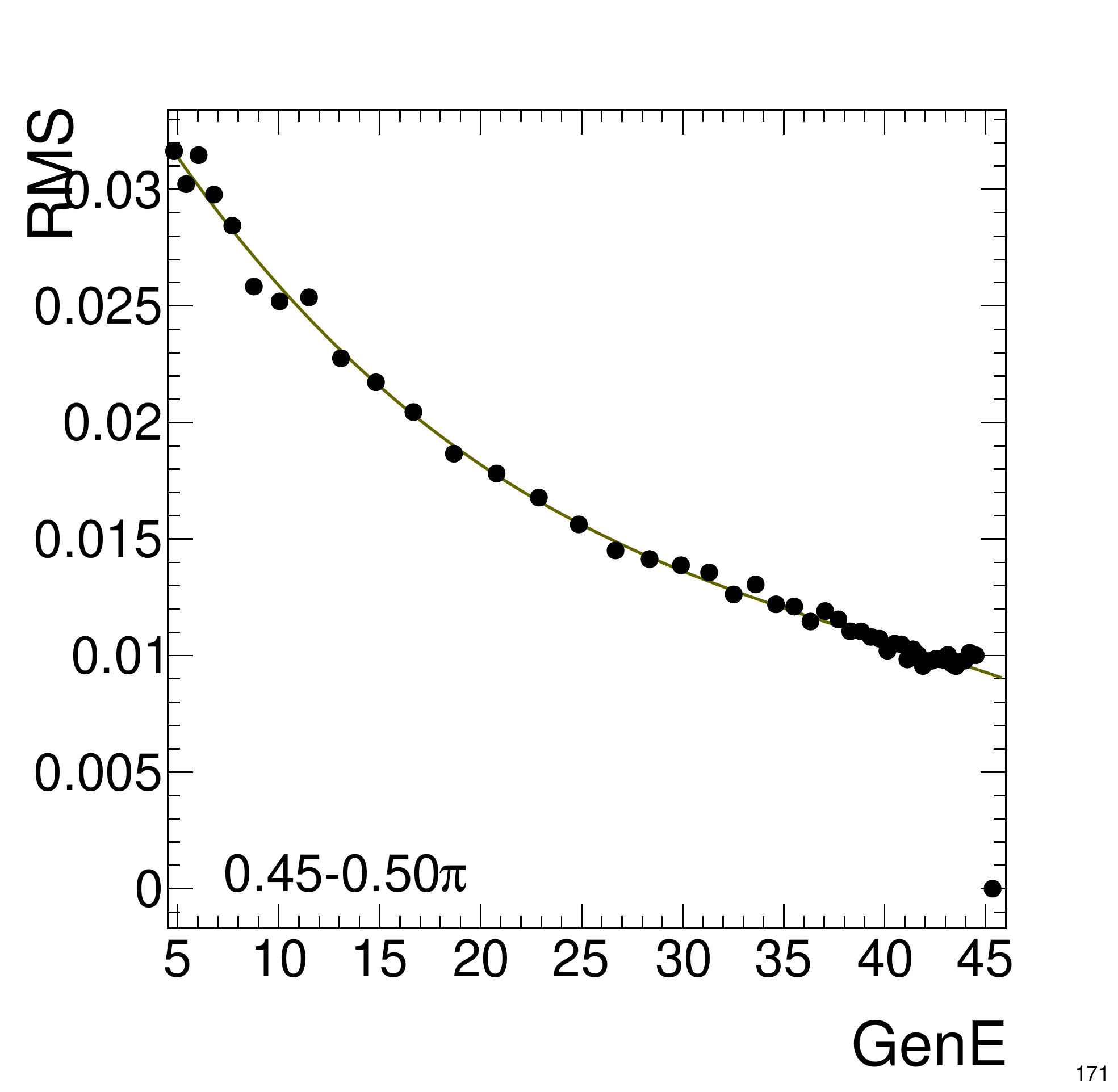}
    \includegraphicsfour{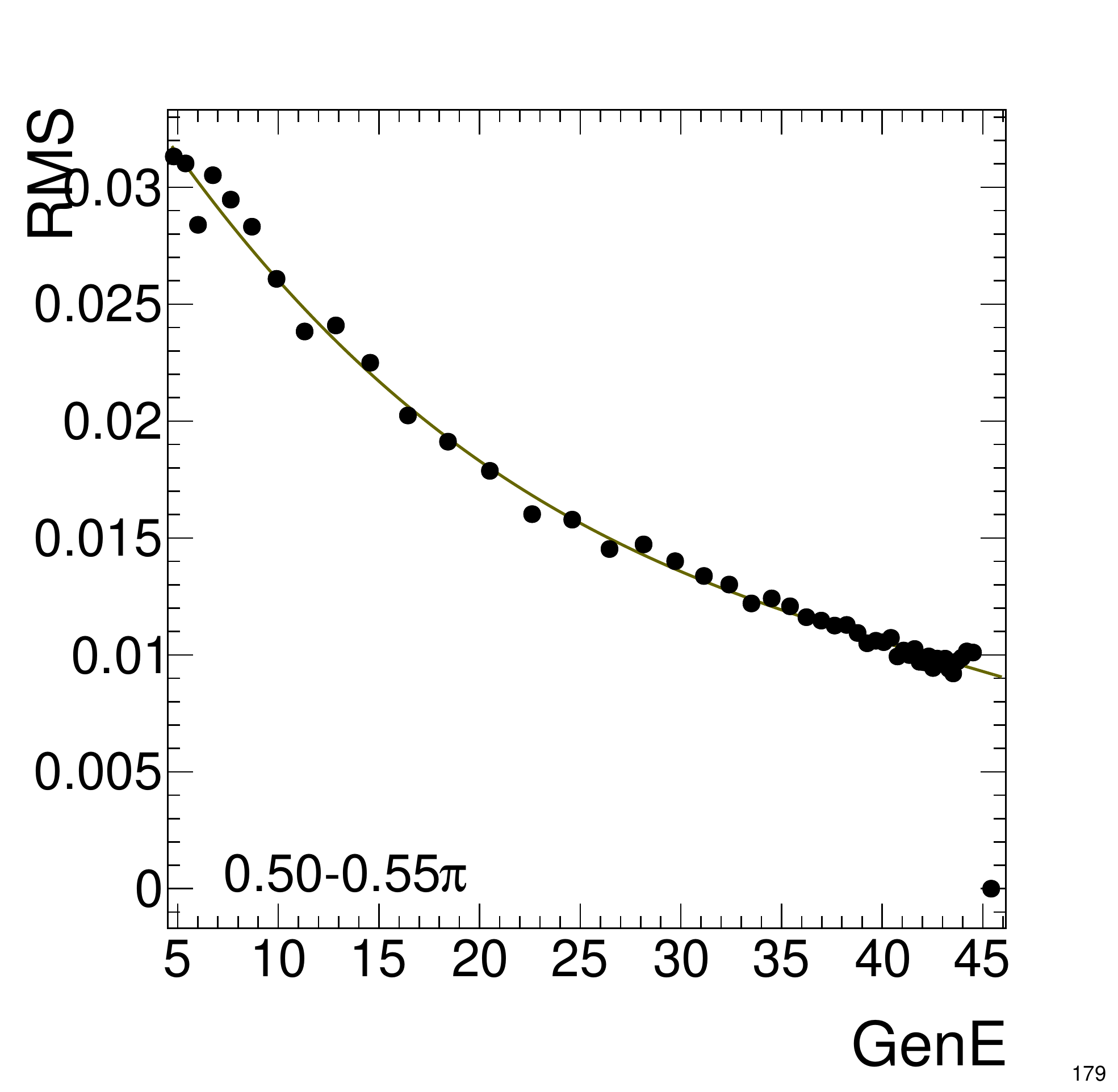}
    \includegraphicsfour{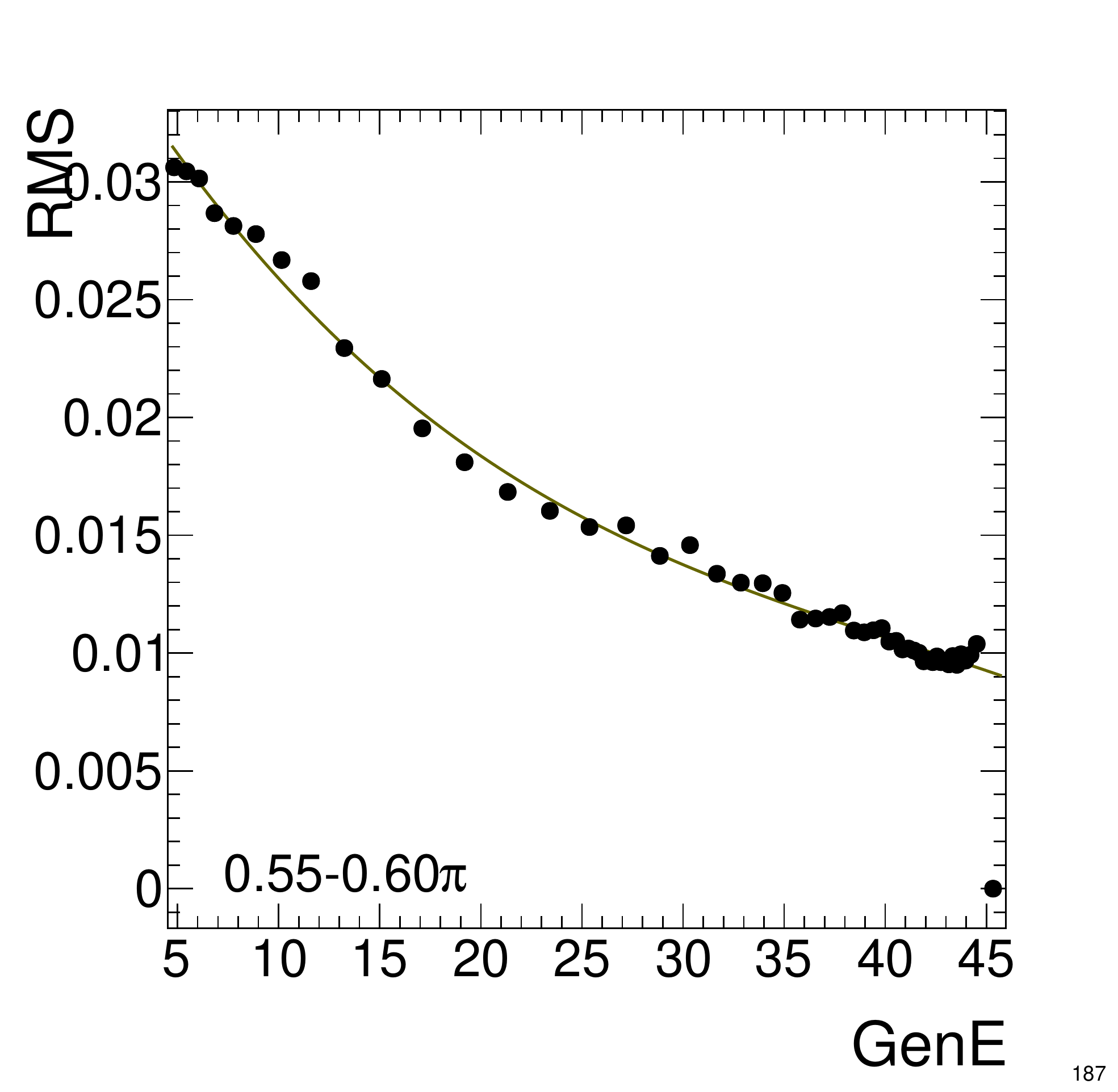}
    \includegraphicsfour{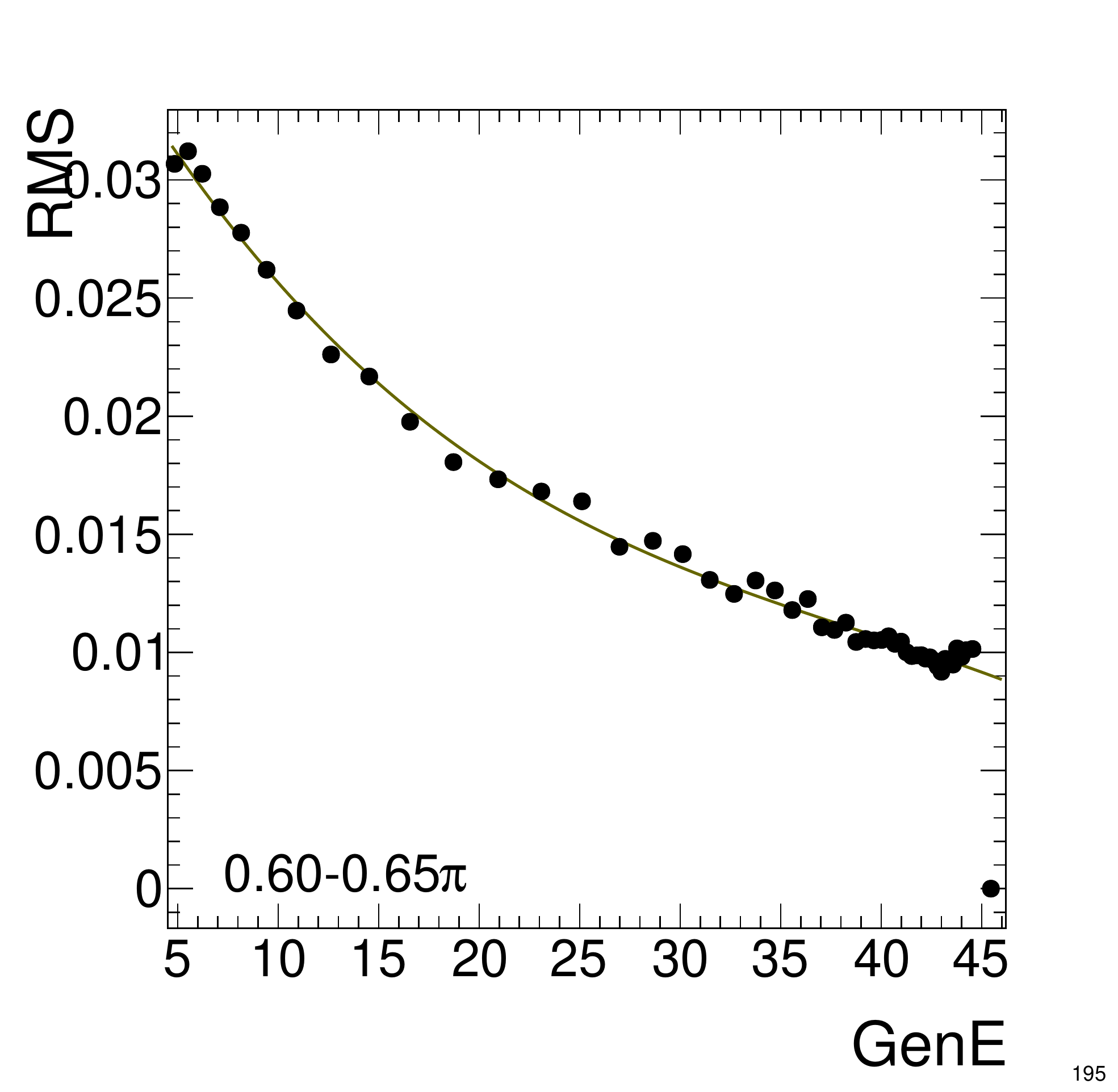}
    \includegraphicsfour{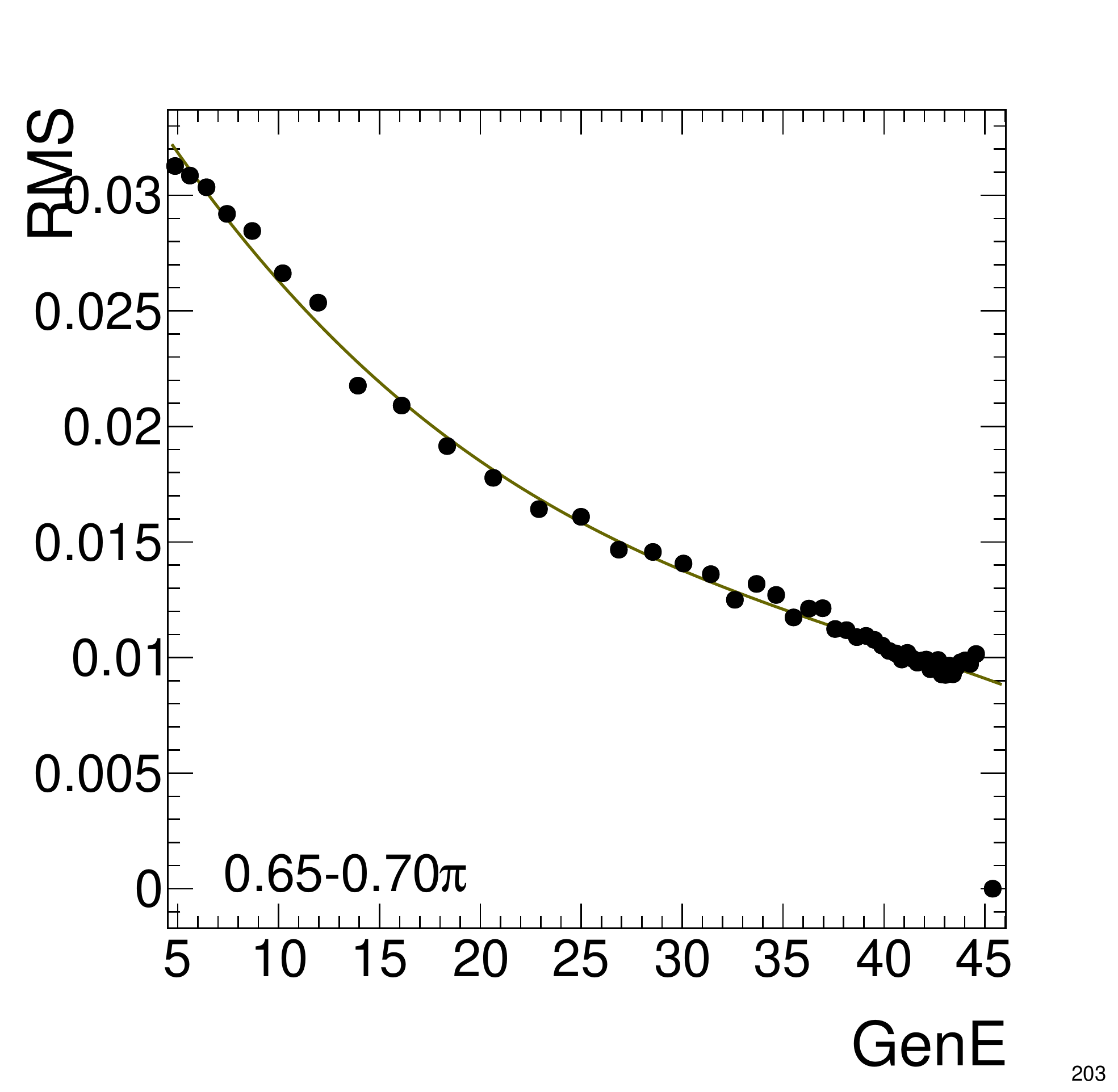}
    \includegraphicsfour{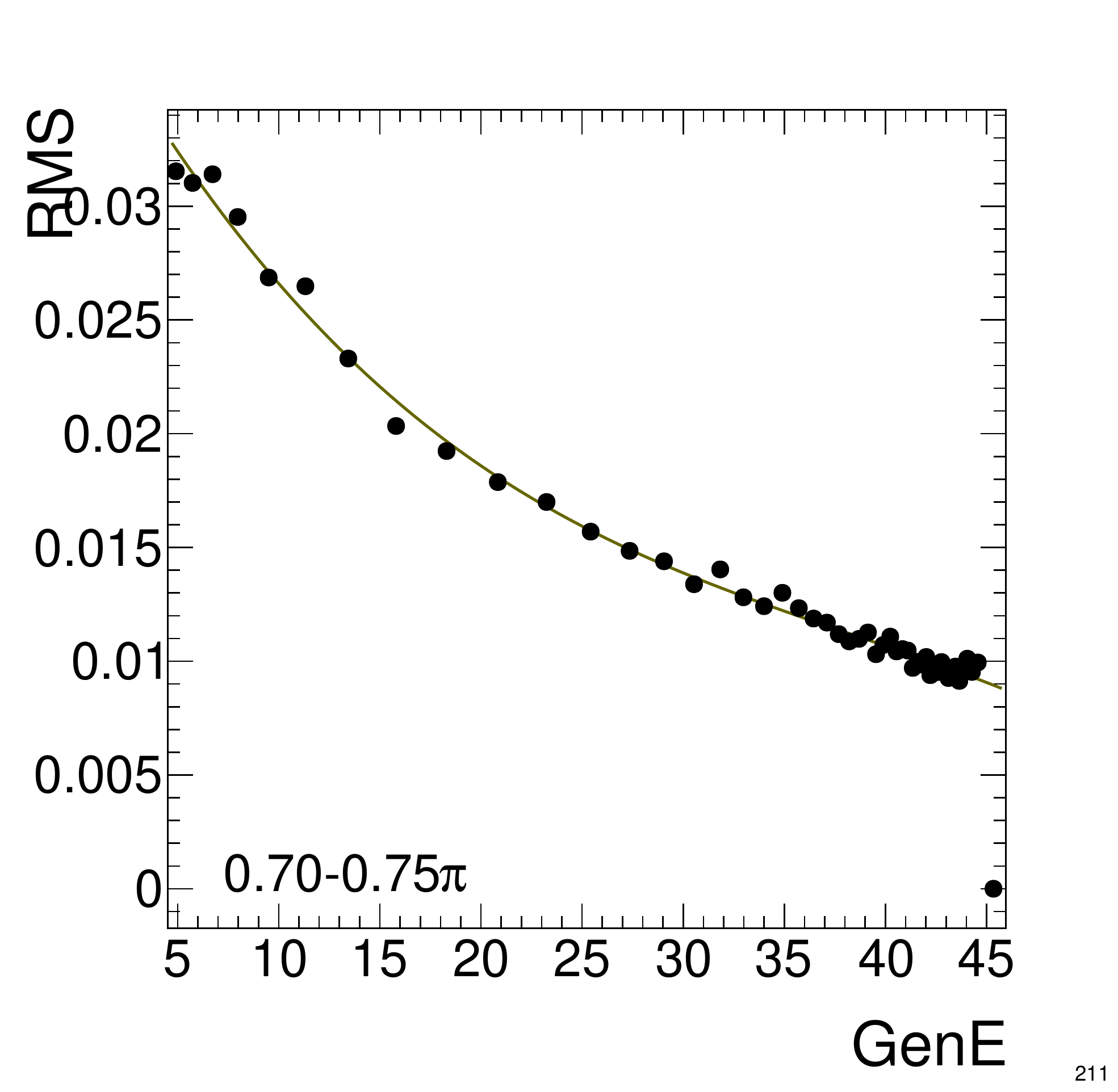}
    \includegraphicsfour{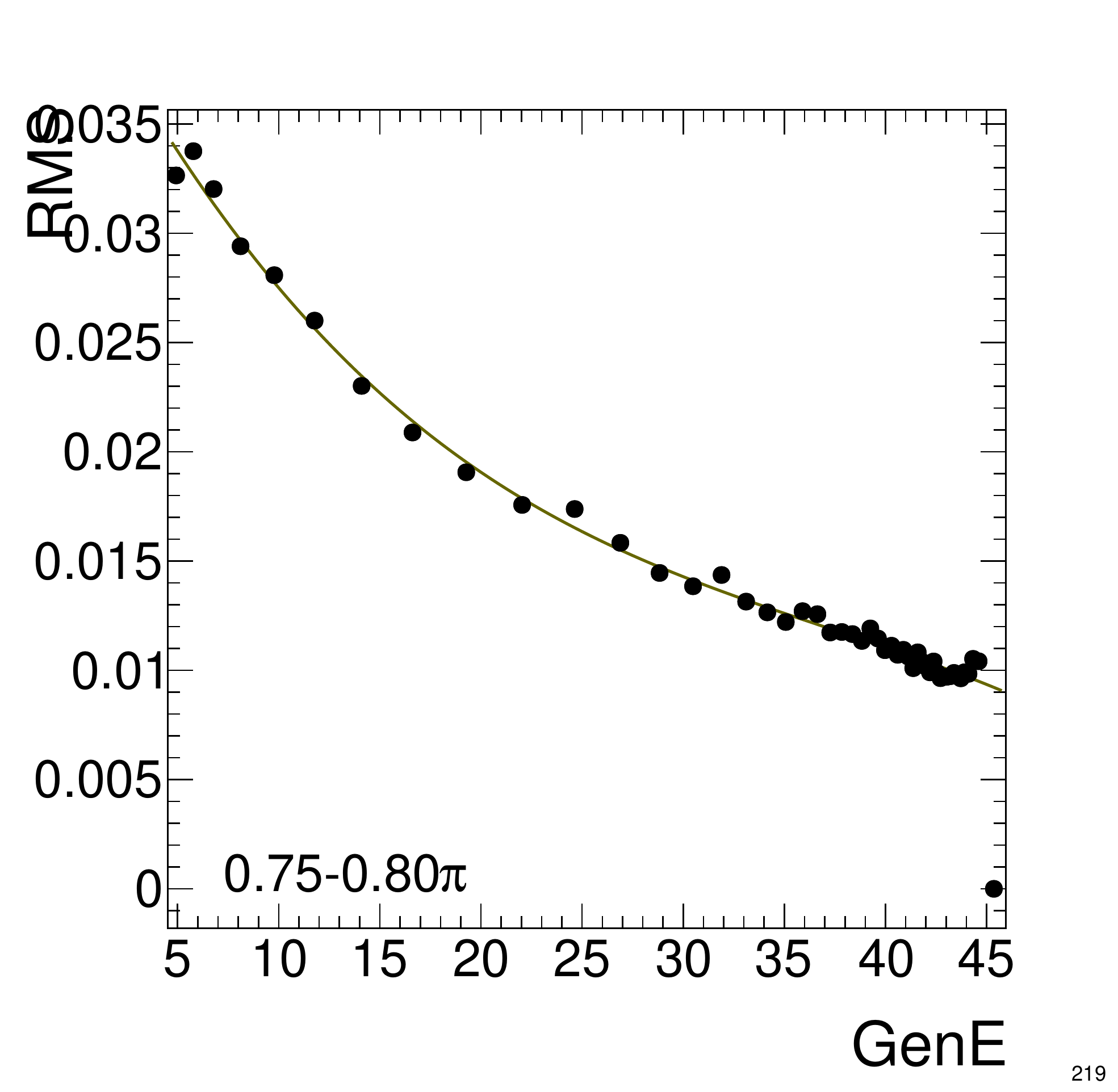}
    \includegraphicsfour{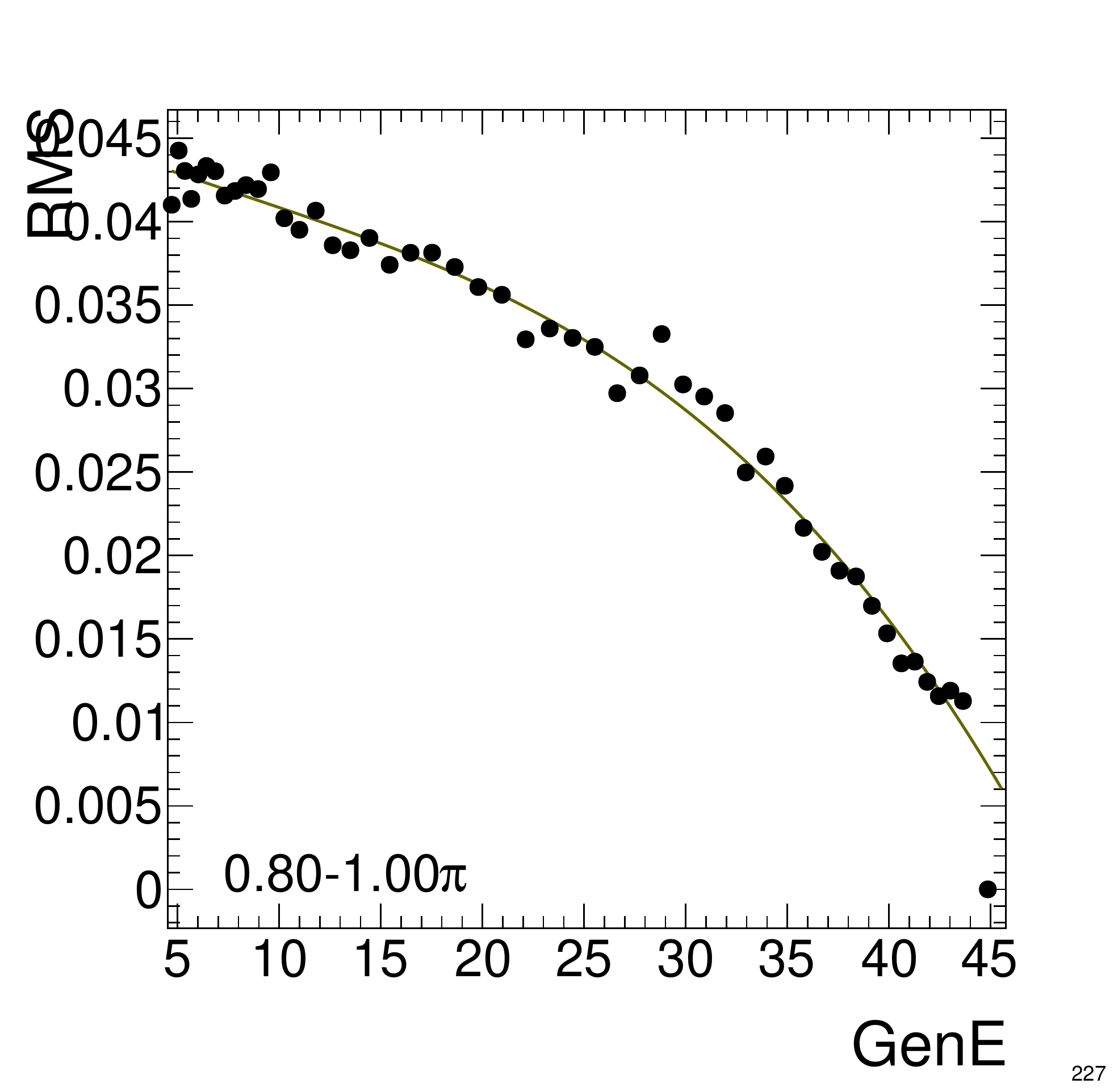}
    \caption{Jet $\theta$ resolution in simulated samples as a function of generated jet momentum in different bins of jet $\theta$.}
    \label{Figure:JetResolution-MCThetaResolution}
\end{figure}

\begin{figure}[htp!]
    \centering
    \includegraphicsfour{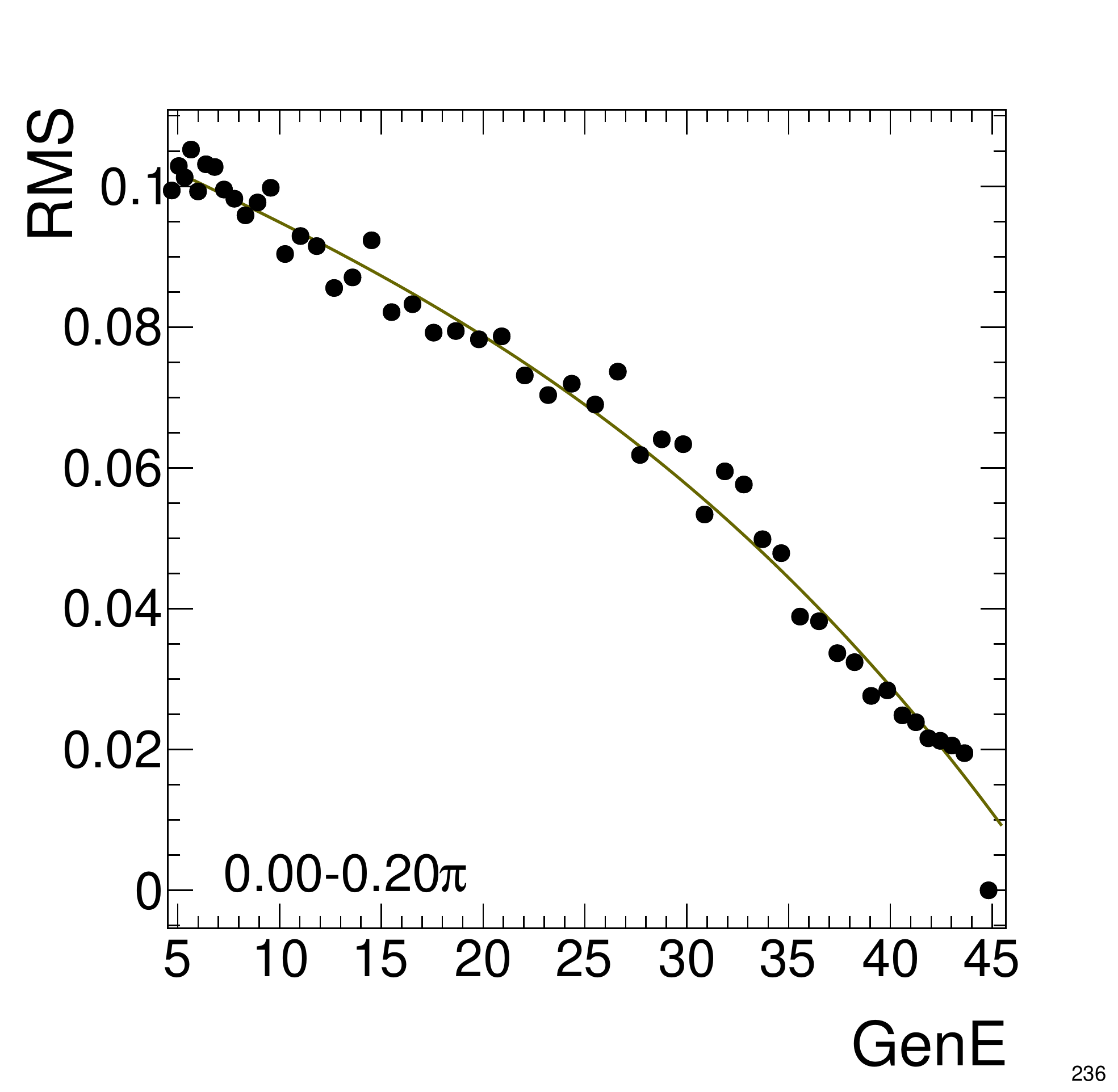}
    \includegraphicsfour{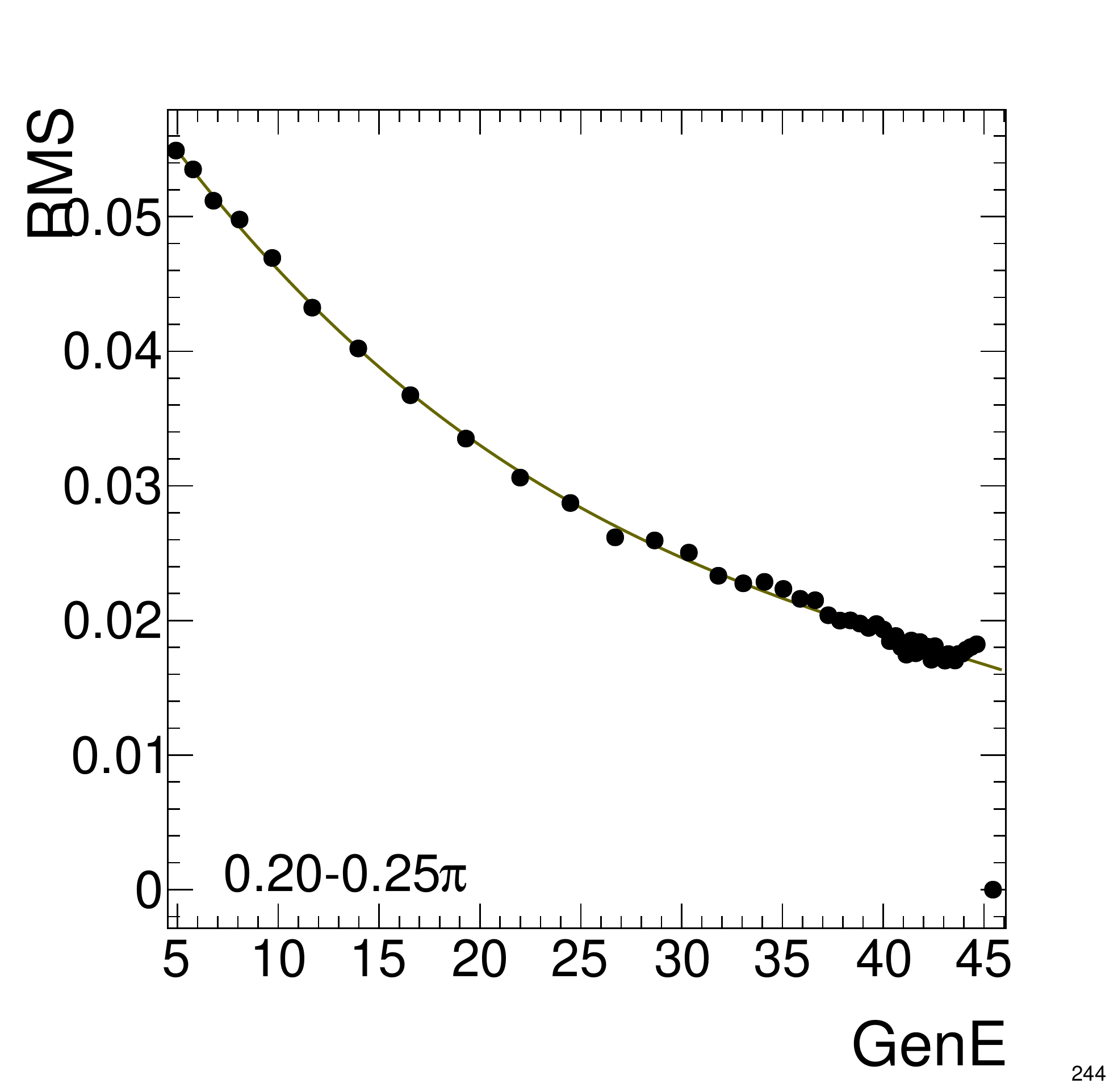}
    \includegraphicsfour{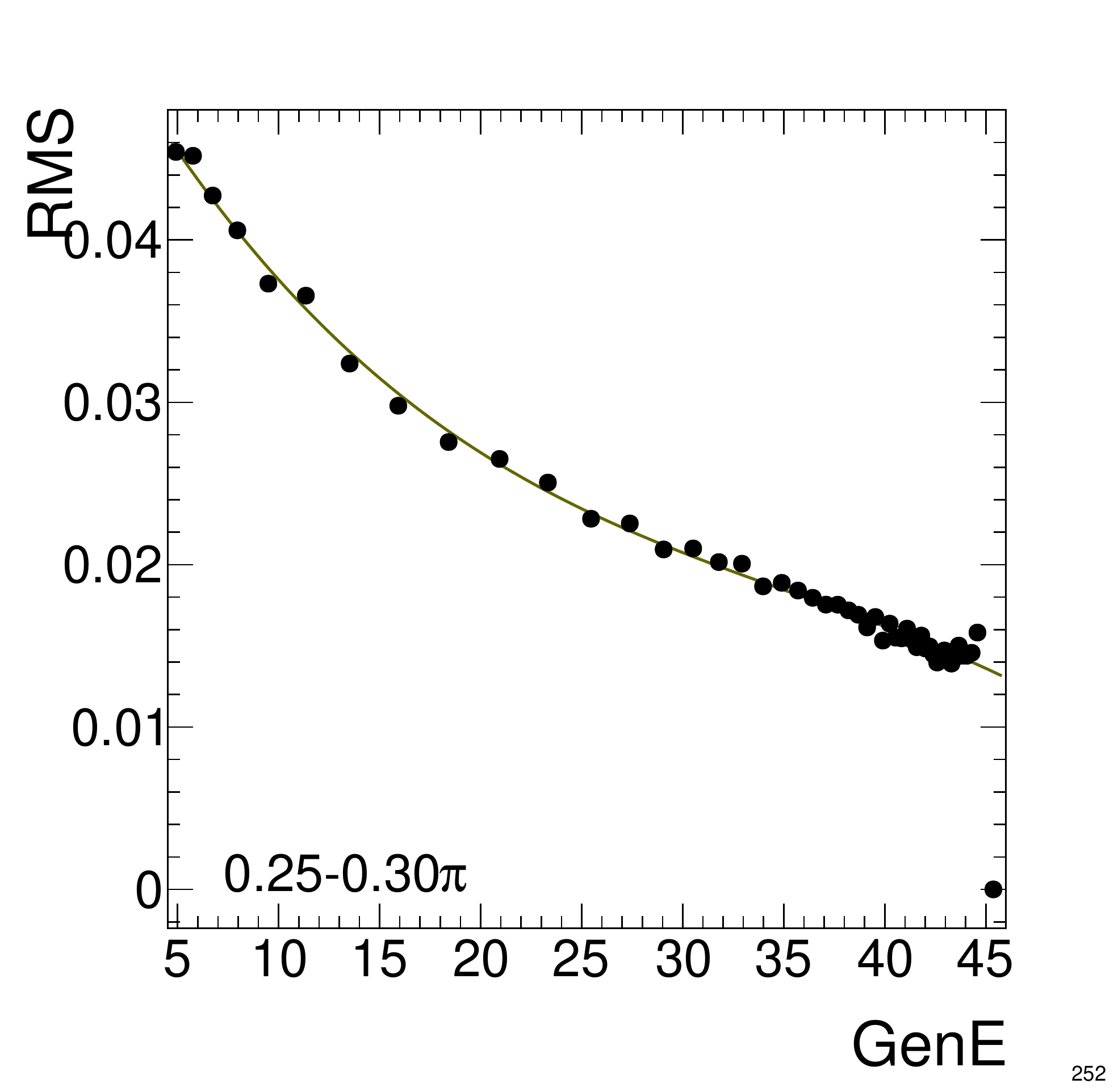}
    \includegraphicsfour{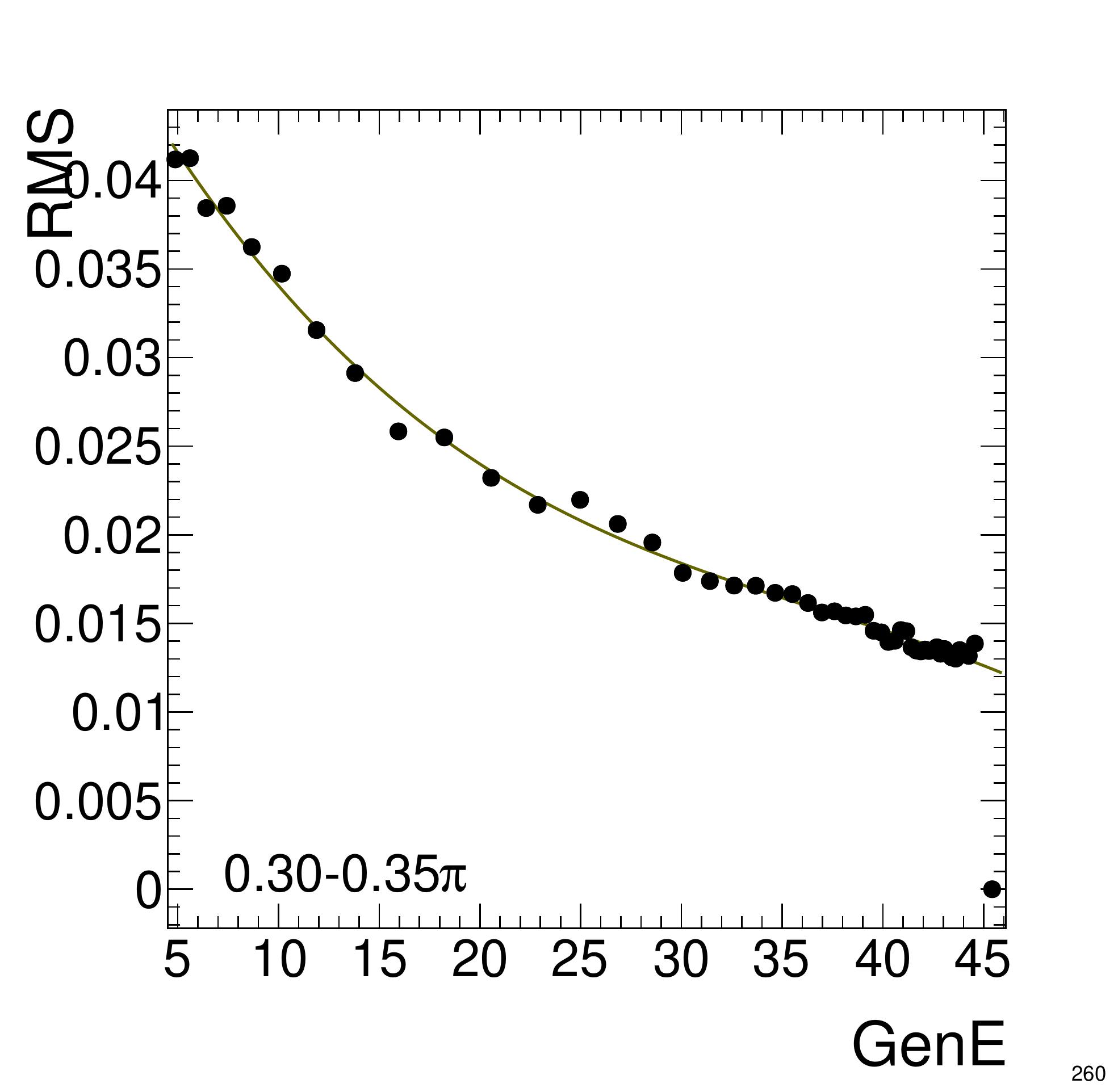}
    \includegraphicsfour{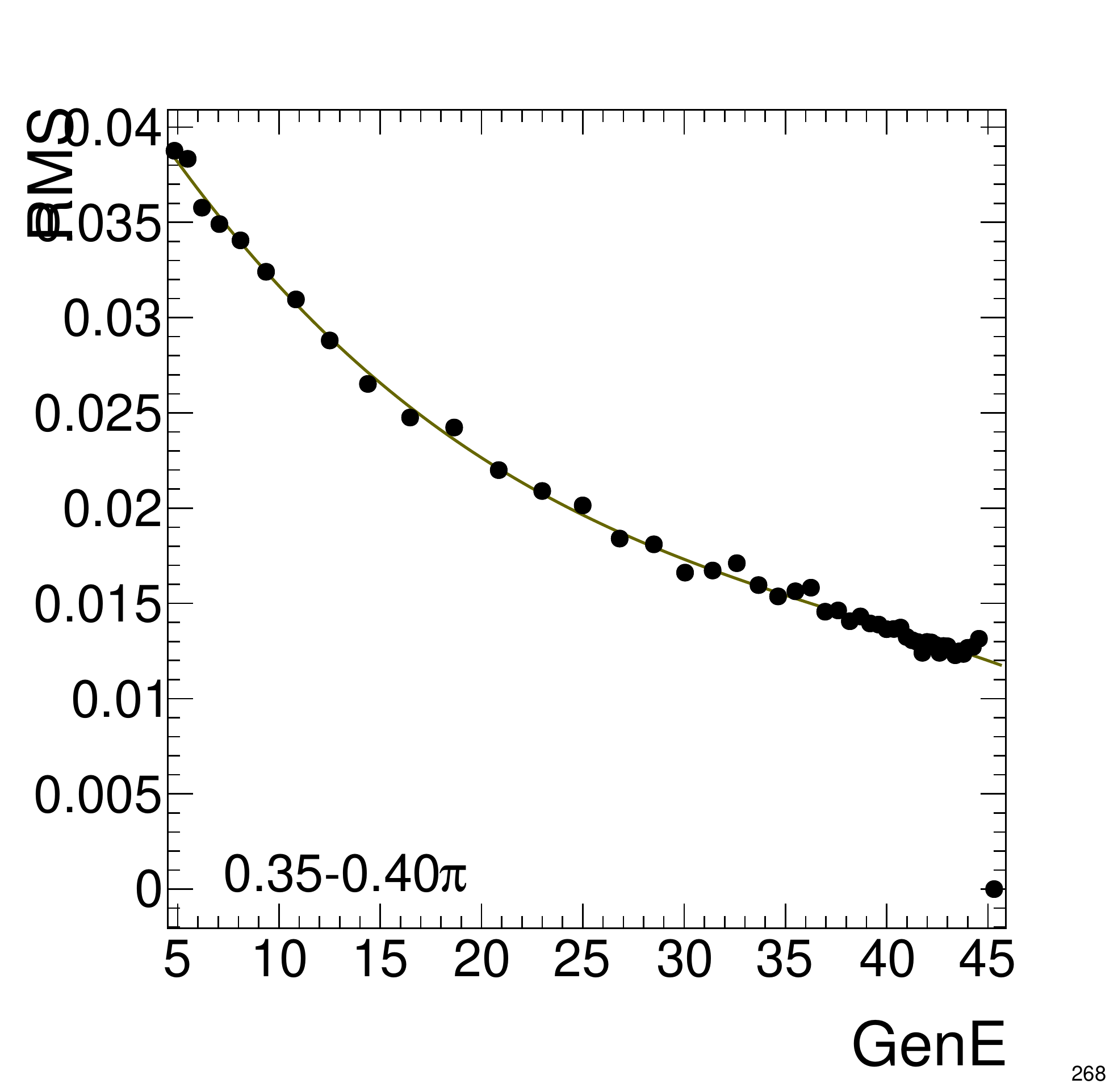}
    \includegraphicsfour{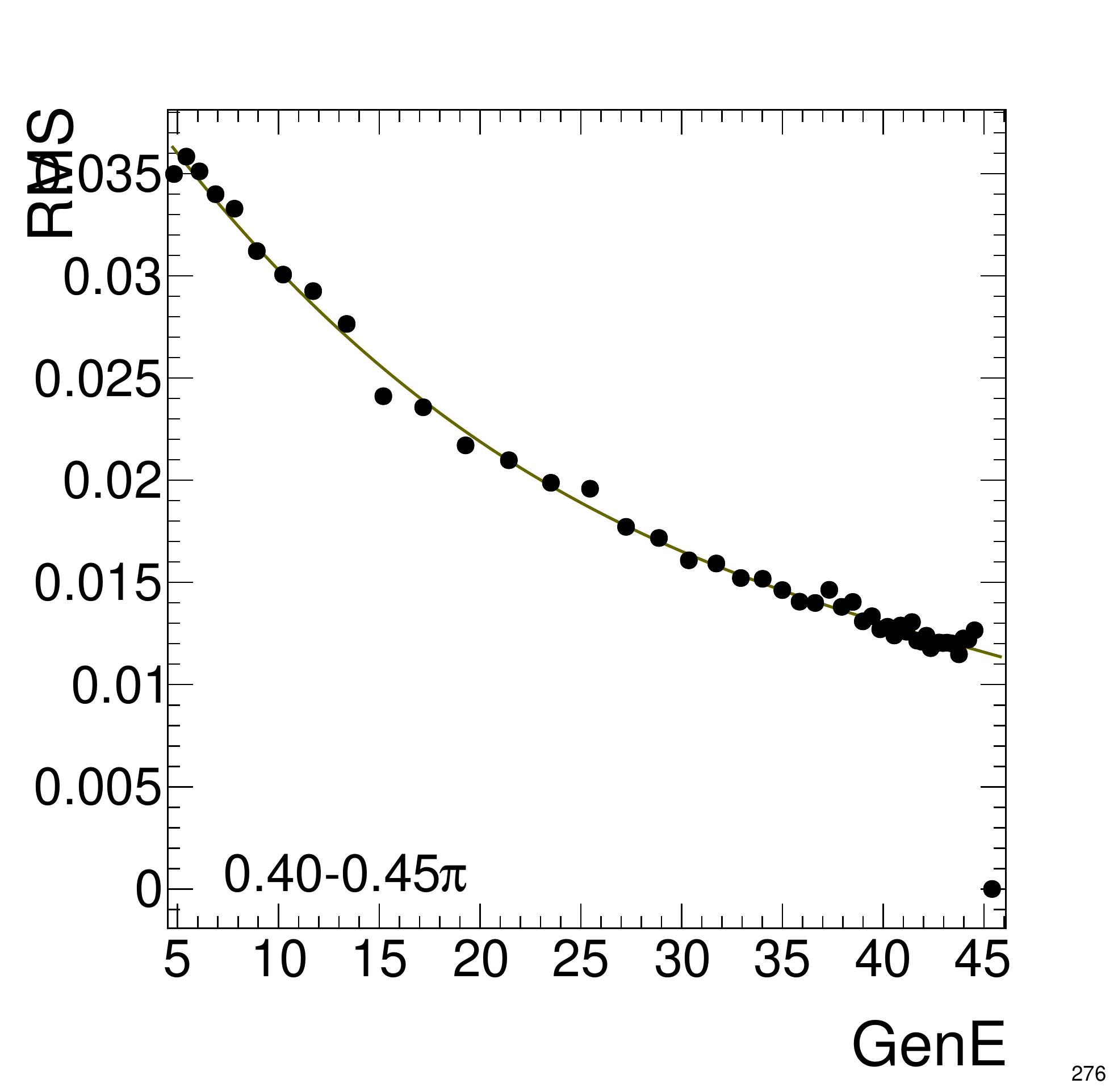}
    \includegraphicsfour{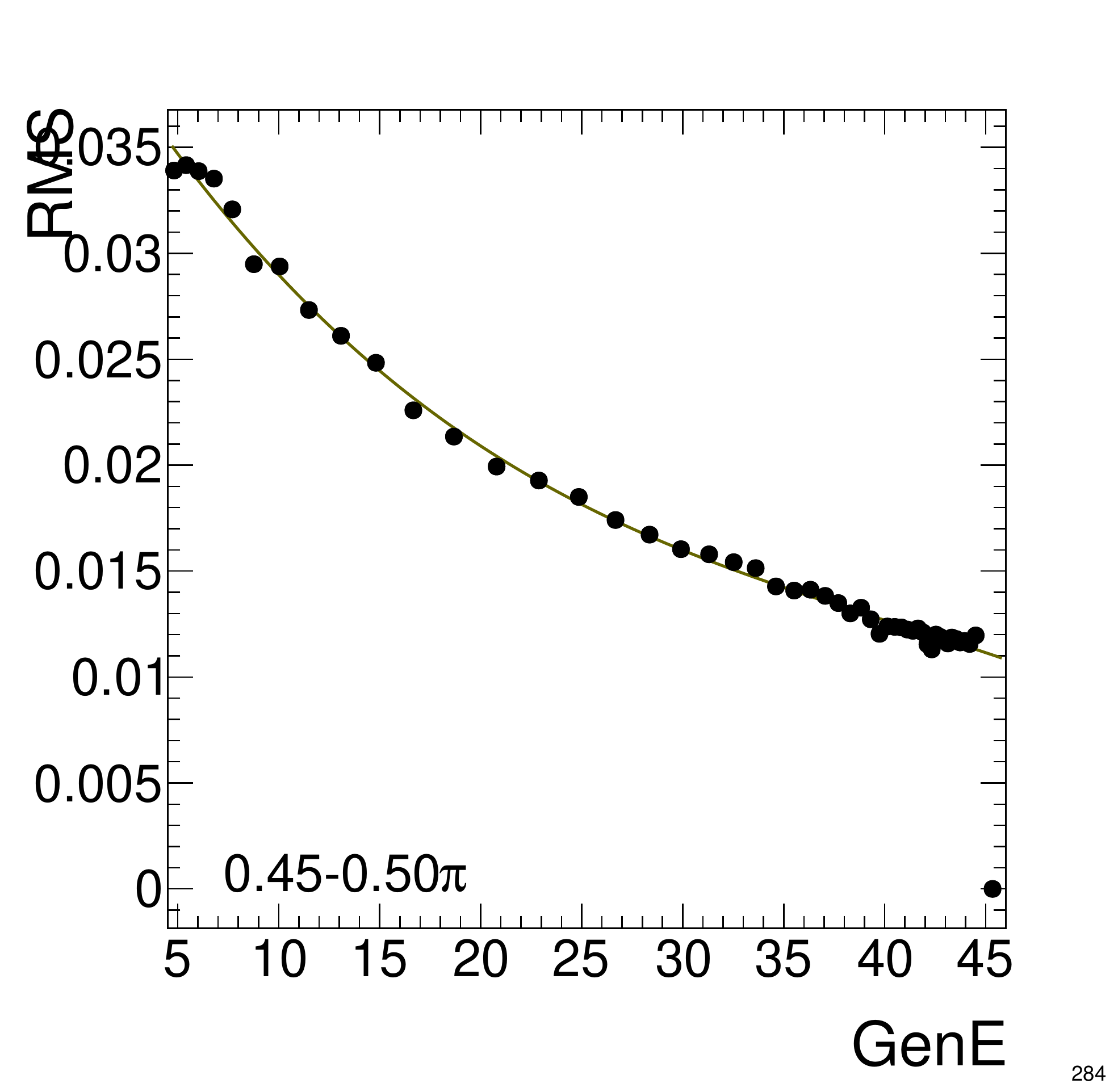}
    \includegraphicsfour{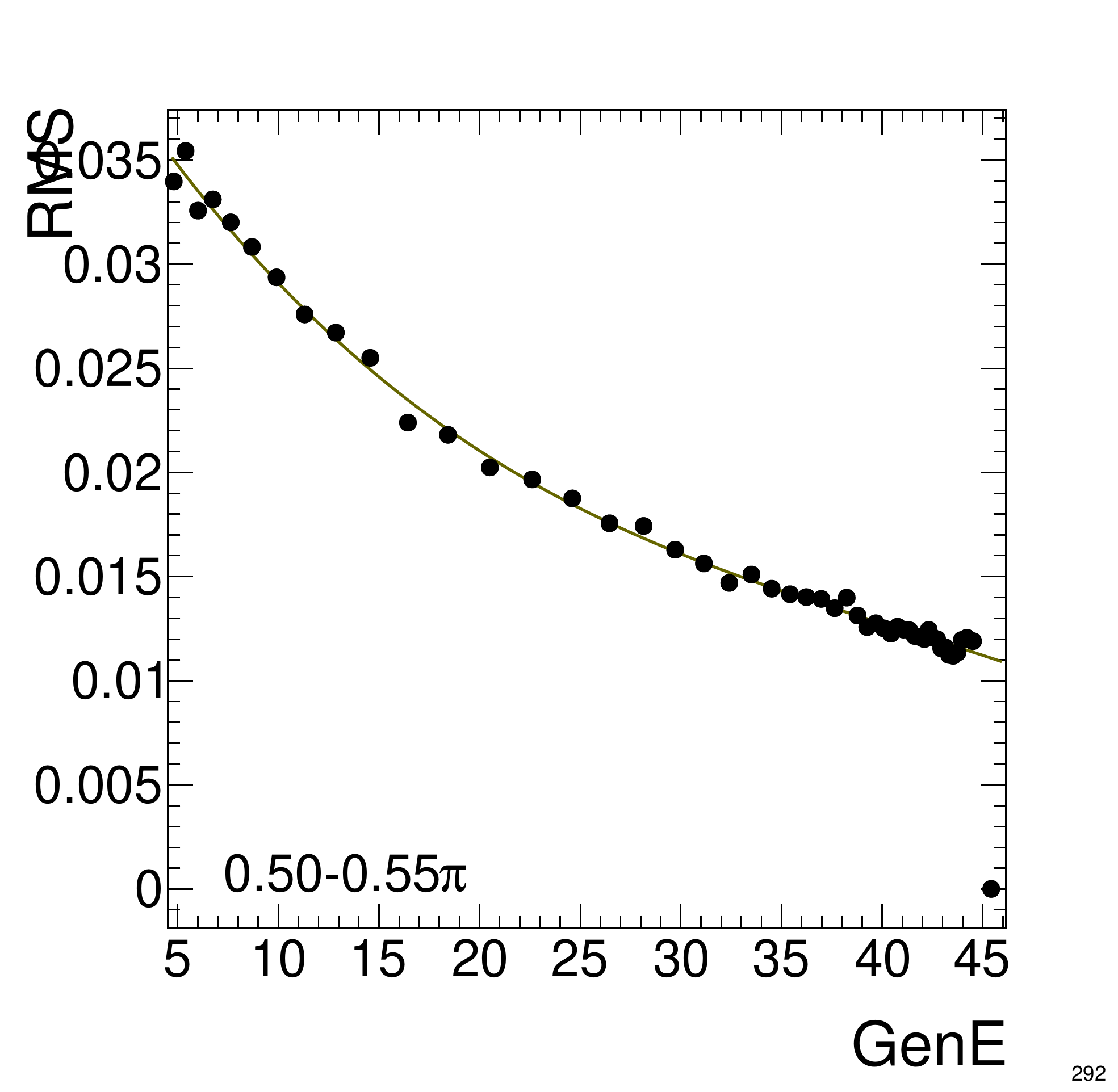}
    \includegraphicsfour{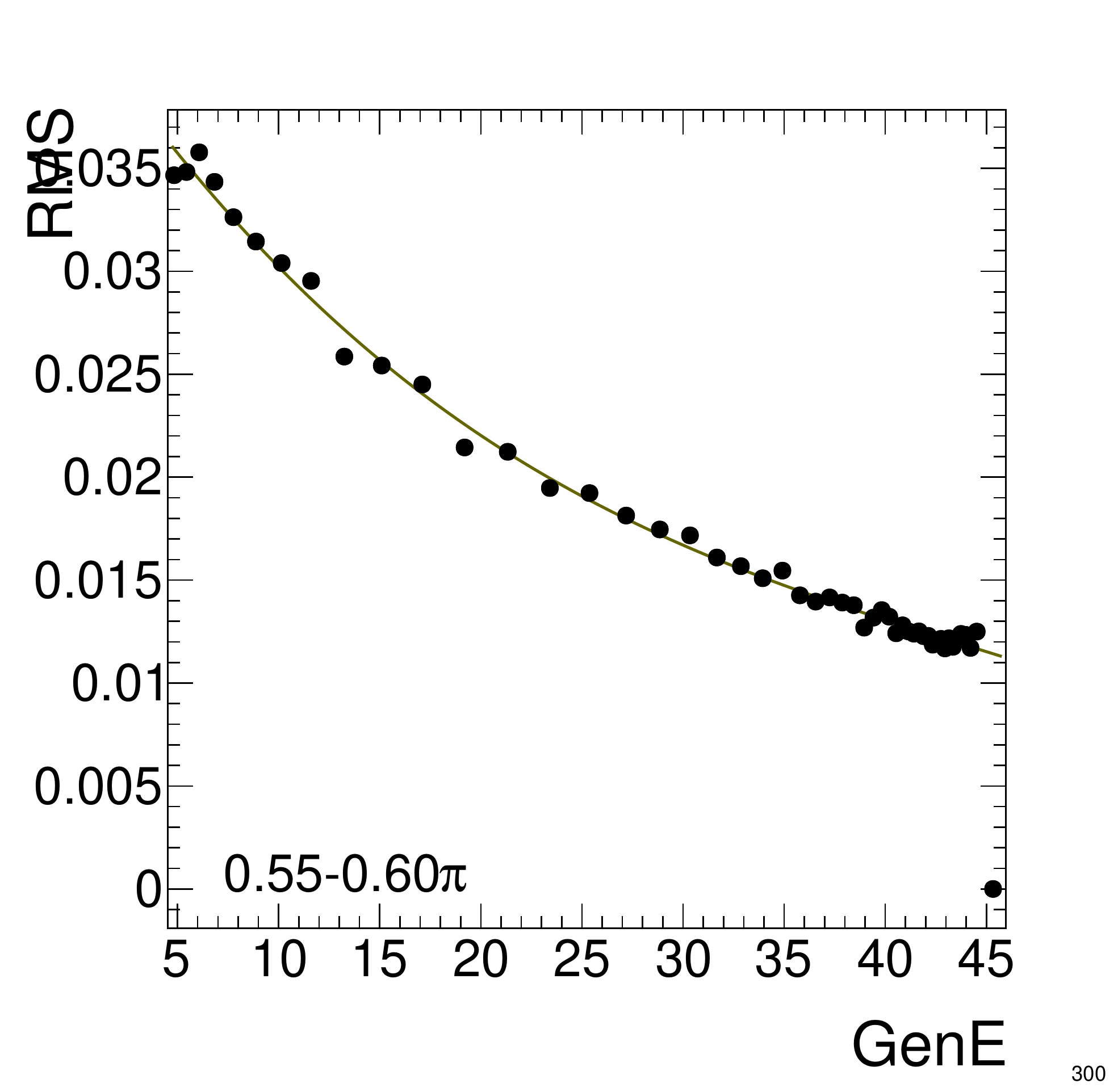}
    \includegraphicsfour{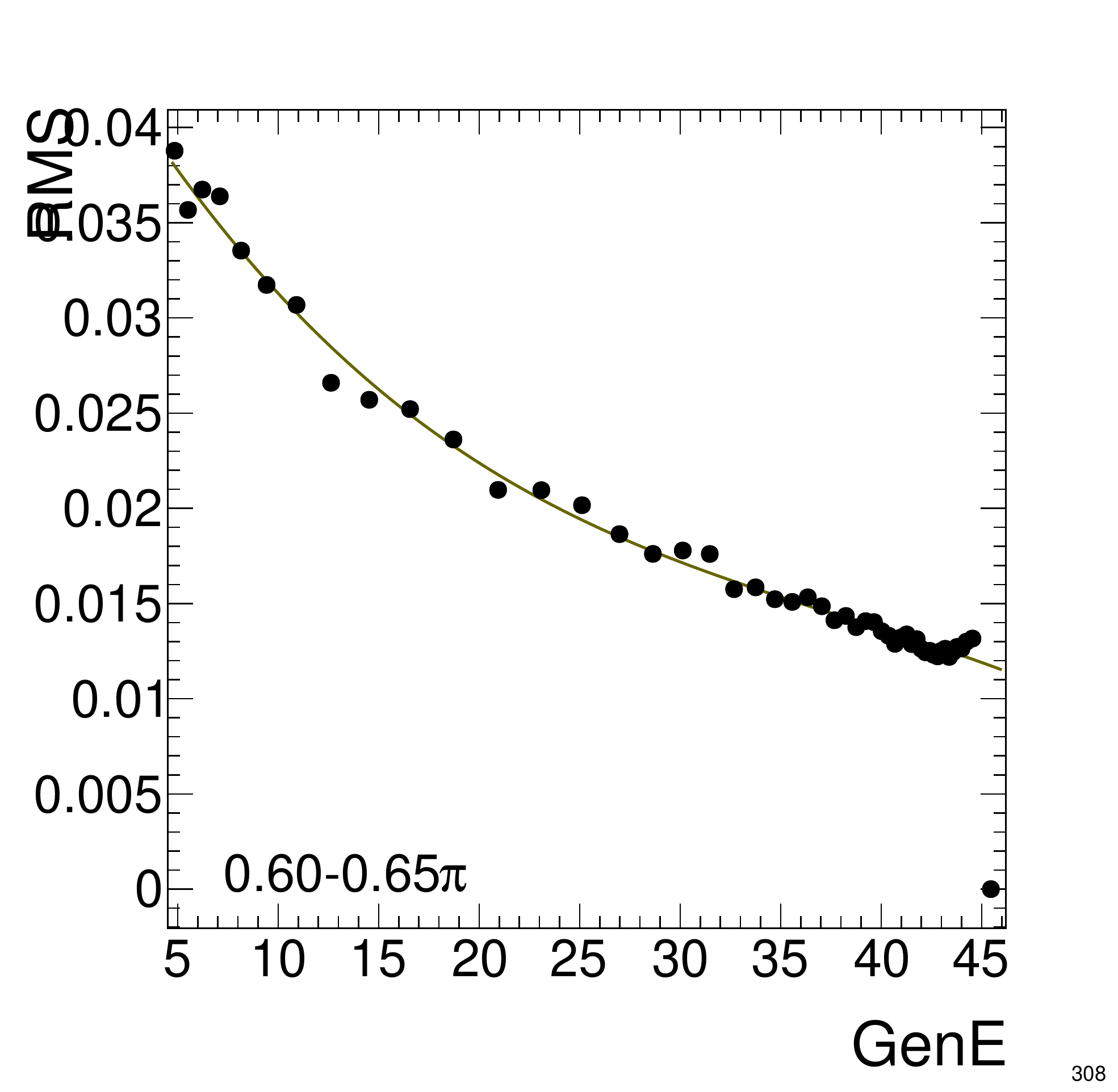}
    \includegraphicsfour{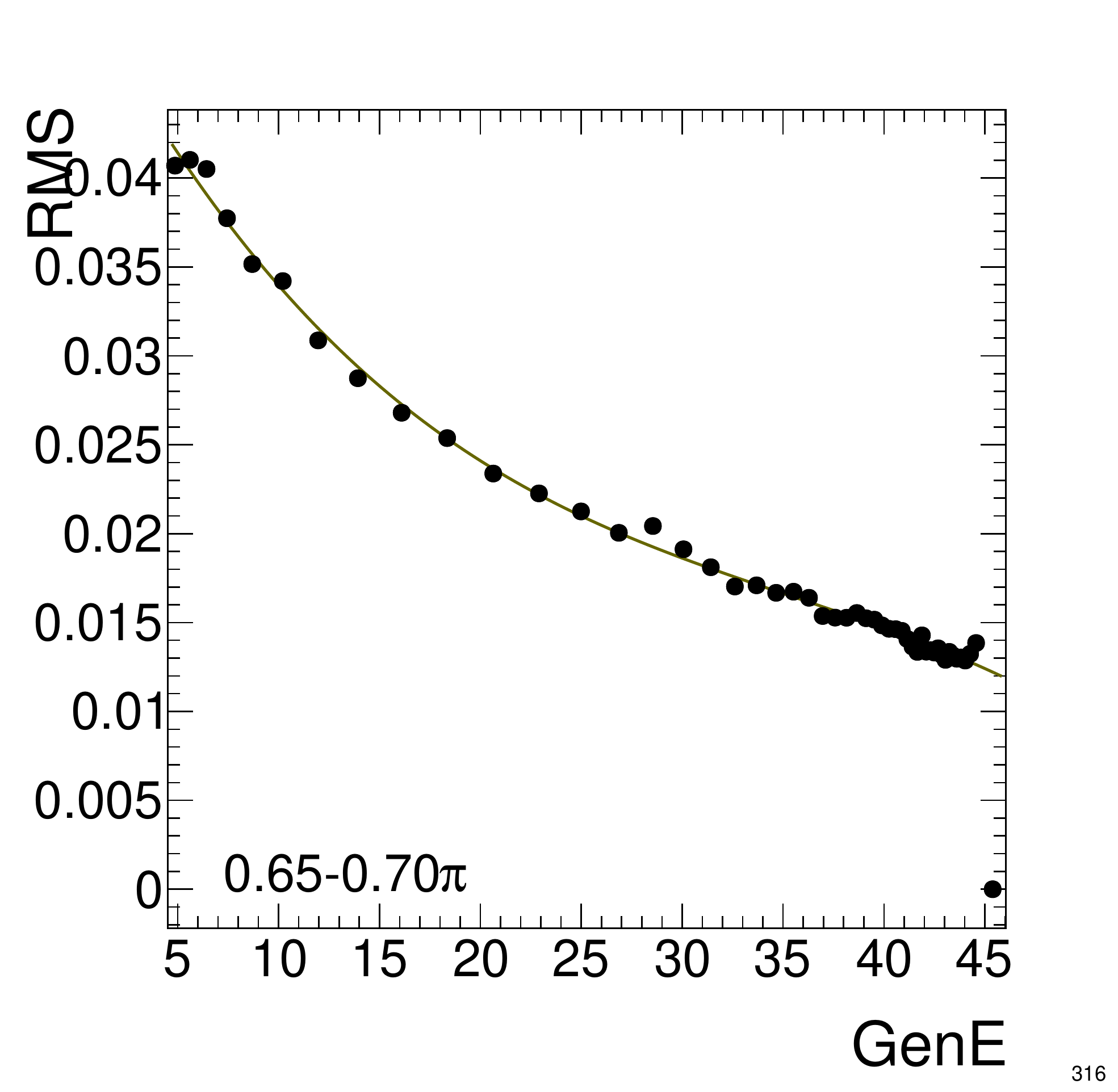}
    \includegraphicsfour{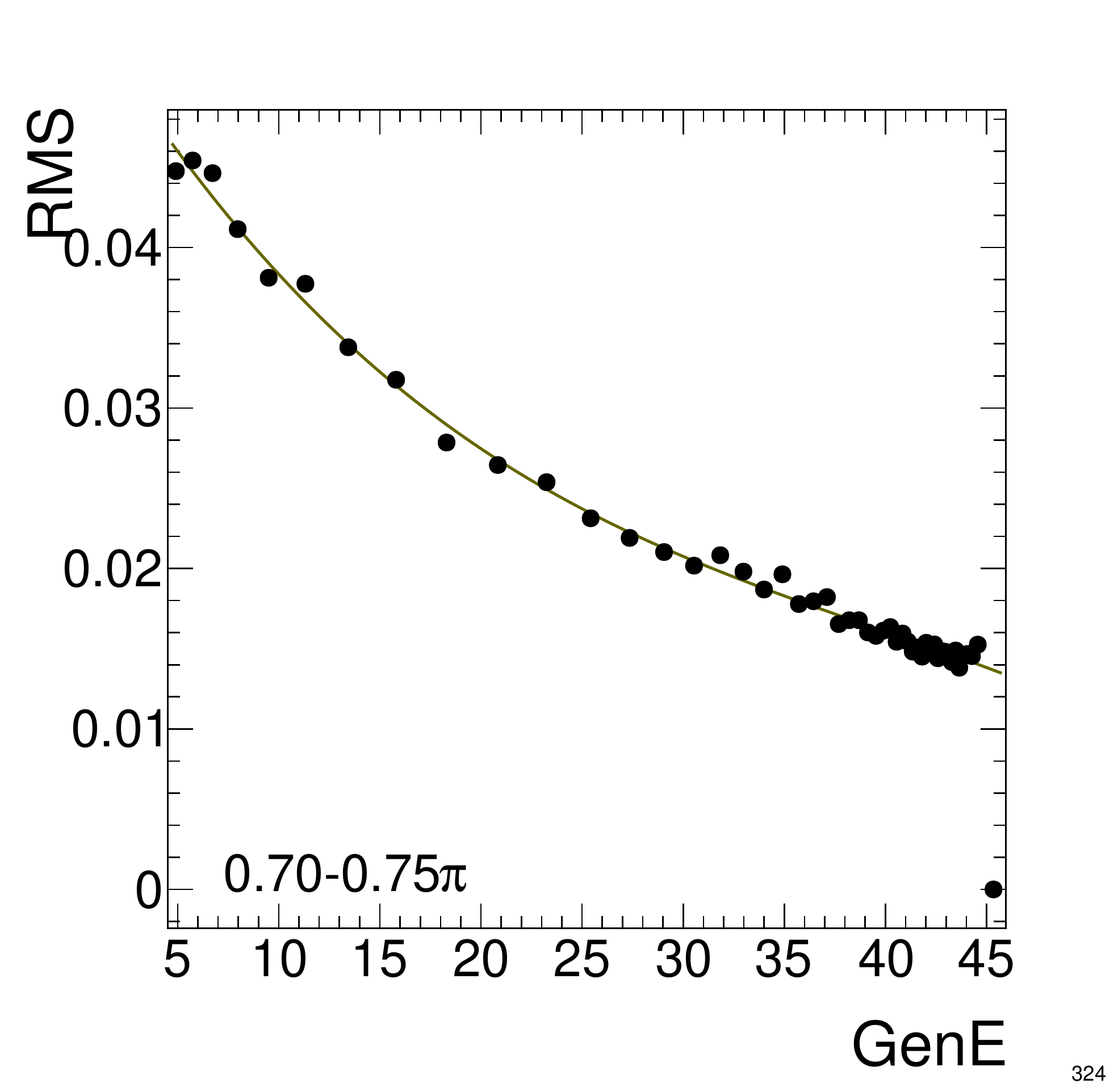}
    \includegraphicsfour{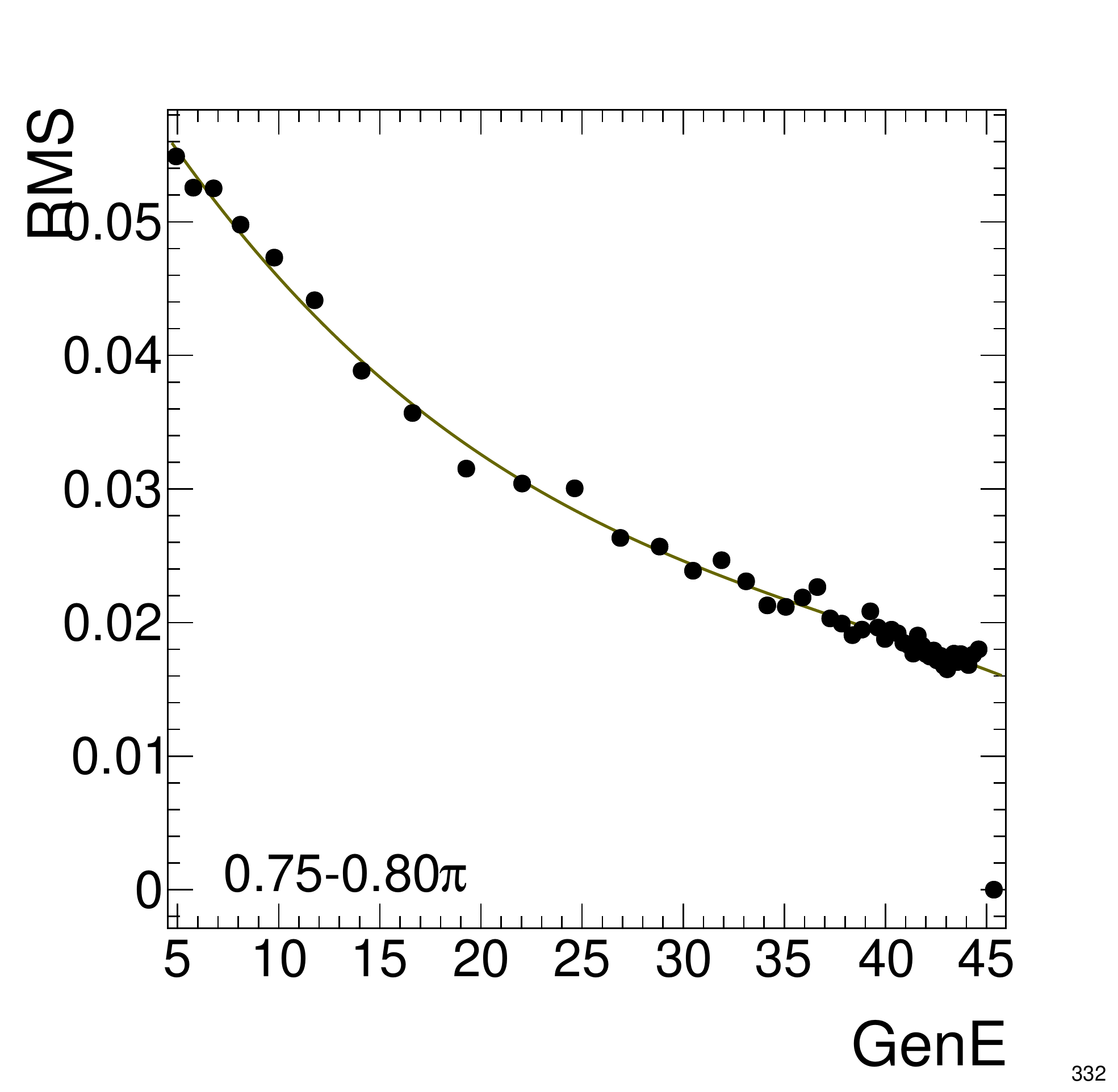}
    \includegraphicsfour{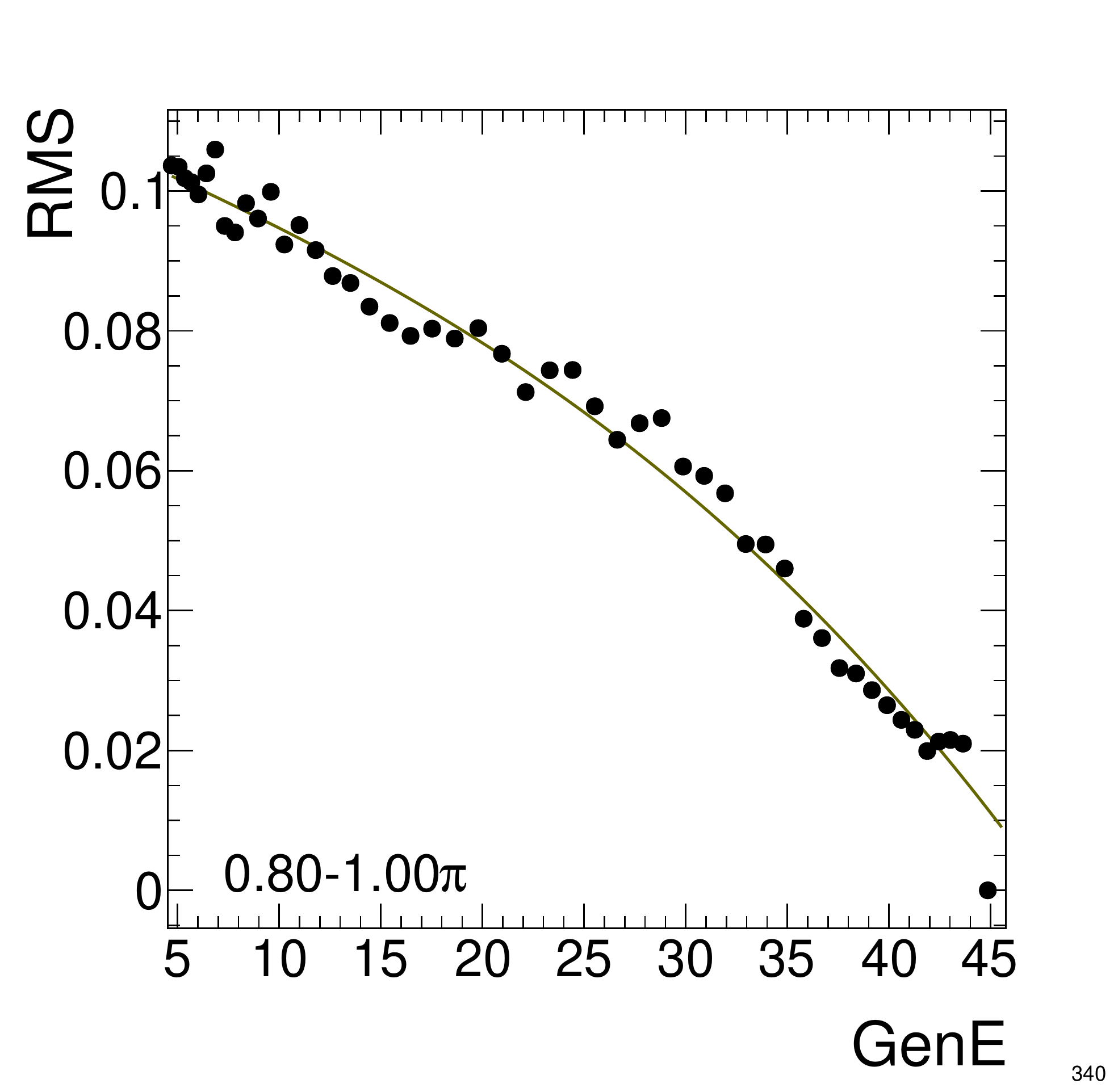}
    \caption{Jet $\phi$ resolution in simulated samples as a function of generated jet momentum in different bins of jet $\theta$.}
    \label{Figure:JetResolution-MCPhiResolution}
\end{figure}

The $\theta$ resolution is shown in Fig.~\ref{Figure:JetResolution-MCThetaResolution} and the $\phi$ resolution is shown in Fig.~\ref{Figure:JetResolution-MCPhiResolution}.  They are fitted with a simple third-order polynomial function.  Again when the jets are close to the beam line, the behavior is significantly different.  Away from the beam line, the resolution does not depend too much on the jet $\theta$.

\subsection{Jet Resolution Difference Between Data and Simulation}

The resolution in data is examined through leading dijets.  The effective resolution is extracted for data and simulation with the exact same selection, and the ratio between them is taken as the scale factor needed to smear simulated jets to match the resolution in data.  Due to the available amount of statistics, only a constant scale factor (as a function of momentum) is extracted, and it applies primarily to jets with energy around the kinematic peak.

\begin{figure}[htp!]
    \centering
    \includegraphicsone{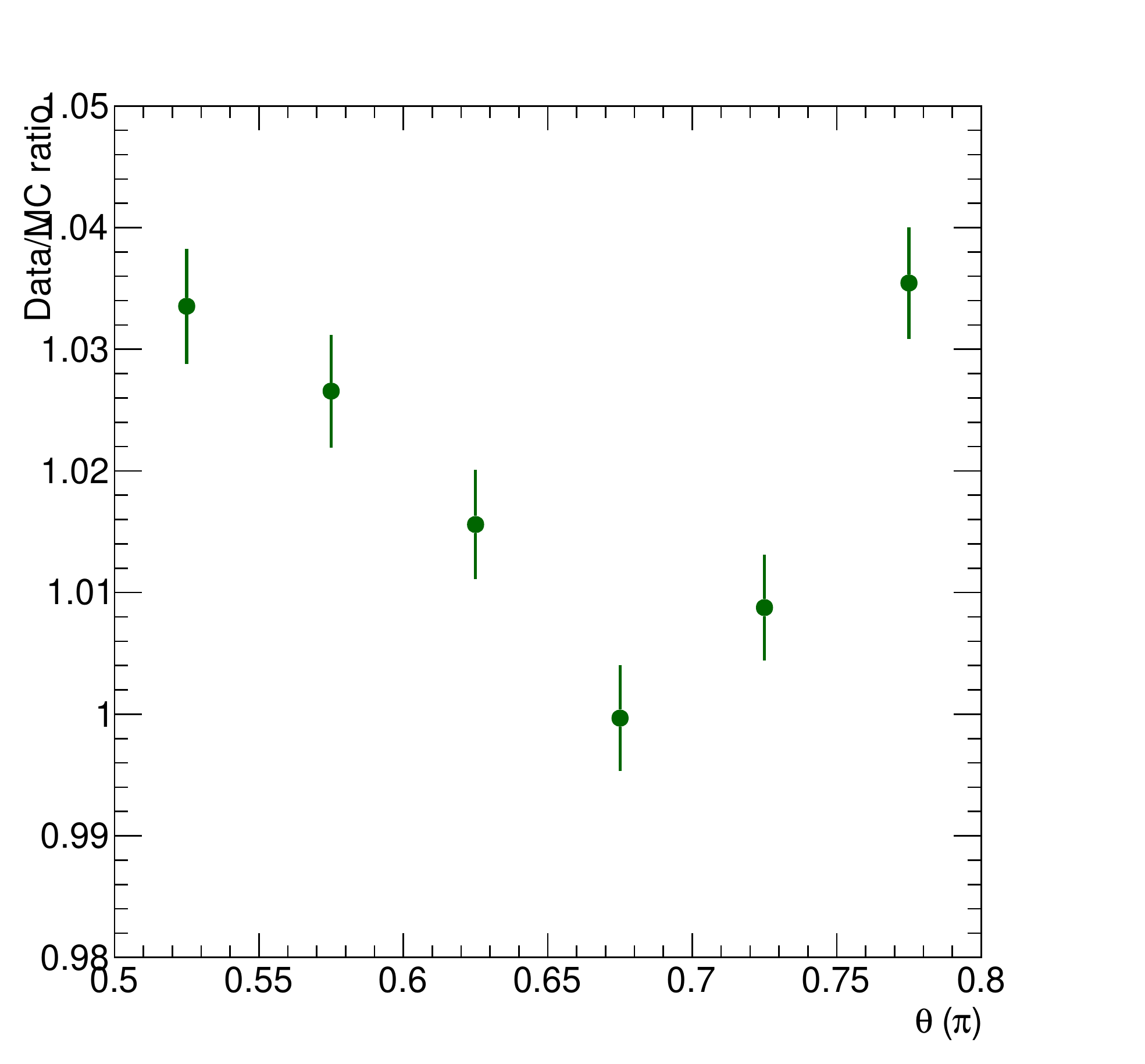}
    \caption{Derived jet resolution ratio between data and simulation in different bins of jet $\theta$.  Only the plus side is shown, but it is symmetrized to the negative side as well.}
    \label{Figure:JetResolution-ResolutionScaleFactor}
\end{figure}

The result is shown in Fig.~\ref{Figure:JetResolution-ResolutionScaleFactor} for $N = 9, X = 3$ GeV.  The resolution scale factor needed ranges from 0--5\%, depending on the jet direction.  Varying $X$ from 3 to 5 GeV changes the scale factor by around 1\%.

\clearpage
\section{Leading Jet Selection}\label{Section:LeadingJet}

When the ``global leading jets'', the leading jets in the event without detector acceptance requirement, overlap with the dead region (close to the beam), some of the jet energy will not be detected, and they will appear as lower energy jets.  Therefore additional selection is designed in order to ensure that the leading jets within acceptance ($0.2\pi < \theta_\text{jet} < 0.8\pi$) are the global leading jets, without actually requiring the event-wide leading jet direction.

\subsection{Total Visible Energy}

The simplest quantity that one can think of is the total energy of all visible particles within the acceptance ($0.2\pi < \theta < 0.8\pi$).  Since in an \ee collision, the total energy is known, a low value of total visible momentum implies that there is a higher chance of a jet overlapping with the dead region.

\begin{figure}[htp!]
    \centering
    \includegraphicstwo{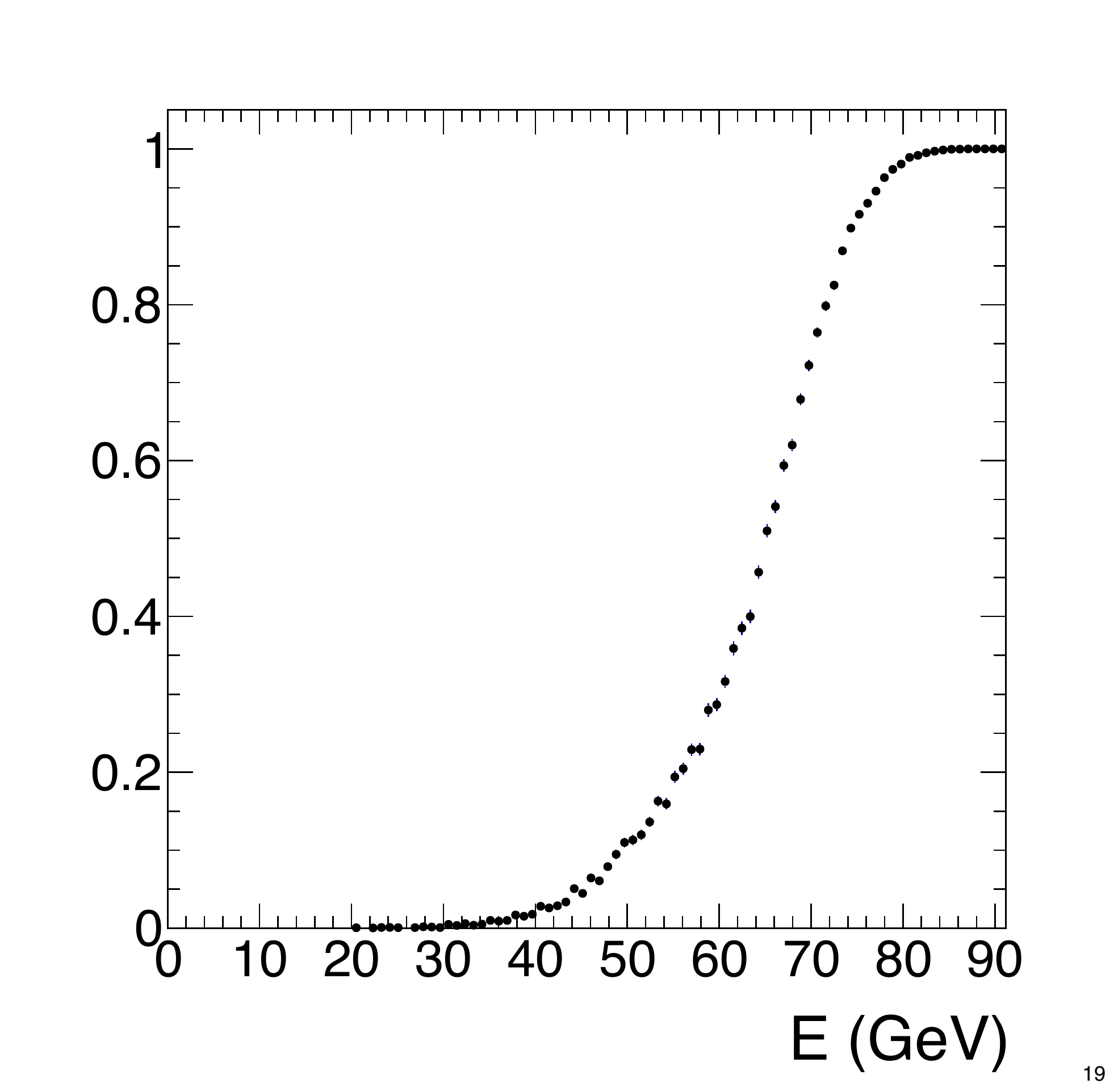}
    \includegraphicstwo{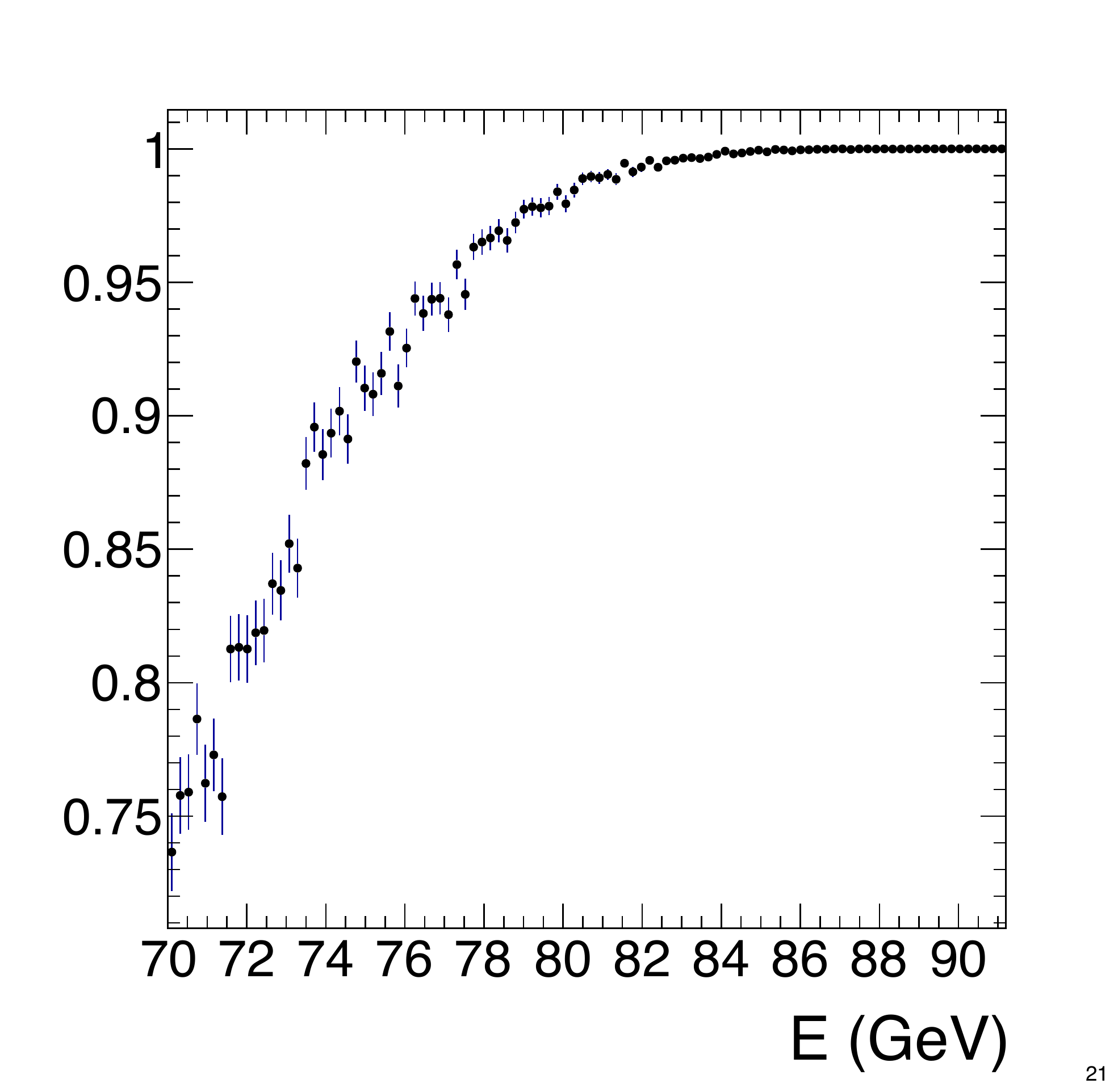}
    \caption{Left: purity of events where leading dijets are inside the acceptance as a function of the total visible energy.  Right: zoomed-in version of the same plot.}
    \label{Figure:LeadingJet-SumEFraction}
\end{figure}

\begin{figure}[htp!]
    \centering
    \includegraphicsonesmall{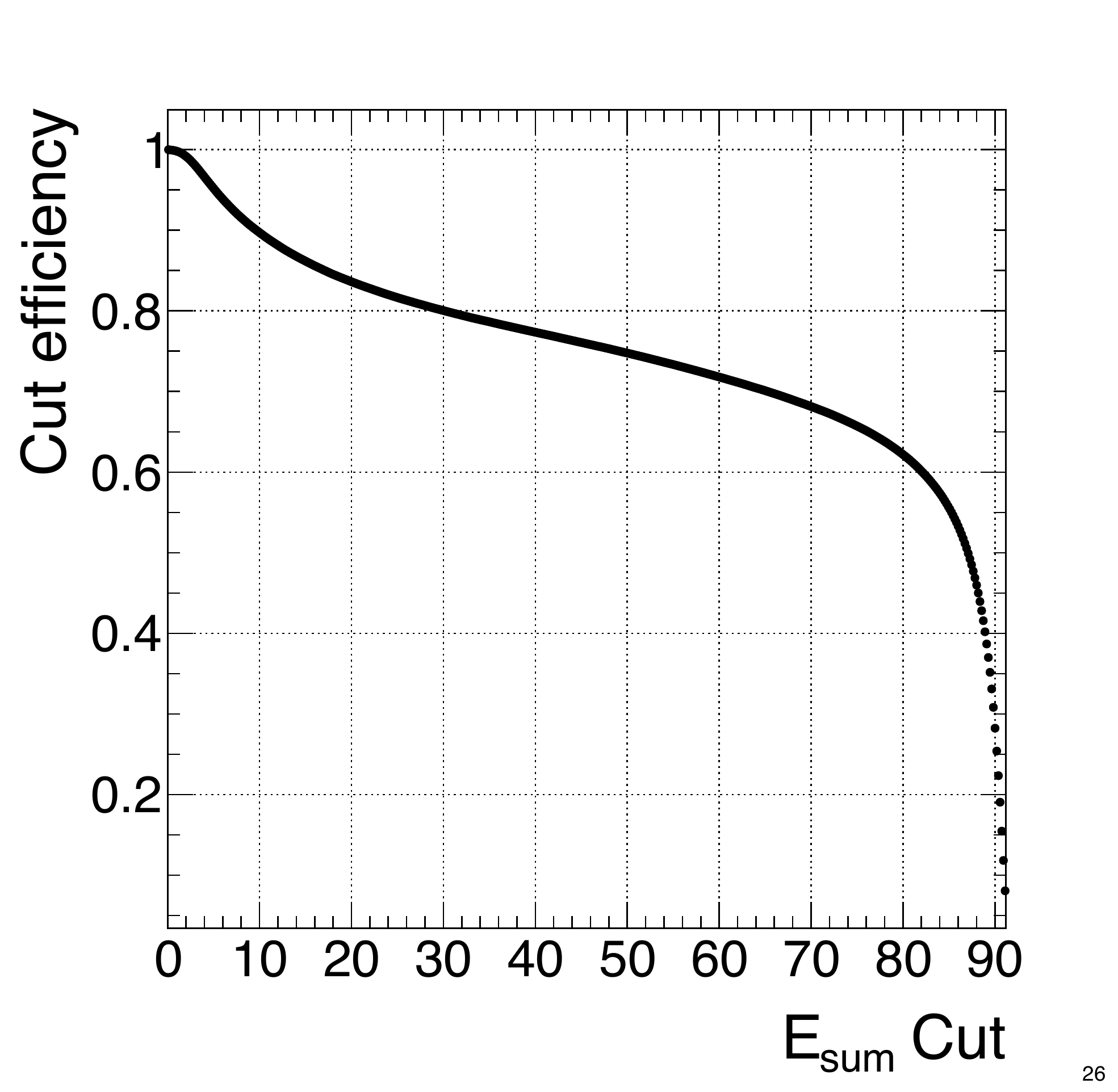}
    \caption{Cut efficiency as a function of the total visible energy cut position.}
    \label{Figure:LeadingJet-SumEEfficiency}
\end{figure}

Figure~\ref{Figure:LeadingJet-SumEFraction} shows the fraction of events with both leading jets inside the acceptance as a function of the total visible particle energy within the same acceptance.  The efficiency of the cut, defined as the percentage of events passing said total energy cut, can be found in Fig.~\ref{Figure:LeadingJet-SumEEfficiency}.  A cut at 83~GeV, which corresponds to about 99\% purity, is about 60\% efficient.  

Since jets are extended objects, if a jet is too close to the edge of the acceptance, some of the jet energy will leak out of the acceptance region and effectively lowers the total visible energy.  Therefore \textit{a priori} we expect the cut efficiency to be between $1-2\times \frac{2\pi(1-\cos(0.2\pi))}{4\pi} = 80.9\%$ and $1-2\times\frac{2\pi(1-\cos(0.2\pi+0.4))}{4\pi} = 51.6\%$, which is obtained by assuming perfect back to back dijets and a spherically uniform distribution.

\begin{figure}[htp!]
    \centering
    \includegraphicsone{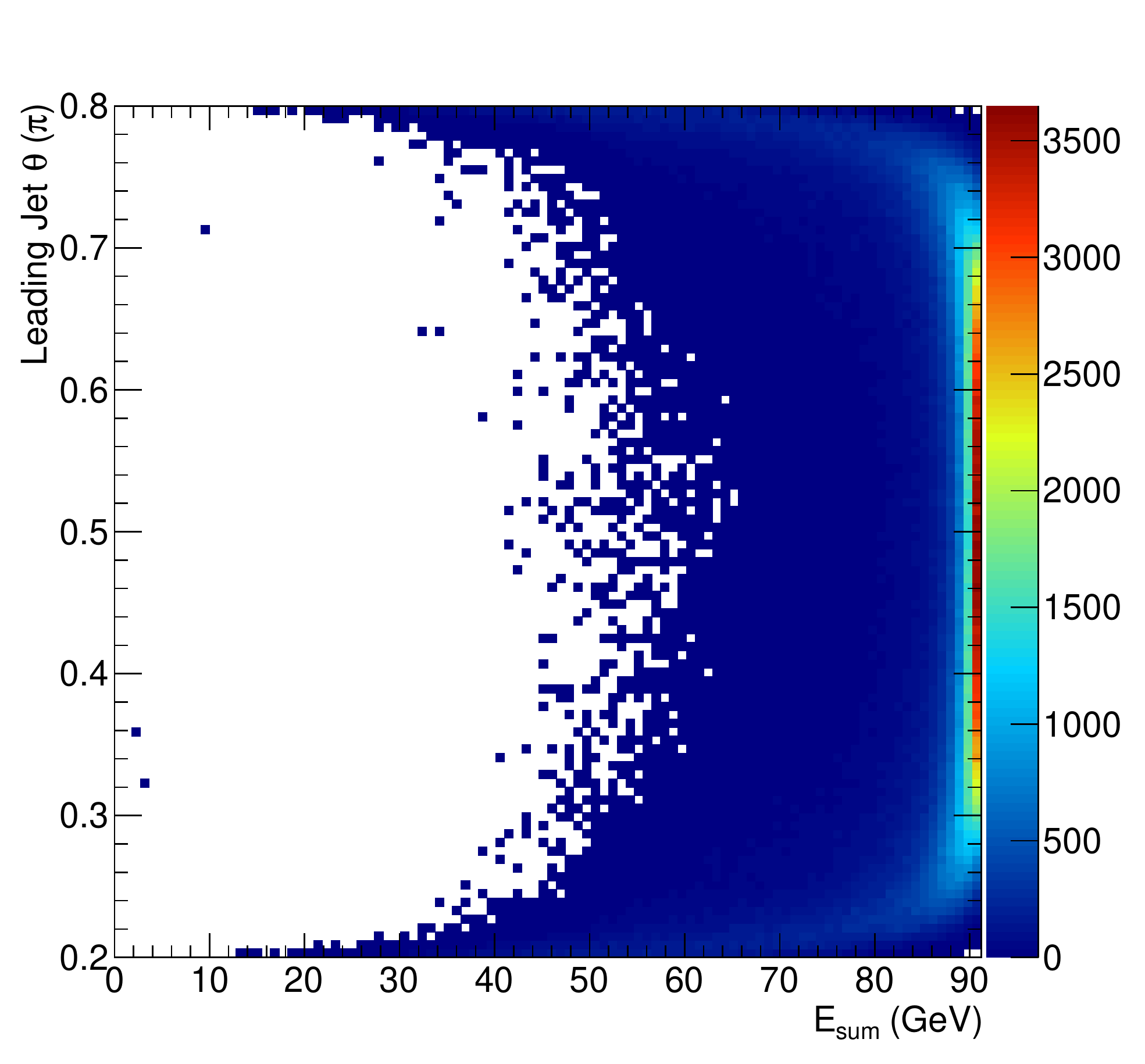}
    \caption{Correlation between leading jet $\theta$ and total visible energy.  When the leading jet is close to the edge of the acceptance, a visible distortion is seen on the total visible energy due to the leaked energy.}
    \label{Figure:LeadingJet-SumEJetTheta}
\end{figure}

This effect can be more clearly seen in Fig.~\ref{Figure:LeadingJet-SumEJetTheta}, where the total visible energy within acceptance is correlated with the leading jet direction.  A clear dependence of the total visible energy when the leading jet is close to the edge is observed.

It can be improved by using a wider range of acceptance for particles compared to the jets.  This however has a negative side effect of significantly lowering the purity of the cut.  In order to maintain similar purity, the cut position needs to be raised, thereby lowering the efficiency.

Due to this reason, a refined quantity, the hybrid total energy, is considered.

\subsection{Hybrid Total Energy}

\begin{figure}[htp!]
    \centering
    \includegraphicsone{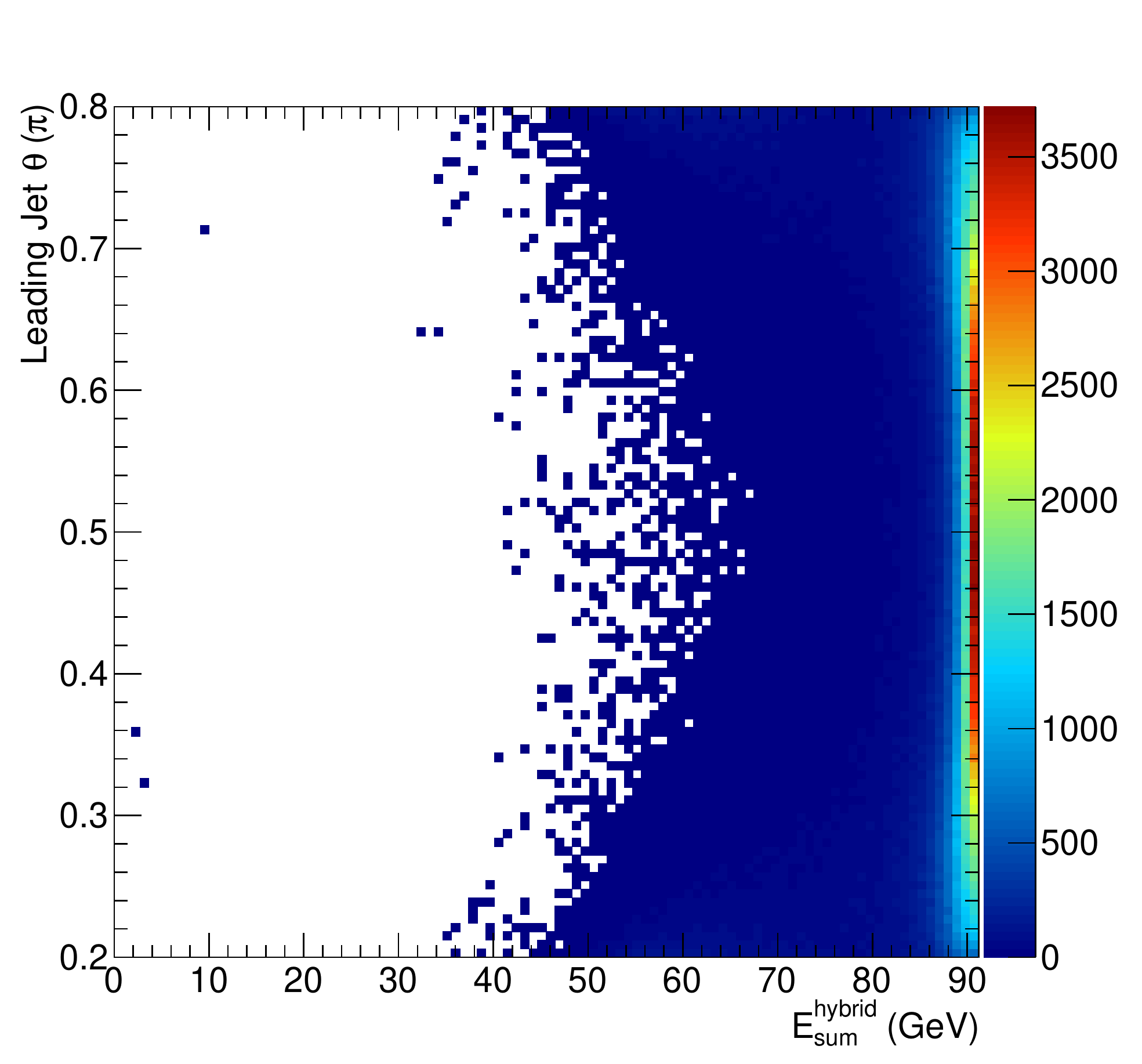}
    \caption{Correlation between leading jet $\theta$ and total visible energy.  The distortion on the \HybridE is greatly reduced compared to the simpler total visible energy.}
    \label{Figure:LeadingJet-HybridEJetTheta}
\end{figure}

In order to reduce edge effects on the total energy calculation, we define a hybrid total energy for this analysis, which is the energy sum of the following set of particles (\HybridE):
\begin{align}
    &\{\text{Particles within acceptance}\} \\\nonumber
    \cup \;\; &\{\text{Particles with angle $< 0.4$ to axes of any jet above $X$ GeV inside the acceptance}\},
\end{align}
%
%
where the acceptance is defined as $0.2\pi < \theta < 0.8\pi$. Note that the angle here is referring to a true Great-Circle Distance. The nominal jet threshold is 5 GeV.  It has been checked by varying it to 1 GeV, and there is negligible effect on the purity and efficiency.  This can be understood since the leading jets are typically higher in energy, and regardless of this threshold, the leaked energy from the leading jets will be included.
The correlation between the leading jet $\theta$ and \HybridE is shown in Fig.~\ref{Figure:LeadingJet-HybridEJetTheta}.  The edge effect is greatly reduced with this new variable definition.

\begin{figure}[htp!]
    \centering
    \includegraphicstwo{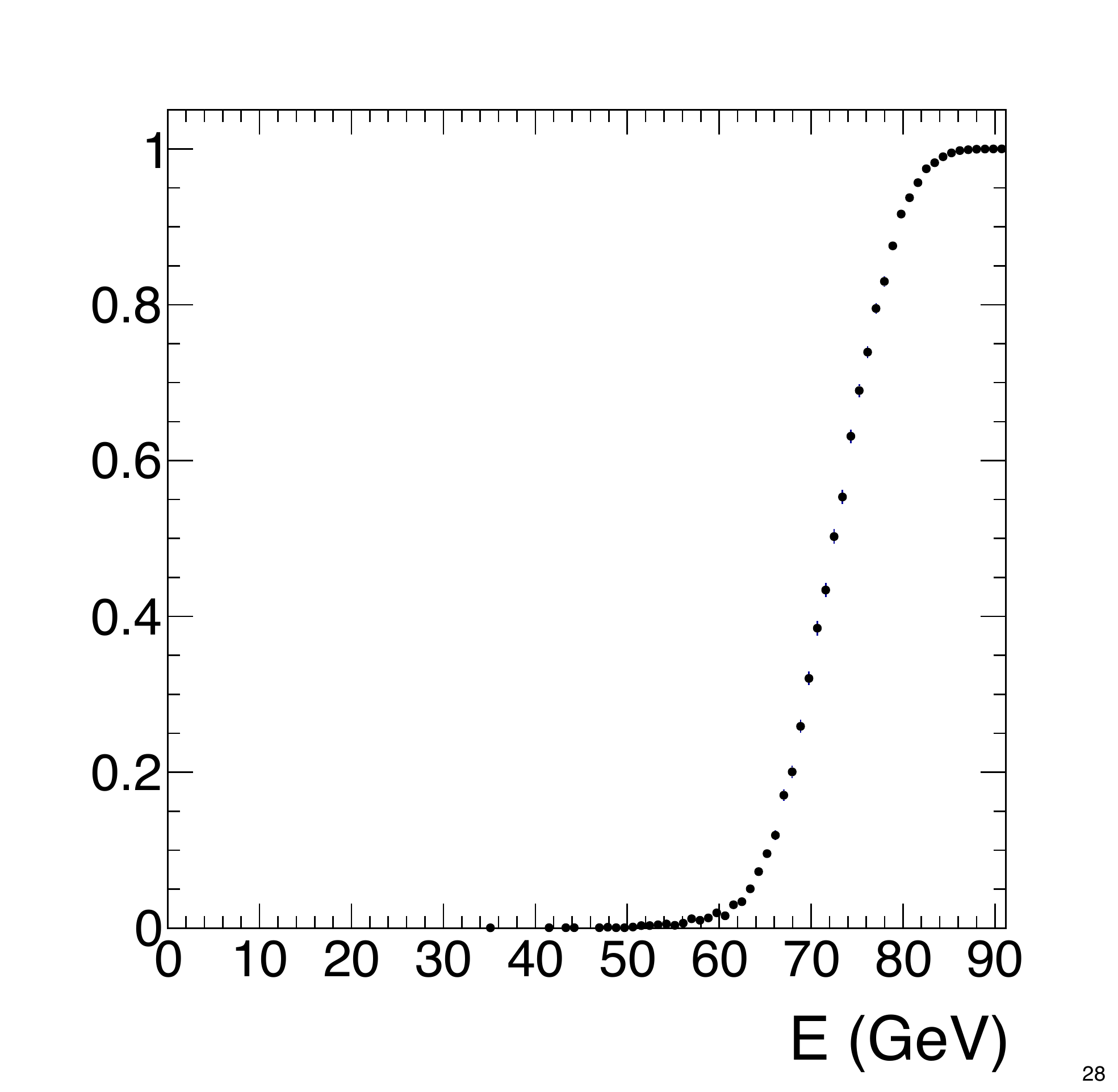}
    \includegraphicstwo{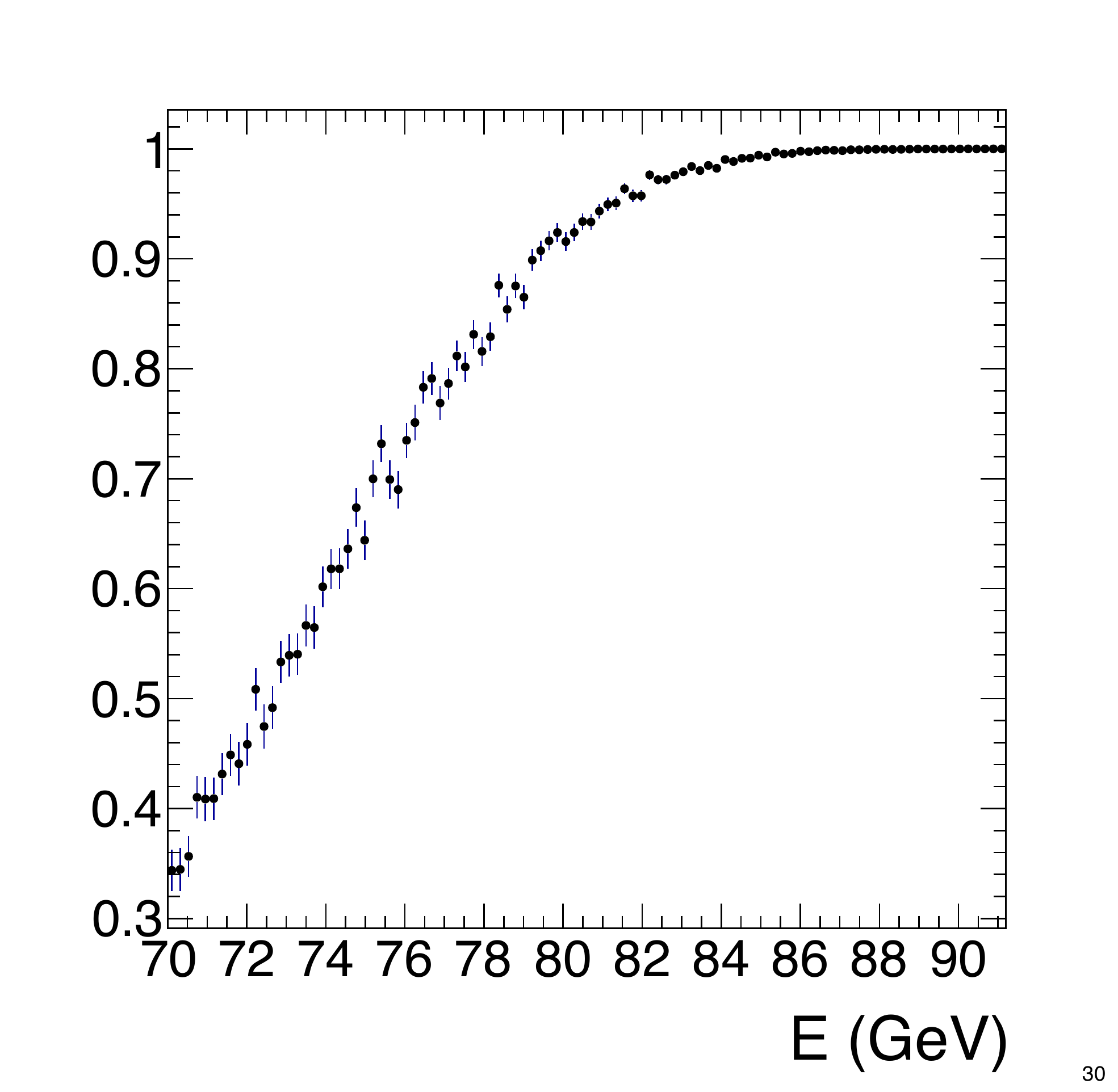}
    \caption{Left: Purity of the selection as a function of \HybridE.  Right: Zoomed in version.}
    \label{Figure:LeadingJet-HybridEFraction}
\end{figure}

\begin{figure}[htp!]
    \centering
    \includegraphicsonesmall{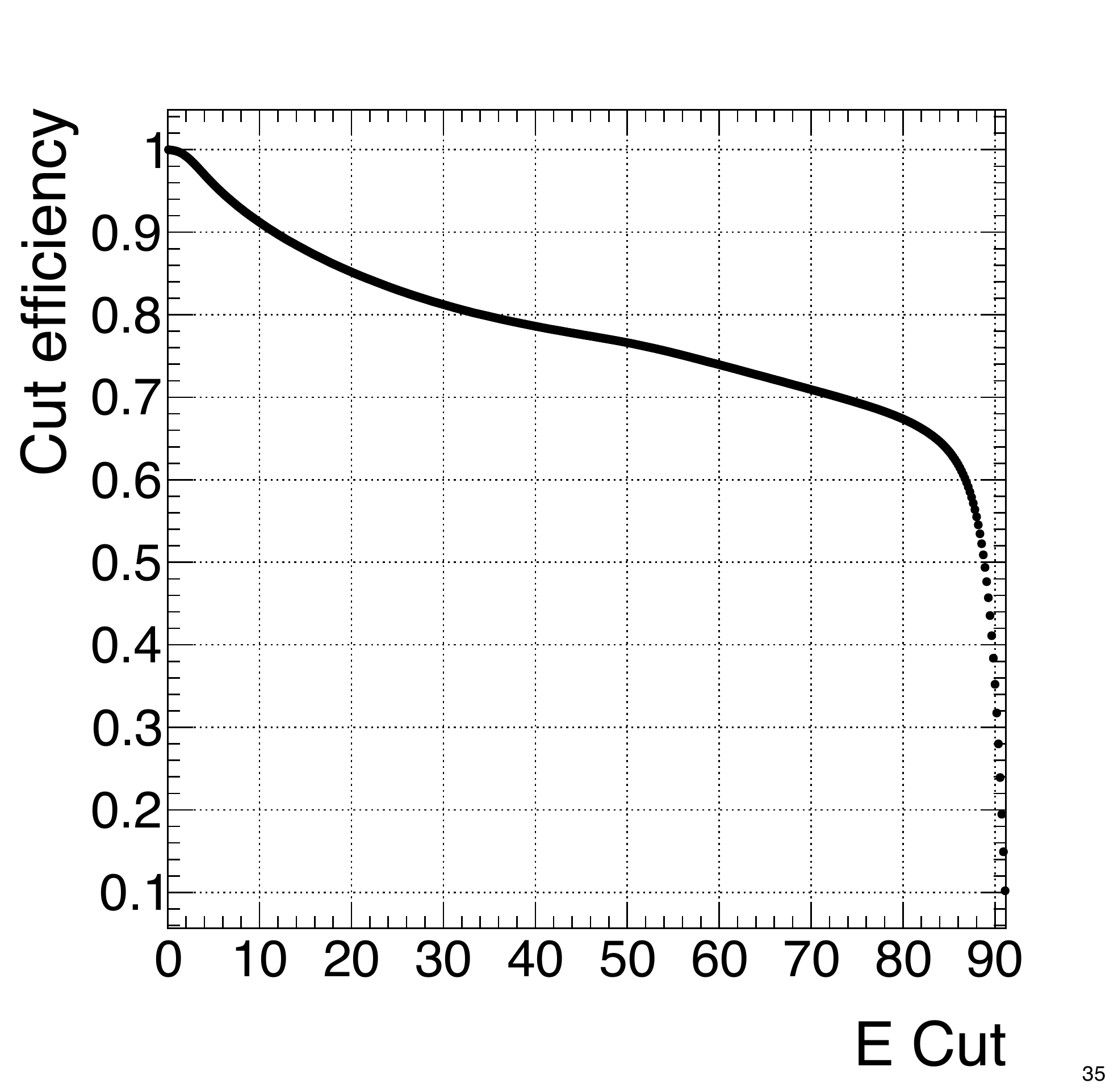}
    \caption{Cut efficiency as a function of \HybridE cut location.}
    \label{Figure:LeadingJet-HybridEEfficiency}
\end{figure}

The purity and efficiency are shown in Fig.~\ref{Figure:LeadingJet-HybridEFraction} and \ref{Figure:LeadingJet-HybridEEfficiency}, respectively.  There is an increase in cut efficiency compared to simple total visible energy due to the recovery of leaked energy.  The nominal cut correspond to 99\% purity, which is 83 GeV.

\begin{figure}[htp!]
    \centering
    \includegraphicsonesmall{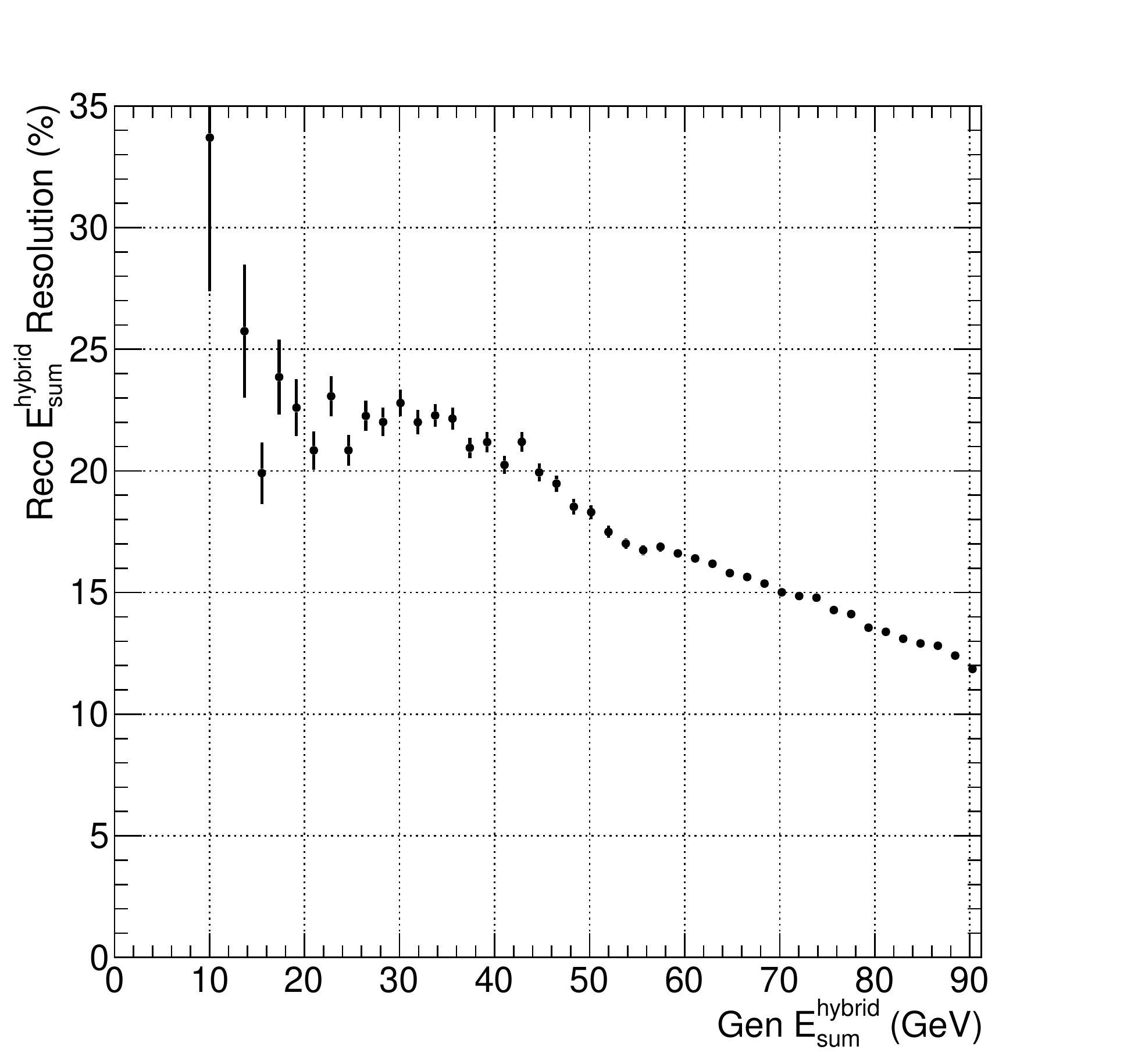}
    \caption{Resolution of reconstructed \HybridE as a function of generated \HybridE.}
    \label{Figure:LeadingJet-HybridEResolution}
\end{figure}

The resolution of reconstructed \HybridE as a function of generated \HybridE can be found in Fig.~\ref{Figure:LeadingJet-HybridEResolution}.  Around the places of the cut (80-85 GeV), the resolution is 12.5\%.

\subsection{Correction Factor}\label{Subsection:LeadingJetCorrection}

The efficiency of the cut depends on the leading dijet energies, since \HybridE includes the energies of the jets.  When the jet energies are high, it is more likely that \HybridE is high as well.  On the other hand, when leading jet energies are lower, there is a chance that there is a significant third jet out of the acceptance, and lowers the \HybridE, even though both leading dijets are inside the acceptance.  Therefore a correction is needed to account for this effect.

The current approach is to derive correction factors from simulation.  It is not perfect, however, since it depends not only on the kinematics, but also jet spectra.  It is necessary to quote systematic uncertainties on the imperfection of the model dependence.

The correction factor is defined as follows:
\begin{align}
    \text{Correction} = \dfrac{N(\text{Both dijet within acceptance})}{N(\text{Both dijet within acceptance and pass \HybridE selection})}.
\end{align}
Separate corrections are derived as a function of leading dijet jet energy and as a function of leading dijet total energy.  The unfolding, described in Sec.~\ref{Section:Unfolding}, unfolds detector smearing effects, and this correction factor is applied on top of the unfolded spectra, as illustrated below:
\begin{align}
    &\text{Detector-level spectra with \HybridE selection}\nonumber\\
    \xrightarrow{\text{Unfolding}}\; &\text{Truth-level spectra with \HybridE selection} \nonumber\\
    \xrightarrow{\text{Correction}}\; &\text{Truth-level spectra with no \HybridE selection}.
\end{align}
A potential alternative strategy is to take care of both steps with the unfolding.  This however has an adverse side effect where the unfolding step fills in jet spectra from simulation for events which fail \HybridE selection.  By doing it in two steps, we only take the ratio of spectra from simulation, and reduce (somewhat) the model dependence.

\begin{figure}[htp!]
    \centering
    \includegraphicstwo{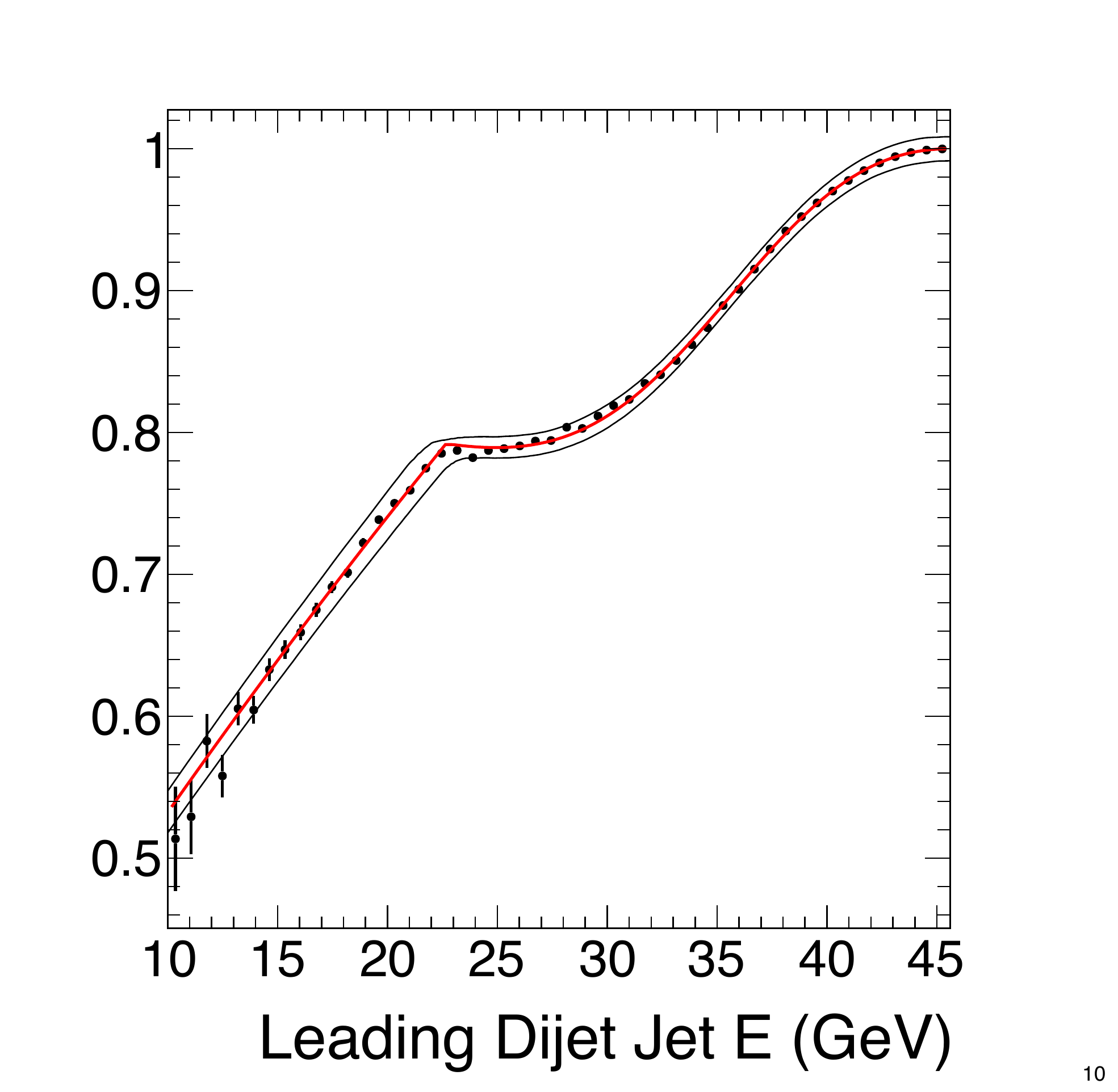}
    \includegraphicstwo{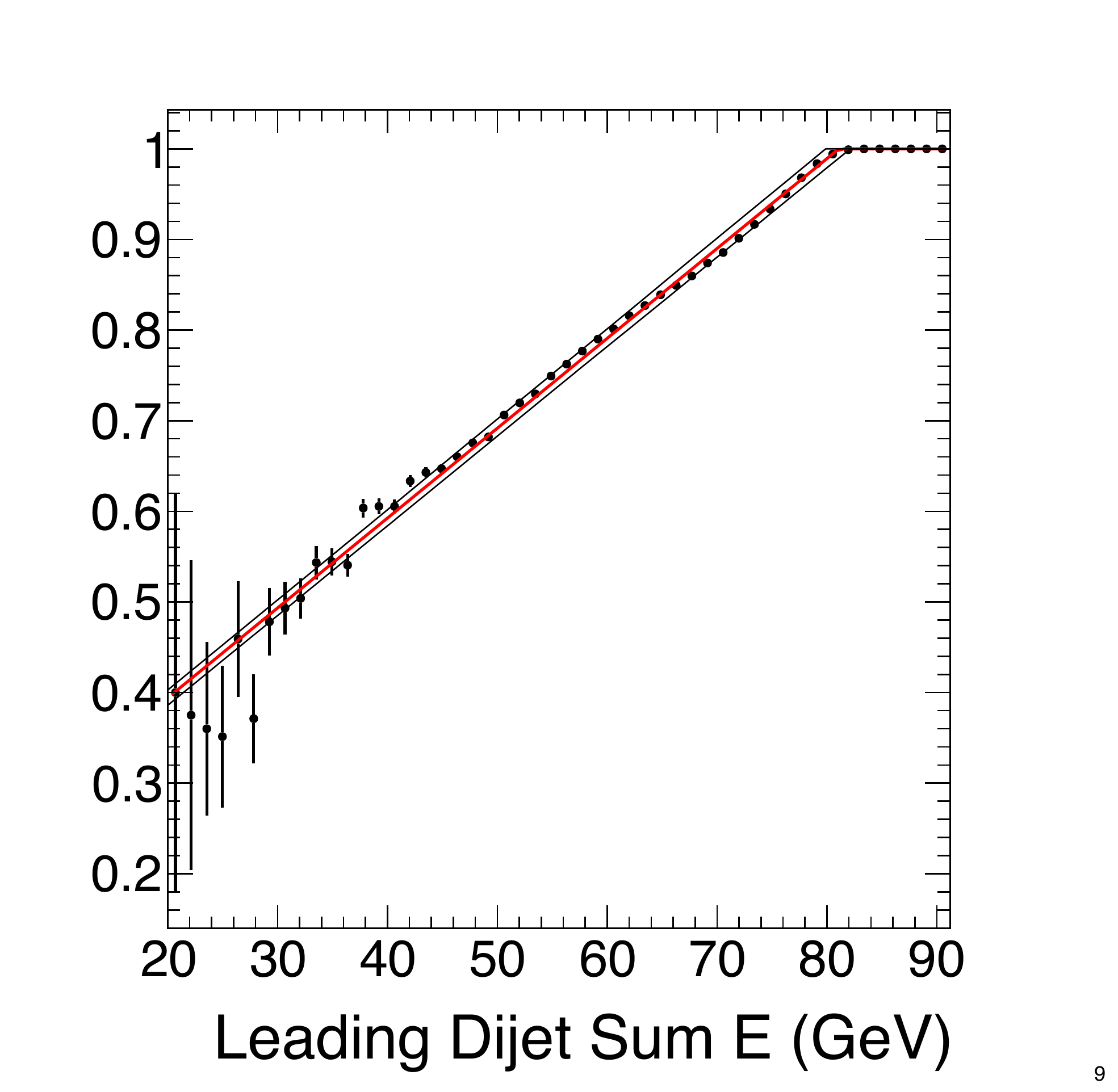}
    \caption{Efficiency (inverse of the correction factor) as a function of leading dijet jet energy (left) and dijet total energy (right).  The nominal fit function is also shown, together with the uncertainty band from the fit.}
    \label{Figure:LeadingJet-HybridECorrection}
\end{figure}

The selection efficiency (inverse of the correction factor) as functions of leading dijet jet energy and total energy are shown in Fig.~\ref{Figure:LeadingJet-HybridECorrection}.  They are fitted with
\begin{align}
    f(E) = \min\left(a_0 + a_1 E, a_2 + a_3 \times  \text{erf}\left(\dfrac{E - a_4}{a_5}\right) \times  (1+a_6 E)\right),
\end{align}
and
\begin{align}
    f(E) = \min(a_0 + a_1 E, 1.0)
\end{align}
respectively.  The nominal parameters are reported in Tab.~\ref{Table:LeadingJet-HybridECorrection}.

\begin{table}[htp!]
    \centering
    \begin{tabular}{|c|c|c|}
        \hline
 & Leading Dijet Jet Energy & Leading Dijet Total Energy \\\hline
$a_0$ & $0.300085 \pm 0.006641$ & $0.195666 \pm 0.001712$ \\\hline
$a_1$ & $0.025771 \pm 0.000328$ & $0.009920 \pm 0.000022$ \\\hline
$a_2$ & $1.082800 \pm 0.000280$ & - \\\hline
$a_3$ & $0.203061 \pm 0.000304$ & - \\\hline
$a_4$ & $36.013658 \pm 0.011172$ & - \\\hline
$a_5$ & $8.273973 \pm 0.019290$ & - \\\hline
$a_6$ & $-0.004598 \pm 0.000004$ & - \\\hline
    \end{tabular}
    \caption{The nominal parameters for the efficiency fit.}
    \label{Table:LeadingJet-HybridECorrection}
\end{table}

\subsection{Variations to the Correction Factors}\label{Subsection:LeadingJetCorrectionVariation}

\begin{figure}[htp!]
    \centering
    \includegraphicsone{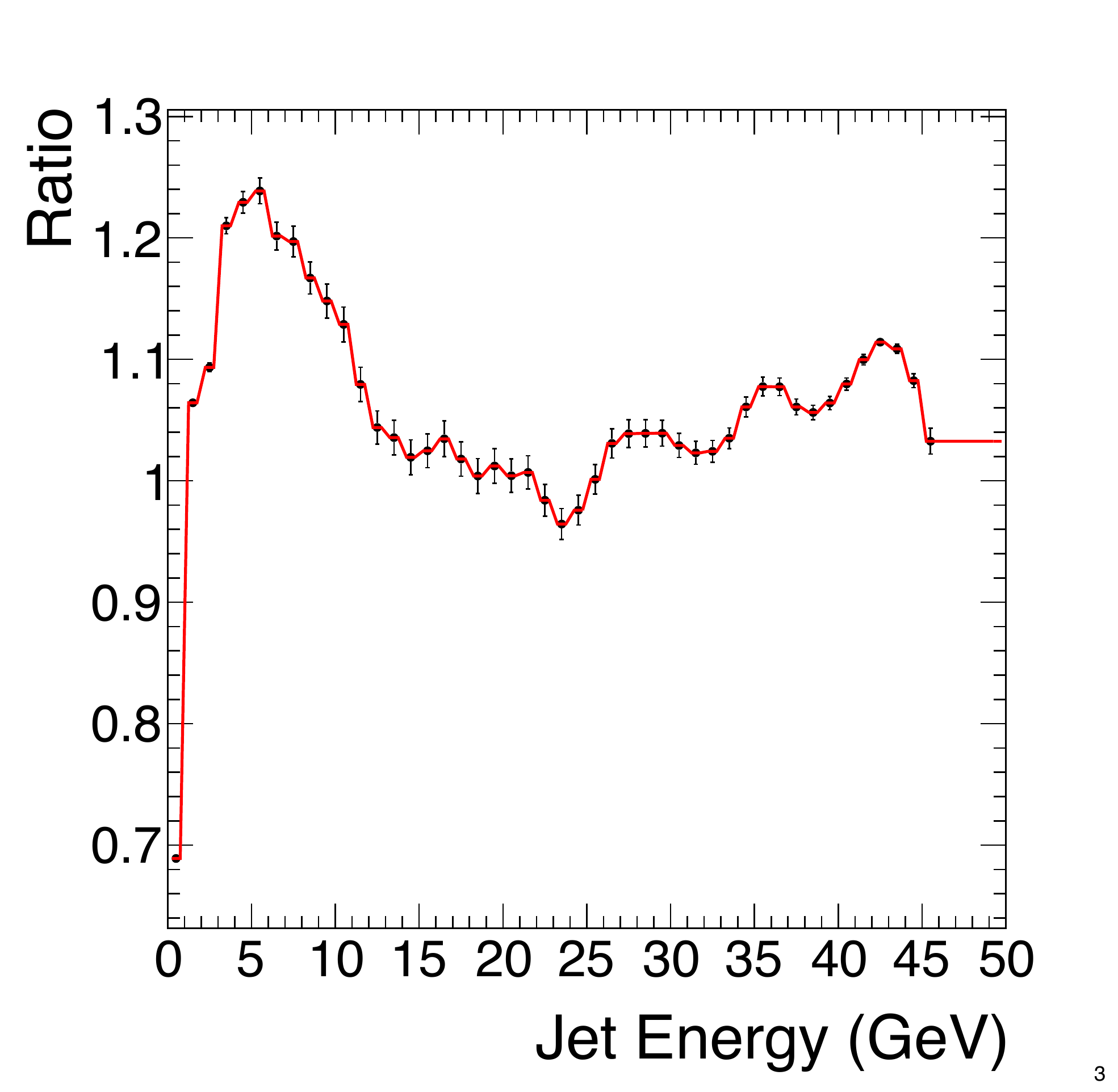}
    \caption{Jet weights based on the inclusive spectra in acceptance}
    \label{Figure:LeadingJet-CorrectionVariationWeight}
\end{figure}

The nominal correction factor is derived from the simulation, and therefore it is model-dependent.  In order to address the imperfect modeling, the simulated sample is reweighted based on the inclusive jet energy spectrum.  The ratio of the inclusive jets within nominal acceptance ($0.2\pi < \theta < 0.8\pi$) between the unfolded data and generator level MC distribution is first derived in fine bins, as shown in figure~\ref{Figure:LeadingJet-CorrectionVariationWeight}.

The weight is then applied to all generated jets (also outside of acceptance) in a PYTHIA8 sample where there is no generator-level cuts on the final state particle directions.  Each event is weighted by the product of the weights for all jets in the event.  The underlying assumption for this procedure is that the degree of disagreement between simulation and data is similar inside the acceptance and outside acceptance, and therefore by reweighting also the jets out of the acceptance, we can estimate how much the mismodeling between data and simulation can change the leading jet correction.

\begin{figure}[htp!]
    \centering
    \includegraphicstwo{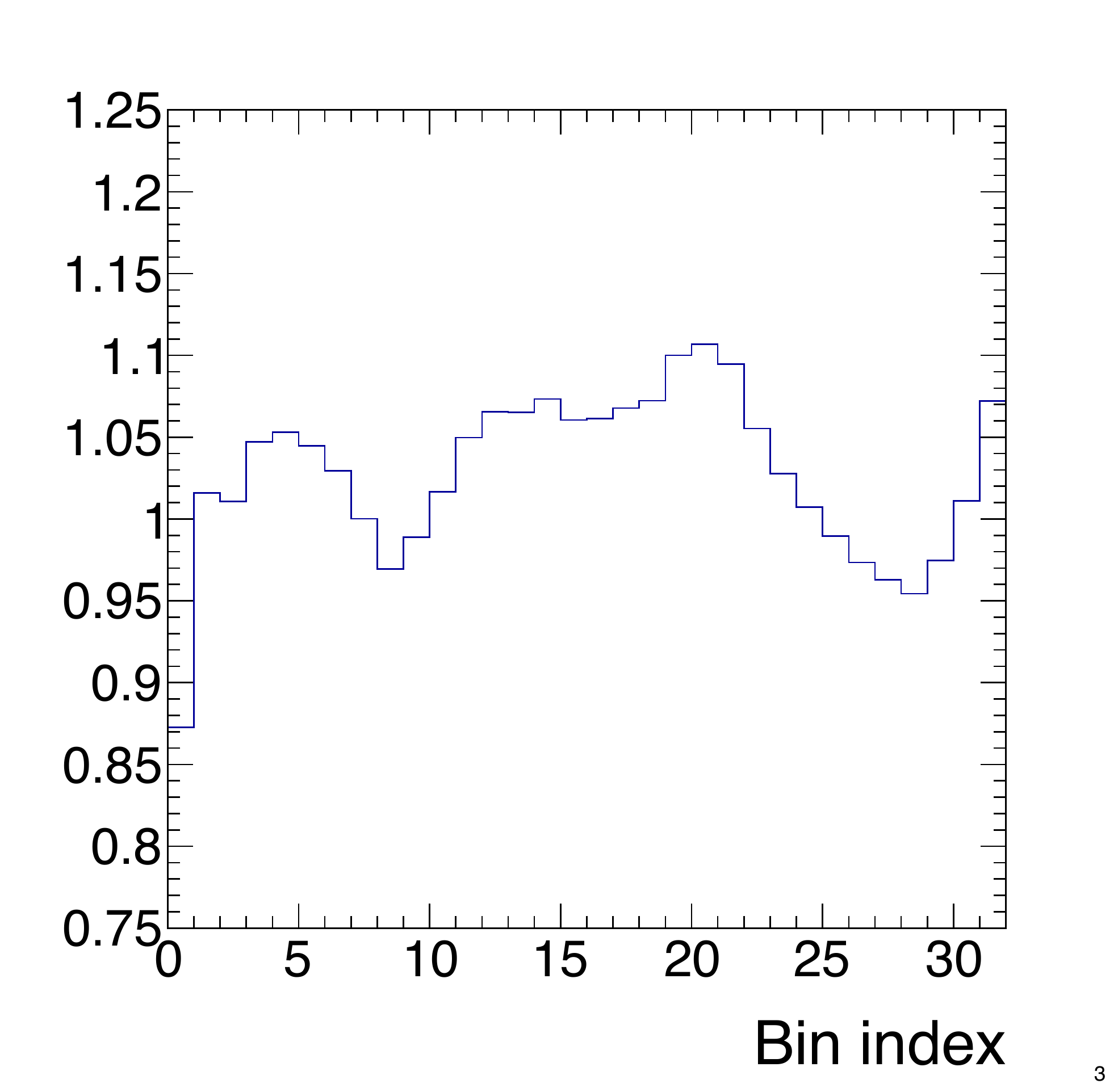}
    \includegraphicstwo{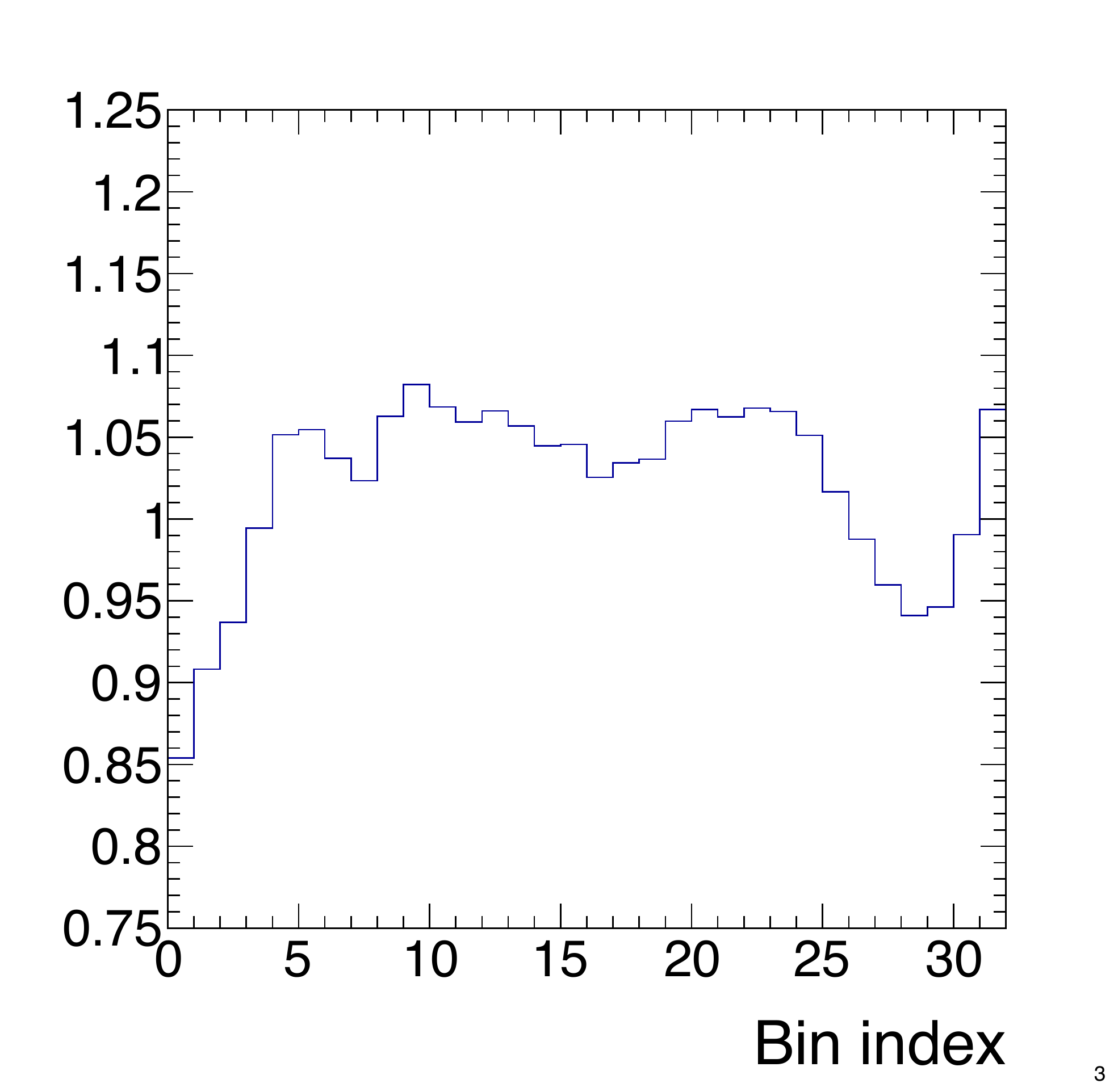}
    \caption{Variation in correction based on reweighted simulation for (left) leading dijet energy and (right) leading dijet total energy.}
    \label{Figure:LeadingJet-CorrectionVariation}
\end{figure}

The result is shown in figure~\ref{Figure:LeadingJet-CorrectionVariation}, where we compare the derived correction factor using original and the reweighted simulated samples.  There is up to $\pm 10\%$ variation observed.

\clearpage

\section{Unfolding}\label{Section:Unfolding}

\subsection{Unfolding Overview}

In order to remove detector effects, an unfolding is performed using the \textsc{RooUnfold} package (v2.0.0).  As nominal, the \Bayes method is used, with the \SVD as cross check.

For jet energy, leading dijet energy, and leading dijet total energy, a 1D unfolding is performed.  For the other jet structure and substructure (mass, groomed $\zg$, groomed angle, groomed mass), because of the significant jet energy migration, a 2D unfolding of the observables in bins of jet energy is done.

\subsection{Jet Ranking Studies}

\begin{figure}[htp!]
    \centering
    \includegraphicsonesmall{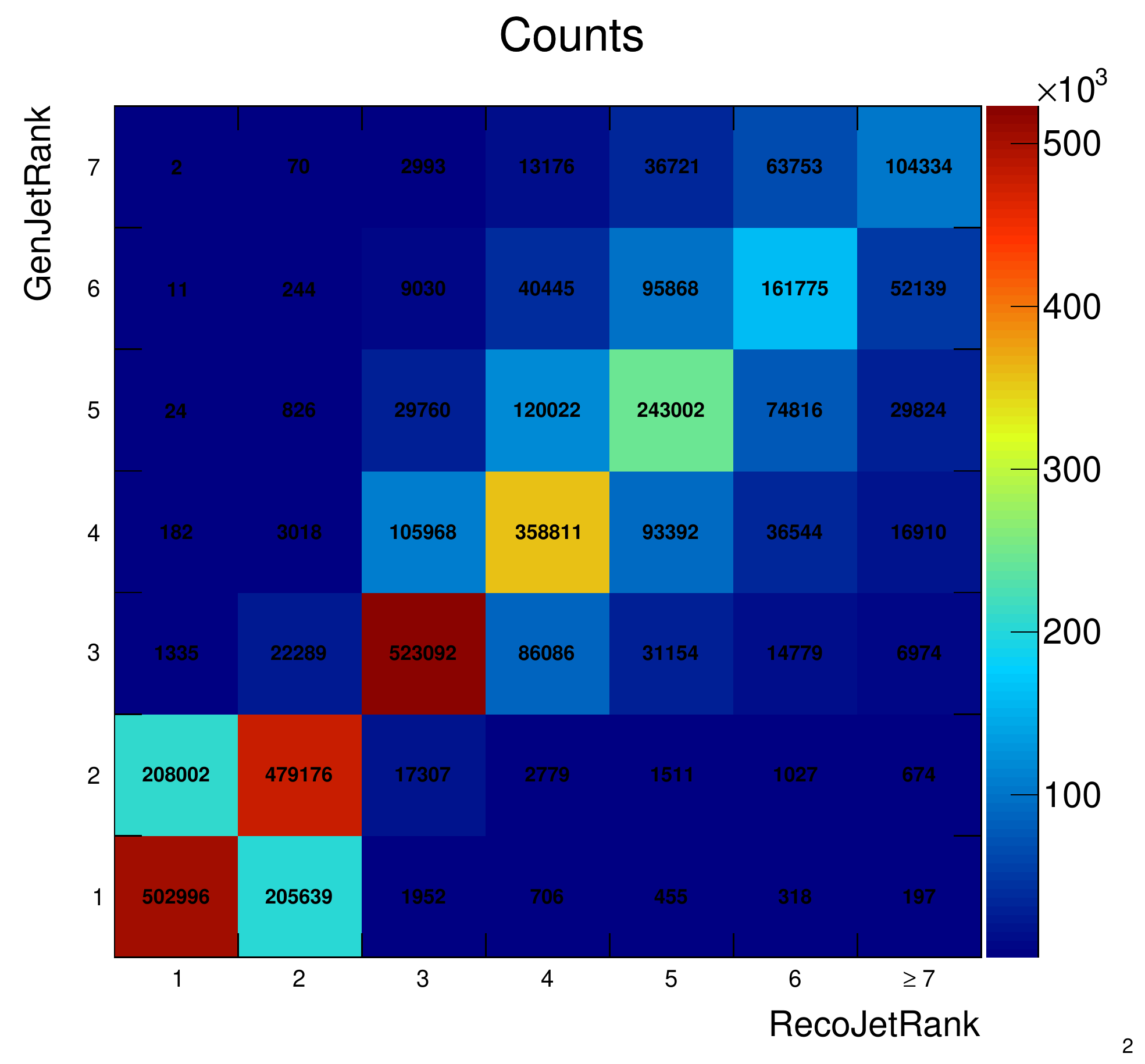}
    \caption{Correlation between generated jet rank and reconstructed jet rank.}
    \label{Figure:Unfolding-JetRankingCount}
\end{figure}

\begin{figure}[htp!]
    \centering
    \includegraphicstwo{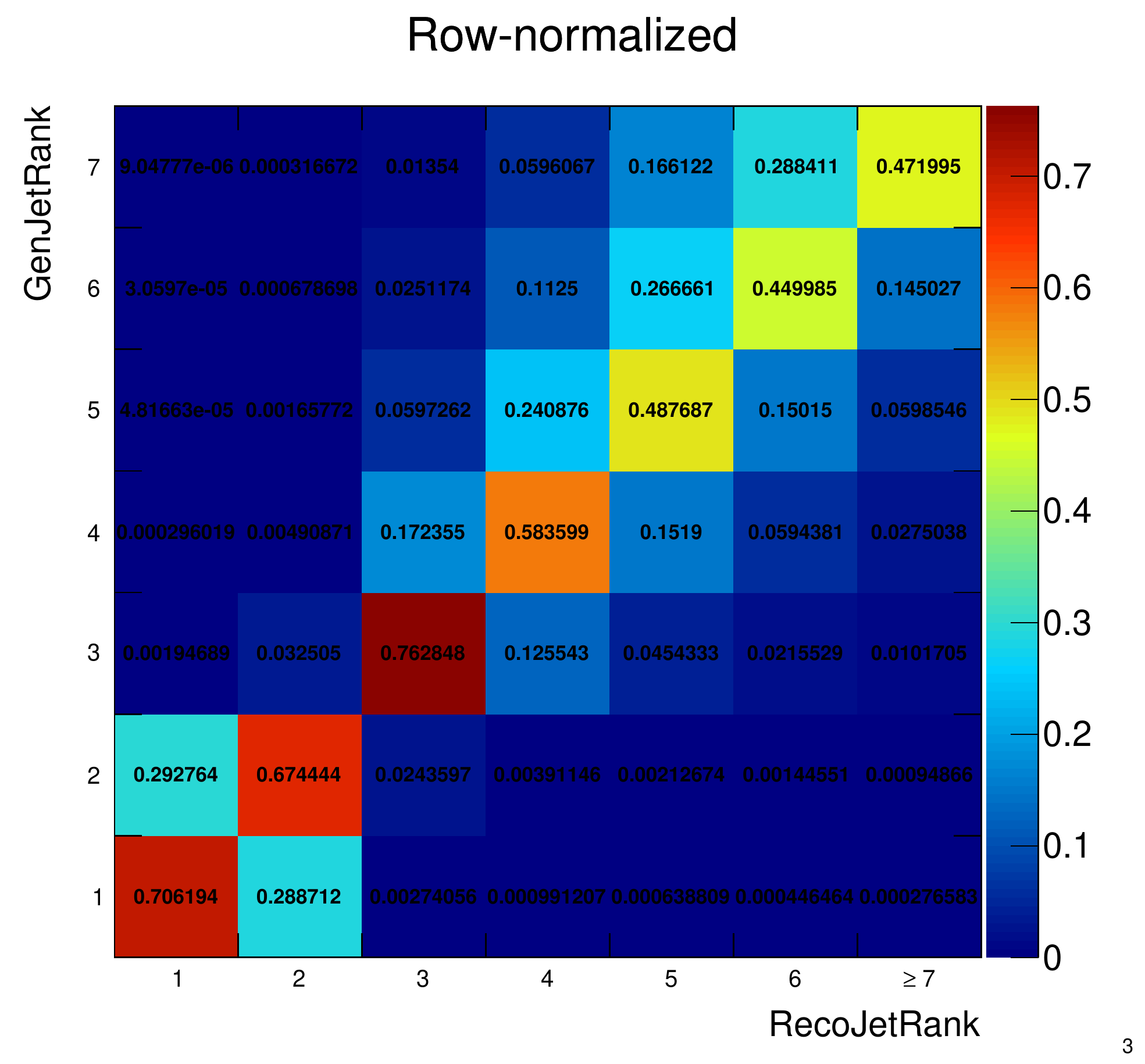}
    \includegraphicstwo{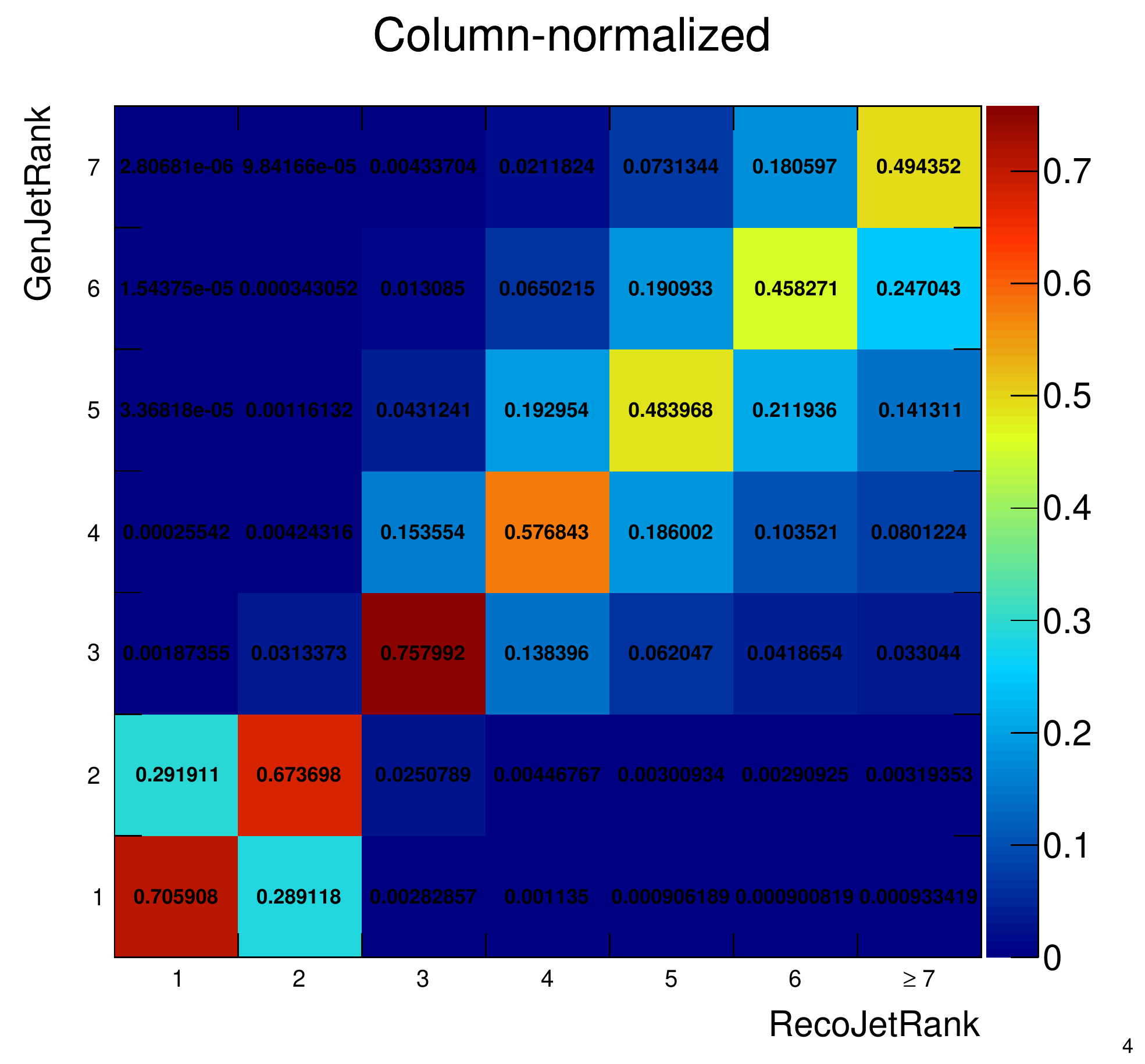}
    \caption{Normalized jet ranking matrices.  Left: row-normalized matrix showing the probability of a generated jet with a given rank being reconstructed into various different ranks.  Right: column-normalized matrix showing the probability of a reconstructed jet with a given rank coming from various ranks of generated jets.}
    \label{Figure:Unfolding-JetRankingPercentage}
\end{figure}

In order to design the unfolding strategy for leading jets, a jet ranking study is performed.  Generated and reconstructed jets are sorted according to their energies and matched to each other.  The frequency of an $m$-th ranked generated jet matched to an $n$-th ranked reconstructed jet is then tabulated for all $m$ and $n$ pairs.  The number of times each pairing occurs is shown in Fig.~\ref{Figure:Unfolding-JetRankingCount} for the archived ALEPH MC sample.  The same table is then normalized either by row, giving the percentage of a given rank generated jet being reconstructed into a different rank reconstructed jet, or by column, giving the percentage of given rank reconstructed jet coming from different rank generated jets.  They are shown in Fig.~\ref{Figure:Unfolding-JetRankingPercentage}.

As can be seen, there is a significant cross-talk between the two leading jets, which is expected as one of the dominant processes in \ee collisions involving jets is the back-to-back production of dijet of similar energy through a $Z$ resonance.

The chance of a generated jet with rank $> 2$ being reconstructed into one of the leading reconstructed dijets is generally small.  The largest is the percentage of a \nth{3}-leading generated jet being reconstructed as a second-leading jet at around 3\%.  In this analysis, it is decided that we measure the leading two jets together to avoid the large cross-talk between the two jets, and the effect of the \nth{3} jet is included in the result as systematic uncertainties.

\subsection{Smearing Matrices}

The matching between generator level jet and reconstructed level jets is done through opening angle criteria.  To ensure a good matching quality, an opening angle of less than half the jet resolution parameter (0.4) is required for a successful match.

\begin{figure}[htp!]
    \centering
    \includegraphicsonesmall{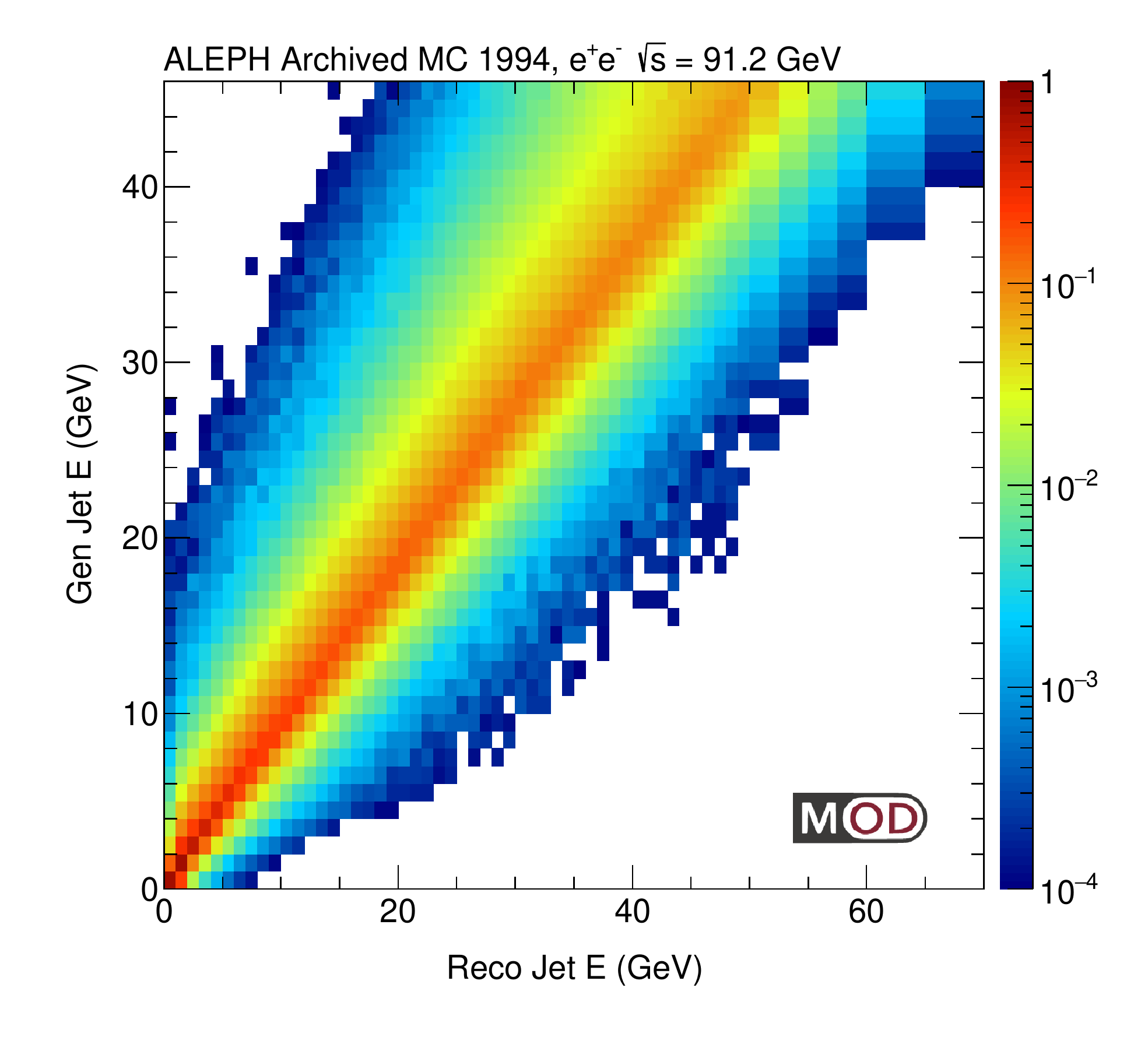}
    \caption{Smearing matrix for jet energy.}
    \label{Figure:Unfolding-SmearingMatrixE}
\end{figure}

The smearing matrix for the inclusive jet energy is shown in Fig.~\ref{Figure:Unfolding-SmearingMatrixE}.  The figure is normalized such that each row integrates to unity (accounting for bin widths).  A nice diagonal matrix is observed, with some off-diagonal jet energy migration according to the jet resolution.  The response is slightly asymmetrical with a long tail toward the lower energy side.  Jets with energy lower than the lower bound of 10 GeV are included in order to serve as underflow bins and stabilize the unfolding procedure.

\begin{figure}[htp!]
    \centering
    \includegraphicstwo{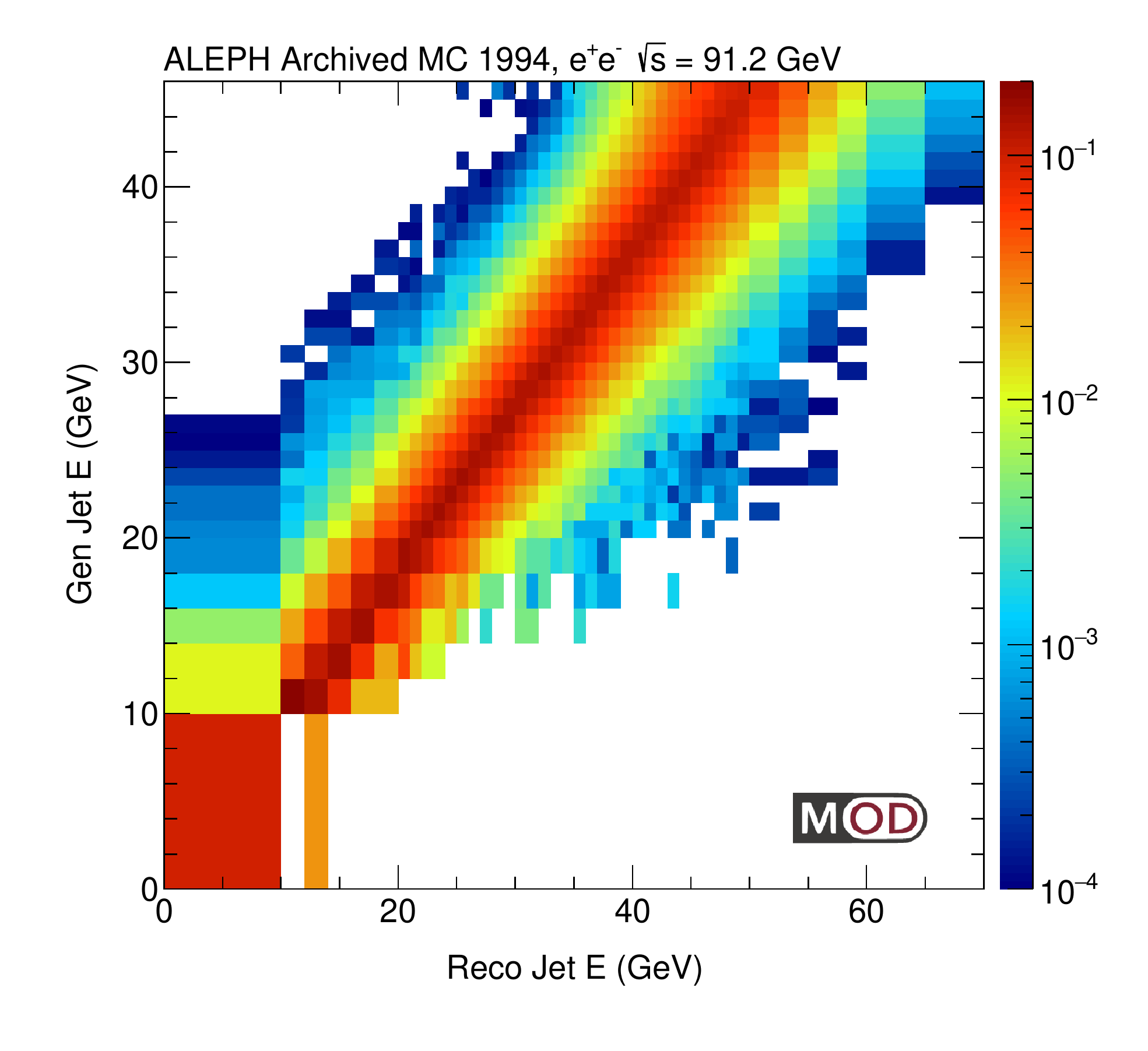}
    \includegraphicstwo{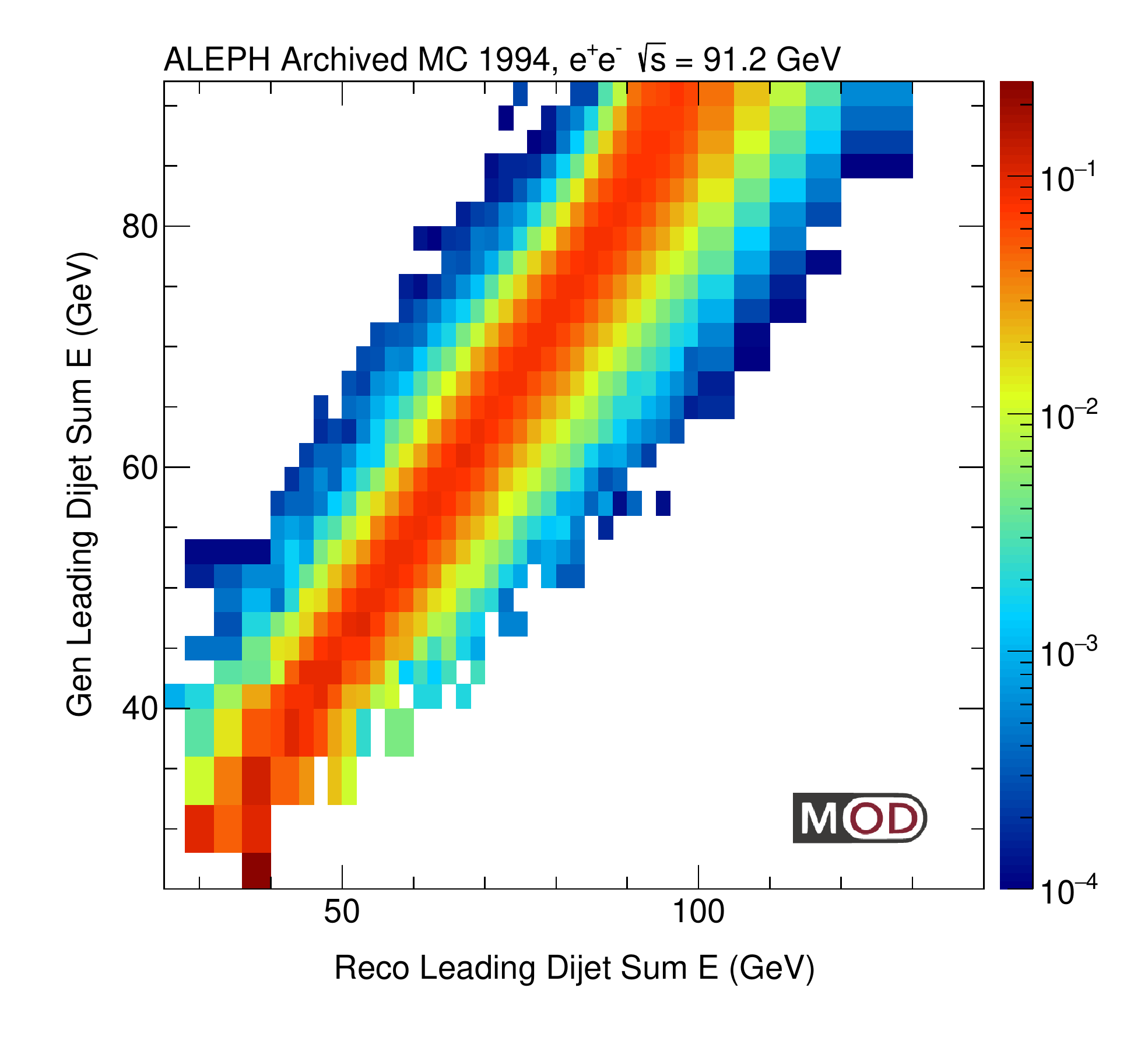}
    \caption{Smearing matrices for leading dijet jet energy (left) and leading dijet total energy (right).}
    \label{Figure:Unfolding-SmearingMatrixLeadingDiJet}
\end{figure}

The smearing matrix for leading dijet jet energy is shown in the left panel of Fig.~\ref{Figure:Unfolding-SmearingMatrixLeadingDiJet}.  It is nearly identical to the inclusive jet one except a larger binning at the low energy end where the statistics is much lower due to the observable definition.  The smearing matrix for the leading dijet sum energy is shown in the right panel.  It is nearly identical to the inclusive jet smearing matrix except a factor of 2 in the energies.

\begin{figure}[htp!]
    \centering
    \includegraphicsone{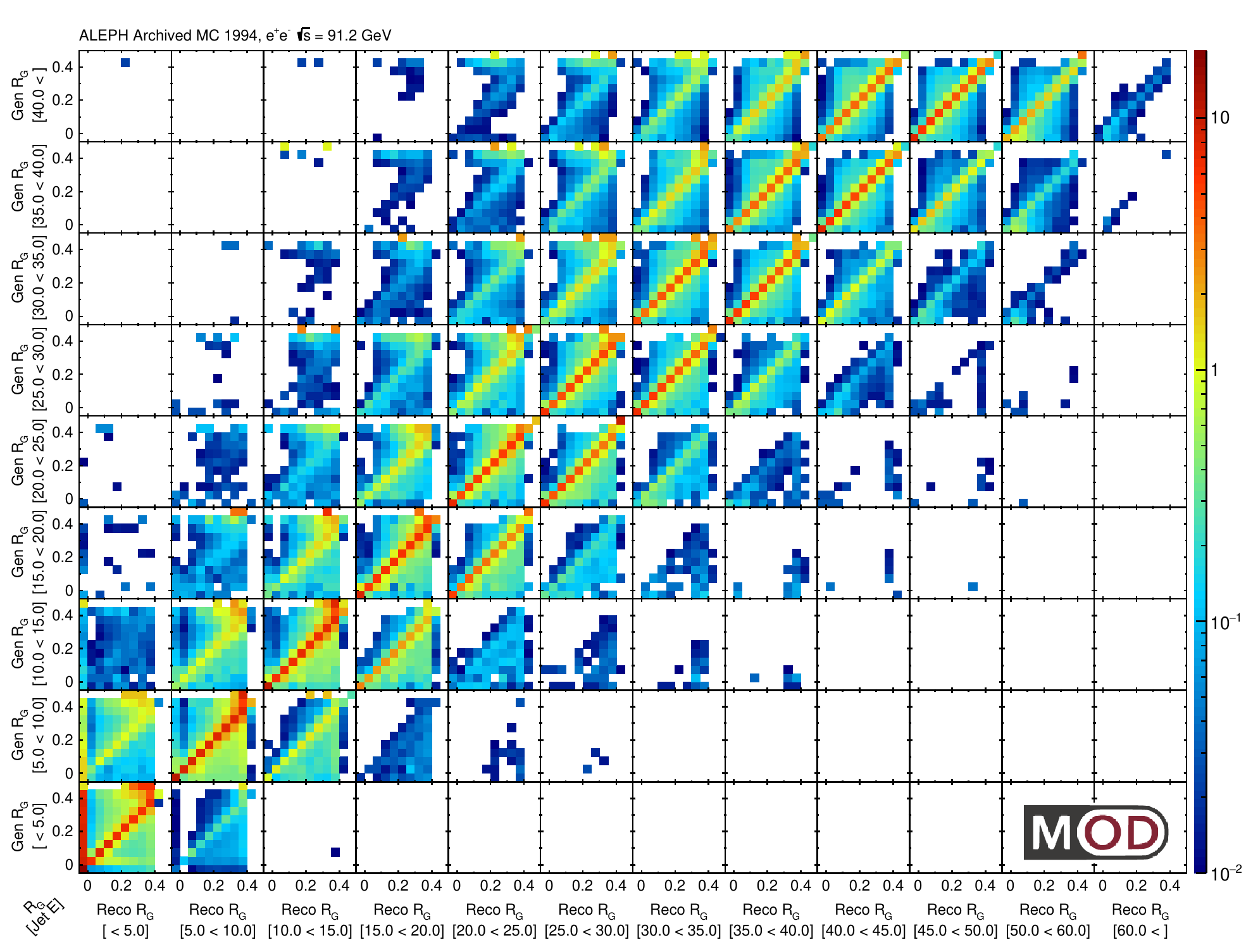}
    \caption{Smearing matrix for \Rg.  Each block represents a different jet energy interval.}
    \label{Figure:Unfolding-SmearingMatrixRG}
\end{figure}

\begin{figure}[htp!]
    \centering
    \includegraphicsone{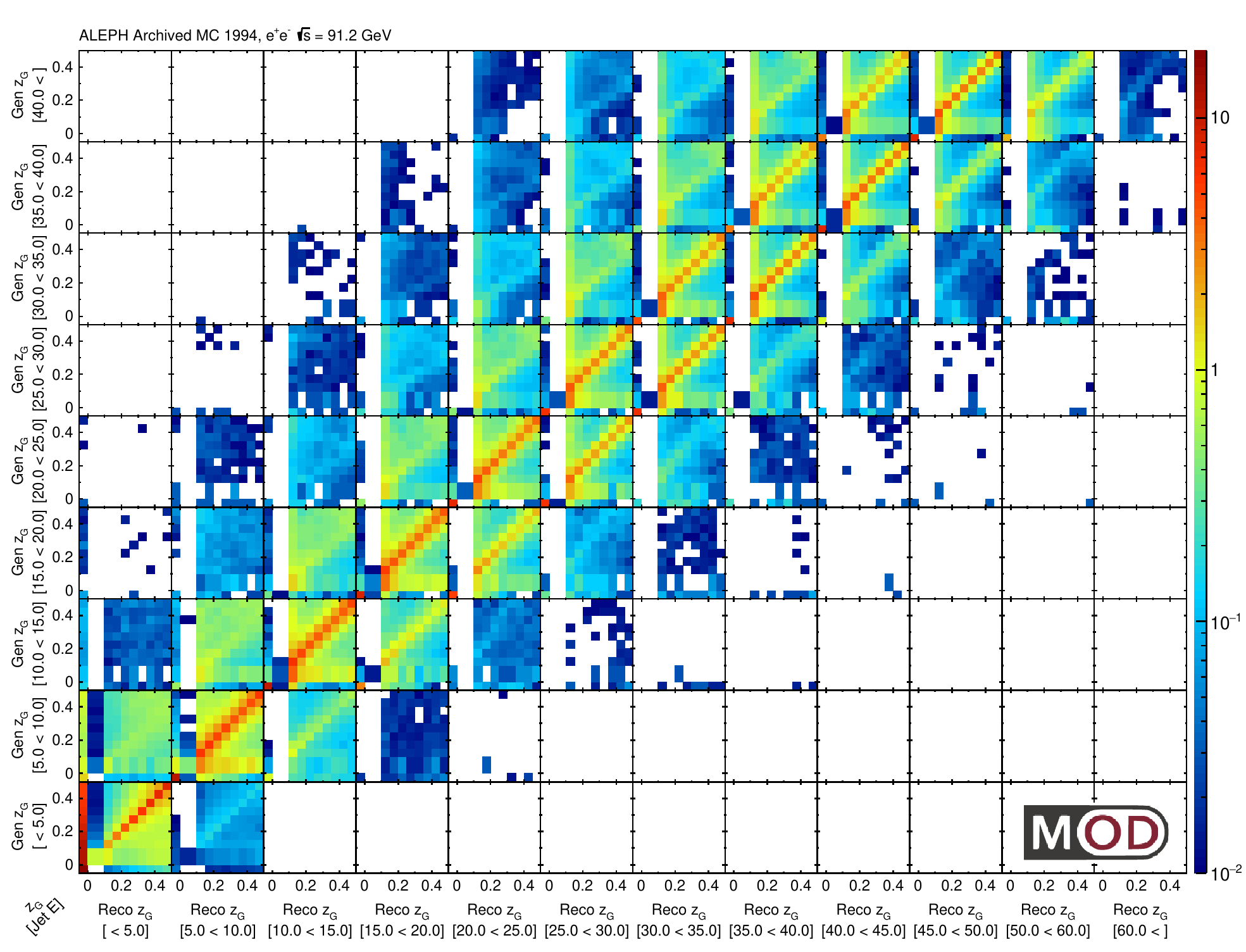}
    \caption{Smearing matrix for \zg.  Each block represents a different jet energy interval.}
    \label{Figure:Unfolding-SmearingMatrixZG}
\end{figure}

The smearing matrices for $\Rg$ and $\zg$ can be found in Fig.~\ref{Figure:Unfolding-SmearingMatrixRG} and Fig.~\ref{Figure:Unfolding-SmearingMatrixZG}, respectively.  Each of the subpanel shows the matrix for given generated and reconstructed jet energy intervals.  As expected, there are significant contribution in the off-diagonal blocks, encoding the energy smearing.  The normalization is done for each row across all reconstructed energy blocks.  The first bin in each block includes jets that are completely groomed away by the grooming procedure. Note that if the grooming procedure ends at different number of declustering between RECO and gen level (due to particle reconstruction efficiencies and energy resolution), it creates vertical and horizontal components in the smearing matrix located at either Gen $z_G=0$ or RECO $z_G=0$. For instance, the line at RECO $Z_G=0$ is created by a RECO jet that failed the grooming procedure while in generator level a subjet pair can be identified in the grooming procedure.

\begin{figure}[htp!]
    \centering
    \includegraphicsone{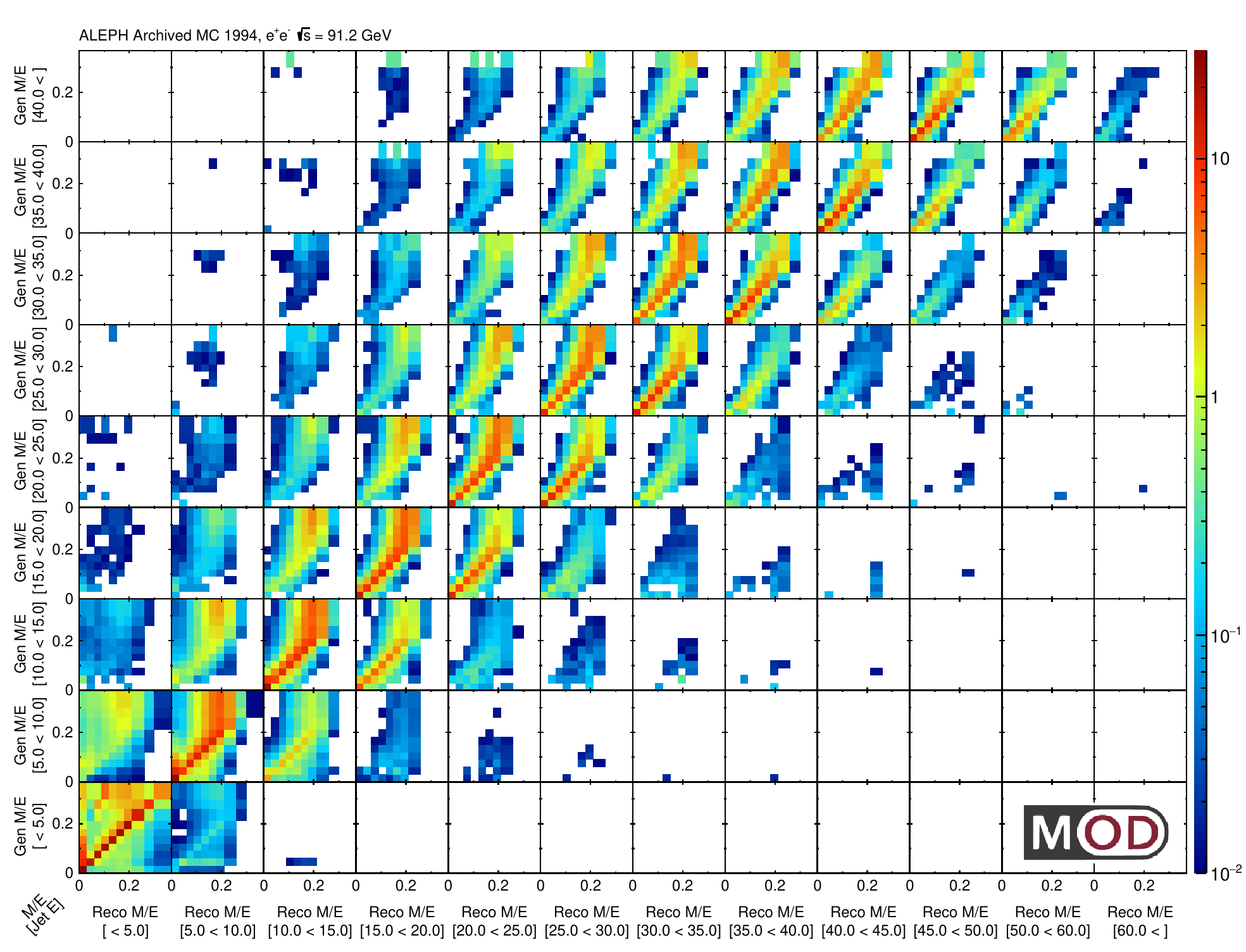}
    \includegraphicsone{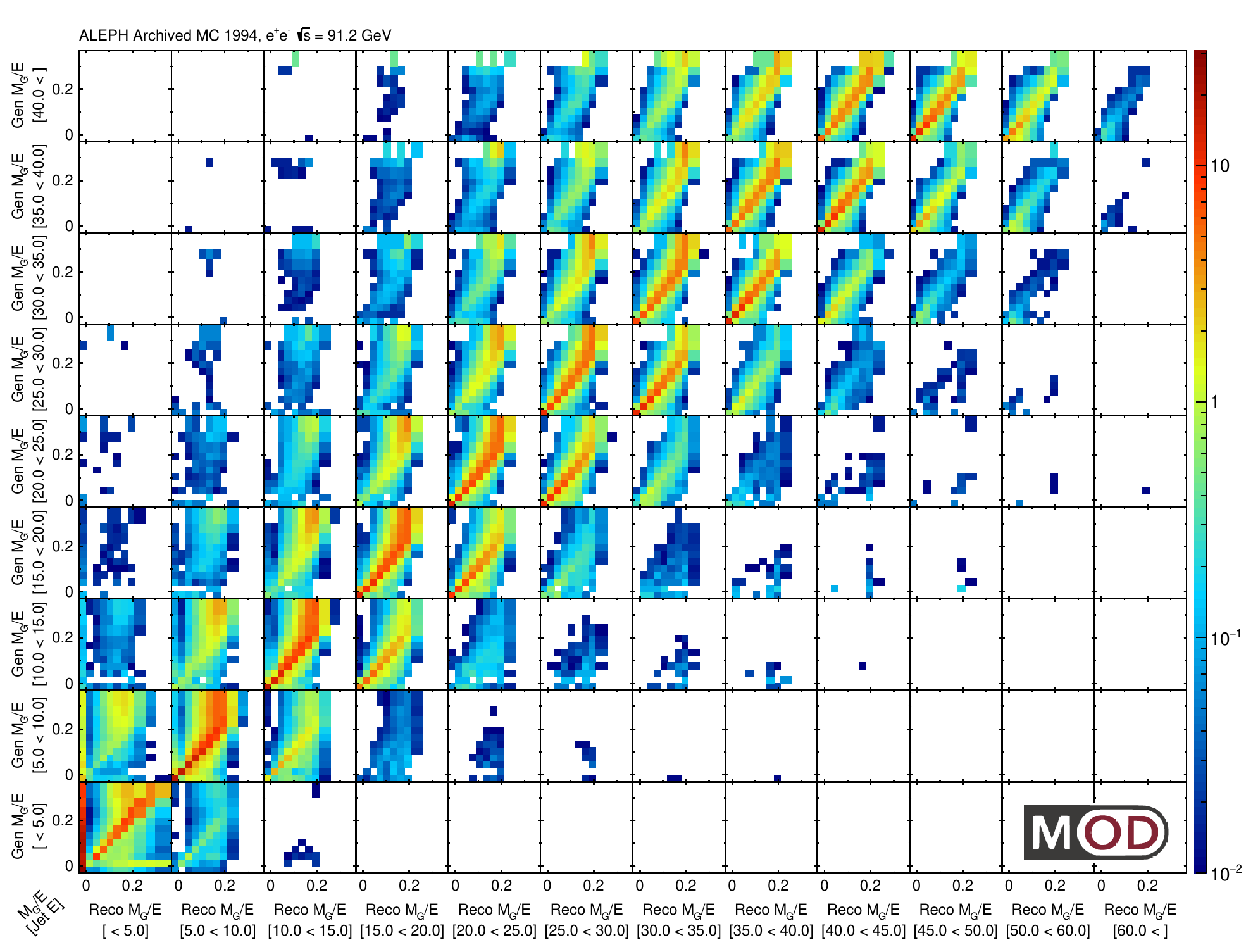}
    \caption{Smearing matrix for ungroomed and groomed jet mass, normalized by the jet energy.  Each block represents a different jet energy interval.}
    \label{Figure:Unfolding-SmearingMatrixMass}
\end{figure}

The smearing matrices for mass and groomed mass can be found in Fig.~\ref{Figure:Unfolding-SmearingMatrixMass}.  Each of the subpanel shows the matrix for given generated and reconstructed jet energy intervals.  The correlation is high in each subpanel, but there is a trend where the reconstructed mass is lower than the generated mass.

\subsection{Toy Data Generation}

For studies of the unfolding, toy data allows fuller control of various tests compared to regular simulated events.  If we take the smearing matrix as given, one can compute the ``no extra statistical fluctuation'' version of the reconstructed spectrum by a matrix multiplication of the smearing matrix and a ``truth'' spectrum.  Based on this perfect reconstructed spectrum, realistic toy datasets can be generated by use of Poisson statistics and which is compatible with the amount of statistics in data.

The toy datasets are used in various studies, described in later sections:
\begin{enumerate}
    \item Evaluate and validate the statistical uncertainty reported by the unfolding procedure
    \item Evaluate the ideal regularization parameter (number of iterations for the case of \Bayes)
\end{enumerate}

\subsection{Statistical Uncertainty Validation}

For each of the observables, 500 statistically independent toy datasets are generated with the generator-level spectrum, the smearing matrix, and amount of statistics in data, as described in the previous section.  All toy datasets are then unfolded, and the results are compared with the input ``truth'' distribution.  Since both the input ``truth'' distribution and the smearing matrix are held constant across all toy datasets, the only variation comes from the statistical nature of the toy datasets sampling from the perfect reconstructed spectrum.  Therefore the spread in the unfolded results is a measure of the statistical uncertainty.  It should be equal to the unfolding uncertainty (for the ensemble of toy datasets), which also quantifies the statistical uncertainty.

In order to quantify the agreement between the spread in unfolded toy datasets and the uncertainty reported by the unfolding, the pull distribution is calculated, where the pull is defined as the difference between the unfolded result and the input ``truth'' bin by bin, divided by the unfolding uncertainty.  

\begin{figure}[htp!]
    \centering
    \includegraphicsthree{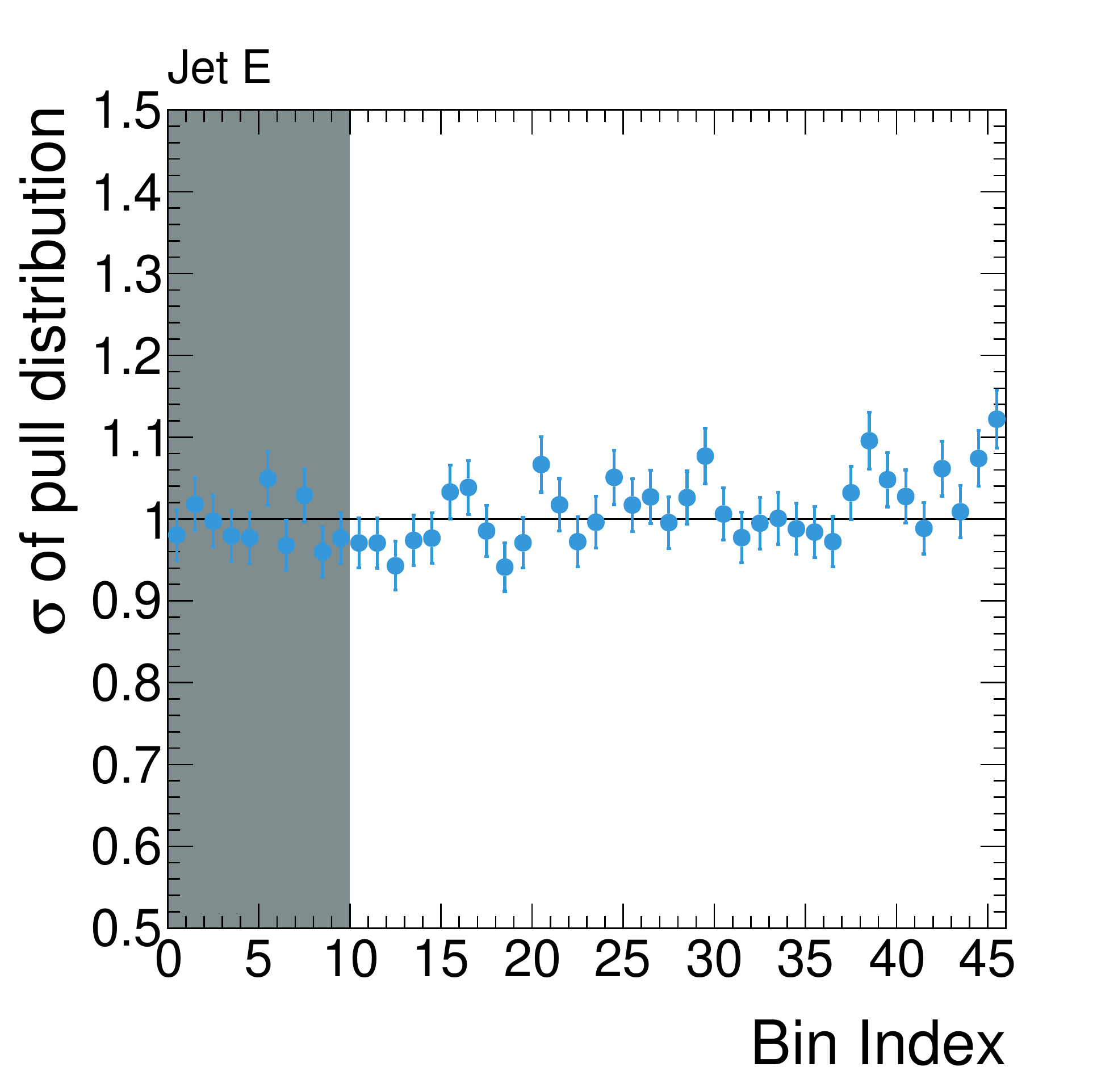}
    \includegraphicsthree{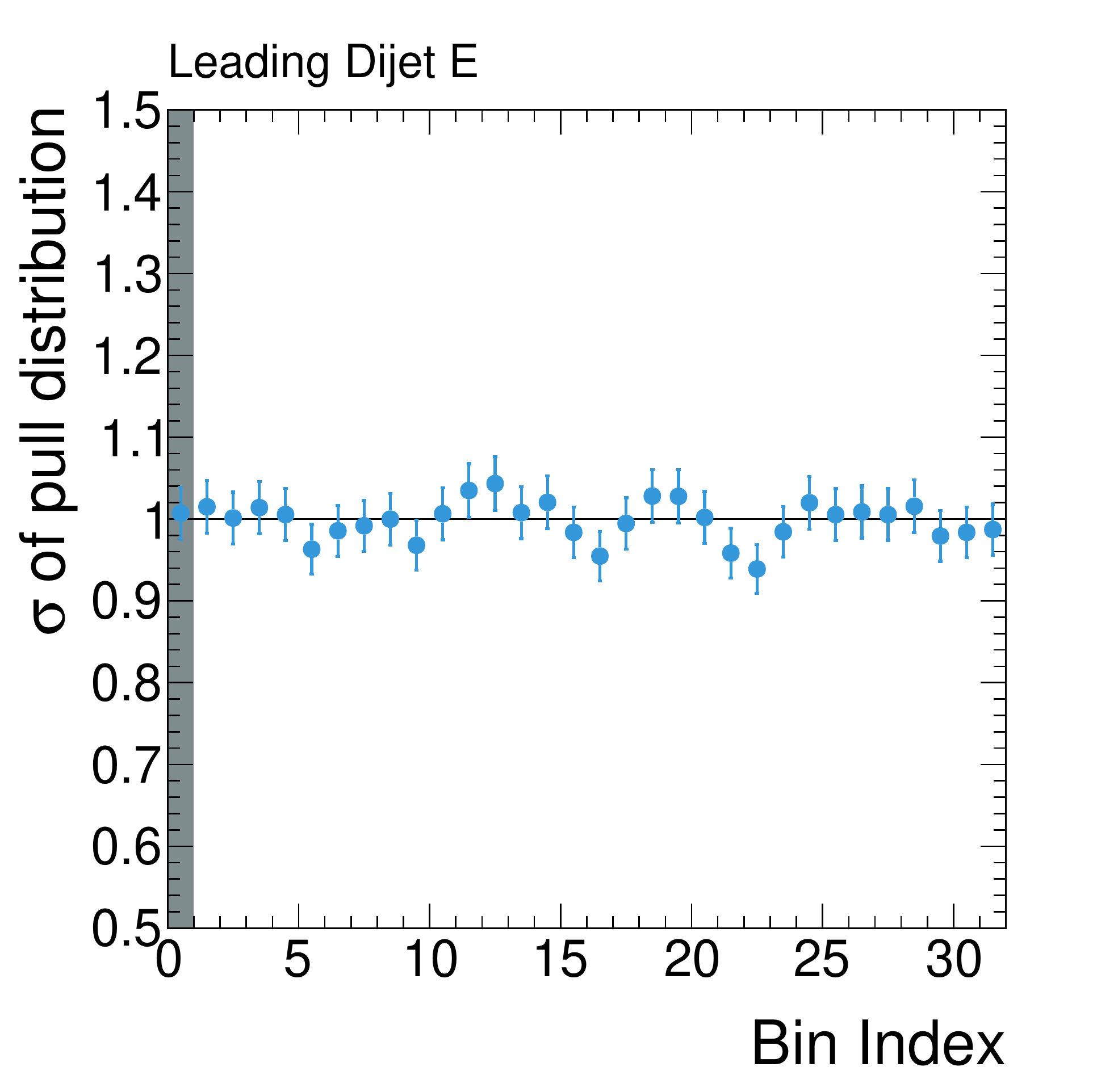}
    \includegraphicsthree{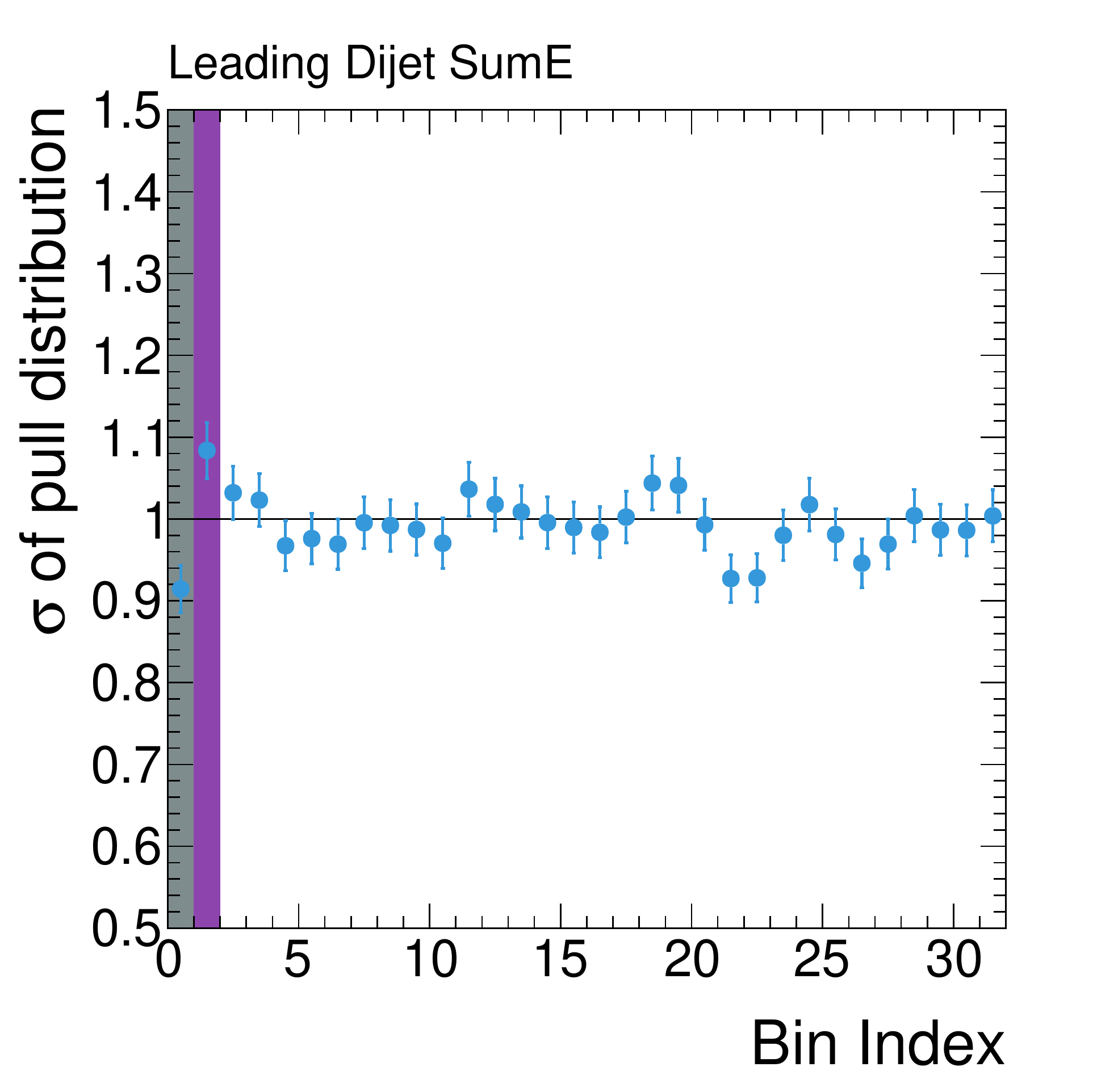}
    \includegraphicsthree{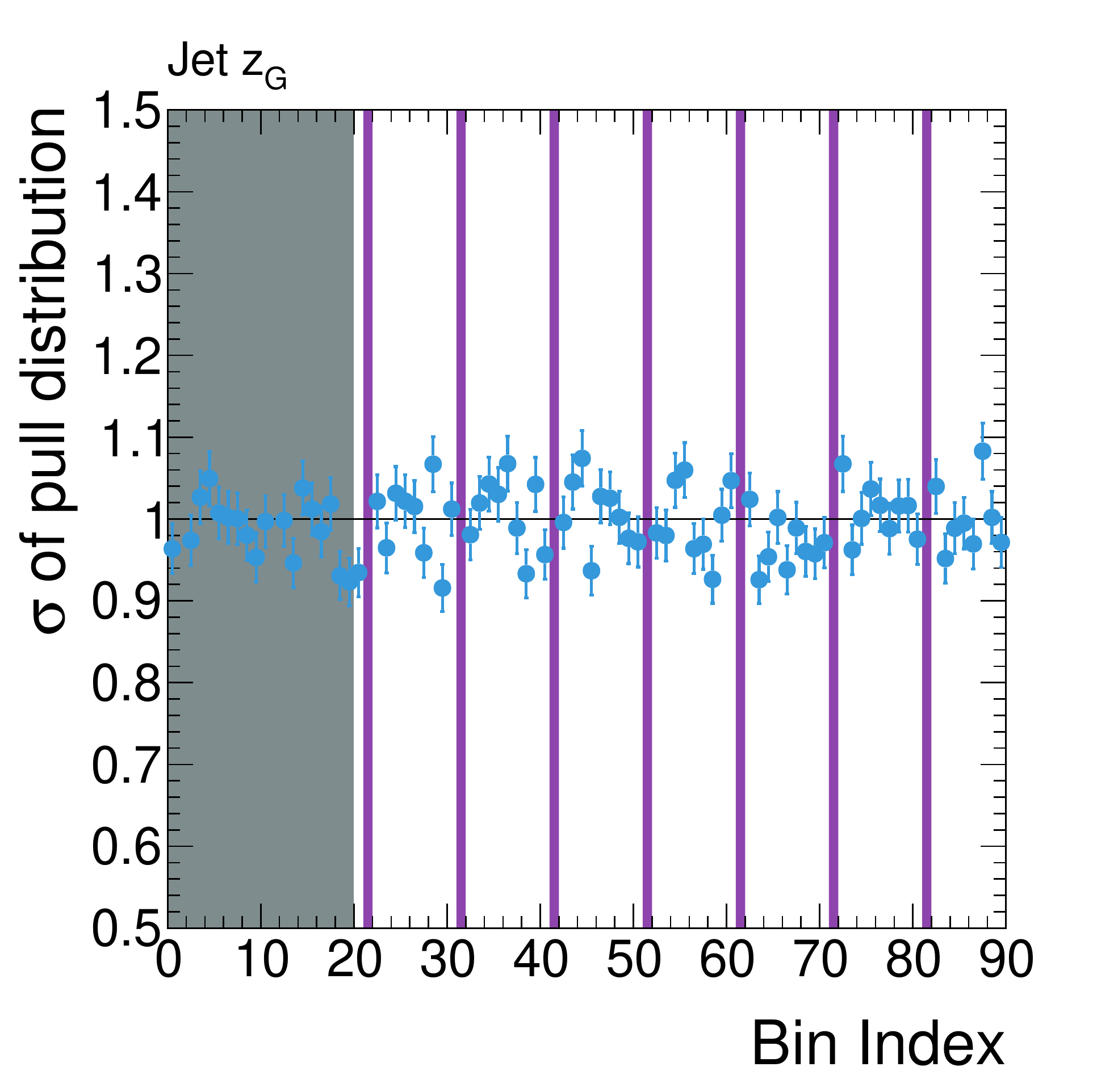}
    \includegraphicsthree{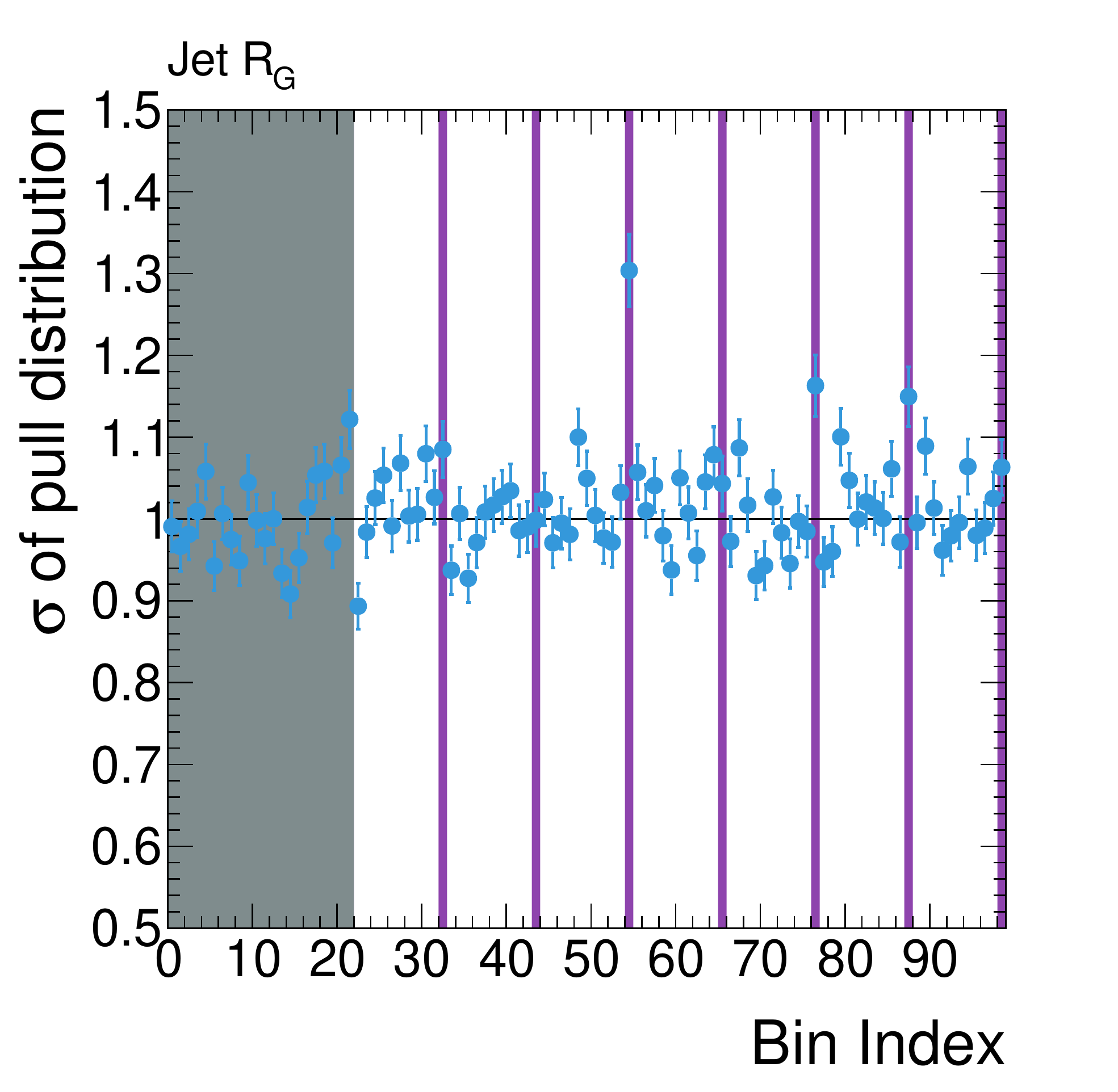}\\
    \includegraphicsthree{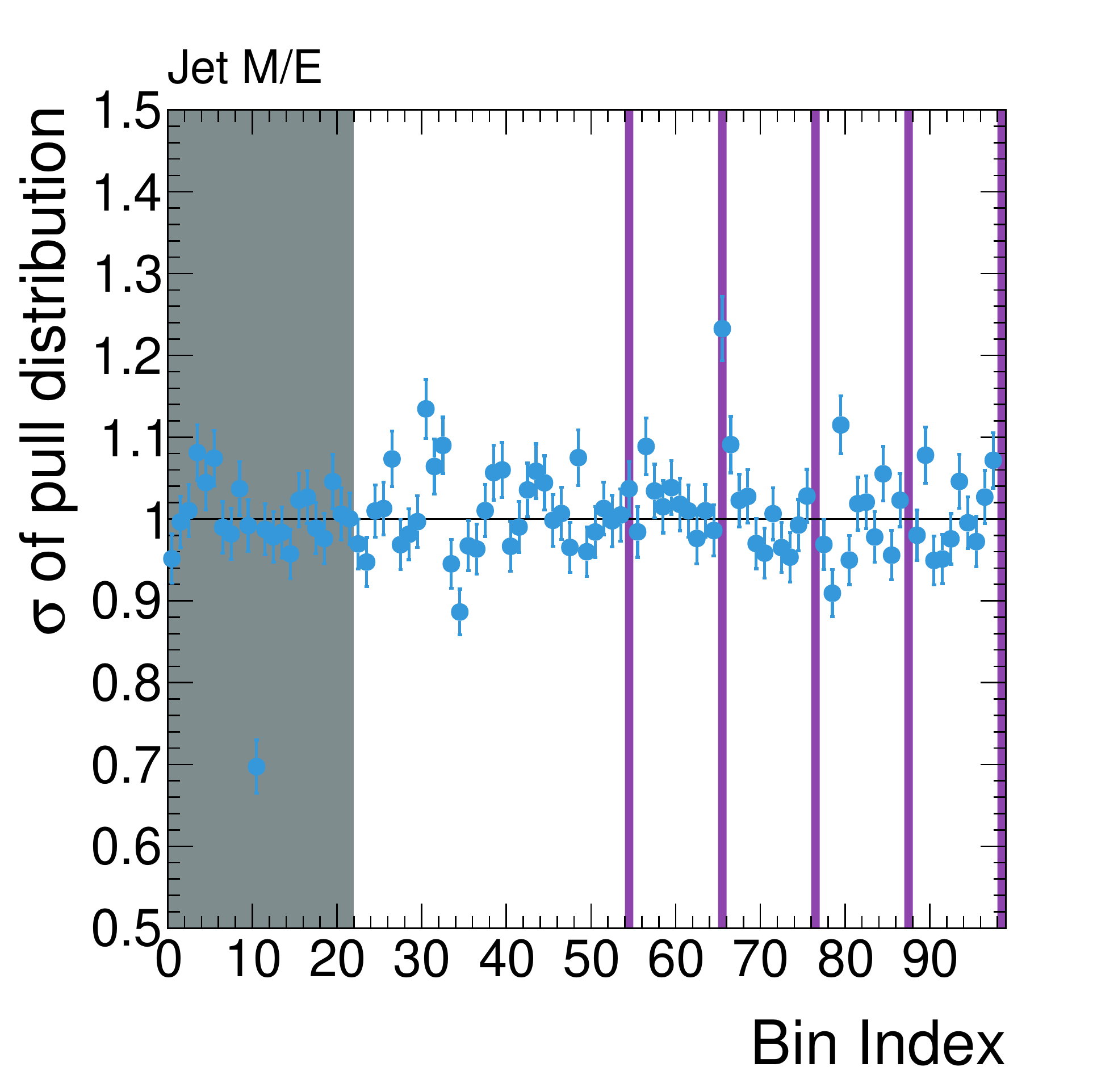}
    \includegraphicsthree{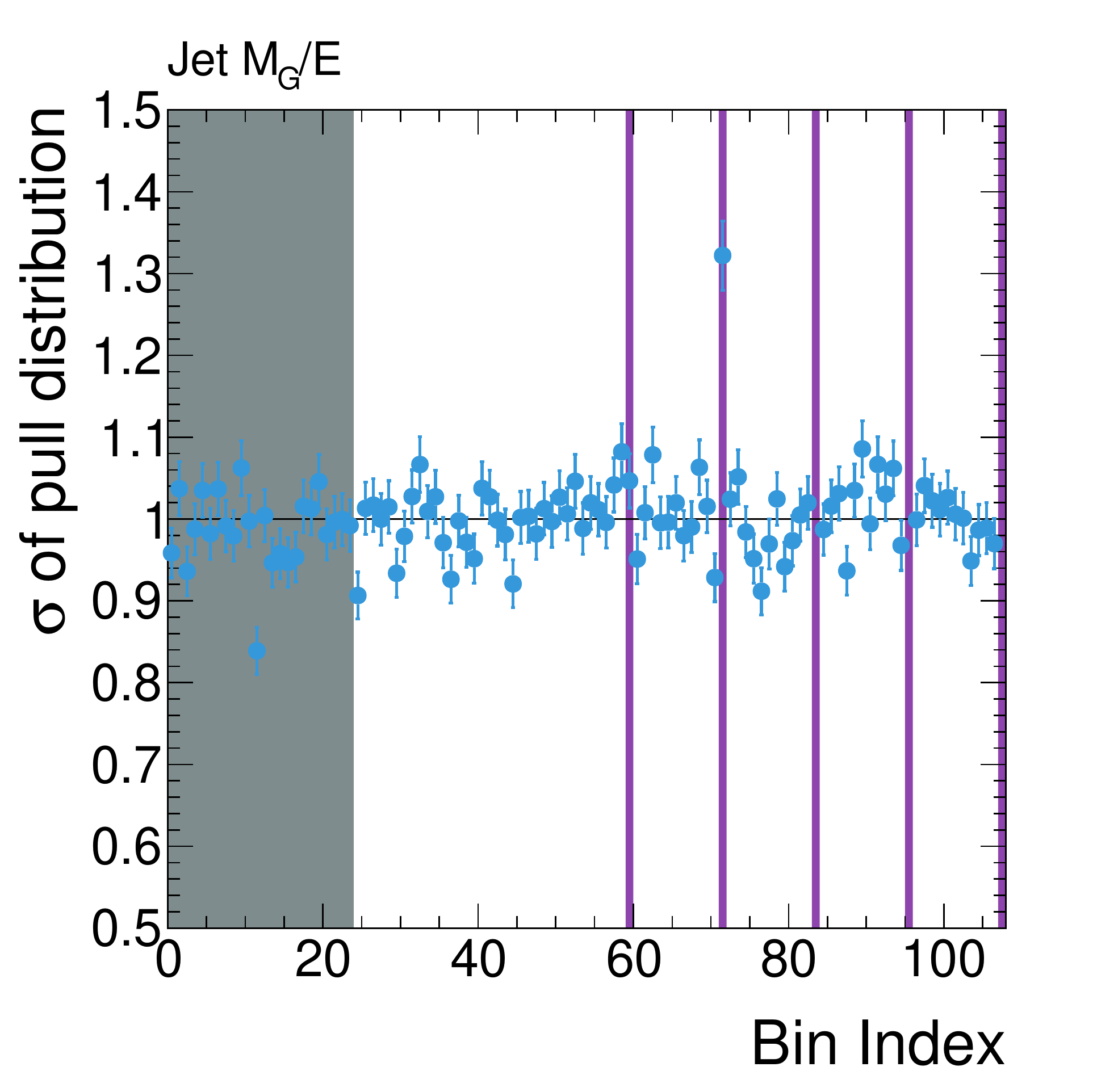}
    \caption{Width of the bin by bin pull distribution for jet energy, leading dijet energy, leading dijet total energy (top row), \zg, \Rg (middle row), and ungroomed and groomed jet mass (bottom row).  Underflow bins are shaded in gray, and bins with low statistics are shaded in purple.}
    \label{Figure:Unfolding-StatisticsValidation}
\end{figure}

The result is summarized in Fig.~\ref{Figure:Unfolding-StatisticsValidation}.  The results are consistent with one for the most part.  The lower bin indices are underflow bins (marked by gray background), and they are not considered as those bins do not appear in the final result.  In the calculation of pull, Gaussian statistics is assumed, and it is inaccurate when the statistics in the bin is low (marked as purple background), where a Poisson-like adaptation is needed.  Therefore in the tail of the jet substructure quantities (mass, groomed observables), occasionally we see the width of the pull deviating from one.

\subsection{Regularization Parameter Determination}

Another important aspect of the unfolding is the regularization parameter for the unfolding.  In terms of \Bayes, the regularization parameter is the number of iterations.  For \SVD, there is also a regularization parameter as input to the unfolding.

A two-step procedure is employed.  In the first step, the optimal regularization parameter for simulated sample is found through the use of the toy datasets, with the generated spectrum as the ``truth'' input.  A scan through different regularization parameters is performed, and the sum of the square differences across all bins between the unfolded spectrum and the input ``truth'' is used as the metric.

The data is then unfolded using the optimal parameter setting for the simulation.  This unfolded spectrum is used as the ``truth'' input for the second step of the regularization parameter optimization, and the same procedure as the first step is repeated.

\begin{figure}[htp!]
    \centering
    \includegraphicstwo{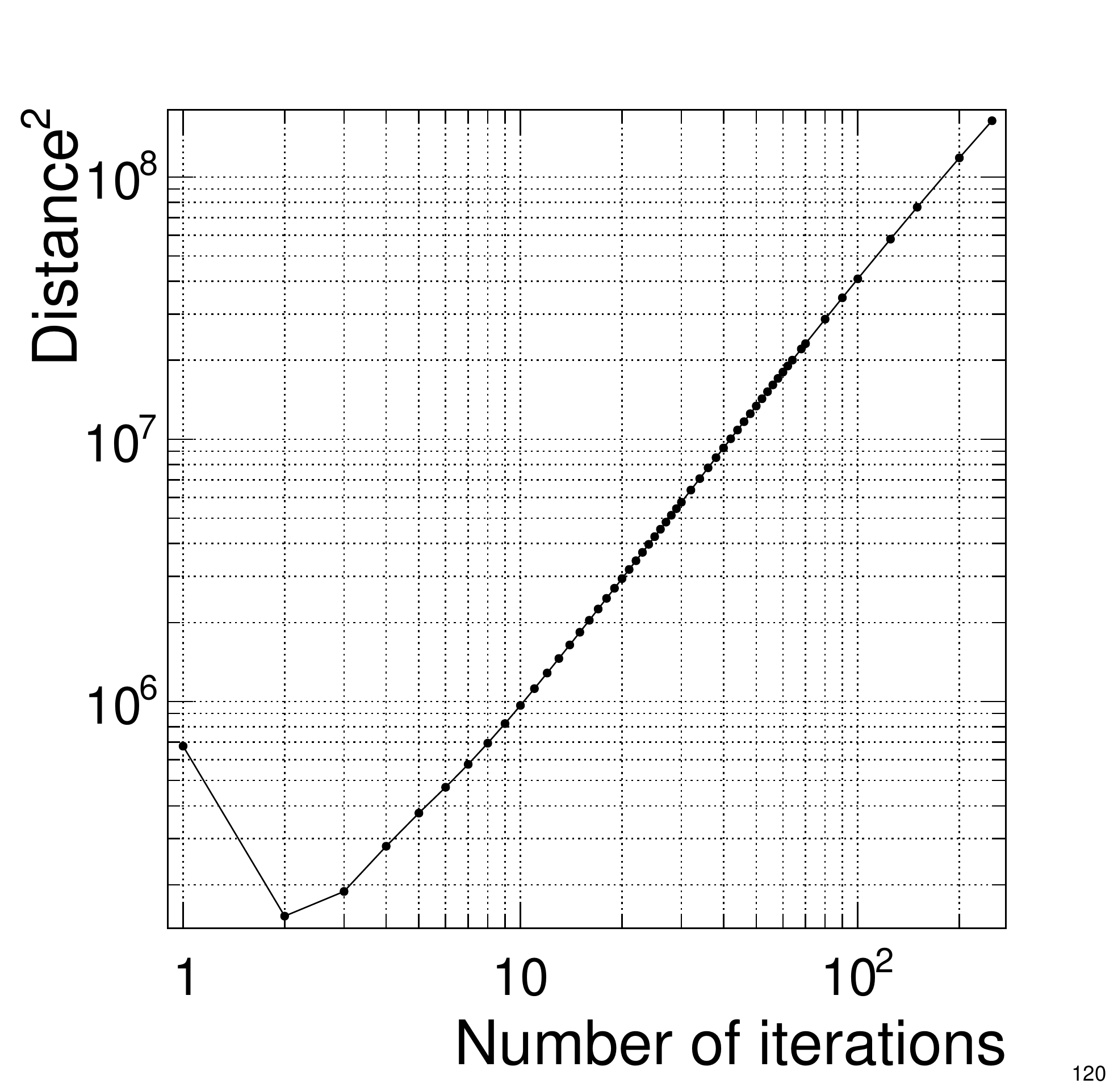}
    \includegraphicstwo{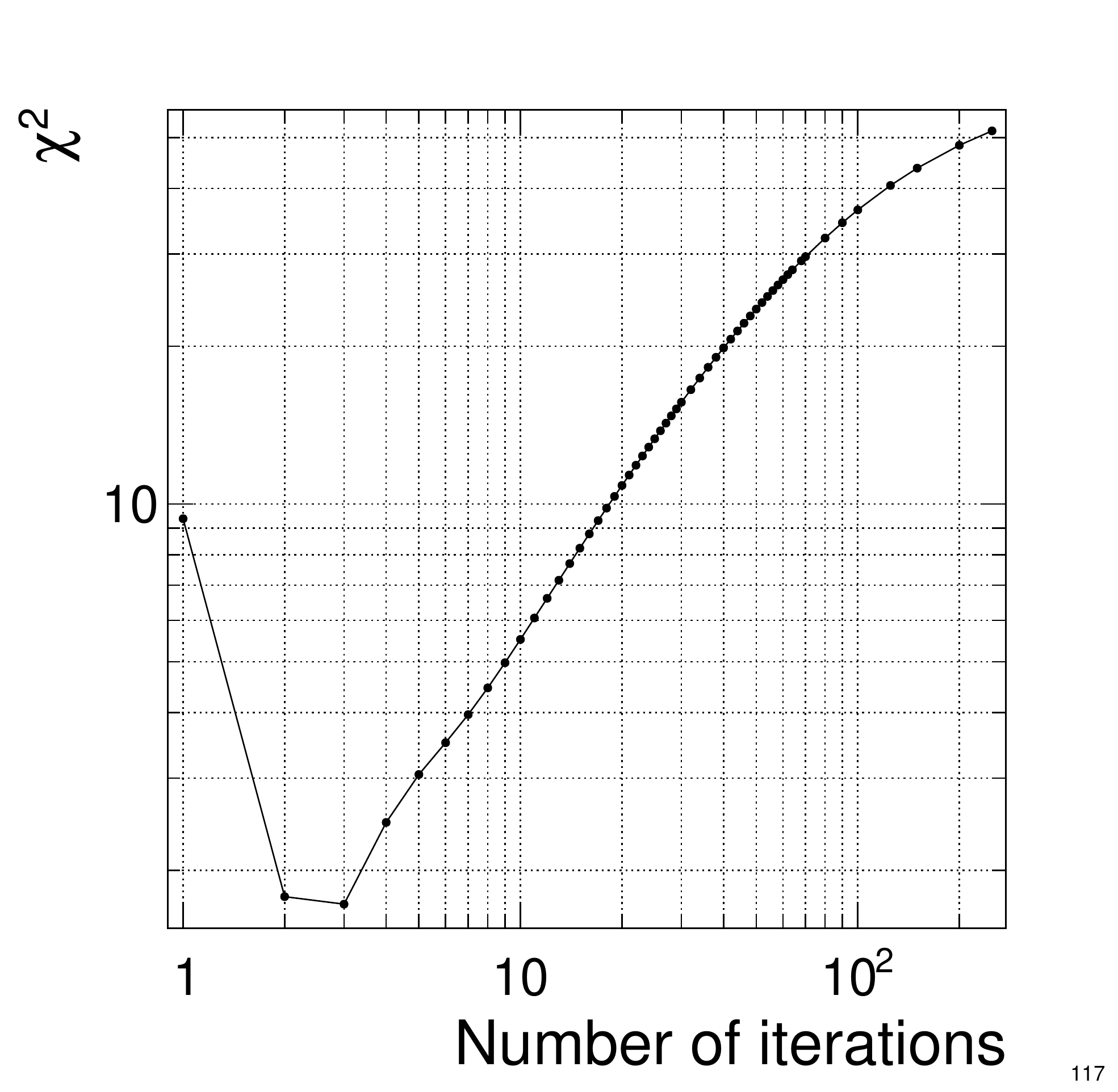}
    \includegraphicstwo{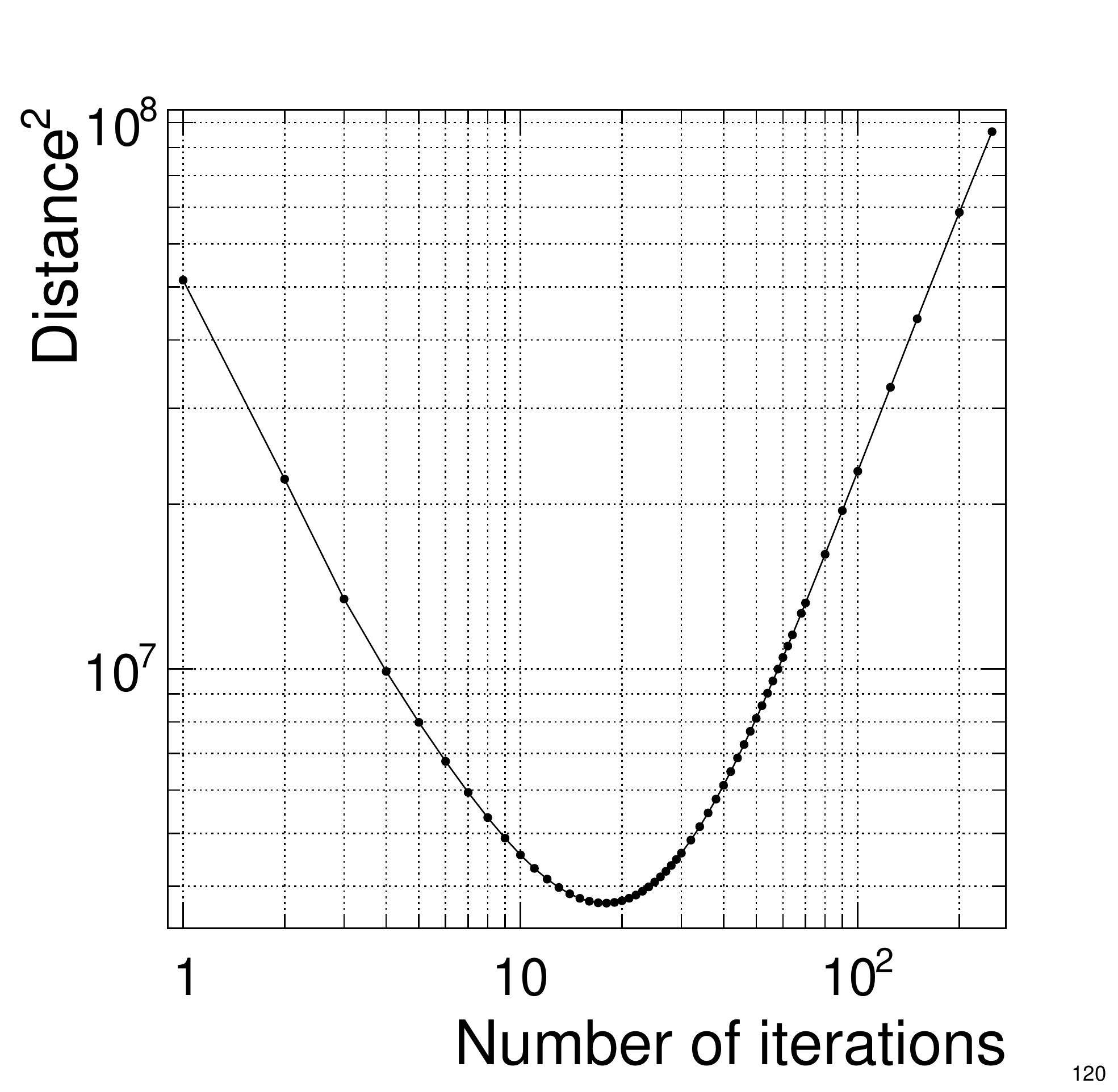}
    \includegraphicstwo{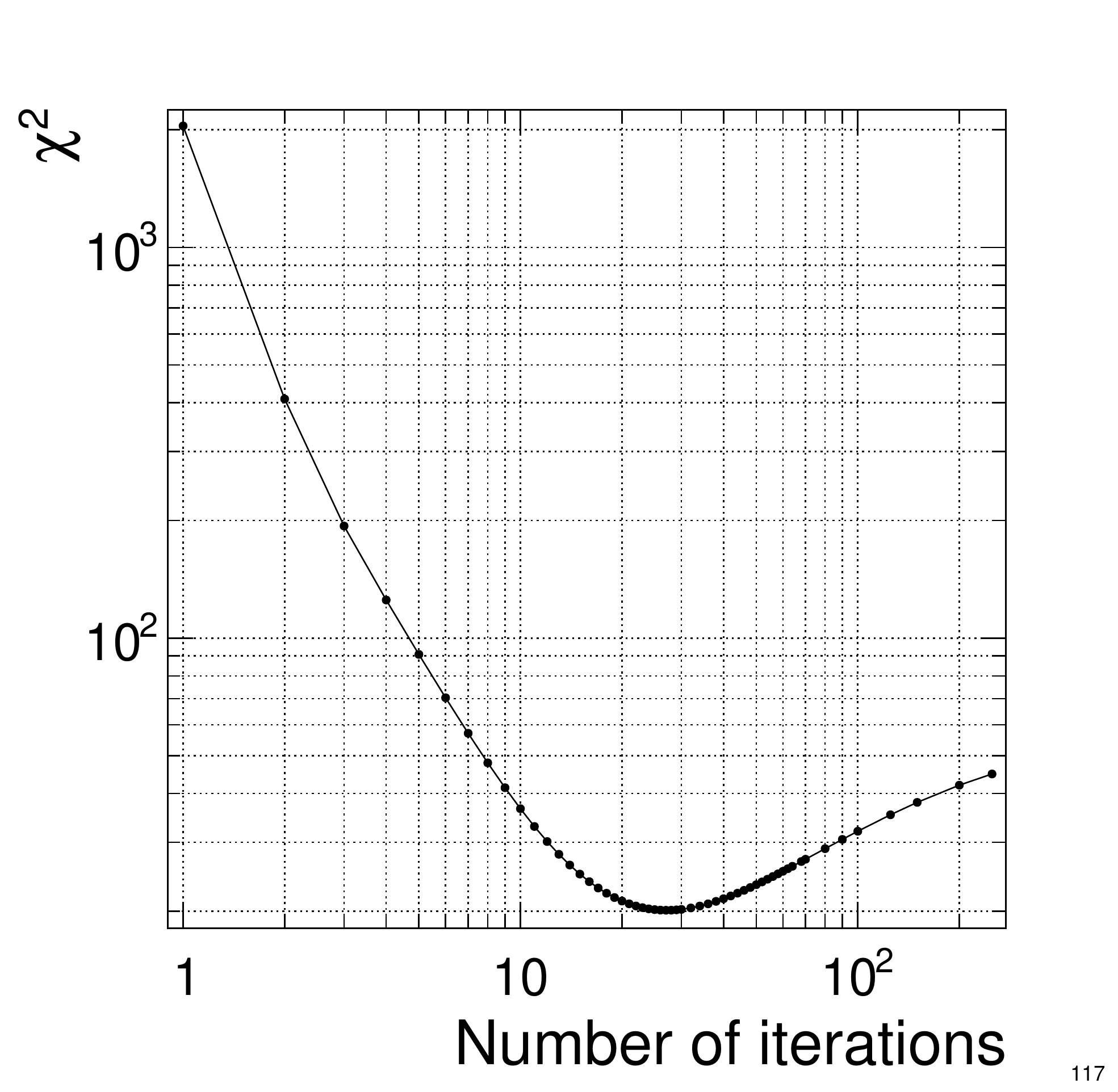}
    \caption{Scan of number of iterations for simulation (upper row) and data (lower row).  Both the squared distance (left) and $\chi^2$ (right) are shown.}
    \label{Figure:Unfolding-IterationScanJetE}
\end{figure}

An example is shown in Fig.~\ref{Figure:Unfolding-IterationScanJetE} for the determination for the jet energy unfolding.  The result of the first step on simulated spectrum is shown in the first row for both $\chi^2$ and distance squared.  The best iteration (2) is the used as input for the second step on data-like input, and the result is shown in the bottom row.  In this case 17 iterations works the best, after which the unfolding starts to magnify fluctuations and settle into one of the modes, as can be seen in the growing total squared distance.  The $\chi^2$ exhibit a similar trend on a qualitative level.  The $\chi^2$ decreases with increasing iterations both because of the convergence of the unfolded result to the input, and also because of the growing unfolding uncertainty, which becomes dominant when the number of iterations is large.  This is the reason why the number of iterations with the smallest $\chi^2$ is larger than the one using distance as the metric.  The growth in $\chi^2$ is also slower than its distance counterpart at large number of iterations, indicating the inflation of unfolding uncertainty.  For some other observables, the growth of uncertainty is so fast that the $\chi^2$ never increases up to few hundred iterations, well after the squared distance starts to grow rapidly.

For SVD unfolding regularization optimization, the input to the second step is taken from the same input as the second step optimization for \Bayes.

The determined number of iterations, and the SVD regularization parameters are summarized in Tab.~\ref{Table:Unfolding-IterationCount}.

\begin{table}[htp!]
    \centering
    \begin{tabular}{|c|c|c|c|c|}
        \hline
        \multirow{2}{*}{Observable} & \multicolumn{2}{c|}{\Bayes} & \multicolumn{2}{c|}{\SVD} \\\cline{2-5}
        & Simulated & Data & Simulated & Data \\\hline
        Jet Energy & 2 & 17 & 8 & 12 \\\hline
        Leading Dijet Energy & 1 & 7 & 3 & 4 \\\hline
        Leading Dijet Total Energy & 1 & 15 & 4 & 6 \\\hline
        \zg & 6 & 17 & 7 & 29 \\\hline
        \Rg & 2 & 8 & 12 & 42 \\\hline
        $M/E$ & 2 & 7 & 5 & 29 \\\hline
        $M_G/E$ & 2 & 16 & 12 & 34 \\\hline
    \end{tabular}
    \caption{Optimized regularization parameter for simulation and data for \Bayes and \SVD.}
    \label{Table:Unfolding-IterationCount}
\end{table}

\clearpage

\section{Systematic Uncertainties}\label{Section:Systematics}

A common set of systematic uncertainty sources are considered for all observables, and in addition for the leading jet observables, uncertainty sources related to the leading dijet selection are included.

The common systematic sources can be categorized into the following categories, to be described in subsequent subsections: jet energy scale and resolution uncertainties, fake jets and unfolding-related uncertainties.

\subsection{Jet Calibration and Resolution Uncertainties}

The uncertainties related to the jet calibrations are split into different sources: the non-closure to deal with imperfections of simulation-based jet calibration, and the jet energy scale/resolution variation for difference between data and simulation.

\subsubsection{Residual Jet Energy Scale}

The jet energy scale uncertainty targets the potential difference between data and simulation.  The residual jet energy correction is varied by 0.5\% both up and down from the nominal value, and the variation in the unfolded spectrum is taken as the uncertainty associated with the residual jet energy scale.  It is a conservative choice which covers the observed spread in the residual jet energy correction derivation.

\subsubsection{Residual Jet Energy Resolution}

The jet energy resolution varies from 0\% to 5\% worse in data compared to simulation, depending on the jet direction.  Therefore the nominal value of the resolution scale factor is taken as 2.5\%, and a variation from 0\% to 5\% is taken as the systematic variation.  The unfolded spectra using varied jet energy resolution is quoted as the uncertainty.

\subsection{Fake Jets}

Fake jets are defined as the accidental clusters of energy in the final state that do not correspond to an initial high energy parton produced in the hard process.  The contribution for fake jets are estimated by using only reconstructed jets that are matched to a hard process parton.  The unfolded spectrum is then compared with the nominal unfolded result from simulation.  The difference between the two is quoted as the uncertainty for the fake jet contribution.

\subsection{Unfolding Uncertainties}

\subsubsection{Choice of Prior}

In the \Bayes formalism, there is a possibility to supply a prior knowledge of how the unfolded spectrum should look like.  The nominal result is done with a flat prior (i.e., no prior knowledge of the unfolded result).  Since a flat prior is still a choice we make, it is necessary that we test the effects from this assumption.  In order to estimate effect, the unfolding is redone using the simulated spectrum as the prior.  The difference in the unfolded spectra is then taken as the uncertainty from this source.

\subsubsection{Choice of Regularization}

The regularization parameter is varied to assess the impact on the choice.  The number of unfold iteration is varied by 1 on both sides from the optimal number determined with a procedure described in Section~\ref{Section:Unfolding}.  The difference to the nominal result is then quoted as the uncertainty.

\subsubsection{Unfolding Method}

In order to address any potential bias in the \Bayes methodology itself, an alternative unfolding method, \SVD is used.  The unfolded spectra between the two unfolding methods are compared, and the difference is taken as systematic uncertainty.

\subsubsection{Closure in Simulation}

The nonclosure, or the difference between the unfolded simulated spectra and the generated spectra, is quoted as a source of uncertainty.  Any potential miscalibration of simulation-based jet energy correction will be covered by this.


\subsection{Leading Jet Selection Uncertainties}

The uncertainty described in this subsection applies only to the leading jet measurements.

\subsubsection{Choice of Cuts}

The nominal cut is selected for 99\% purity for the leading dijet selection.  The variation is chosen by cuts for 98\% and 99.5\% purity.  Unfolding is performed for each of the variation, and the difference in the unfolded spectra is quoted as the systematic uncertainty.

\subsubsection{Resolution Effects}

The resolution of \HybridE is evaluated in Section~\ref{Section:LeadingJet}.  The value of \HybridE is smeared event-by-event in simulated sample according to the resolution, and the unfolding is performed on the smeared sample, and the difference in the spectra is examined.

\subsubsection{Correction}

The nominal correction is derived from simulation, as described in section~\ref{Subsection:LeadingJetCorrection}.  The reweighting procedure to estimate potential mismodeling of the simulation is described in section~\ref{Subsection:LeadingJetCorrectionVariation}.  The ratio in the derived correction factor is used as the base for the systematic uncertainty, with care taken to bridge the gap where the ratio switches sign to aovid artificial dips in the uncertainty.

\subsection{Summary of Uncertainties}

The summary of all the uncertainties considered is shown in this section.

\begin{figure}[htp!]
    \centering
    \includegraphicsone{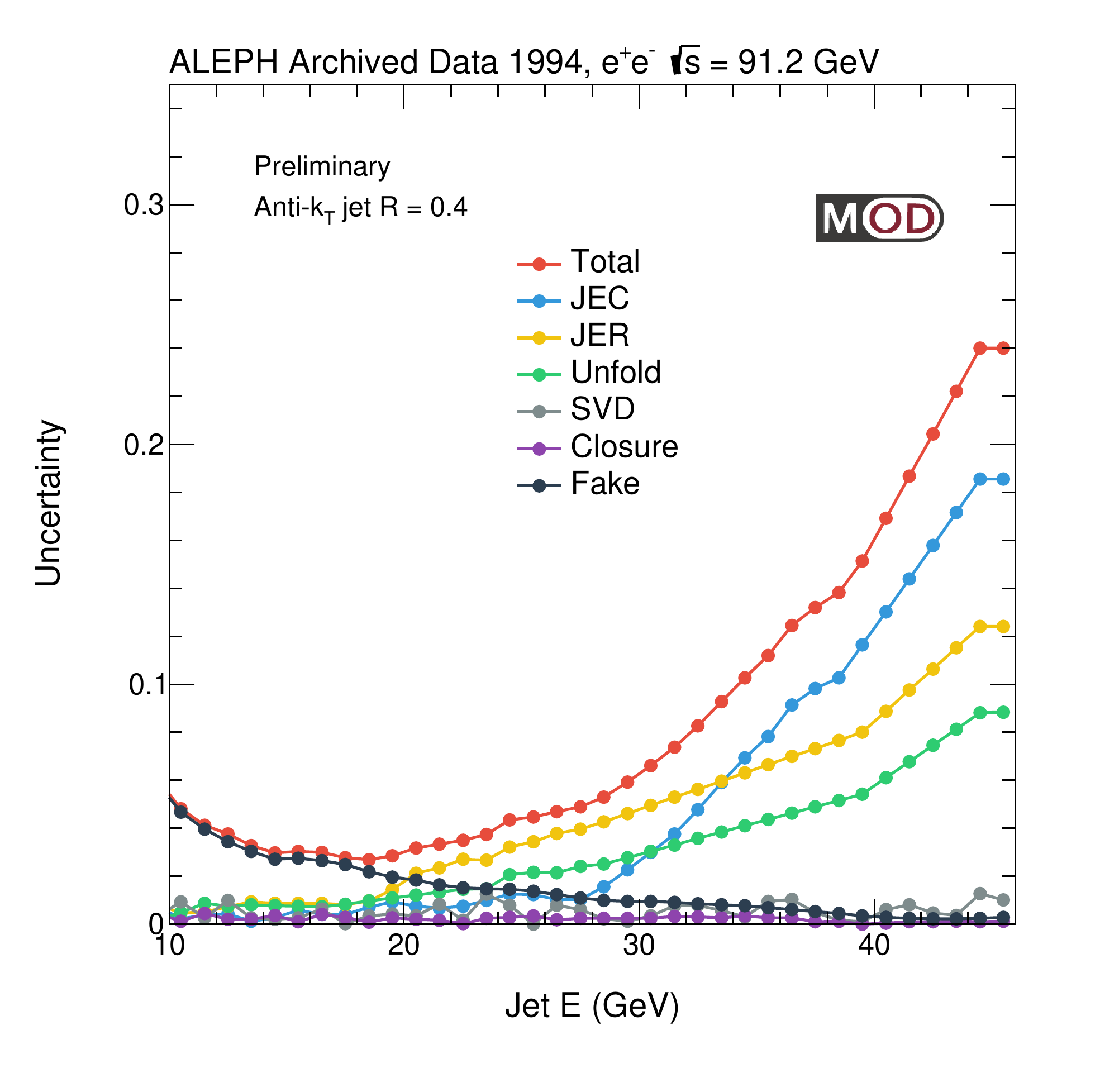}
    \caption{Summary of systematic uncertainty for the jet energy spectrum.}
    \label{Figure:Systematics-JetE}
\end{figure}

The uncertainties for the inclusive jet spectrum is shown in Figure~\ref{Figure:Systematics-JetE}.  It is dominated by the jet energy resolution and correction uncertainties.  For lower jet energies, the fake jets dominates.

\begin{figure}[htp!]
    \centering
    \includegraphicsone{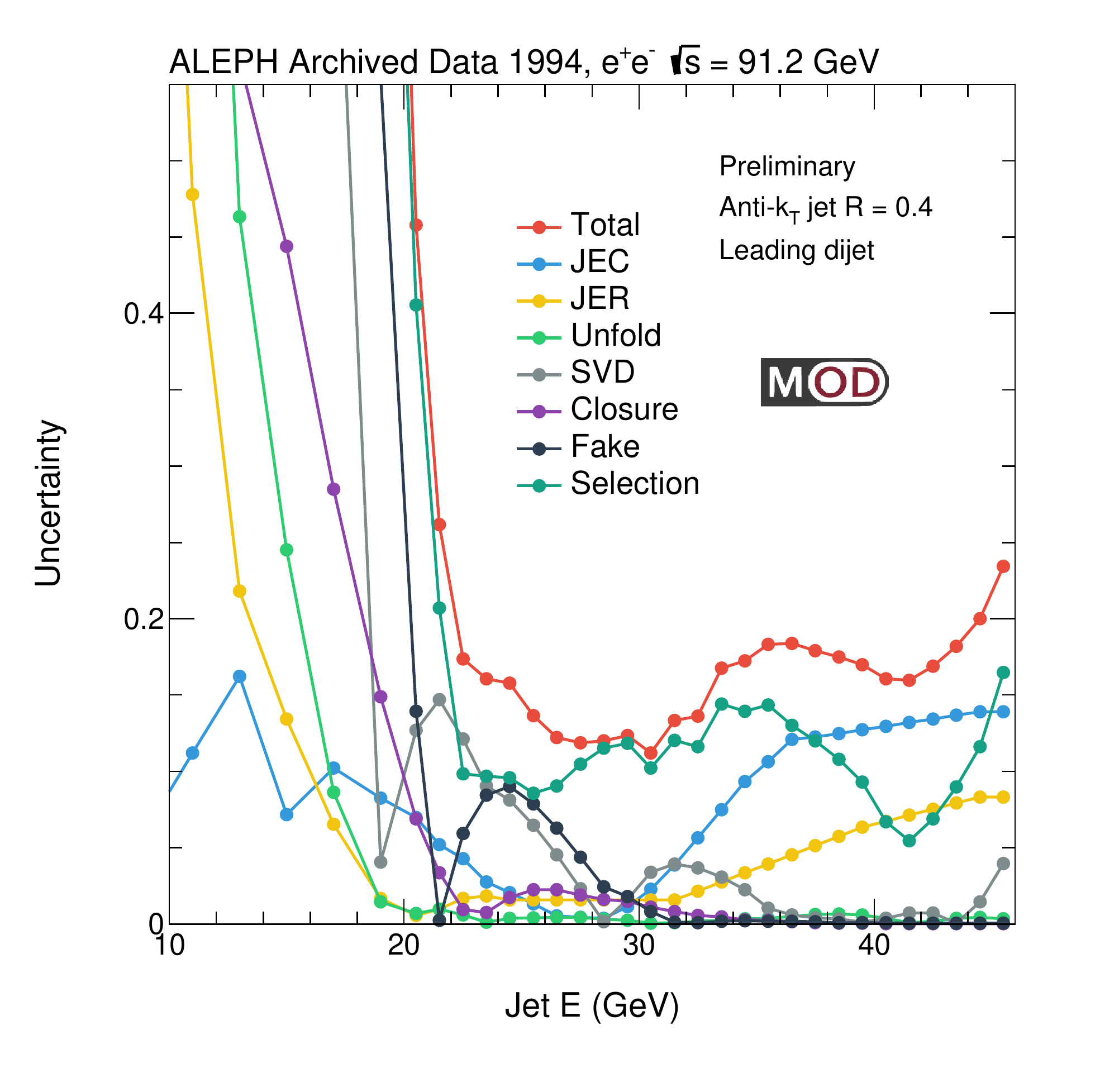}
    \caption{Leading dijet energy systematics}
    \label{Figure:Systematics-DiJetE}
\end{figure}

The uncertainties for leading dijet energy is summarized in Figure~\ref{Figure:Systematics-DiJetE}.  The leading dijet selection-related uncertainties limits the accuracy for lower energy jets, while for higher energy jets the usual jet energy correction and resolution uncertainties dominates.

\begin{figure}[htp!]
    \centering
    \includegraphicsone{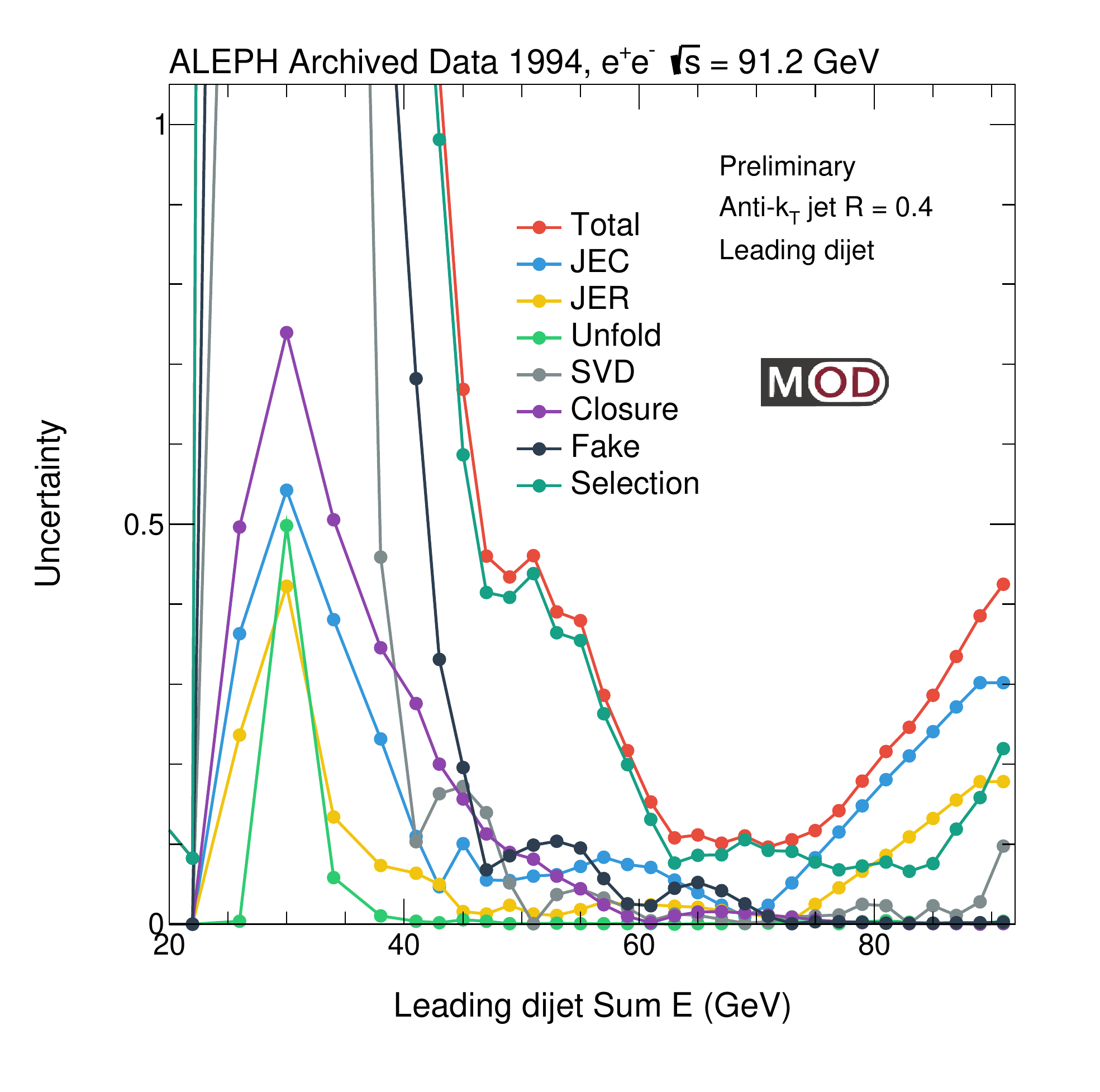}
    \caption{Leading dijet sum energy systematics}
    \label{Figure:Systematics-DiJetSum}
\end{figure}

A similar picture is found for the leading dijet total energy, shown in Figure~\ref{Figure:Systematics-DiJetSum}.  In addition to the selection and jet reconstruction, the \SVD uncertainty is seen not to be completely stable for this observable.

\begin{figure}[htp!]
    \centering
    \includegraphicsonewide{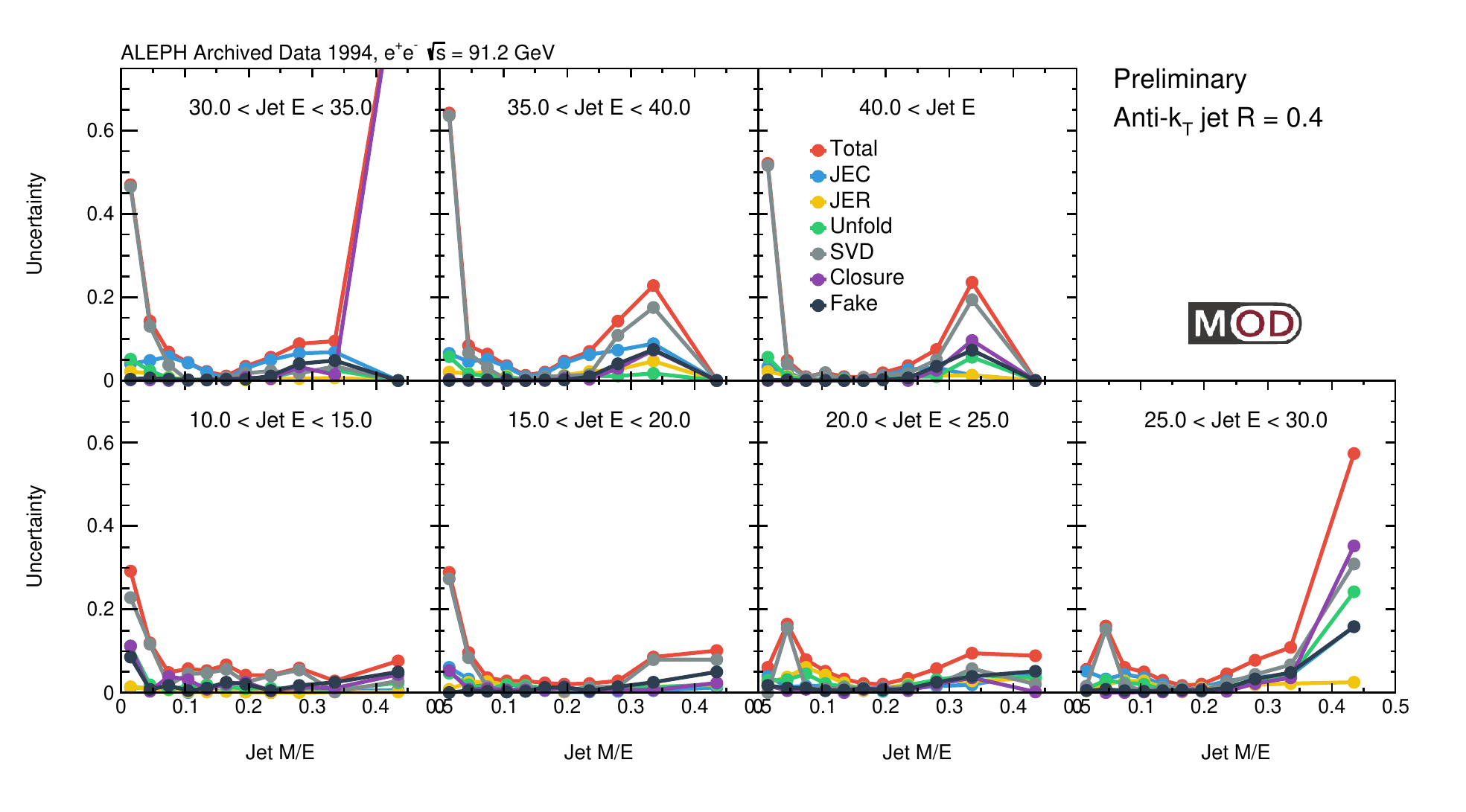}
    \caption{Mass / energy systematics}
    \label{Figure:Systematics-JetME}
\end{figure}

\begin{figure}[htp!]
    \centering
    \includegraphicsonewide{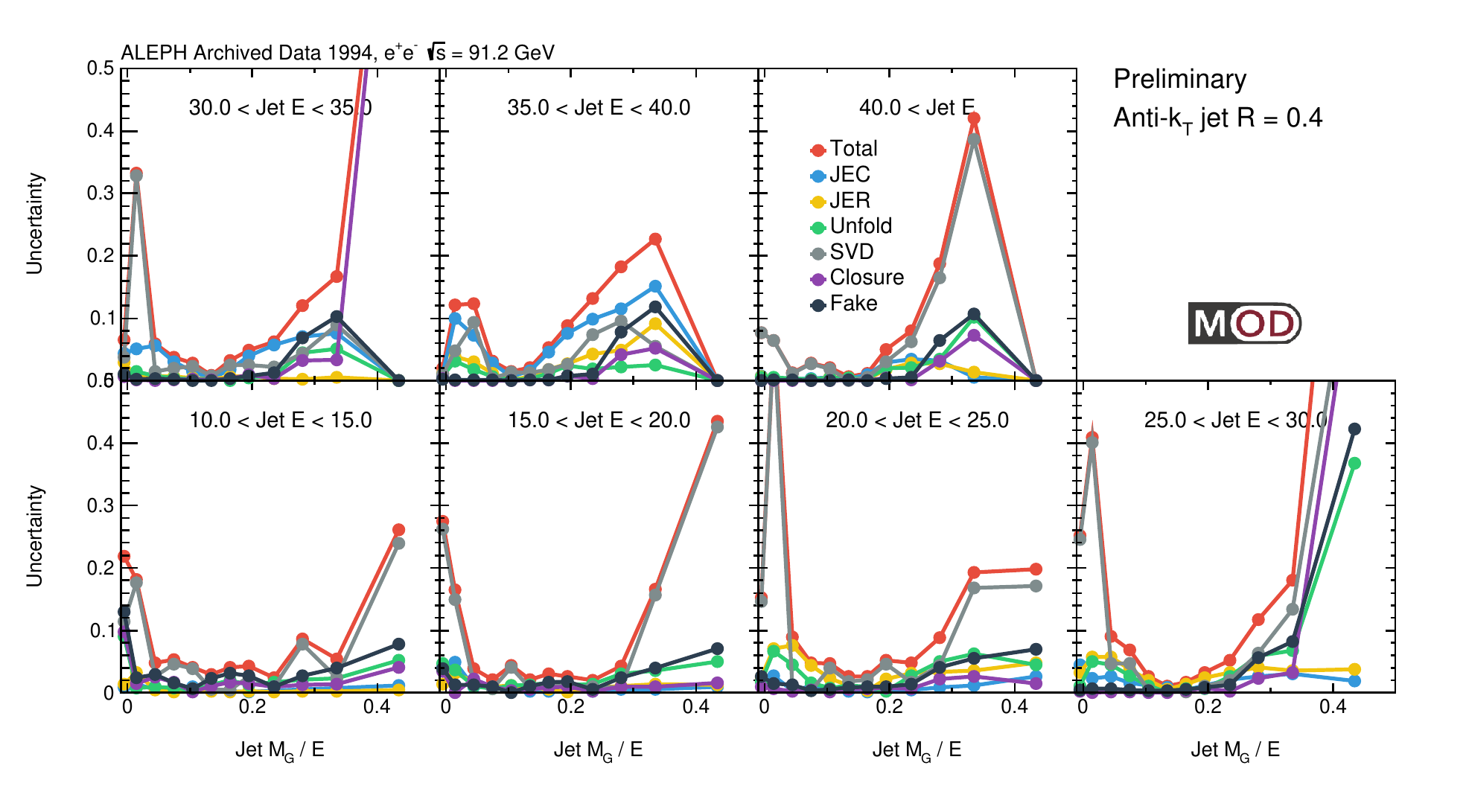}
    \caption{Groomed mass / energy systematics}
    \label{Figure:Systematics-JetMGE}
\end{figure}

The uncertainties related to jet mass are shown in figure~\ref{Figure:Systematics-JetME} for inclusive jet mass, and figure~\ref{Figure:Systematics-JetMGE} for the groomed jet mass.  The relative uncertainty for the last bin in each panel varies greatly mainly due to the diminishing statistics in those bins.  The dominant systematics are jet energy reconstruction related, as well as the \SVD variation.

\begin{figure}[htp!]
    \centering
    \includegraphicsonewide{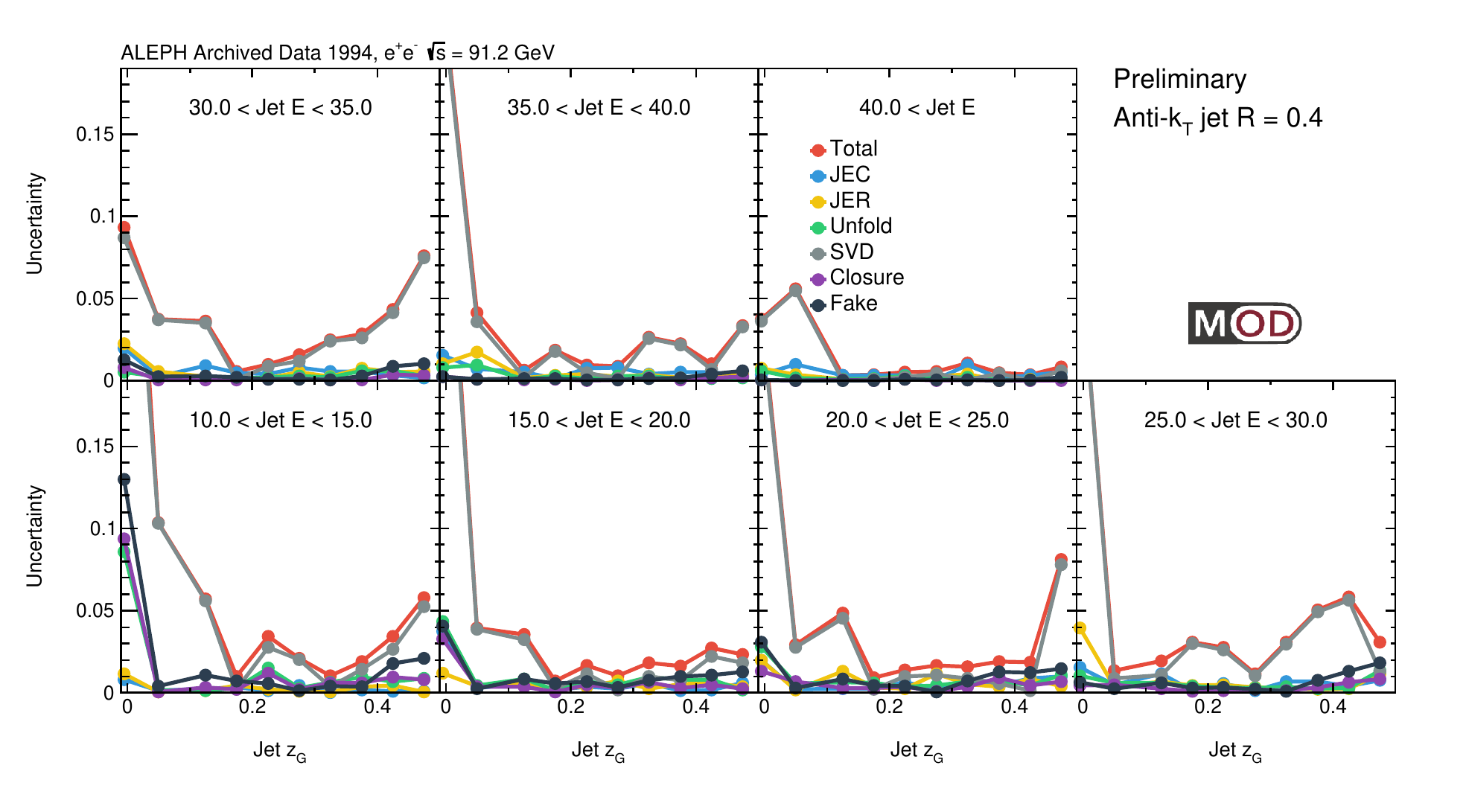}
    \caption{\zg systematics}
    \label{Figure:Systematics-JetZG}
\end{figure}

\begin{figure}[htp!]
    \centering
    \includegraphicsonewide{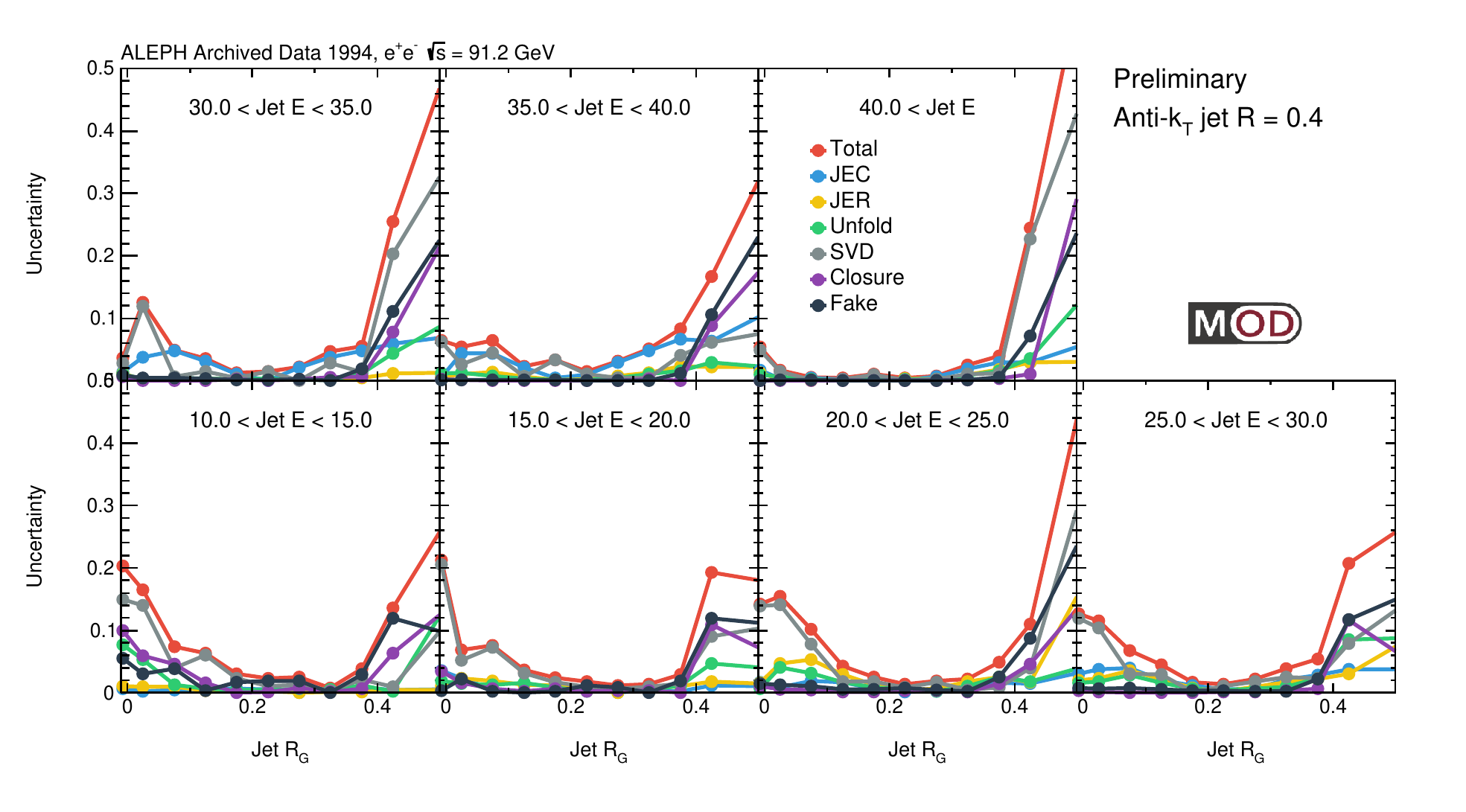}
    \caption{\Rg systematics}
    \label{Figure:Systematics-JetRG}
\end{figure}

Finally, uncertainties for the groomed \zg (Figure~\ref{Figure:Systematics-JetZG}) and \Rg (Figure~\ref{Figure:Systematics-JetRG} are shown.  Here the jet energy related systematics mostly cancel, and the dominant is the \SVD variation.

\clearpage

\section{Cross Checks}\label{Section:CrossCheck}

\subsection{Thrust Cross Check}

The thrust is an event-wide observable which characterizes the overall distribution of particles.  The thrust is defined as
\begin{align}
    T \equiv \max_{\hat{n}} \left[ \dfrac{\sum_i | \vec{p}_i \cdot \hat{n} |}{\sum_i | \vec{p}_i |} \right],
\end{align}
where $\hat{n}$ is a unit vector, and $\vec{p}_i$ is the momentum of the $i$-th particle.  The $\hat{n}$ that maximizes this quantity defines the thrust axis direction.
The thrust is close to 1 for dijet events, and small for events with particles spreading out uniformly in all directions.

The thrust distribution has been measured by the ALEPH collaboration~\cite{ALEPH:2003obs}.  A cross check using the thrust observable is carried out to check if the obtained unfolded thrust distribution is consistent with what has been published.

\begin{figure}[htp!]
    \centering
    \includegraphicsone{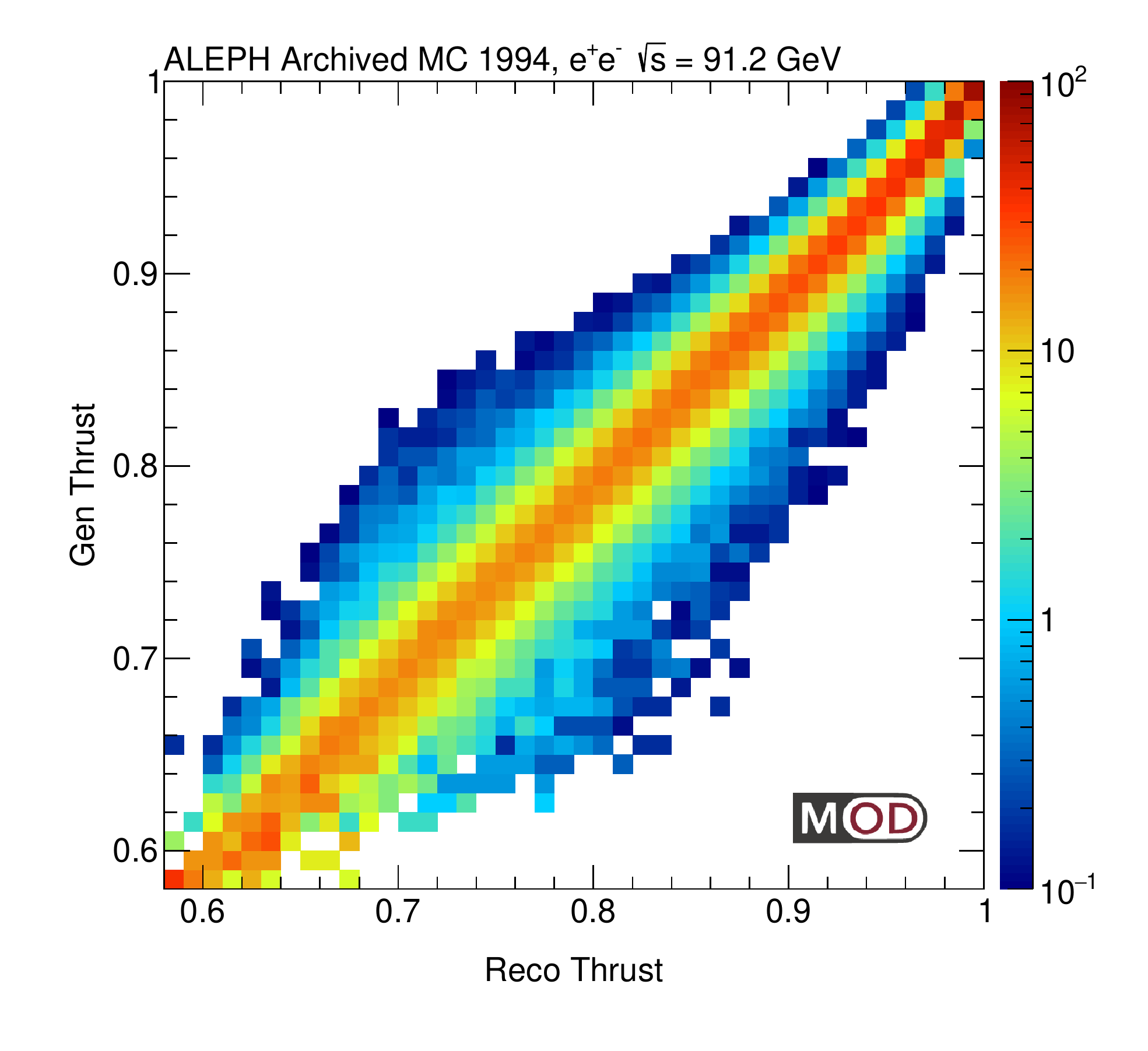}
    \caption{Smearing matrix for the thrust distribution}
    \label{Figure:CrossCheck-MatrixThrust}
\end{figure}

The smearing matrix is shown in figure~\ref{Figure:CrossCheck-MatrixThrust}.  The matrix is row-normalized so that the integrals for all rows are equal to unity.  The bin sizes are uniform and correspond to the same binning used in the published result.  A tight correlation between generated and reconstructed thrust is observed.

\begin{figure}[htp!]
    \centering
    \includegraphicstwo{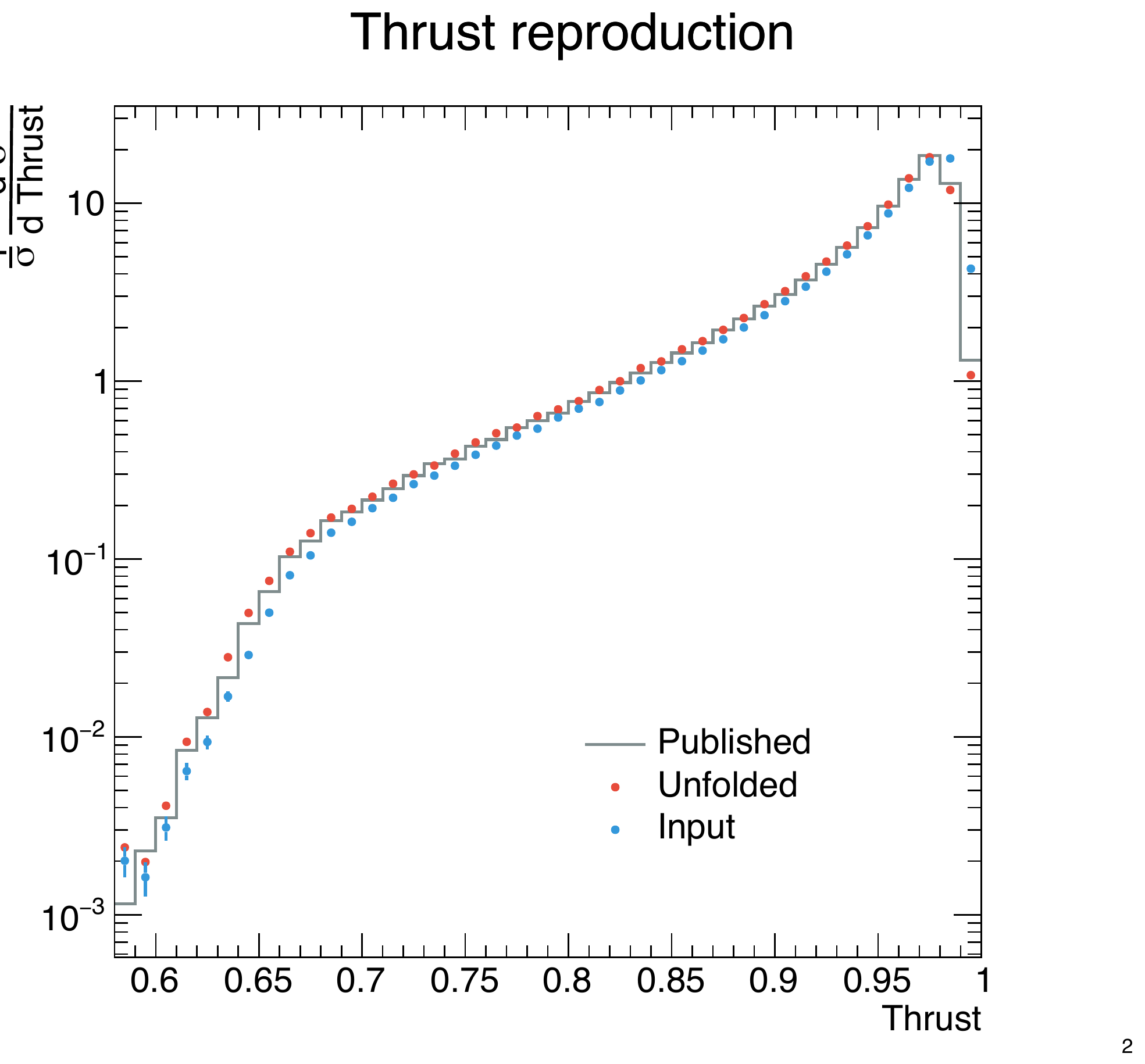}
    \includegraphicstwo{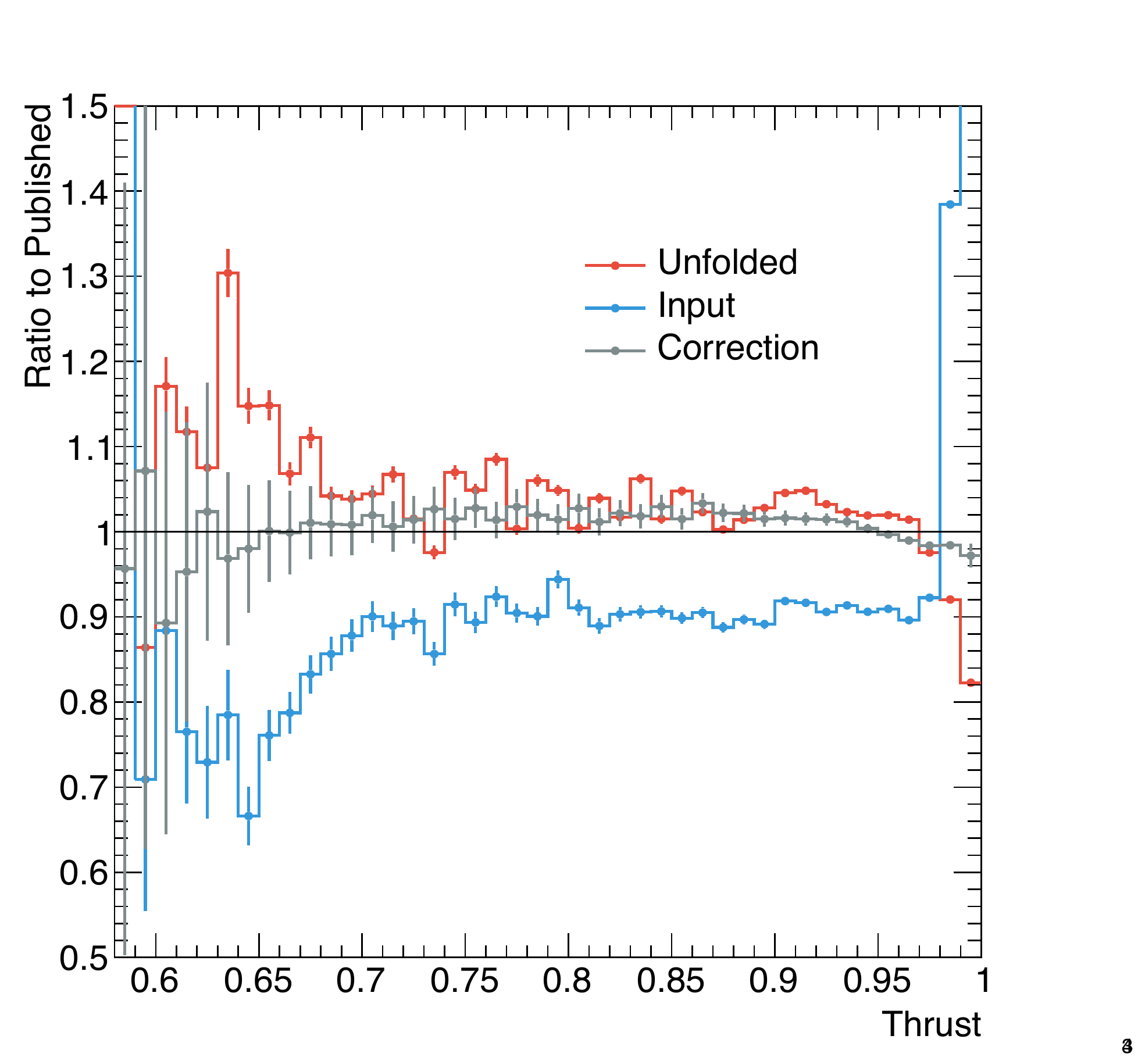}
    \caption{(Left panel) Unfolded thrust distribution (red) compared to the published spectra (gray) and the input distribution before unfolding (blue).  (Right panel) Ratio of the spectra to the published result.  The size of the correction is shown in gray.}
    \label{Figure:CrossCheck-UnfoldedThrust}
\end{figure}

The unfolded result is compared to the published result in the left panel of figure~\ref{Figure:CrossCheck-UnfoldedThrust}, and the ratio in the right.  After unfolding, the distribution agrees much better with the published result.  There is also a correction needed coming from the event selection, which is not applied by default in the unfolded (red) distribution.  The effect of the correction is shown as gray in the ratio plot.  For the majority of ranges, the unfolded distribution agrees with the published result.  There is, however, some remaining discrepancy.

\begin{figure}[htp!]
    \centering
    \includegraphicstwo{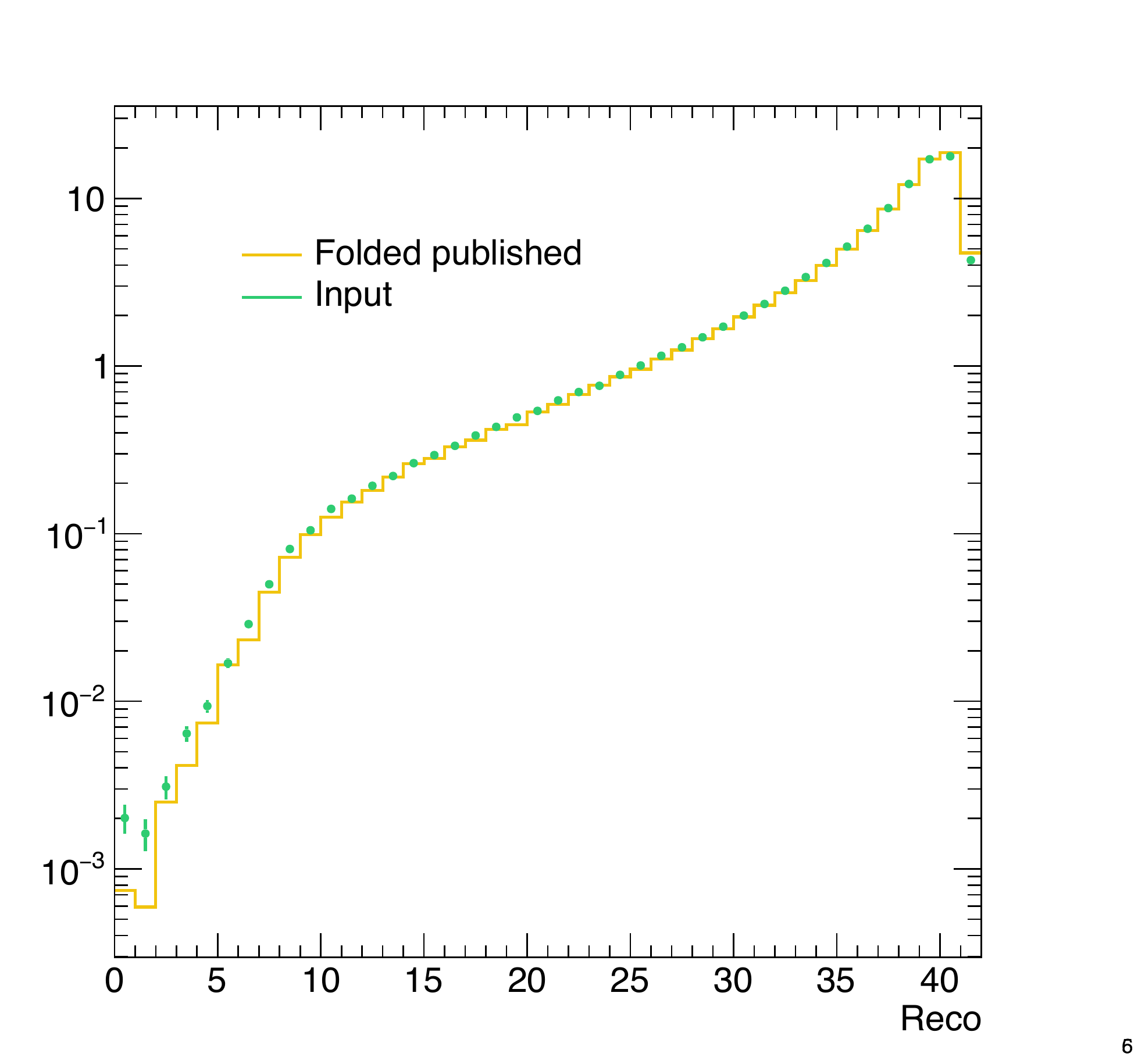}
    \includegraphicstwo{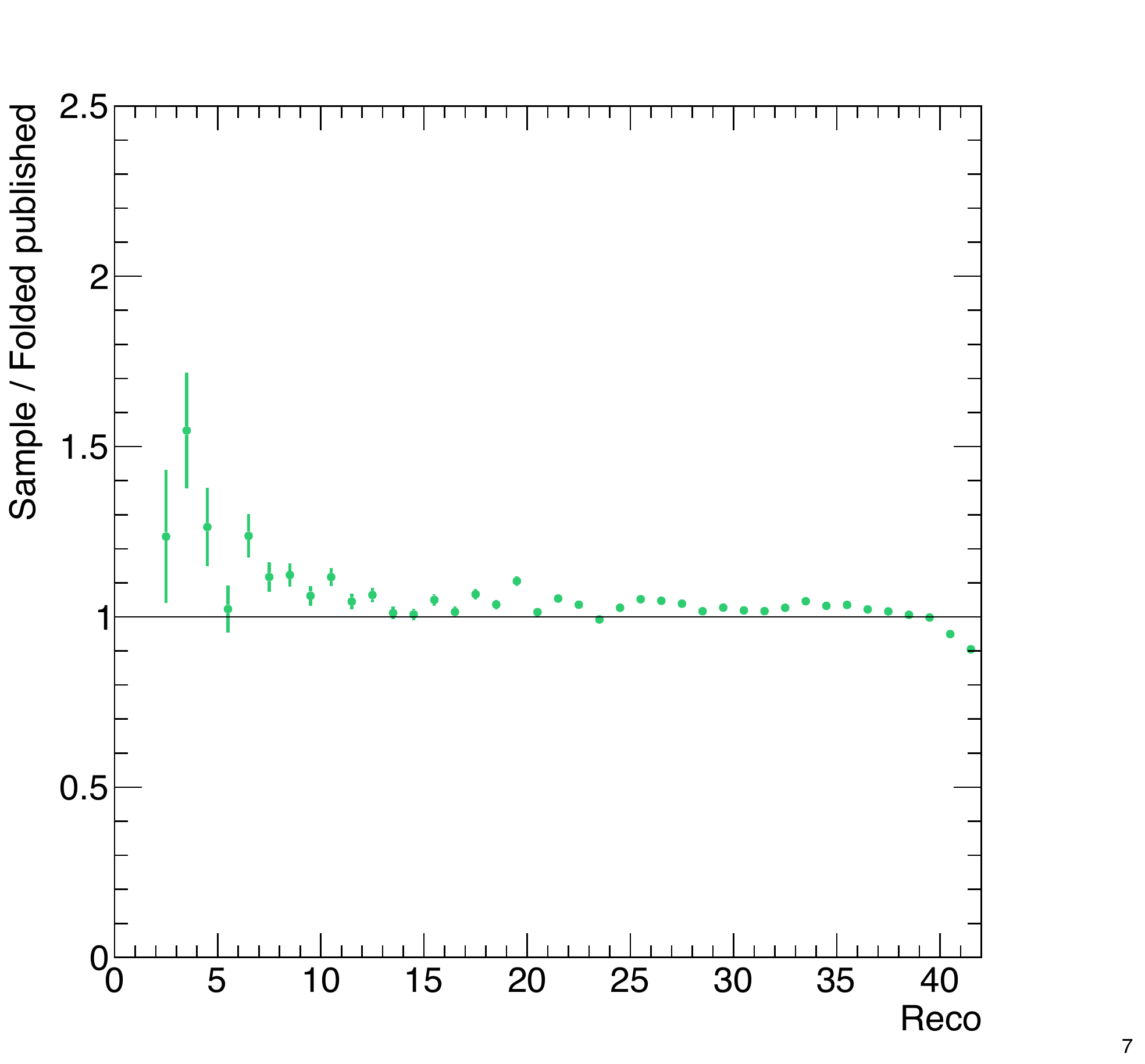}
    \caption{Comparison of the published spectrum folded with the response matrix (yellow), with the raw observed thrust spectra before unfolding (green).  The ratio is shown on the right, showing a similar amount of disagreement as in the unfolded case.}
    \label{Figure:CrossCheck-FoldedThrust}
\end{figure}

In order to understand if the discrepancy comes from the unfolding procedure or from some other sources, a forward folding comparison is done by convoluting the published spectrum with the smearing matrix, and compare to the pre-unfolding input.  The result is shown in figure~\ref{Figure:CrossCheck-FoldedThrust}.  A similar level of discrepancy is observed compared to the unfolded result, indicating that the source of the nonclosure is not from the unfolding procedure, but most likely from the some other corrections that are not applied in this cross check.

\clearpage

\section{Result}\label{Section:Result}

In this section, we present the fully corrected jet spectra and compare them with \textsc{pythia} 6.1 (from archived MC), \textsc{pythia} 8, \textsc{herwig} 7, and \textsc{sherpa} event generators. The results are also compared to analytical calculations with perturbative QCD. Finally, predictions from the \textsc{pyquen} event generator, which added jet quenching effect to the simulated $e^+e^-$ events, are also overlapped to illustrate the possible modifications due to the presence of a strongly interacting medium.

\begin{figure}[htp!]
    \centering
    \includegraphicsone{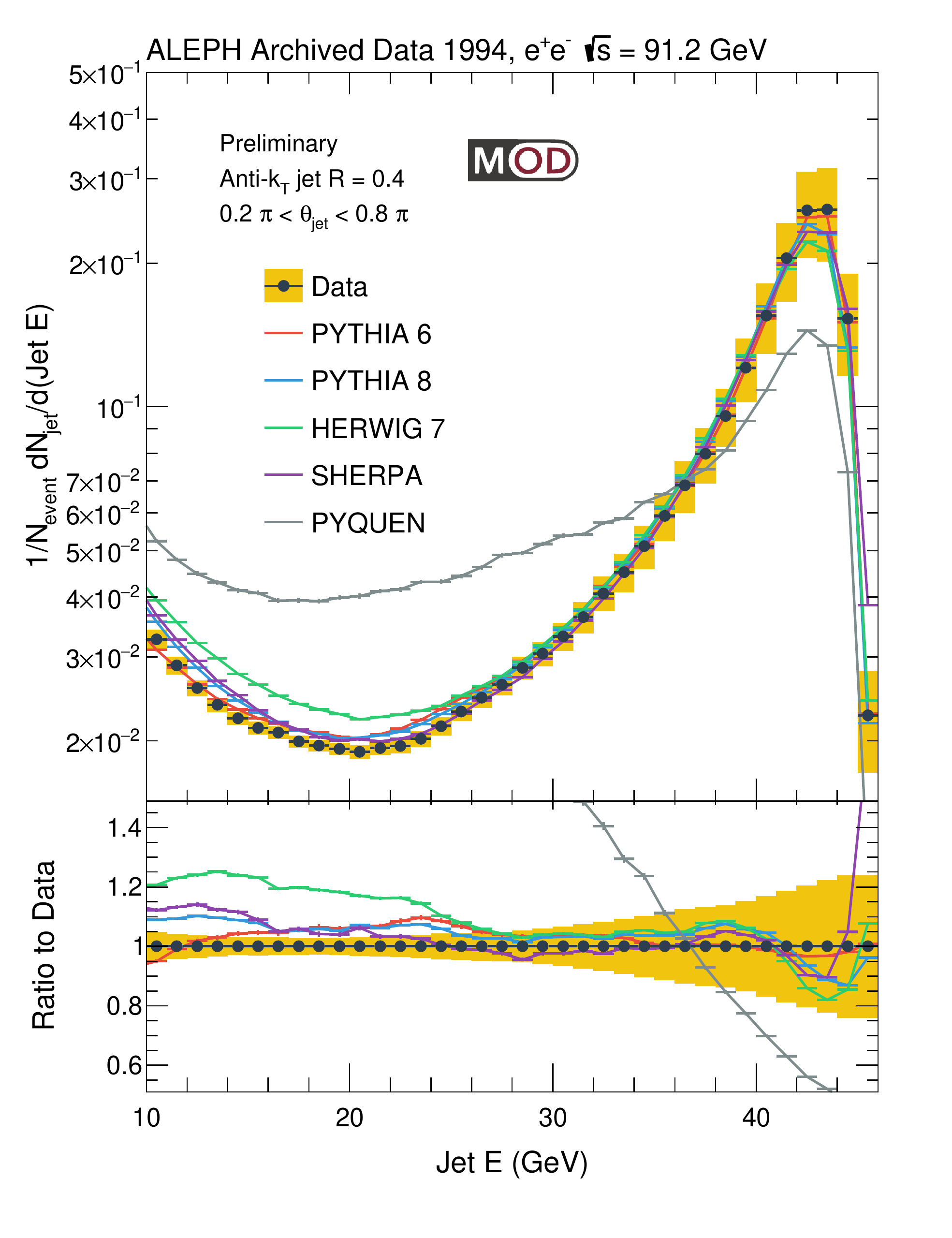}
    \caption{(Upper panel) Inclusive jet energy distribution in $0.2\pi<\theta_{\rm jet}<0.8 \pi$. The yellow boxes are the systematical uncertainties. The predictions from event generators are shown as colored curves (Lower panel) The ratio of theoretical predictions to data}
    \label{Figure:Result-JetEnergy}
\end{figure}

The first unfolded inclusive jet energy spectrum in hadronic Z decay with ALEPH archived data collected in 1994 is shown in Figure~\ref{Figure:Result-JetEnergy}. A peak structure could be observed at around half of the $Z$ boson mass. This is mainly coming from $Z\rightarrow q\bar{q}$ and parton shower of one of the outgoing (anti-)quark is almost fully captured by the anti-$k_T$ algorithm with a resolution parameter of 0.4. 

The spectrum decreases rapidly as one moves to lower and lower jet energy and reaching a minimum at around 20 GeV. Below that, the spectrum increases as we go to even lower energy. The spectrum shape is captured by most of the event generators, although at low jet energy, \textsc{herwig} 7, which has the worse description of the result, over-predicts the jet spectrum at low jet energy. 

\textsc{pyquen} generator which include a large jet quenching effect predicts a large reduction of the population at around 45 GeV and a significant increase on the number of jet at low jet energy.

\begin{figure}[htp!]
    \centering
    \includegraphicsone{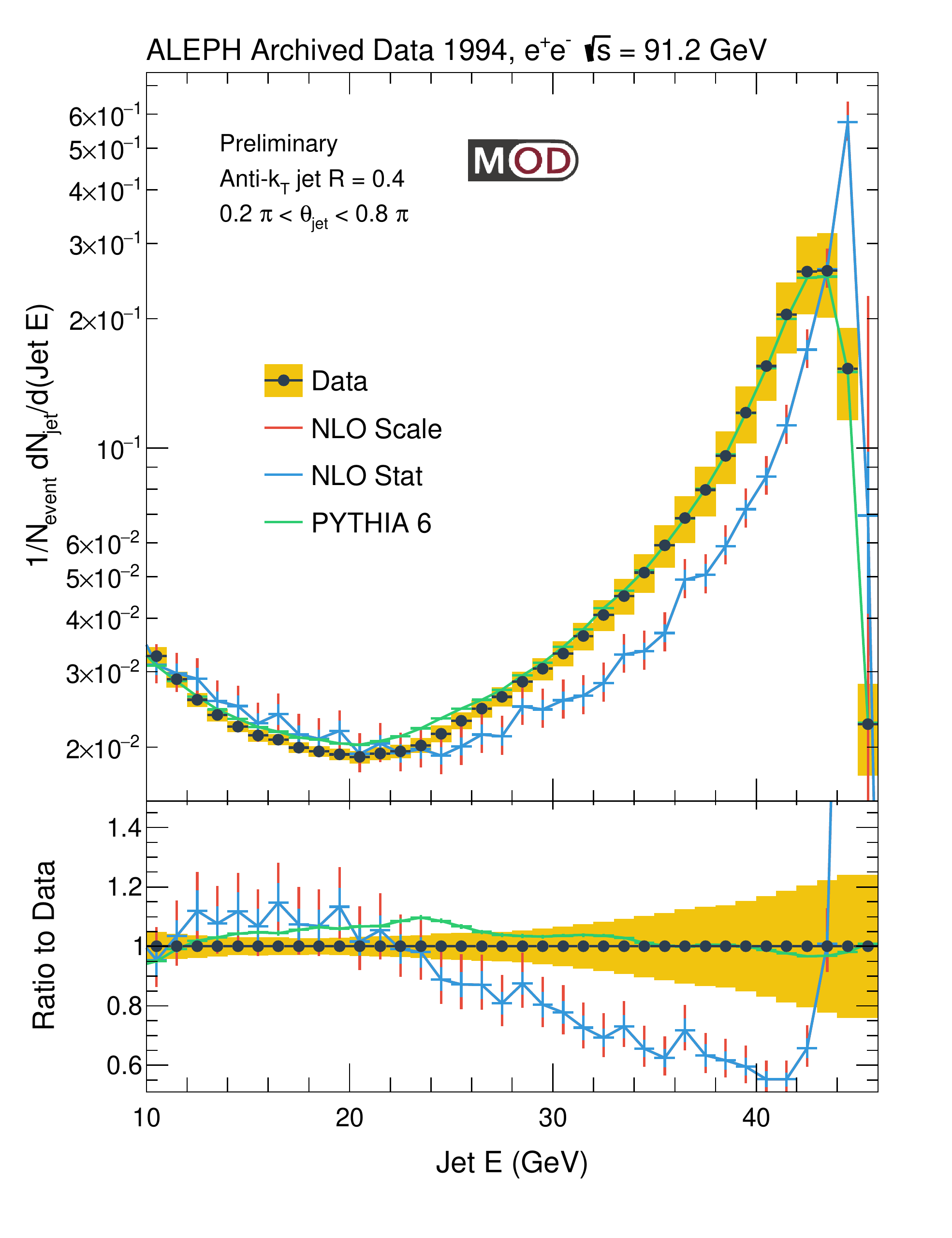}
    \caption{Inclusive jet energy distribution in $0.2\pi<\theta_{\rm jet}<0.8 \pi$ compared to a next-to-leading order perturbative QCD calculation}
    \label{Figure:Result-JetEnergyJoao}
\end{figure}

The same data could also be compared to perturbative QCD calculations at parton level. In LO, the QCD calculation would give a delta function at half of the Z mass which will not describe the data. As shown in Figure~\ref{Figure:Result-JetEnergyJoao}, the Next-to-Leading Order (NLO) calculation predicts a sharper peak at large jet energy. This motivates the calculations beyond NLO and highlights the importance of the threshold resummation ($z=2E/Q\rightarrow 1$)~\cite{Neill:2021std}.

\clearpage

\begin{figure}[htp!]
    \centering
    \includegraphicsone{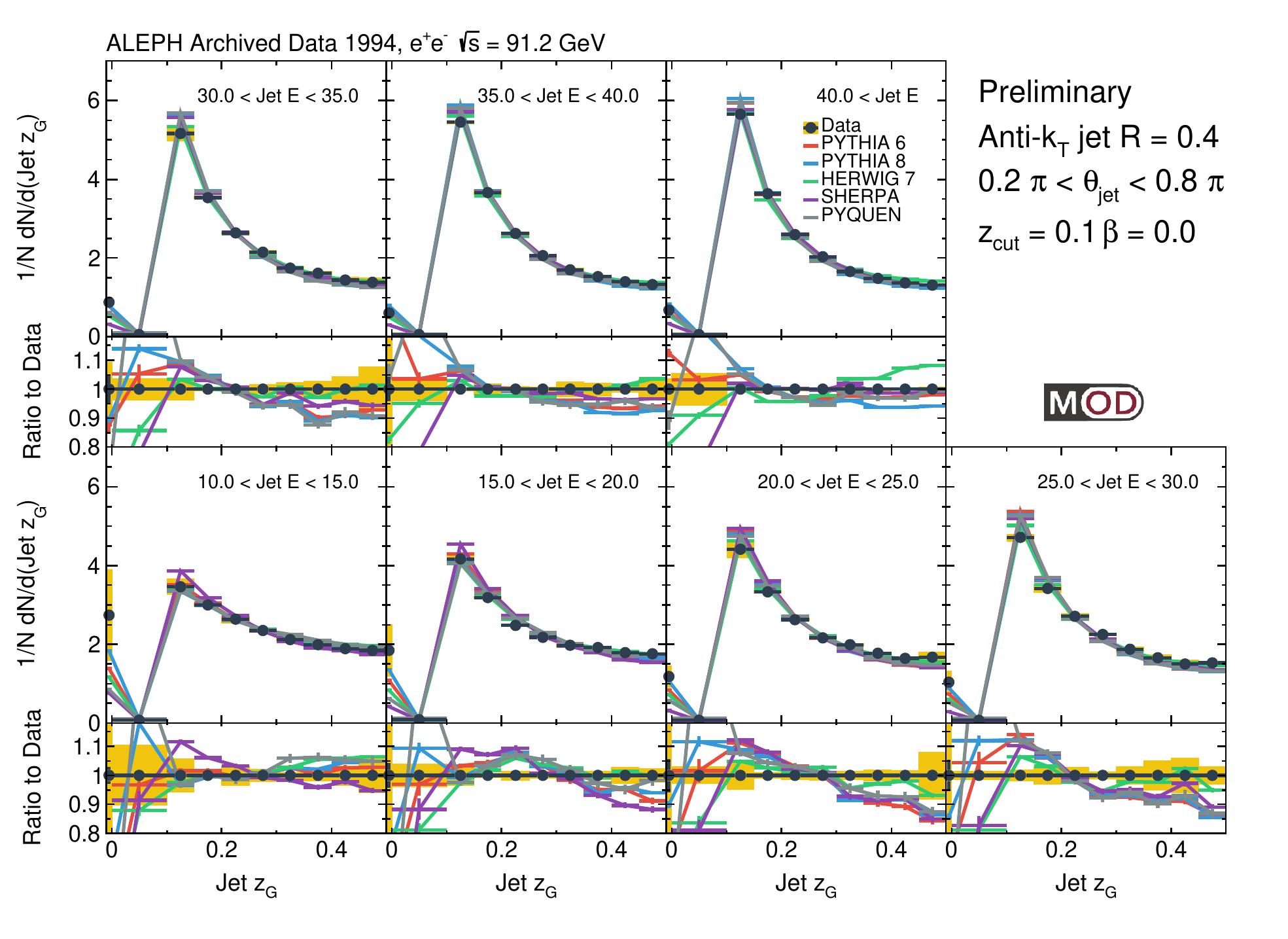}
    \caption{The groomed momentum sharing $\zg$ spectra for jets in 7 different energy intervals.}
    \label{Figure:Result-JetZG}
\end{figure}

In order to characterize the substructure inside the anti-$k_T$ jets, the groomed momentum sharing $\zg$ spectra and the groomed jet radius $\Rg$ are presented in bins of jet energy. As shown in Figure~\ref{Figure:Result-JetZG}, the $\zg$ spectrum is falling as a function of $\zg$ value, reaching a minimum at $\zg\sim 0.5$, which is similar to the data from proton-proton and heavy-ion collisions. At high jet energy, \textsc{herwig} 7 over-predicts the jets with $\zg$ close to 0.5. In most of the jet energy intervals, \textsc{pythia} 6, \textsc{pythia} 8 and \textsc{sherpa} under-predict the number of jets at large $\zg$ by roughly 10\% while they over-predict the $\zg$ spectra at low $\zg$. 

\begin{figure}[htp!]
    \centering
    \includegraphicsone{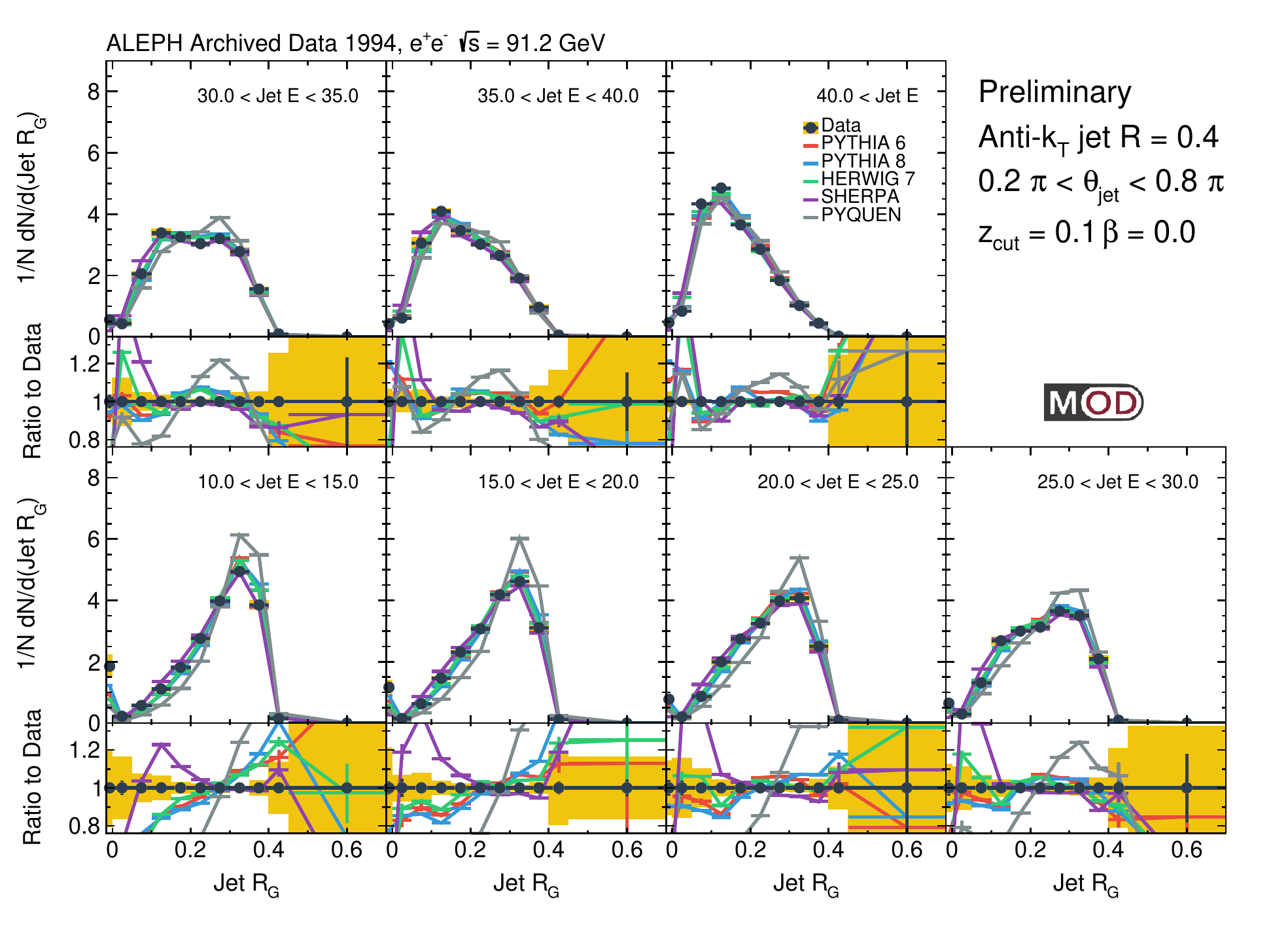}
    \caption{The groomed jet radius $\Rg$ spectra for jets in 7 different energy intervals}
    \label{Figure:Result-JetRG}
\end{figure}

The agreement between $\Rg$ spectra and event generators, as shown in Figure~\ref{Figure:Result-JetRG} is worse than that observed in the comparison of $\zg$ spectra. At high jet energy, event generators predicts a slightly narrower $\Rg$ spectra compared to data. At low jet energy, the event generators predict on average a larger separation between subjects (larger $\Rg$) than the unfolded data. \textsc{pyquen} generator that include jet quenching effect predicts an even larger fraction of jets with large $\Rg$

\clearpage

The mass of the jet is sensitive to the scale where the initial high energy parton is created.  Due to the large correlation between jet mass and jet energy, we observe similar qualitative behavior: the mass/E is smaller at higher jet energy, and increases progressively with decreasing jet energy.

The effect of a potential jet energy loss effect as modeled by the \textsc{pyquen} generator is the strongest with a small jet energy, and diminishes with higher jet energy.

\begin{figure}[htp!]
    \centering
    \includegraphicsone{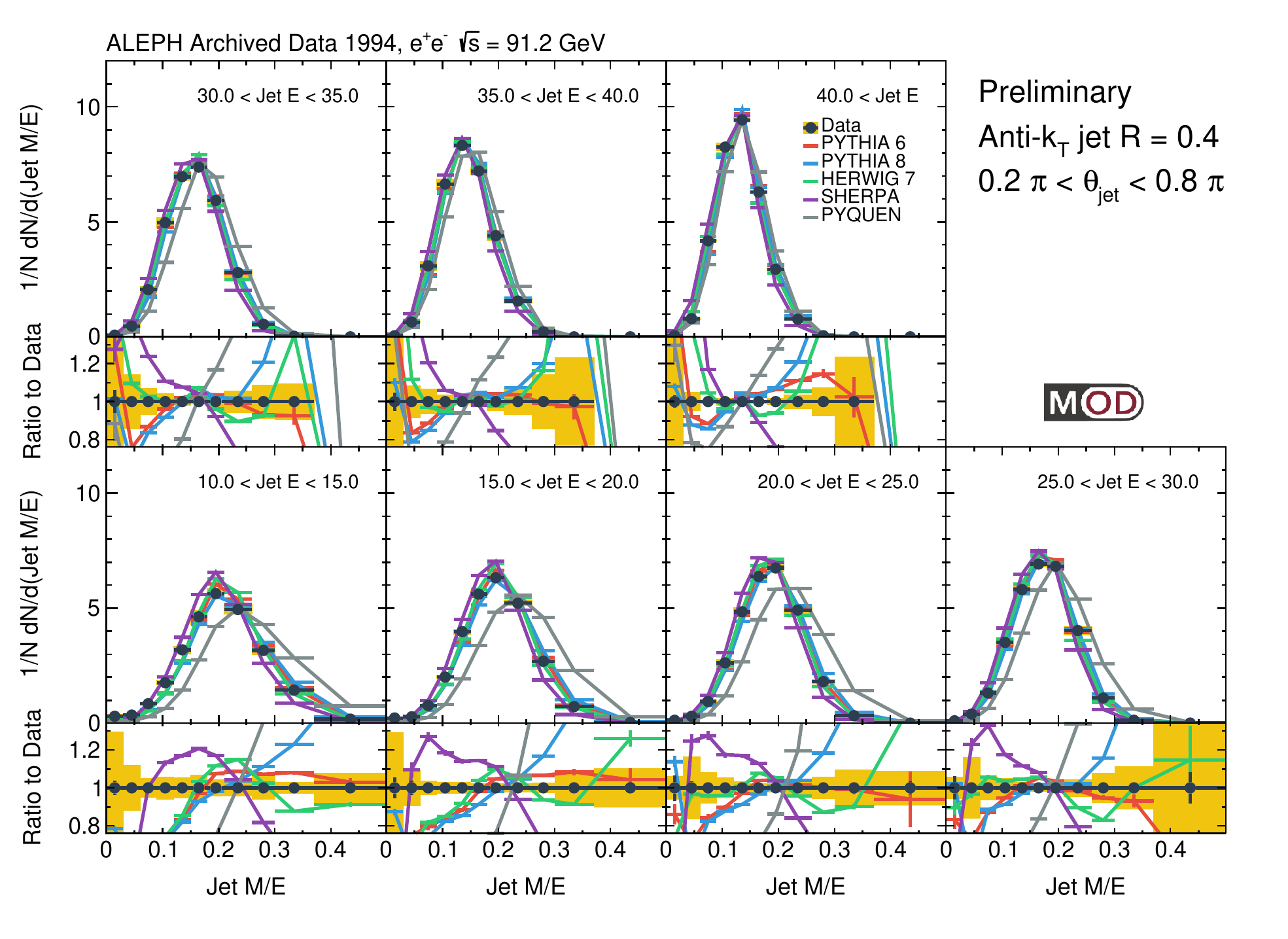}
    \caption{The jet mass to energy ratio $M/E$ distributions for jets in 7 different energy intervals}
    \label{Figure:Result-JetME}
\end{figure}

\begin{figure}[htp!]
    \centering
    \includegraphicsone{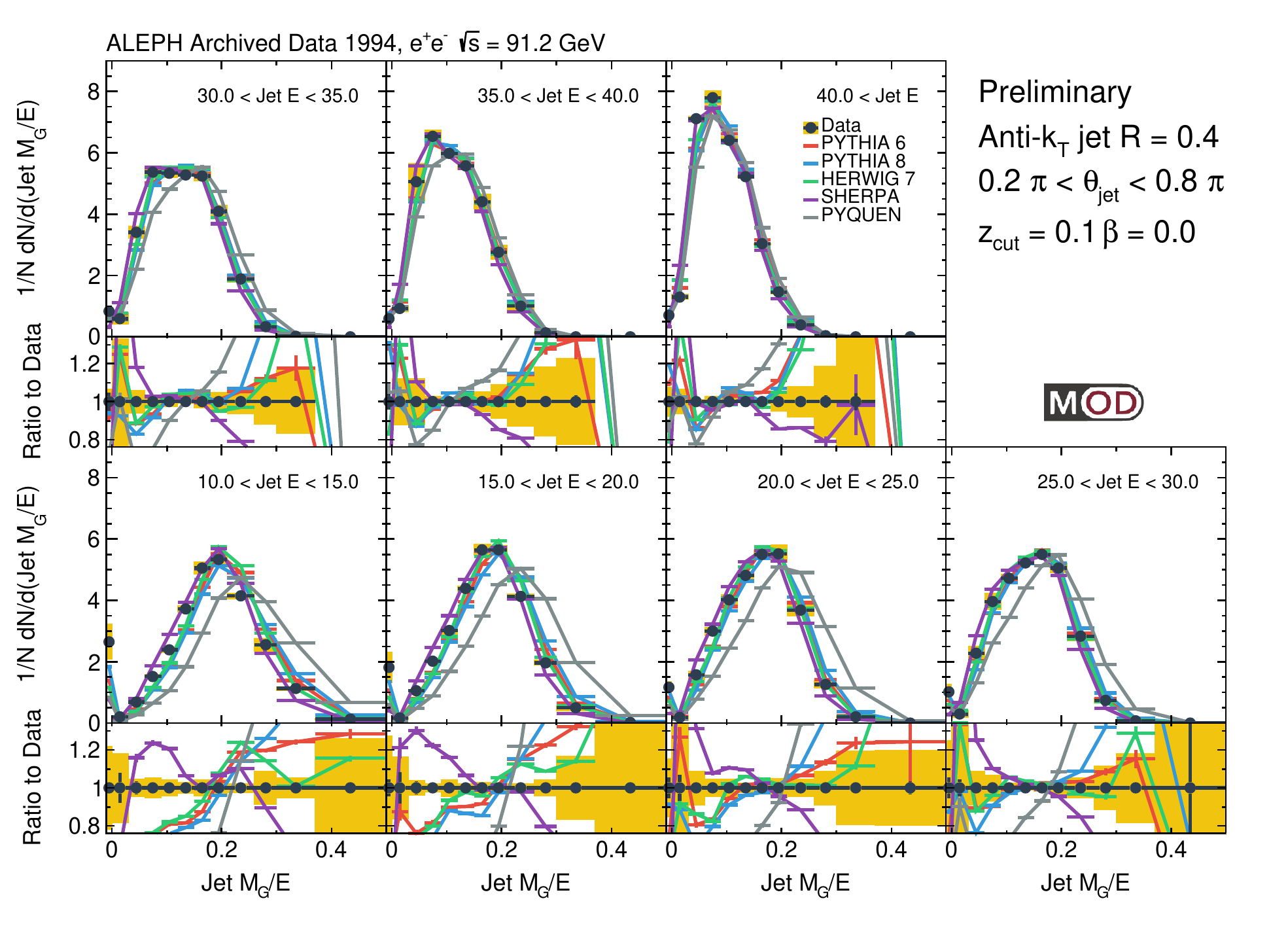}
    \caption{The groomed jet mass to energy ratio $M_G/E$ distributions for jets in 7 different energy intervals}
    \label{Figure:Result-JetMGE}
\end{figure}

By comparing the groomed mass (Figure~\ref{Figure:Result-JetMGE}) with the ungroomed mass (Figure~\ref{Figure:Result-JetME}) it is evident that there is a systematic shift to lower values for the groomed mass, as expected by the grooming procedure, which removes large angle soft particles, whereby lowering the mass.  The difference between the two masses is sensitive to the amount of (relatively) large angle radiation of the jet.

\clearpage

\begin{figure}[htp!]
    \centering
    \includegraphicsone{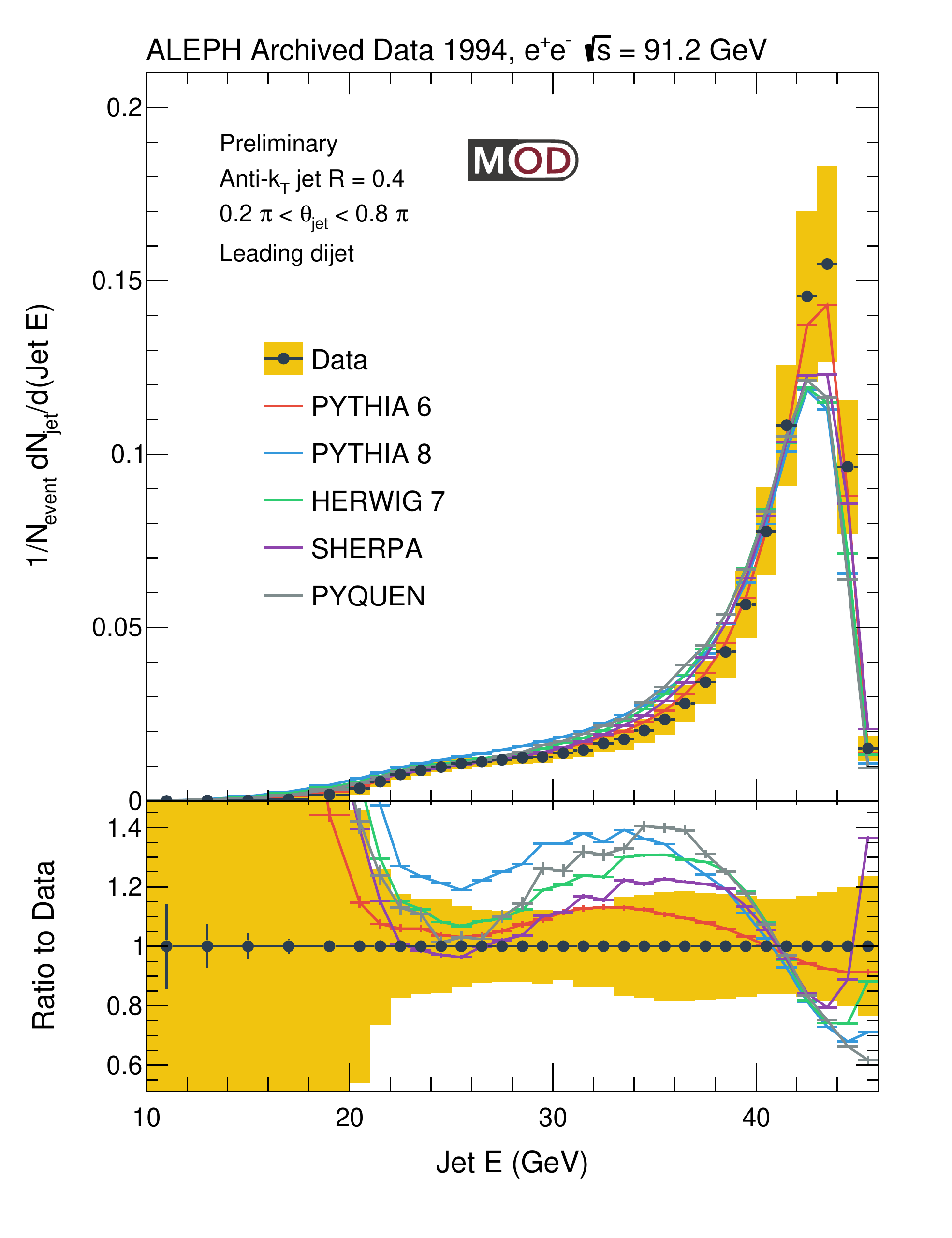}
    \caption{Leading dijet energy spectra.}
    \label{Figure:Result-DiJetE}
\end{figure}

\begin{figure}[htp!]
    \centering
    \includegraphicsone{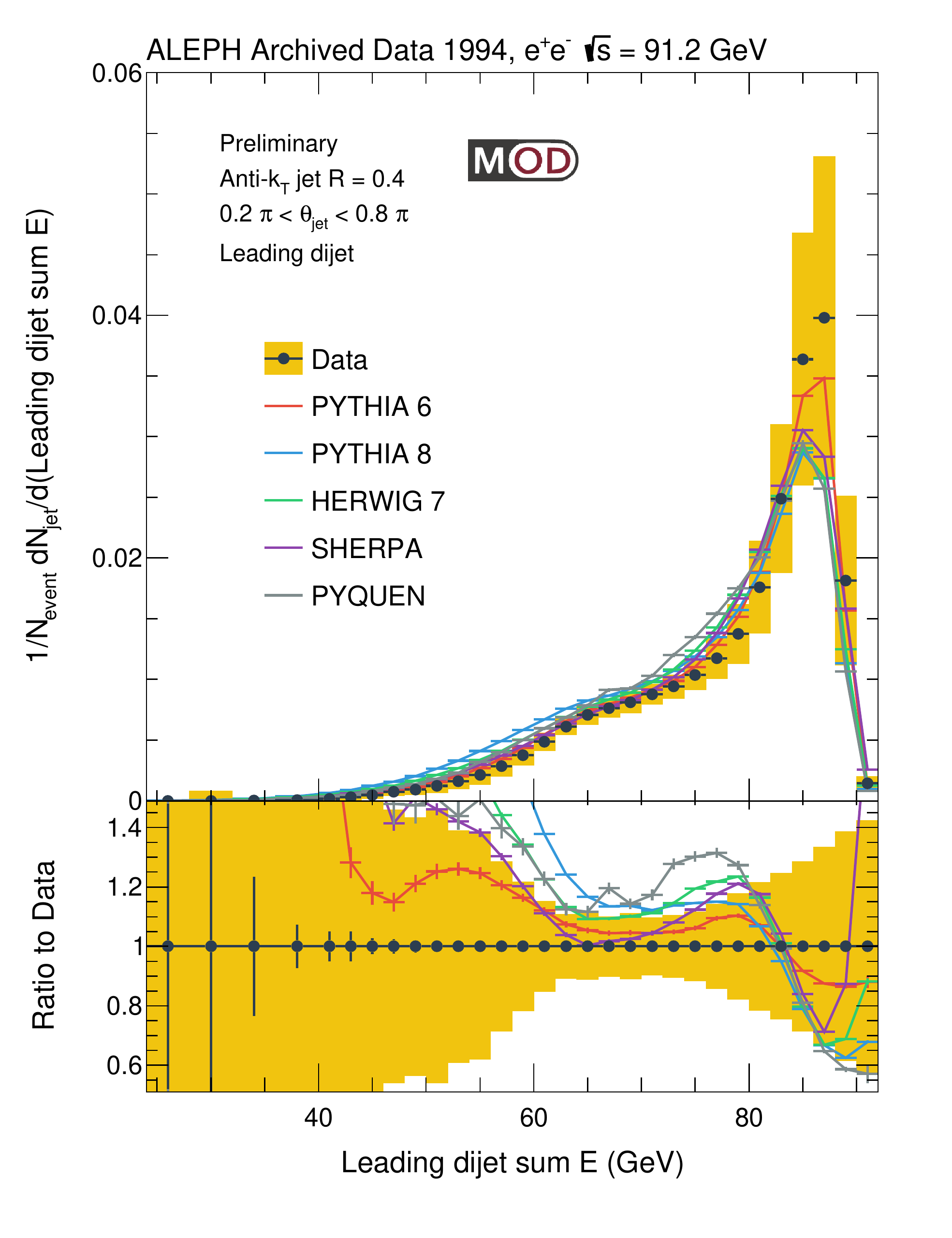}
    \caption{Leading dijet total energy spectra.}
    \label{Figure:Result-DiJetSumE}
\end{figure}

The leading dijet energy is shown in Figure~\ref{Figure:Result-DiJetE}.  The spectra is normalized to the number of events passing the baseline selection described in earlier sections.  A decent agreement between the generators and the unfolded data is observed.  Due to the leading dijet selection, the rise at low jet energy is suppressed.  The sum of the two leading dijet is shown in Figure~\ref{Figure:Result-DiJetSumE}.  The levels of (dis)agreement between the for simulation and data for leading dijet energy and the leading dijet total energy is similar.  The total energy, which is equivalent to $\sqrt{s}$ minus the sum of all the small jets, is more sensitive to modeling of subleading jets.

\clearpage



\clearpage

\section*{Acknowledgement}
The authors would like to thank the ALEPH Collaboration for their support and foresight in archiving their data. We thank Joao Pires for providing the NLO calculations of the jet energy spectrum. We would like to thank the useful comments and suggestions from Roberto Tenchini, Guenther Dissertori, Felix Ringer, Jesse Thaler, Andrew Lakoski, Liliana Apolinario, Ben Nachman, Camelia Mironov, Jing Wang. This work has been supported by the Department of Energy, Office of Science.

\bibliographystyle{unsrt}
\bibliography{references}
\end{document}